\documentclass{jfm}
\usepackage{graphicx}
\usepackage{natbib}
\usepackage{epstopdf, epsfig}
\usepackage{amsmath}
\usepackage{amssymb}
\usepackage{latexsym}
\usepackage{wasysym}
\usepackage{multirow}
\usepackage{color}
\usepackage{subfigure}
\usepackage{pdflscape}
\usepackage{multirow}
\usepackage{dashrule}
\usepackage{tikz}
\usepackage{cancel}
\usetikzlibrary{arrows}
\usepackage{mathtools}
\usepackage[utf8]{inputenc}
\usepackage{float}
\usepackage[version=4]{mhchem}
\usepackage{siunitx}
\usepackage{longtable,tabularx}
\usepackage{gensymb}
\DeclareGraphicsExtensions{.png}

\usepackage [english]{babel}
\usepackage [autostyle, english = american]{csquotes}
\MakeOuterQuote{"}

\newcommand{\bea}{\begin{equation}\begin{aligned}}
\newcommand{\eea}{\end{aligned}\end{equation}}

\shorttitle{Guidelines for authors}
\shortauthor{R. Ma and K. Mahesh}

\title{Global stability analysis and direct numerical simulation of boundary layers with an isolated roughness element}

\author{Rong Ma\aff{1} 
 \and Krishnan Mahesh\aff{1}
 \corresp{\email{kmahesh@umn.edu}}}

\affiliation{\aff{1}Department of Aerospace Engineering and Mechanics, University of Minnesota,
Minneapolis, MN 55455, USA}

\begin{document}

\maketitle

\begin{abstract}
Global stability analysis and direct numerical simulation (DNS) are performed to study boundary layer flows with an isolated roughness element. Wall-attached cuboids with aspect ratios $\eta=1$ and $\eta=0.5$ are investigated for fixed ratio of roughness height to displacement boundary layer thickness $h/\delta^*=2.86$. 
%The global stability analysis is able to capture the growth rates and frequencies of the unstable modes when using the base flow computed by selective frequency damping method. 
%Either using base flow or mean flow for global stability analysis is able to capture the frequency of the primary vortical structures, but the mean flow is marginally stable due to the non-linear saturation farther downstream. Larger $\eta$ and higher $Re_h$ result in a thicker and more sustainable central streak%whose unstable nature plays important roles in the transition process
%. When the ratio of the roughness height to the displacement boundary layer thickness is sufficiently high ($h/\delta^*=2.86$), 
 Global stability analysis is able to capture the frequency of the primary vortical structures.
%Varicose instability is dominant for small aspect ratios ($\eta \le 1$).
For $\eta=1$, only varicose instability is seen. %The varicose mode has its root in the center of the reversed flow region and extracts most of energy from both the wall-normal and spanwise shear of the central streak. 
For the thinner roughness element ($\eta=0.5$), the varicose instability dominates the sinuous instability, and the sinuous instability becomes more pronounced as $Re_h$ increases, due to increased spanwise shear in the near-wake region. %In contrast, the sinuous instability has its root in the lateral parts of the reversed flow region, experiences spatial growth along the lateral parts of the central streak and extracts its energy from the spanwise shear. 
The unstable modes mainly extract energy from the central streak, although the lateral streaks also contribute. %when $h/\delta^*$ is large. %These results highlight that $h/\delta^*$, $\eta$ and $Re_h$ have the joint effects on the instability characteristics. 
The DNS results show that different instability features lead to different behavior and development of vortical structures in the nonlinear transition process. For $\eta=1$, the varicose mode is associated with the shedding of hairpin vortices. %the varicose mode associated with the shedding of hairpin vortices is obtained from the dynamic mode decomposition (DMD) analysis, consistent with the global stability results.
As $Re_h$ increases, the breakdown of hairpin vortices occurs closer to the roughness and sinuous breakdown behavior promoting transition to turbulence is seen in the farther wake. %the wake flow is more homogeneous. 
A fully-developed turbulent flow is established in both the inner and outer layers farther downstream when $Re_h$ is sufficiently high. For $\eta=0.5$, the sinuous wiggling of hairpin vortices is prominent at higher $Re_h$, leading to stronger interactions in the near wake, as a result of combined varicose and sinuous instabilities. A sinuous mode captured by dynamic mode decomposition (DMD) analysis, and associated with the `wiggling' of streaks persists far downstream. %This sinuous like breakdown could destabilize the shear layer and promote transition to turbulence.
%The sinuous oscillations lead to stronger interactions among different fluid motions in the near-wake region, and has a persistent effect on the transition process. %A sinuous mode captured by the analysis of dynamic mode decomposition (DMD) is associated with the wiggling of the streaks farther downstream. The sinuous instability leads to stronger interactions among different fluid motions in the near-wake region, and has a persistent effect on the transition process. %The present cases are compared to the von Doenhoff-Braslow transiton diagram and are well fitted into it.

%The varicose instability has its root in the center of the reversed flow region, experiences spatial transient growth along the central low-speed streak and extracts its energy from the whole 3-D shear layer, contributing to the birth of hairpin vortices.
%The sinuous instability has its root in the lateral parts of the reversed flow region, experiences spatial growth along the lateral parts of the central streak and extracts its energy from the spanwise shear, contributing to the sinuous wiggling of hairpin vortices.

\end{abstract}


\section{Introduction}
%isolated roughness, induce unsteady structures
%linear stability analysis
%global stability theory, literature review, both numerical and experimental

Surface roughness has important effects on boundary layers. The laminar-turbulent transition in boundary layer flows can be greatly modified by the presence of localized or distributed roughness. %The transition to turbulence induced by surface roughness can lead to a significant increase of the skin friction.
Understanding the roughness-induced transition process is therefore important for the control of flows in an engineering context. 

%Important parameters have been identified and empirical correlations have been developed for predicting the transition onset. The roughness Reynolds number is one of the important parameters and different definitions exist. 
The roughness Reynolds number is one of the important parameters in roughness-induced transition, and empirical correlations have been developed to predict the transition onset. One commonly used definition is $Re_h=U_eh/\nu$, where $U_e$ is the boundary layer edge velocity, $h$ is the roughness height and $\nu$ is the kinetic viscosity of the fluid. This definition, however, does not account for the impact of the relative location of the roughness element in a boundary layer. Another definition $Re_{hh}=u_{h}h/\nu$ based on the Blasius velocity solution at the roughness tip location $u_h$ has been suggested to best characterize roughness-induced transition \citep{klanfer1953effect}. \cite{von1961effect} have suggested a transition diagram that correlates the roughness aspect ratio $\eta=d/h$ (where $d$ is the roughness width and $h$ is the roughness height) with $Re_{hh}$, using experimental dataset from different types of distributed surface roughness. Their diagram highlights the crucial roles of $\eta$ and $Re_{hh}$ on the transition onset. %and demonstrates a general trend that as the aspect ratio decreases, the transition Reynolds number $Re_{hh,tr}$ increases. 

Isolated, three-dimensional (3-D) roughness elements may be considered as the primary models to be generalized and extended for more complex roughness geometries. The effects of isolated roughness on transition have been investigated experimentally by \cite{gregory1956effect}. The main flow pattern is observed to be horseshoe vortices that wrap around the roughness element, and whose legs trail downstream and give birth to the streamwise vortices farther downstream. %the spanwise vortices wrapping around the front of the roughness element, creating horseshoe vortices that trail downstream and give birth to counter-rotating streamwise vortices. 
\cite{baker1979laminar} has experimentally studied the vortex system around an isolated cylindrical roughness element, and shown the dependency of the horseshoe system dynamics on $Re_D=U_eD/\nu$ and $D/\delta^*$, where $D$ is the cylinder diameter and $\delta^*$ is the displacement boundary layer thickness. The streamwise vortices induced by the 3-D roughness elements create longitudinal streaks downstream that are lifted upwards \citep{landahl1980note,reshotko2001transient}. These streamwise longitudinal streaks are related to the disturbance transient growth, which can be unstable and cause transition downstream of the roughness \citep{fransson2004experimental,fransson2005experimental}. The concept of optimal perturbation was introduced by \cite{boberg1988onset} and \cite{butler1992three} to define these `most dangerous' initial perturbations that generate the maximum energy growth. \cite{luchini2000reynolds} provided a numerical method to determine the optimal perturbation and explain that the linear growth of initially small disturbances can excite nonlinear interactions and cause transition. 
%which trigger the strong transient growth of the boundary layer streaks \citep{joslin1995growth}. %\cite{cossu2004tollmien} and \cite{fransson2004experimental,fransson2005experimental} found that the streamwise streaks induced by the fully 3-D roughness can stabilize the Tollmien-Schlichting (T-S) waves and delay the transition. However, some physical mechanisms of the transition process are still not fully understood. The relation between the roughness characteristics and the transition remains to be further investigated. 

%The streamwise vortices induce disturbance transient growth by creating longitudinal streaks driven up by the lift-up phenomenon \citep{landahl1980note,reshotko2001transient}. 

Both symmetric (termed `varicose') and anti-symmetric (termed `sinuous') streak instabilities have been detected and are of importance in turbulent boundary layers. The varicose type is associated with horseshoe vortices that originate from a normal inflectional instability in the streamwise velocity profile \citep{robinson1991kinematics,asai2002instability,skote2002varicose}. The sinuous streak instability is correlated with a base state with a spanwise inflection and contributes to the regeneration of near-wall turbulence \citep{jimenez1991minimal,skote2002varicose}.  \cite{asai2002instability} observed that wider streaks more easily undergo varicose breakdown while narrower streaks are more likely to undergo the sinuous breakdown. 

The strength and stability properties of the streamwise streaks also play important roles in roughness-induced transition, and are dependent on roughness characteristics, such as its shape, height and aspect ratio. \cite{white2005receptivity} conducted experiments to investigate the effects of surface roughness characteristics on transient growth and suggested a strong dependence of the resulting transient growth on roughness diameters. \cite{piot2008stability} used bi-global stability approach and DNS to investigate the local stability of streamwise streaks past a single smooth roughness element. A stabilizing effect of a `pre-streaky' structure associated with the counter-rotating streamwise vortices is identified in the near wake. \cite{cherubini2013transient} performed a global optimization analysis to search for the optimal perturbation inducing the largest transient growth for a boundary layer past smooth 3-D roughness elements. They investigated bumps with different heights at different target times, and their results indicate that when the bump is sufficiently large, it can destabilize the wake flow on a short time scale. 

With large-scale linear algebra computations now being possible, global linear stability theory \citep{theofilis2011global} has been performed on roughness-induced transition. Global stability is especially useful for non-parallel flows such as roughness wakes, thus is a promising tool to predict and analyze roughness-induced transition. \cite{loiseau2014investigation} used global stability theory to investigate the flow past a cylindrical roughness element. They suggested that the frequencies associated with the dominant fluid dynamics are well predicted by global stability analyses, and that the unstable nature of the central low-speed streak is of crucial importance in the transition process. \cite{citro2015global} presented the direct and adjoint global eigenmodes for boundary layer flows past a hemispherical roughness element, and found that %the core of the instability is symmetric and localized in the immediate downstream region of the roughness element. They also found that 
the critical Reynolds number is constant when the ratio of the roughness height and the displacement boundary layer thickness $h/\delta^*$ is less than $1.5$. \cite{kurz2016mechanisms} used DNS and global stability analysis to investigate the effects of discrete surface roughness with various roughness height and background disturbance on a swept-wing boundary layer. Their results suggest that large elements are able to trip turbulence by either a convective or a global instability in the near-wake region. \cite{puckert2018experiments} conducted experiments to detect global instability in the 3-D isolated cylindrical roughness cases of \cite{loiseau2014investigation}. %Their results have shown good agreement with \cite{loiseau2014investigation} for the critical Reynolds number of global instability. However, t
They report that the critical Reynolds number is higher than the transition Reynolds number and suggest that global instability may not be the decisive mechanism for transition. 
%They also found that the roughness can amplify the external disturbances in the subcritical regime.

While linear instability is usually detected in the near wake of roughness elements using local or global stability analysis, secondary instability could appear farther downstream if the streak amplitude is sufficiently large. \cite{andersson2001breakdown} found that secondary instabilities appear at large amplitudes of the primary streaks and suggested that the sinuous modes of instability are dominant and most often reported for streak breakdown, while the varicose instability only occurs for larger streak amplitude and thus is barely observed in natural transition. \cite{denissen2013secondary} studied the stability of steady roughness-induced transient growth to unsteady fluctuations, and found that the transition could be caused via secondary instability in the mid-wake region when the roughness size is sufficiently large. \cite{vadlamani2018distributed} found that secondary sinuous instability developed on the low-speed streak is evident in the transition process induced by distributed roughness. They also suggested that this sinuous like breakdown is reminiscent with the sinuous instability observed in the context of transition under the effects of free-stream turbulence \citep{brandt2004transition,hack2014streak}. 

%transition vs instability
%linear stability
%varicose vs sinuous 

Understanding global varicose and sinuous instabilities in roughness-induced transition and their dependency on roughness configuration is important. \cite{de2013laminar} conducted bi-global and three-dimensional parabolized stability analyses to investigate the transition induced by a sharp-edged isolated roughness element in a supersonic boundary layer. Their results suggest that the varicose mode is associated with the entire 3-D shear layer while the sinuous mode is a consequence of the lateral streaks. \cite{loiseau2014investigation} suggest that the sinuous global mode is similar to the von K{\'a}rm{\'a}n vortex street in the 2-D cylinder flow and the varicose mode is associated with the hairpin vortices. They also investigate the dependence of instability types on aspect ratios and suggest that varicose instability is observed for wider roughness elements and sinuous instability is observed for thinner roughness elements. %They suggested that the sinuous instability is different from the sinuous instability of optimal streaks underlined by \cite{andersson2001breakdown}.
%\cite{shin2015stability} conducted bi-global linear stability analysis and found that the varicose mode dominated the sinuous mode. Their experimental results in a water channel show agreement with the bi-global stability results. 
\cite{puckert2018experiments} reported experimental observation of sinuous oscillations in roughness-induced transition using PIV. They found that for thin roughness elements (i.e., $\eta=1$), the sinuous mode competes with the varicose mode and becomes dominant in the supercritical regime. %They also found that the sinuous mode can develop convectively when the forcing is strong enough.
\cite{bucci2021influence} fixed the aspect ratio of the cylinder to unity and studied the effect of freestream turbulence on roughness-induced transition. They noted that the roughness Reynolds number and aspect ratio might not be the only important parameters for flow characteristics, the shear ratio also plays a crucial role in the onset and symmetry of the primary global instability. %They found that the leading global mode is characterized by a varicose symmetry when $h/\delta^* \ge 2.04$, but whether this is  

%As Reynolds number increases, the flow unsteadiness is revealed downstream the roughness element. The unsteady structures are associated with the instability. 

The joint effects of the parameters mentioned above make the instability characteristics and transition process highly sensitive to the flow configuration. Although the ratio $h/\delta^*$ seems to play a crucial role in the determination of the dominant instability, the dependency and sensitivity of the onset of sinuous instability on $\eta$ and $Re_h$ need further investigations to complement the current understanding. While it has been observed that the sinuous instability is related to a sinuous wiggling similar to von K{\'a}rm{\'a}n vortices, the influence of sinuous instability and its interplay with the varicose instability on the non-linear patterns and dynamics needs further analysis. To address these points, we perform global stability and adjoint sensitivity analyses to study global instability of boundary layer flows over a cuboid with two aspect ratios. The ratio of the cuboid height to the displacement boundary layer thickness is $2.86$, which is larger than most past work. The thin cuboids with aspect ratios $\eta=1$ and $0.5$ are considered to examine the dominant instability and the sensitivity of the onset of sinuous instability in terms of $\eta$ and $Re_h$ for roughness elements with high $h/\delta^*$. We also perform DNS to examine the dependence of $Re_h$ and $\eta$ on the laminar-turbulence transition process, and use DMD analysis to study the development of vortical structures and associated non-linear dynamics corresponding to different global instability characteristics.

%The general trend higher $Re_h$, larger $\eta$ and higher $h/\delta^*$ lead to a stronger shear and a more sustainable central streak. At smaller $h/\delta^*$, smaller $\eta$ and higher $Re_h$, the sinuous instability is more likely to occur.

%The joint effects of the parameters mentioned above make the instability characteristics and the transition process highly sensitive to the flow configuration. Some current observations may not remain true for a different combination of parameters. Further investigations are needed to examine the influence of different roughness shapes with various parameters (e.g., $h/\delta^*$, $Re_h$ and $\eta$), and to understand how the different types of instability contribute to the transition process. In the present work, we perform DNS, global stability and adjoint sensitivity analyses to study the instability and transition process of boundary layer flows over a cuboid with two aspect ratios. The ratio of the cuboid height to the displacement boundary layer thickness is $2.86$, which is relatively large compared to most of the past work.

The numerical methodology is introduced in \S \ref{sec:numerical} and validations of the global stability and adjoint sensitivity solver are shown in \S \ref{sec:validations}. The flow configuration, base flow computation, grid convergence and domain length sensitivity are demonstrated in \S \ref{sec:setup}. The results and discussions are presented in \S \ref{sec:results}. Finally, the paper is summarized in \S \ref{sec:conclusions}. 

%The role of instability in the transition process would also be further investigated. As a first step towards these goals, the following questions are to be asked:

%(1) What difference would be in the base flow and the leading eigenmodes for a cuboid roughness element compared to a cylindrical roughness element?

%(2) Would the transition Reynolds number obtained from DNS be lower than the critical Reynolds number obtained from GLSA, as suggested by \cite{puckert2018experiments} from the experimental results?

%(3) Would the changeover from convective instability to global instability observed in experiments also be found in DNS and consistent with the critical Reynolds number obtained from the GLSA theory?

%(4) How does the arrangement of roughness elements affect the transition process and instability mechanisms? (e.g., an array compared to an isolated roughness element)

%(5) Would the instability mechanisms maintain the same in the supercritical regime as Reynolds number increases?

%(6) As suggested by \cite{loiseau2014investigation}, sinuous modes are the dominant instability for thinner roughness elements ($\eta \le 1$), whereas the varicose instability is the dominant one for wider roughness elements ($\eta \ge 2$). However, this is still limited by a certain set-up (i.e., with a larger $\delta^*/h$). Would this conclusion be valid for an isolated roughness element with a smaller $\delta^*/h$? How would the roughness height (i.e., $\delta^*/h$) affect the instability mechanisms?

\section{Numerical methodology}\label{sec:numerical}
The governing equations and numerical method are briefly summarized. An overview of modal linear stability, adjoint sensitivity and details regarding the iterative eigenvalue solver are provided. 

\subsection{Direct numerical simulation}
The incompressible Navier-Stokes (N-S) equations are solved using the finite volume algorithm developed by \cite{mahesh2004numerical}:%. The governing equations for the momentum and continuity equations are given by the Navier-Stokes equations:
\begin{equation}
\frac{\partial u_i}{\partial t} + \frac{\partial}{\partial x_j}(u_iu_j)=-\frac{\partial p}{\partial x_i} + \nu\frac{\partial^2 u_i}{\partial x_j x_j}+K_i,~~\frac{\partial u_i}{\partial x_i} = 0,
\label{eqn:nsme}
\end{equation}
where $u_i$ and $x_i$ are the $i$-th component of the velocity and position vectors respectively, $p$ denotes pressure divided by density, $\nu$ is the kinematic viscosity of the fluid and $K_i$ is a constant pressure gradient (divided by density). Note that the density is absorbed in the pressure and $K_i$.
The algorithm is robust and emphasizes discrete kinetic energy conservation in the inviscid limit which enables it to simulate high-Re flows without adding numerical dissipation. A predictor-corrector methodology is used where the velocities are first predicted using the momentum equation, and then corrected using the pressure gradient obtained from the Poisson equation yielded by the continuity equation. The Poisson equation is solved using a multigrid pre-conditioned conjugate gradient method (CGM) using the Trilinos libraries (Sandia National Labs). \\
The DNS solver has been validated for a variety of problems on wall-bounded flows, including: realistically rough superhydrophobic surfaces \citep{alame2019wall}, random rough surfaces \citep{ma2021direct} and response of a plate in turbulent channel flow \citep{anantharamu2021response}.
%The implicit time advancement uses the second-order Crank-Nicolson discretization:
%\begin{equation}
%     \frac{\hat{u}_i-u_i^n}{\Delta t}=\frac{1}{2}[(NL+VISC)^{n+1}+(NL+VISC)^{n}] \, ,
% \end{equation}
% where the face normal velocities $V^{n+1}_N$ are linearized in time (time-lagged) such that $V^{n}_N$ is used instead; the linearization in time yields an error of $O(\Delta t^2)$, which is the same order as that of the overall scheme. All the terms expressed as $\hat{u}_i$ are taken to the left hand side and a system of equations is solved using SOR until convergence.
\subsection{Linear stability analysis}
Linear stability analysis enables the investigation of the linearized dynamics of infinitesimal perturbations evolving on a base state. In the present work, the incompressible Navier-Stokes equations are linearized about a base state, $\overline{u}_i$ and $\overline{p}$. The flow can be decomposed into a base state subject to a small $O(\epsilon)$ perturbation $\Tilde{u}_i$ and $\Tilde{p}$. The linearized Navier-Stokes (LNS) equations are obtained by subtracting the base state from equation (\ref{eqn:nsme}) and can be written as follows:
\begin{equation}
\frac{\partial \Tilde{u}_i}{\partial t} + \frac{\partial}{\partial x_j}(\Tilde{u}_i\overline{u}_j) + \frac{\partial}{\partial x_j}(\overline{u}_i\Tilde{u}_j)=-\frac{\partial \Tilde{p}}{\partial x_i} + \nu\frac{\partial^2 \Tilde{u}_i}{\partial x_j x_j},~~\frac{\partial \Tilde{u}_i}{\partial x_i} = 0.
\label{eqn:lnsme}
\end{equation}
The same numerical schemes as the N-S equations are used to solve the LNS equations. The LNS equations can be rewritten as a system of linear equations,
\begin{equation}
\frac{\partial \Tilde{u}_i}{\partial t} = A \Tilde{u}_i,
\label{eqn:linear_sys}
\end{equation}
where $A$ is the LNS operator and $\Tilde{u}_i$ is the velocity perturbation. The solutions to the linear system of equations (\ref{eqn:linear_sys}) are:
\begin{equation}
\Tilde{u}_i(x,y,z,t)=\sum_{\omega}\hat{u}_i(x,y,z)e^{\omega t} + c.c.
\label{eqn:linear_sol}
\end{equation}
$Re(\omega)$ is defined as the growth rate and $Im(\omega)$ is the temporal frequency. The linear system of equations can then be transformed into a linear eigenvalue problem:
\begin{equation}
\Omega \hat{U}_i = A \hat{U}_i,
\label{eqn:eigen_direct}
\end{equation}
where $\omega_j = diag(\Omega)_j$ is the $j$-th eigenvalue and $\hat{u}_i^j = U_i[j,:]$ is the $j$-th eigenvector. For the global stability analysis, the computational cost to solve the eigenvalue problem using direct methods is very expensive. Instead, a matrix-free method, the implicitly restarted Arnoldi method (IRAM) is usually used. We make use of the IRAM implemented in the PARPACK library to solve for the leading eigenvalues and eigenmodes.

\subsection{Adjoint sensitivity analysis}
Adjoint sensitivity analysis solves for the dominant eigenvalues and eigenmodes of the adjoint LNS Equations, which yields the dominant sensitivity modes corresponding to the direct modes. According to the definition of the continuous adjoint to the LNS equations by \cite{hill1995adjoint}, the adjoint equations are:
\begin{equation}
\frac{\partial \Tilde{u}_i^{\dagger}}{\partial t} + \frac{\partial}{\partial x_j}(\Tilde{u}_i^{\dagger}\overline{u}_j) - \Tilde{u}_j^{\dagger}\frac{\partial}{\partial x_i}\overline{u}_j=-\frac{\partial \Tilde{p}^{\dagger}}{\partial x_i} - \nu\frac{\partial^2 \Tilde{u}_i^{\dagger}}{\partial x_j x_j},~~\frac{\partial \Tilde{u}_i^{\dagger}}{\partial x_i} = 0.
\label{eqn:ansme}
\end{equation}
Similar to the direct problem, the adjoint systems of linear equations can be simplified to an eigenvalue problem:
\begin{equation}
- \Omega \hat{U}_i^{\dagger} = A^{\dagger} \hat{U}_i^{\dagger}.
\label{eqn:eigen_adjoint}
\end{equation}
\cite{hill1995adjoint} suggested that the adjoint perturbation velocity field would highlight the optimal locations where the largest response to unsteady point forcing occurs. In the present work, our aim is to use the global adjoint sensitivity analysis in conjunction with the global stability analysis to determine the most sensitive flow regions to point forcing and the inception of instability.

\section{Validation}\label{sec:validations}
The global stability solver is developed in the present work on 3-D structured platforms. First, the global stability of a 3-D lid-driven cavity is validated against \cite{regan2017global}. Then, the global stability and adjoint sensitivity analyses are performed for laminar channel flow, where the results are compared to the parallel flow stability of Poiseuille flow conducted by \cite{juniper2014modal}. The global adjoint sensitivity results are also examined.

\subsection{3-D lid-driven cavity}
\begin{figure}
 \includegraphics[height=40mm,trim={1.5cm 0.2cm 1.5cm 0cm},clip]{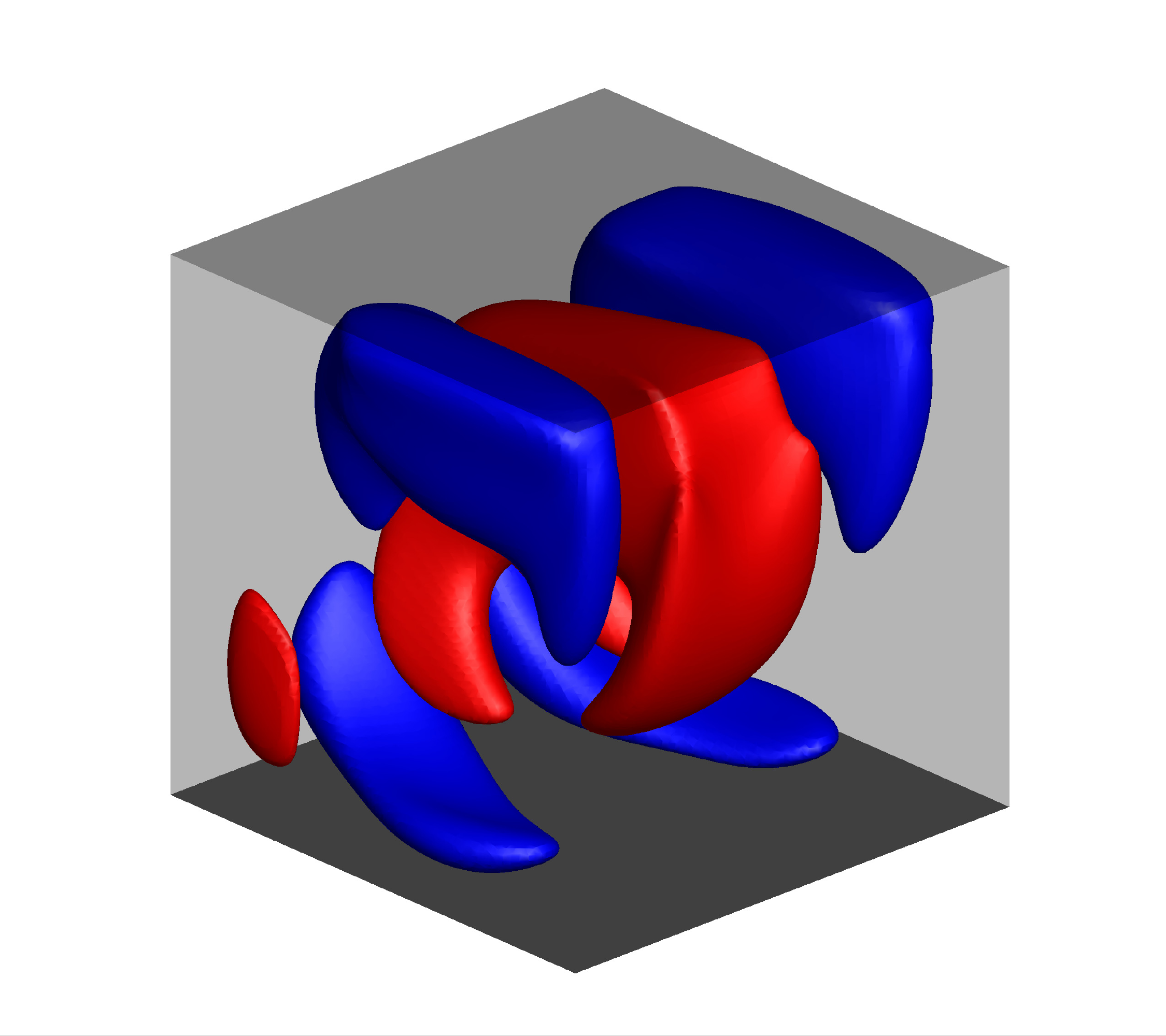}
\put(-130,115){$(a)$}
\put(-110,115){$\omega=-0.1350 \pm i0.294$}
\put(-60,105){$\hat{u}$}
\includegraphics[height=40mm,trim={1.5cm 0.2cm 1.5cm 0cm},clip]{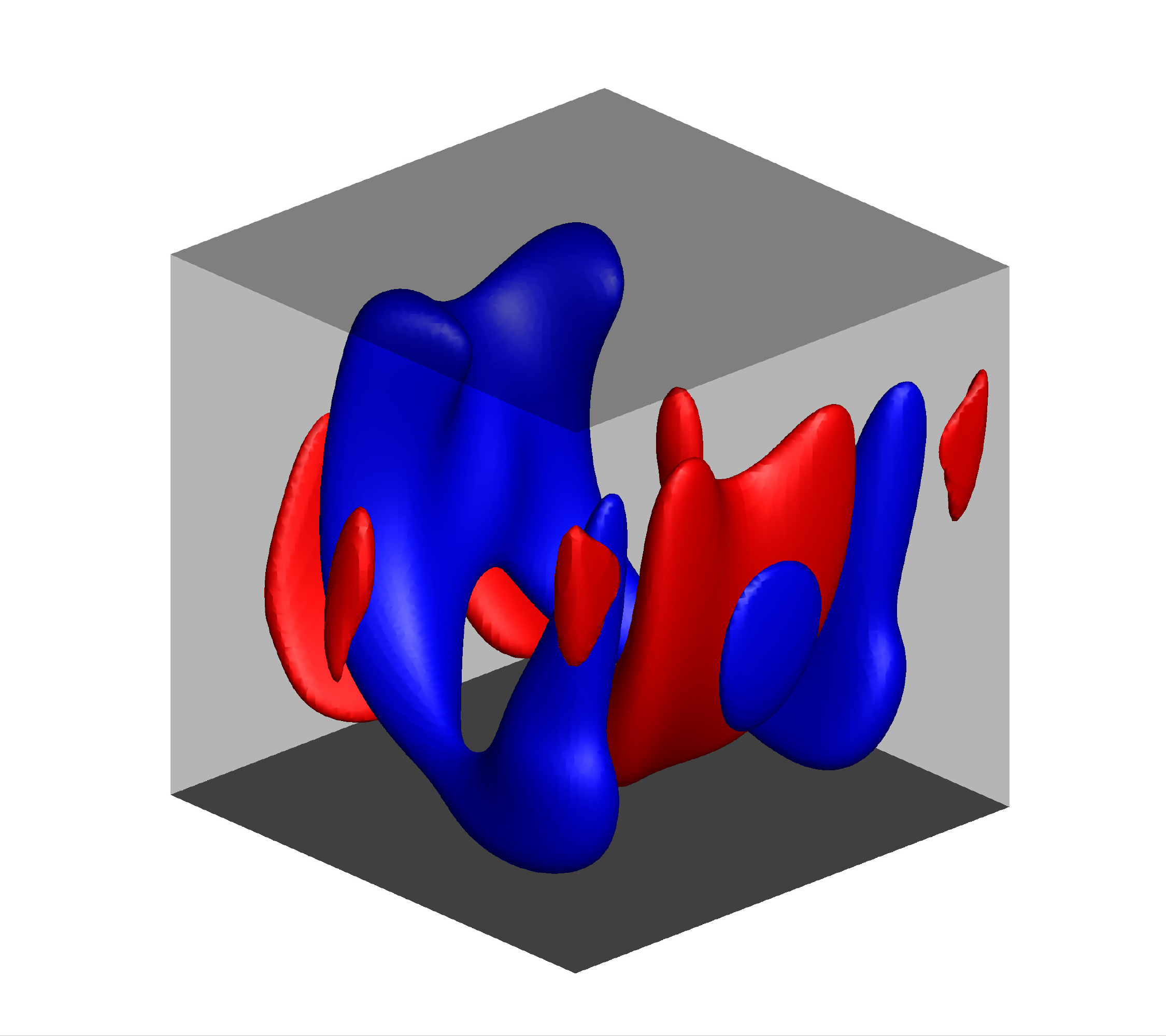}
\put(-60,105){$\hat{v}$}
\includegraphics[height=40mm,trim={1.5cm 0.2cm 0cm 0cm},clip]{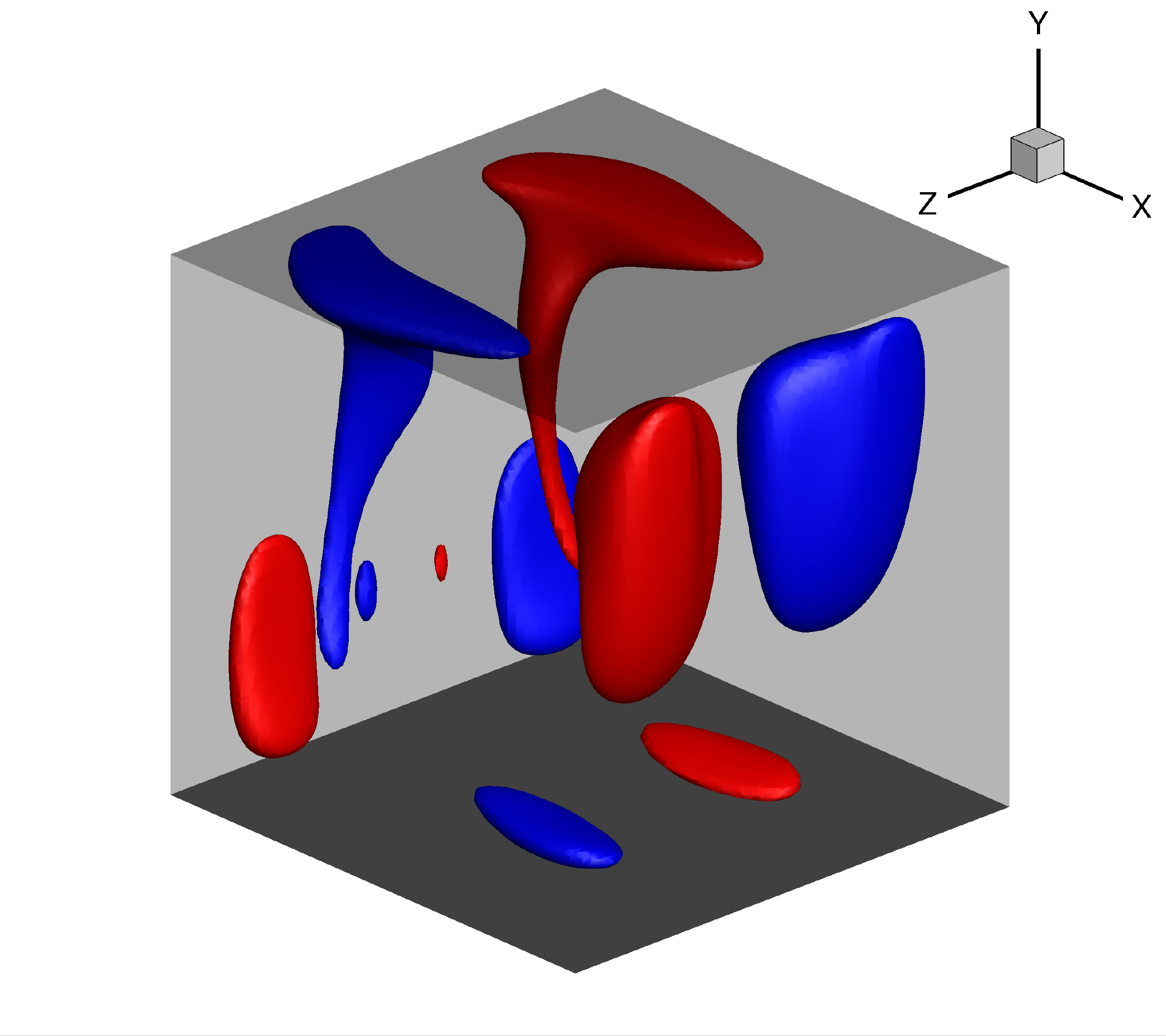}
\put(-70,105){$\hat{w}$}
\hspace{3mm}

\includegraphics[height=40mm,trim={1.5cm 0.2cm 1.5cm 0cm},clip]{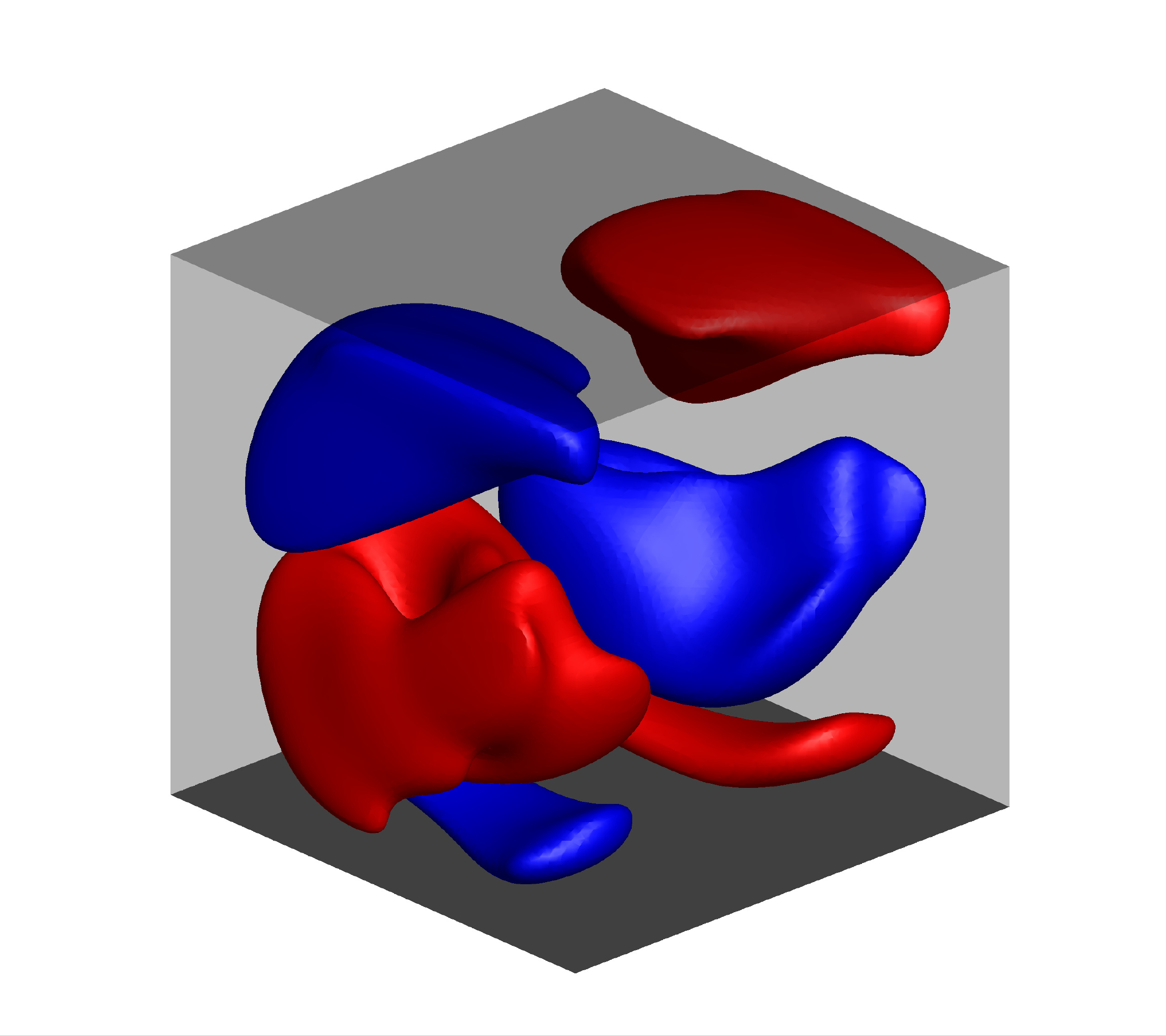}
\put(-130,110){$(b)$}
\put(-110,110){$\omega=-0.1343 \pm i0.485$}
\includegraphics[height=40mm,trim={1.5cm 0.2cm 1.5cm 0cm},clip]{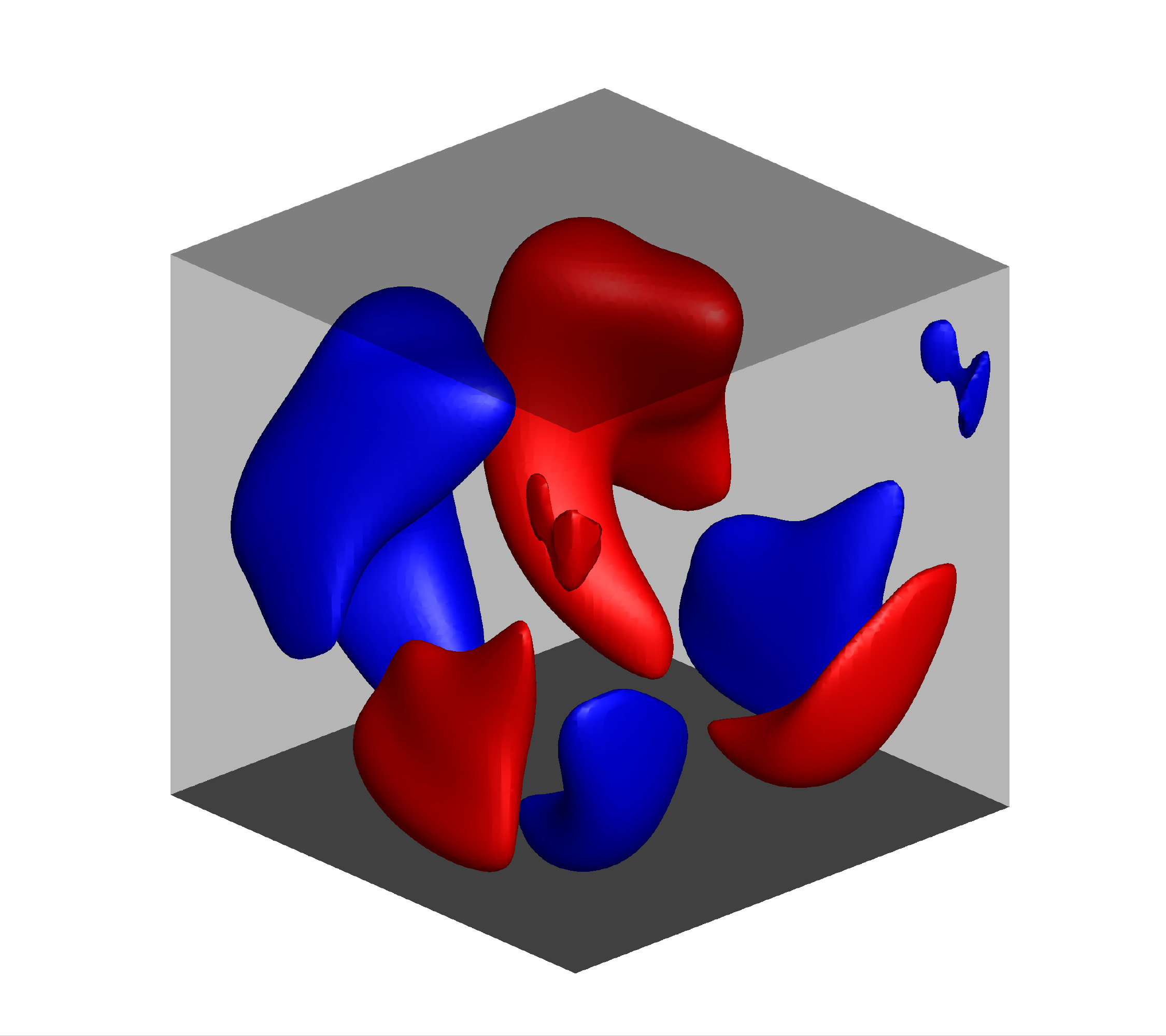}
\includegraphics[height=40mm,trim={1.5cm 0.2cm 1.5cm 0cm},clip]{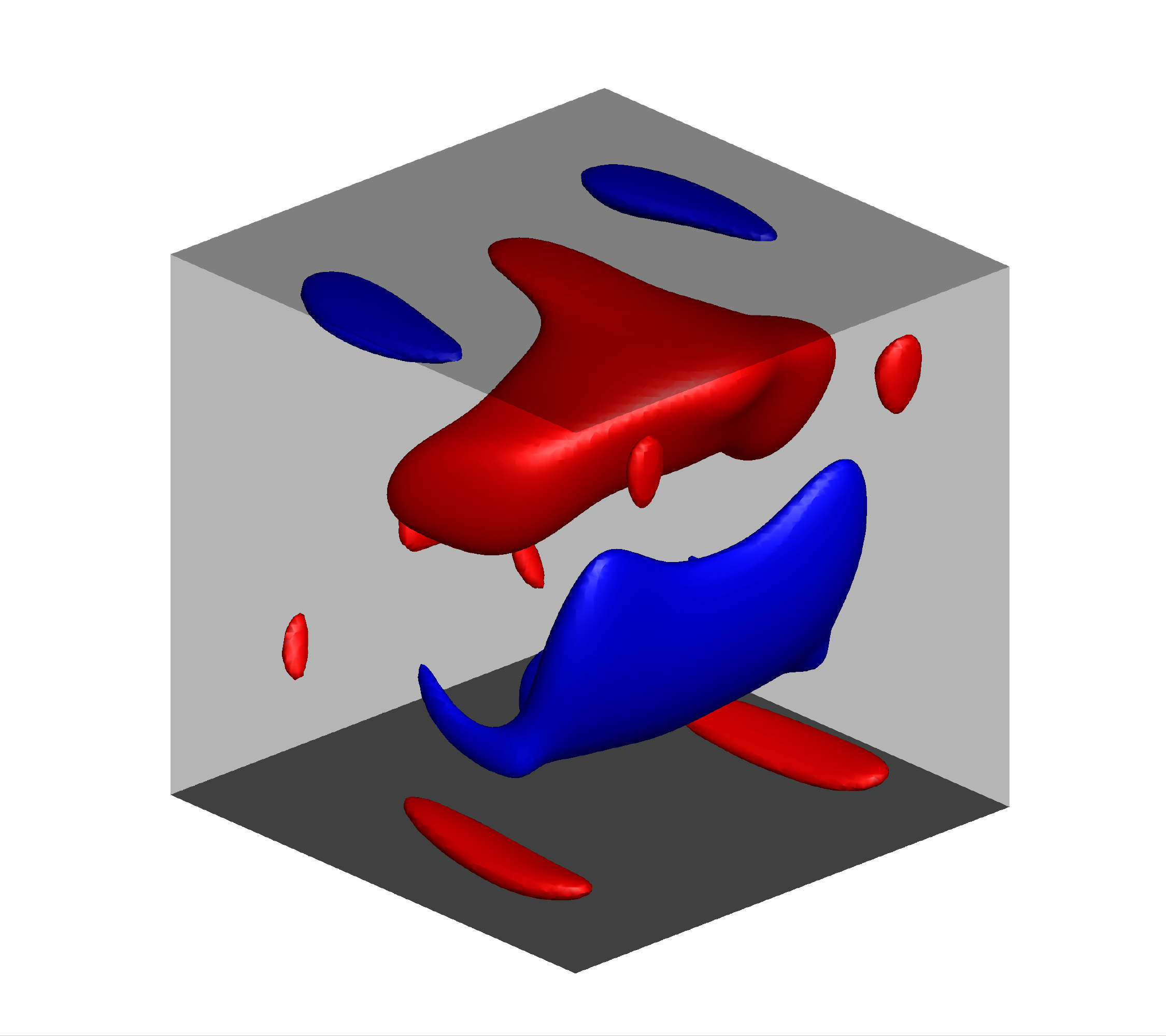}
\hspace{3mm}

\includegraphics[height=40mm,trim={1.5cm 0.2cm 1.5cm 0cm},clip]{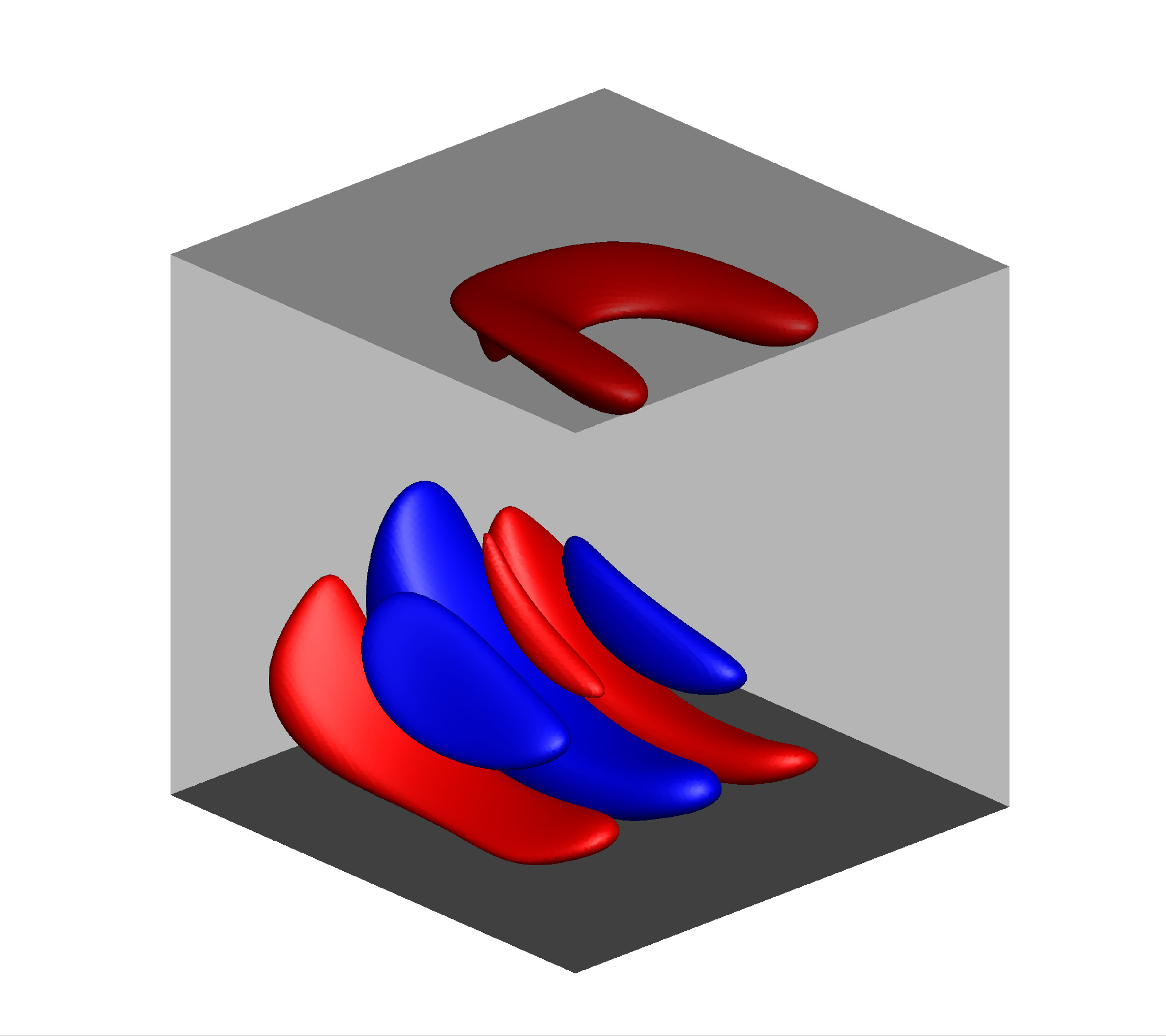}
\put(-130,110){$(c)$}
\put(-110,110){$\omega=-0.1405$}
\includegraphics[height=40mm,trim={1.5cm 0.2cm 1.5cm 0cm},clip]{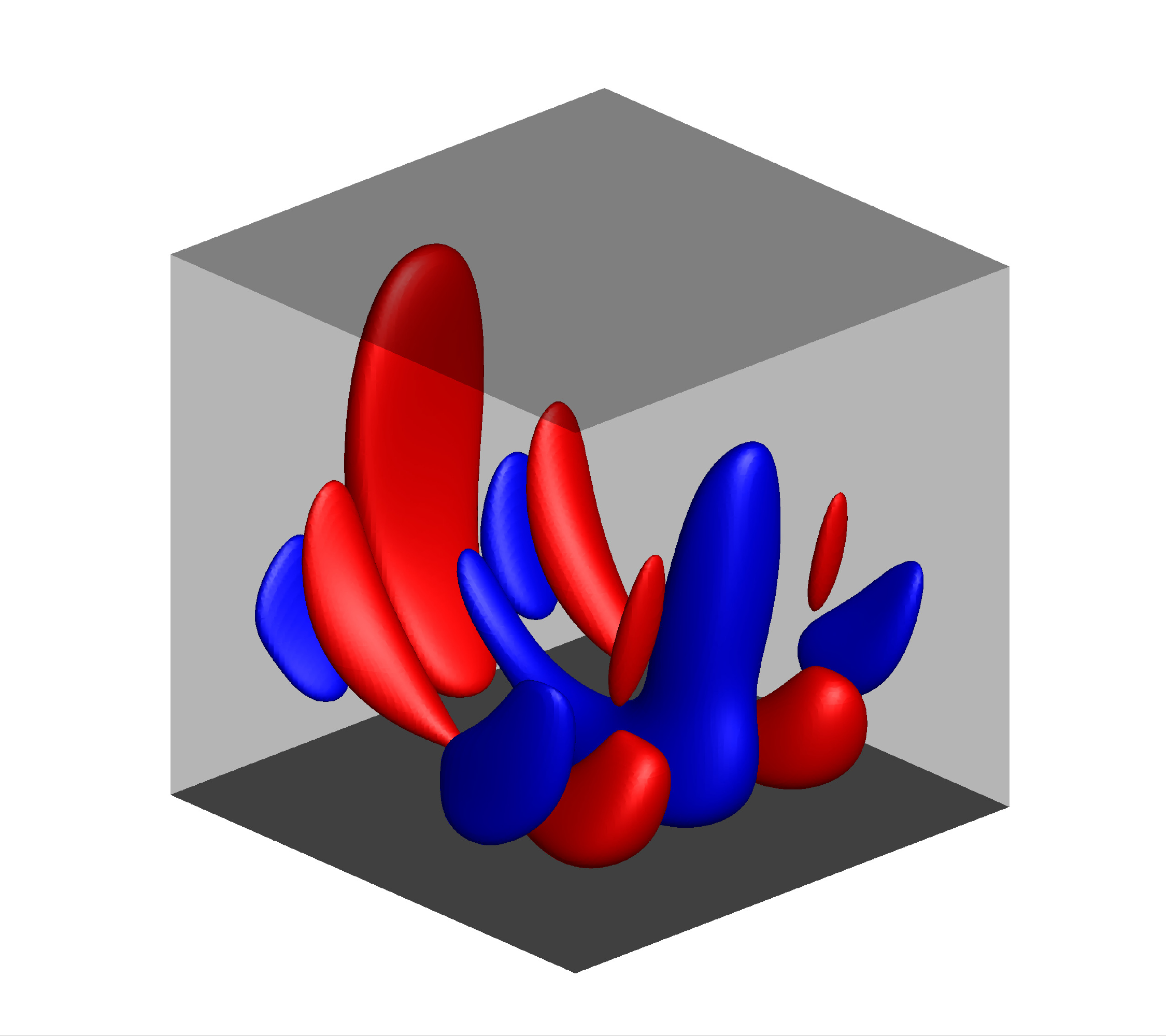}
\includegraphics[height=40mm,trim={1.5cm 0.2cm 1.5cm 0cm},clip]{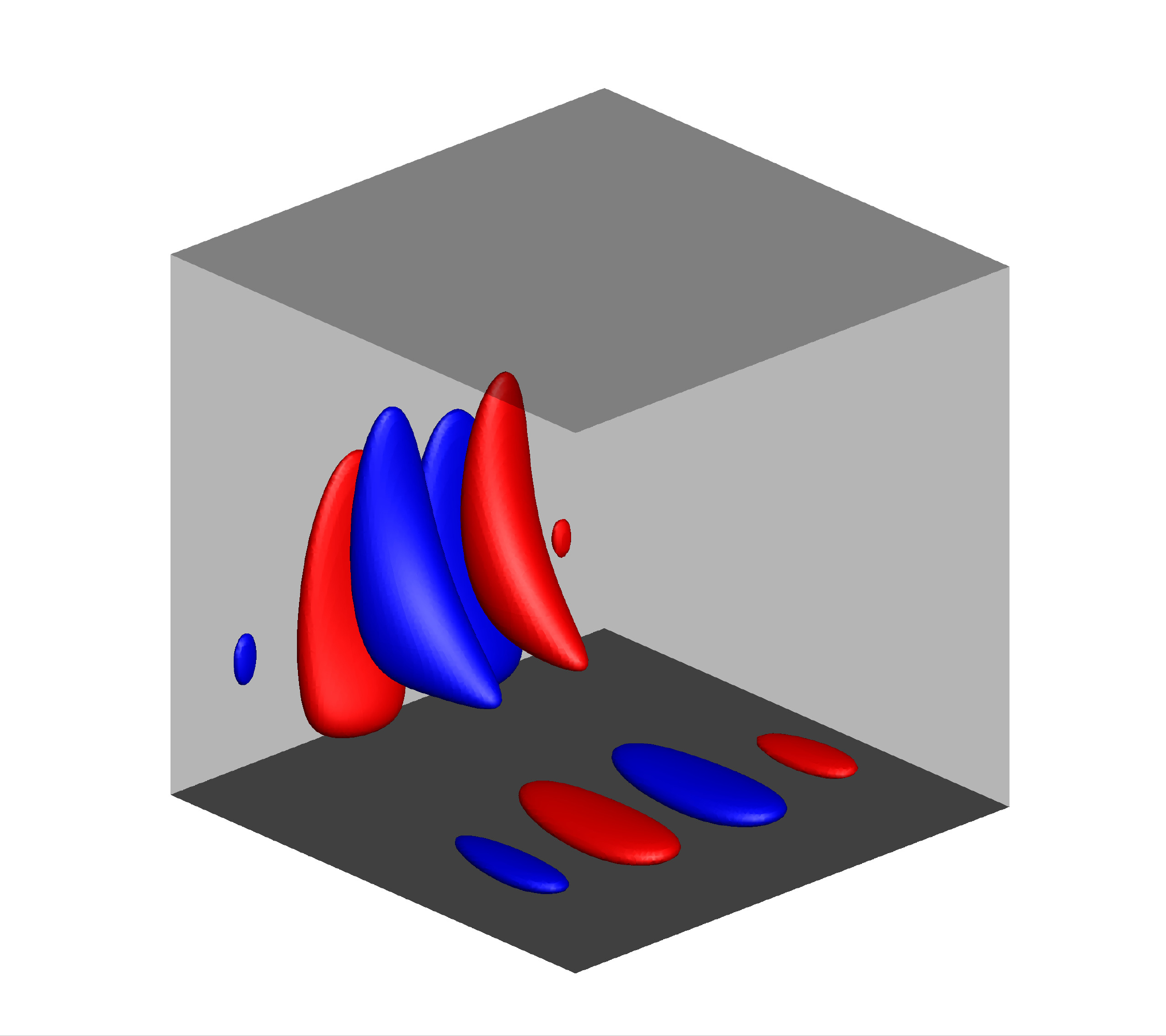}

\caption{Real part of the eigenmodes from global stability analysis of the cubic lid-driven cavity at $Re=1000$. Eigenfunction velocity fields ($\hat{u},\hat{v},\hat{w}$) are normalized with max($\hat{u}$). The results are shown with positive and negative isocontours of $\hat{u},\hat{v},\hat{w}=\pm 0.15$.}
\label{fig:3d_ldc}
\end{figure}

  \begin{table}
  \begin{center}
\def~{\hphantom{0}}
    \begin{tabular}{lc}
    %\cite{gomez2014three} & Present \\[3pt]\hline
    \cite{regan2017global} & Present \\[3pt]\hline
    % $-0.1292 \pm i0.329$ & $-0.1350 \pm i0.294$ \\
    % $-0.1348 \pm i0.485$ & $-0.1343 \pm i0.485$ \\
    % $-0.1382$ & $-0.1405$ \\\hline
    $-0.1352 \pm i0.299$ & $-0.1350 \pm i0.294$ \\
    $-0.1304 \pm i0.487$ & $-0.1343 \pm i0.485$ \\
    $-0.1375$ & $-0.1405$ \\\hline
    \end{tabular}
    \caption{The leading eigenvalues from global stability analysis for a stable 3-D lid-driven cavity at $Re=1000$ compared to \cite{regan2017global}.}
      \label{tab:ldc}
   \end{center}
\end{table}

3-D lid-driven cavity is studied as the first validation case for the global stability solver. The Reynolds number based on the cavity height and the lid velocity is 1000.

%The present work solves LNS and utilizes the IRAM implemented in the parallel PARPACK library. 
The leading eigenvalues solved from the global stability solver show good agreement with \cite{regan2017global} in table \ref{tab:ldc}. The isocontours of the real part of the leading eigenmodes in figure \ref{fig:3d_ldc} show good qualitative agreement with the results in \cite{gomez2014three} and \cite{regan2017global}.

\subsection{Laminar channel flow}

\begin{figure}
\centering
\includegraphics[width=90mm,trim={0.5cm 0.2cm 1.5cm 0cm},clip]{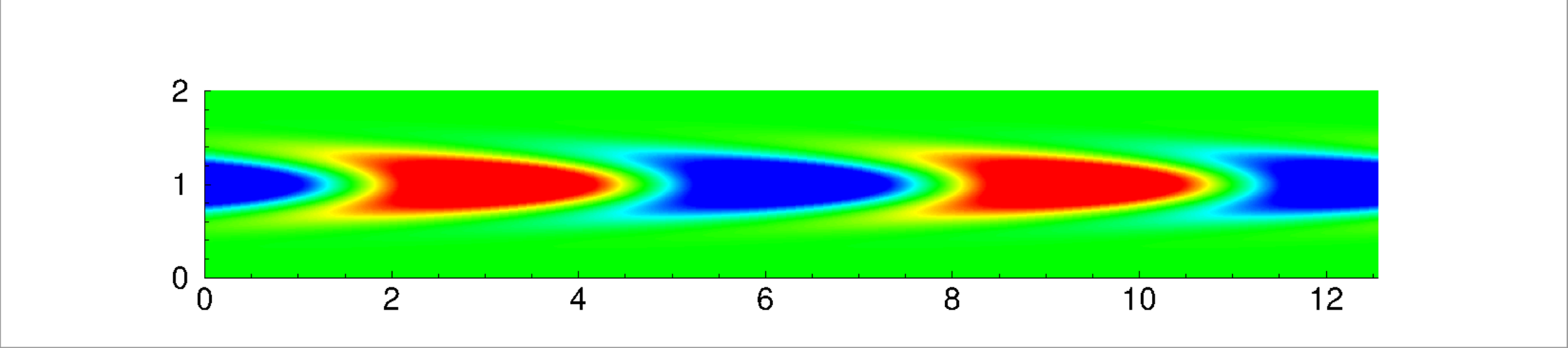}
\put(-125,-5){$x/\delta_{ch}$}
\put(0,25){$\hat{w}$}
\put(-255,25){$y/\delta_{ch}$}
\put(-260,45){$(a)$}
\hspace{5mm}
\includegraphics[width=90mm,trim={0.5cm 0.2cm 1.5cm 0cm},clip]{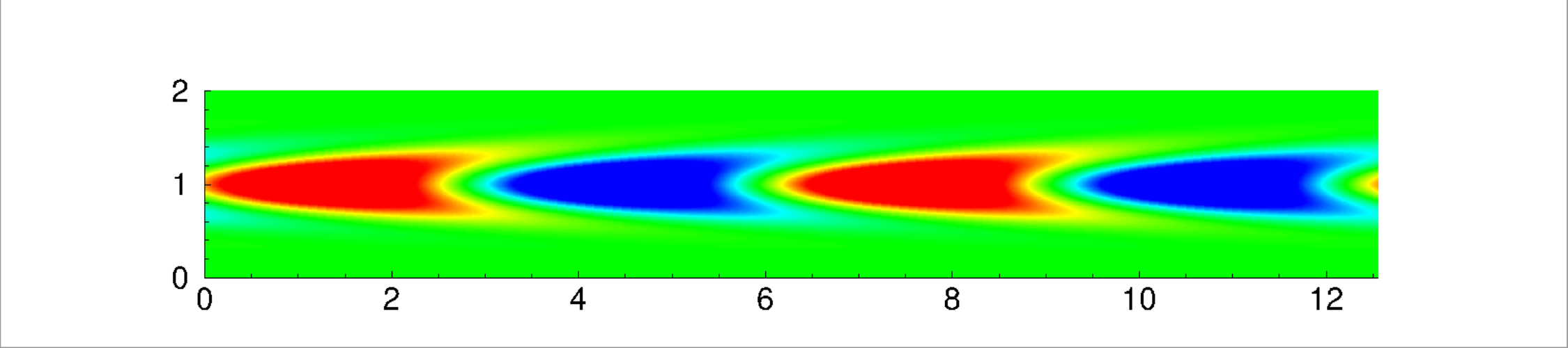}
\put(-125,-5){$x/\delta_{ch}$}
\put(0,25){$\hat{w}^{\dagger}$}
\put(-255,25){$y/\delta_{ch}$}
\put(-260,45){$(b)$}
\hspace{5mm}
\includegraphics[height=40mm,trim={0.5cm 0.2cm 1.5cm 0cm},clip]{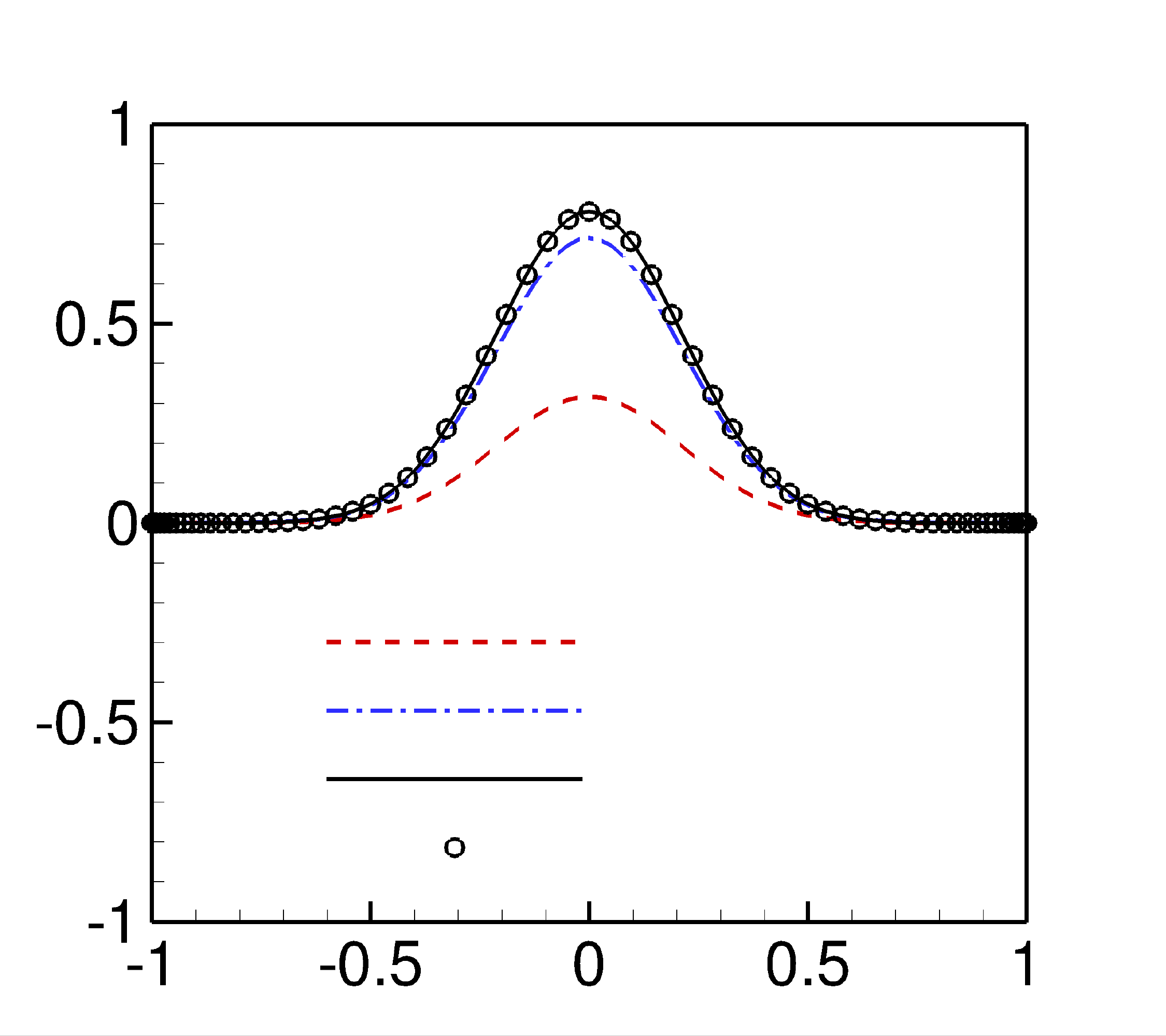}
\put(-125,95){$(c)$}
\put(-125,55){$\hat{w}$}
\put(-60,-3){$y/\delta_{ch}$}
\put(-54,42){\scriptsize{Re($\hat{u}_i$)}}
\put(-54,34){\scriptsize{Im($\hat{u}_i$)}}
\put(-54,26){\scriptsize{$|\hat{u}_i|$}}
\put(-54,18){\scriptsize{Juniper et al.}}
\hspace{3mm}
\includegraphics[height=40mm,trim={0.5cm 0.2cm 1.5cm 0cm},clip]{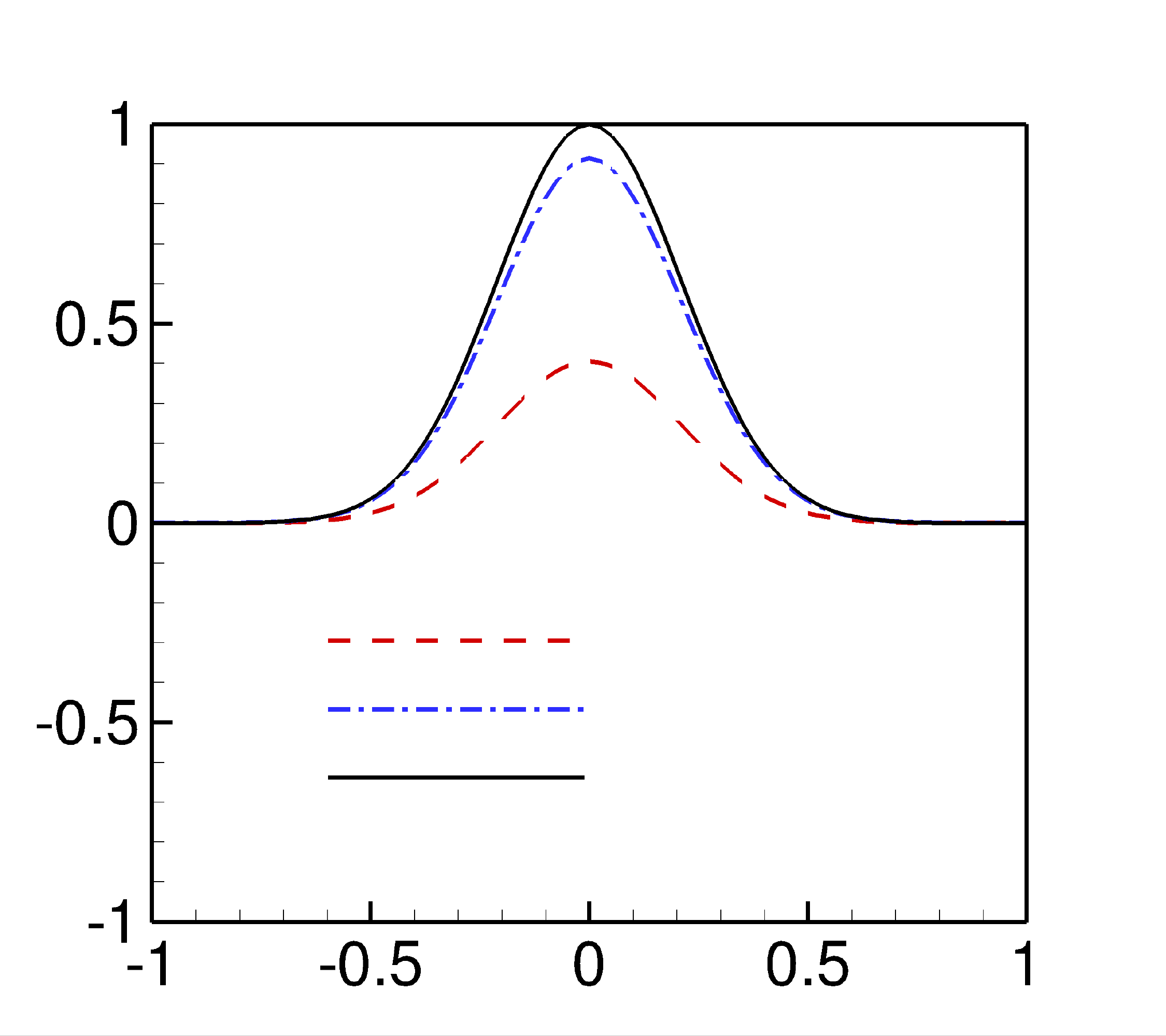}
\put(-125,95){$(d)$}
\put(-125,55){$\hat{w}^{\dagger}$}
\put(-60,-3){$y/\delta_{ch}$}
\put(-54,42){\scriptsize{Re($\hat{u}_i^{\dagger}$)}}
\put(-54,34){\scriptsize{Im($\hat{u}_i^{\dagger}$)}}
\put(-54,26){\scriptsize{$|\hat{u}_i^{\dagger}|$}}
\caption{Real part of $(a)$ the direct and $(b)$ adjoint eigenmodes corresponding to the first leading eigenvalue ($\alpha=1,\beta=0$). Shown with the contour of $\hat{w}$ at the mid-plane $z=(2\pi/3)\delta_{ch}$. The associated Fourier coefficients $(c)$ $\hat{w}$ and $(d)$ $\hat{w}^{\dagger}$ are shown for completeness. %Real part (dash), imaginary part (dash dot) and the absolute value (solid) of the global eigenmode Fourier coefficients $\hat{u}_i$. 
The absolute values of the global eigenmode match the results of $|\hat{w}|$ from parallel flow stability of \cite{juniper2014modal}. Note that $\hat{u}=\hat{v}=\hat{u}^{\dagger}=\hat{v}^{\dagger}=0$.}
\label{fig:mode15}
\end{figure}
%direct: laminar_channel_test_4; laminar_channel_adjoint

\begin{figure}
\centering
\includegraphics[width=85mm,trim={0.5cm 0.2cm 1.5cm 0cm},clip]{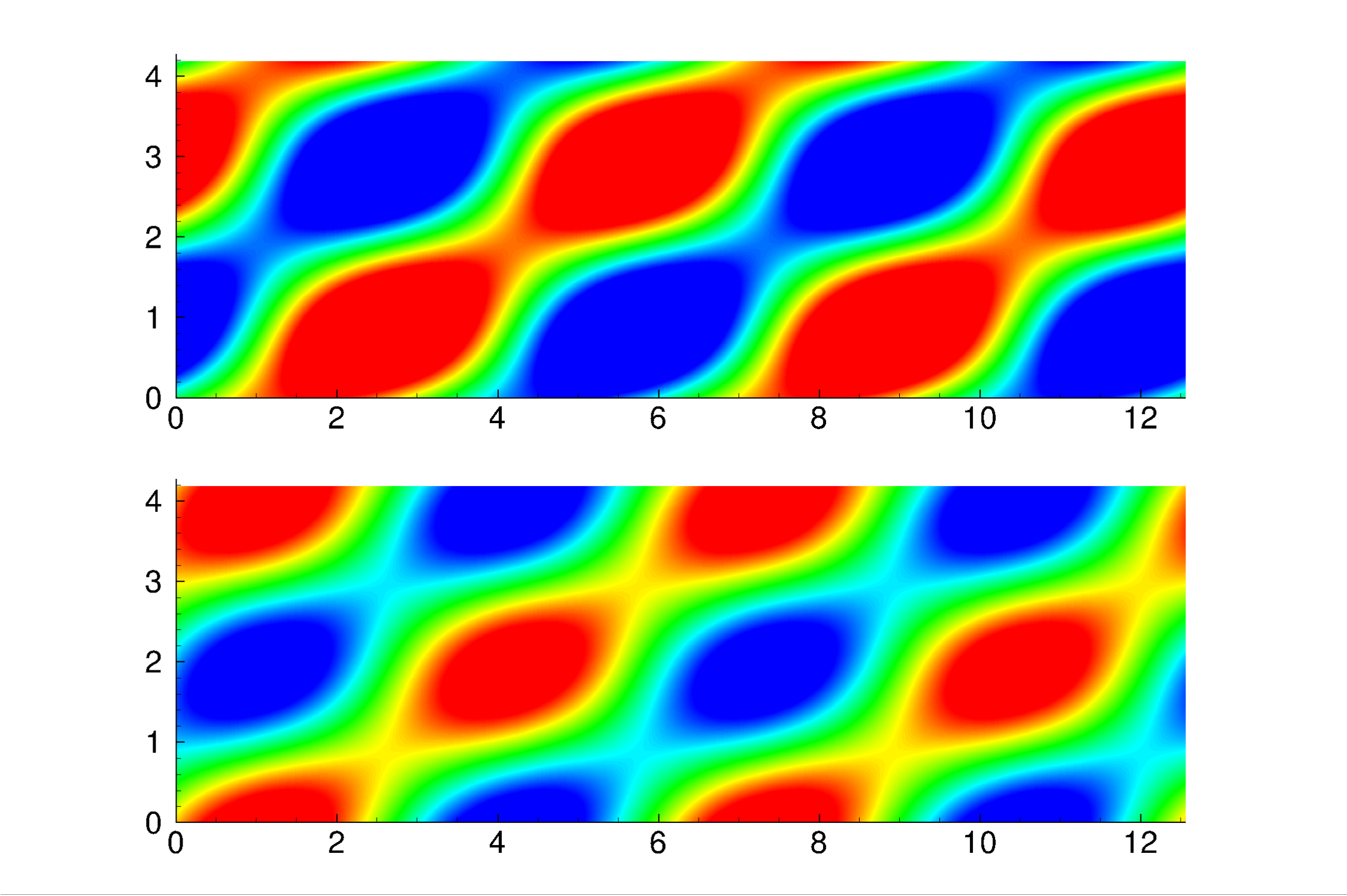}
\put(-115,0){$x/\delta_{ch}$}
\put(0,43){$\hat{w}$}
\put(-115,83){$x/\delta_{ch}$}
\put(0,127){$\hat{u}$}
\put(-245,125){$z/\delta_{ch}$}
\put(-245,40){$z/\delta_{ch}$}
\put(-245,165){$(a)$}
\hspace{8mm}
 \includegraphics[height=40mm,trim={0.5cm 0.2cm 1.5cm 0cm},clip]{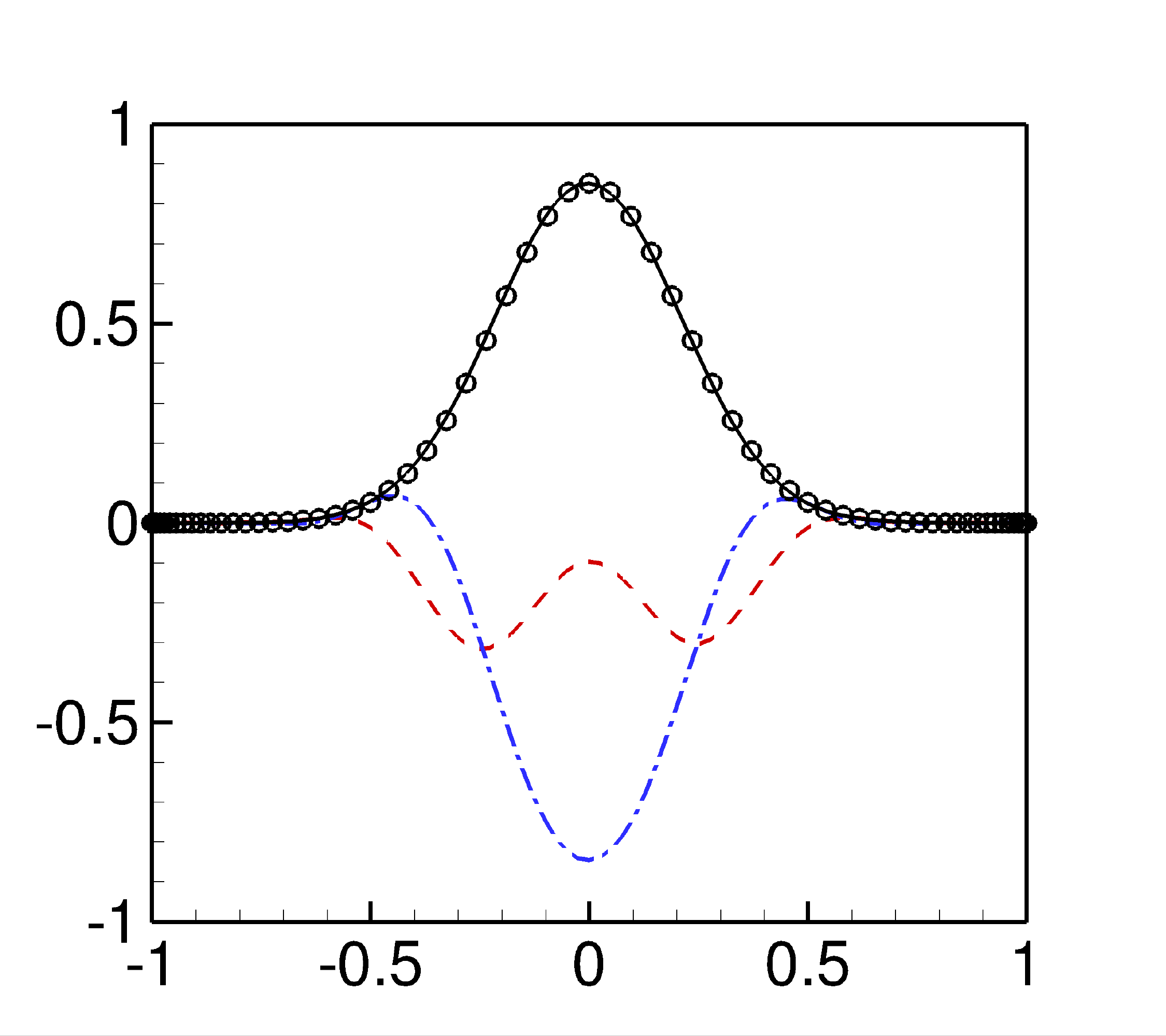}
\put(-125,95){$(b)$}
%\put(-90,115){$\alpha=1,\beta=1.5$}
\put(-125,55){$\hat{u}$}
\put(-60,-3){$y/\delta_{ch}$}
\hspace{3mm}
 \includegraphics[height=40mm,trim={0.5cm 0.2cm 1.5cm 0cm},clip]{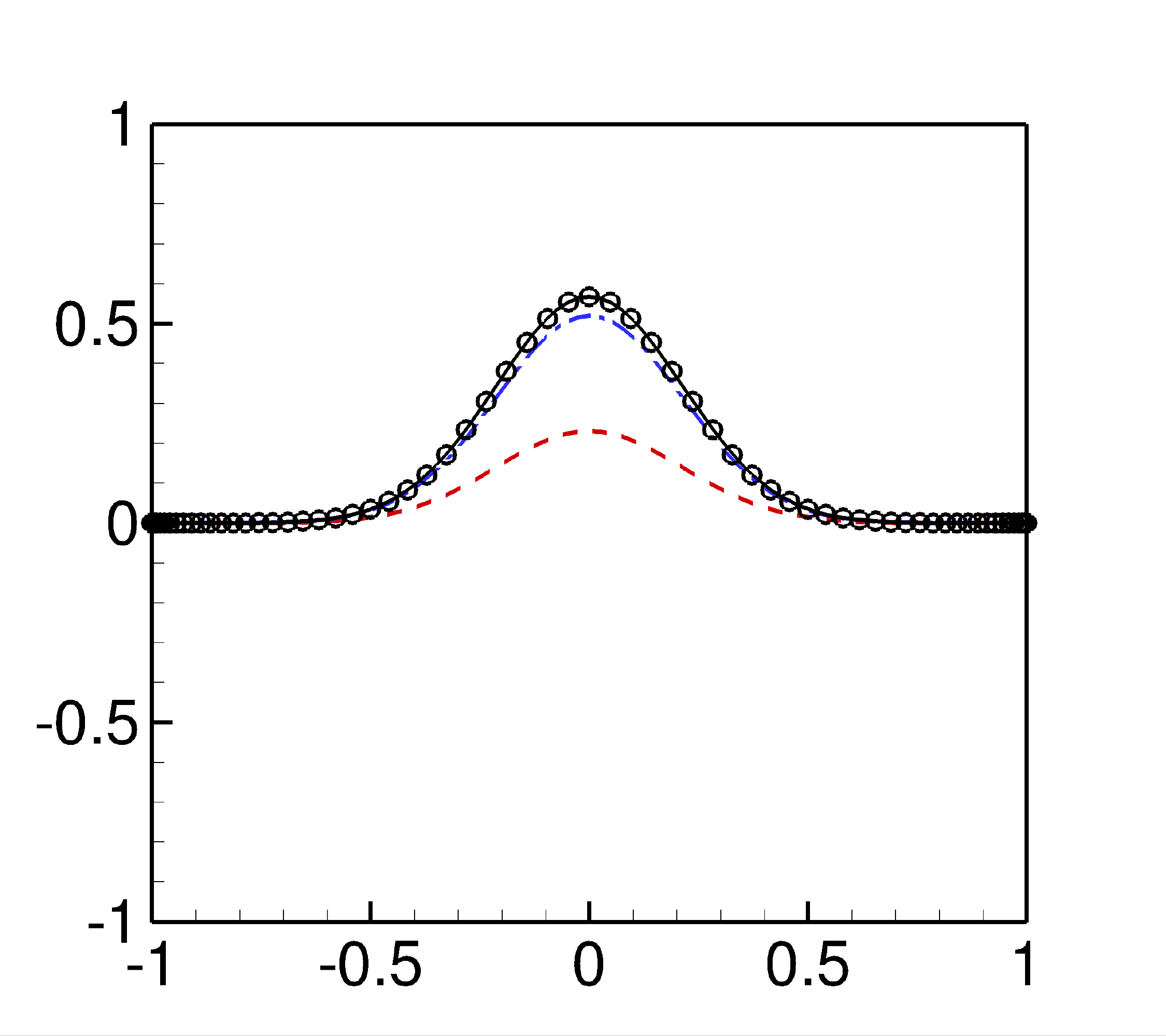}
%\put(-125,95){$(c)$}
%\put(-90,115){$\alpha=1,\beta=1.5$}
\put(-125,55){$\hat{w}$}
\put(-60,-3){$y/\delta_{ch}$}
\caption{$(a)$ Real part of the direct eigenmode corresponding to the second leading eigenvalue ($\alpha=1,\beta=1.5$). Shown with the contour of $\hat{u}$ and $\hat{w}$ at the mid-plane $y=\delta_{ch}$. $(b)$ The associated Fourier coefficients $\hat{u}$ and $\hat{w}$ are shown for completeness. The legend is the same as figure \ref{fig:mode15}$(c)$. Note that $\hat{v}=0$. } 
\label{fig:mode17}
\end{figure}
%LNS_solver/runs/laminar_channel_test2/v/plots

\begin{figure}
\centering
\includegraphics[width=90mm,trim={0.5cm 0.2cm 1.5cm 0cm},clip]{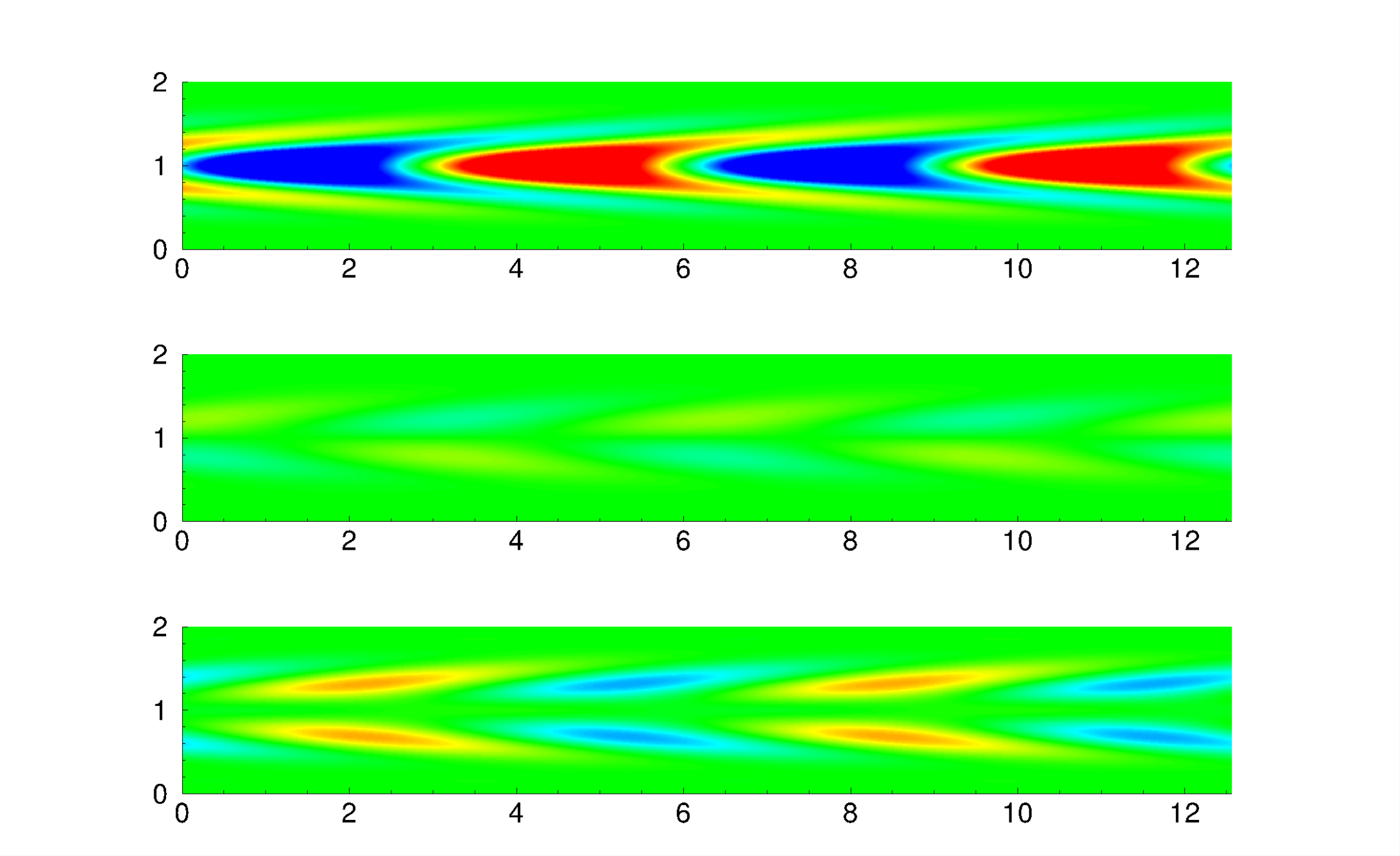}
\put(-250,160){$(a)$}
\put(-125,-2){$x/\delta_{ch}$}
\put(0,25){$\hat{w}^{\dagger}$}
\put(-255,25){$y/\delta_{ch}$}
\put(-125,52){$x/\delta_{ch}$}
\put(0,80){$\hat{v}^{\dagger}$}
\put(-255,80){$y/\delta_{ch}$}
\put(-125,105){$x/\delta_{ch}$}
\put(0,135){$\hat{u}^{\dagger}$}
\put(-255,135){$y/\delta_{ch}$}
\hspace{5mm}
\includegraphics[height=40mm,trim={0.5cm 0.2cm 1.5cm 0cm},clip]{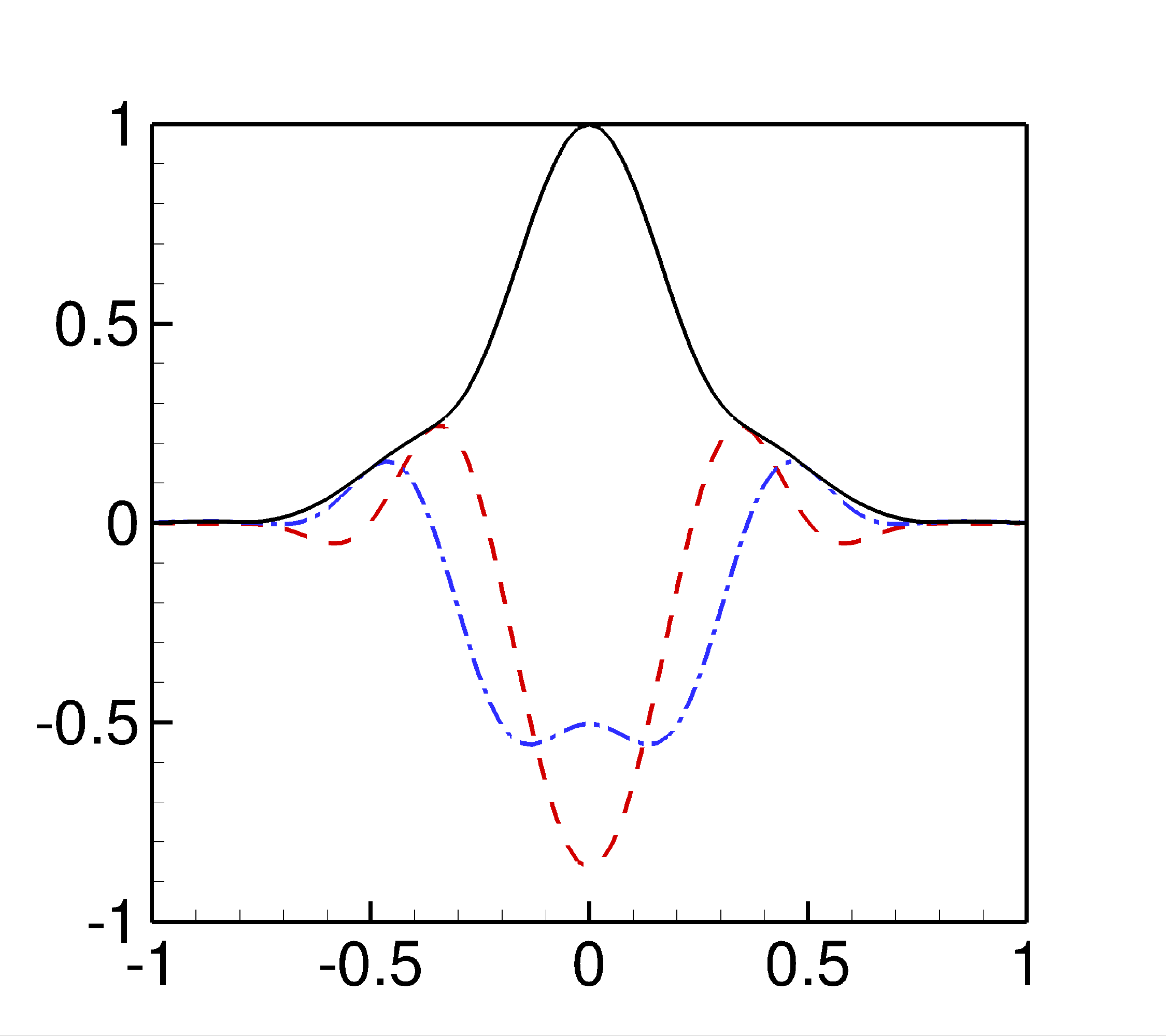}
\put(-125,105){$(b)$}
\put(-125,55){$\hat{u}^{\dagger}$}
\put(-60,-3){$y/\delta_{ch}$}
\hspace{3mm}
 \includegraphics[height=40mm,trim={0.5cm 0.2cm 1.5cm 0cm},clip]{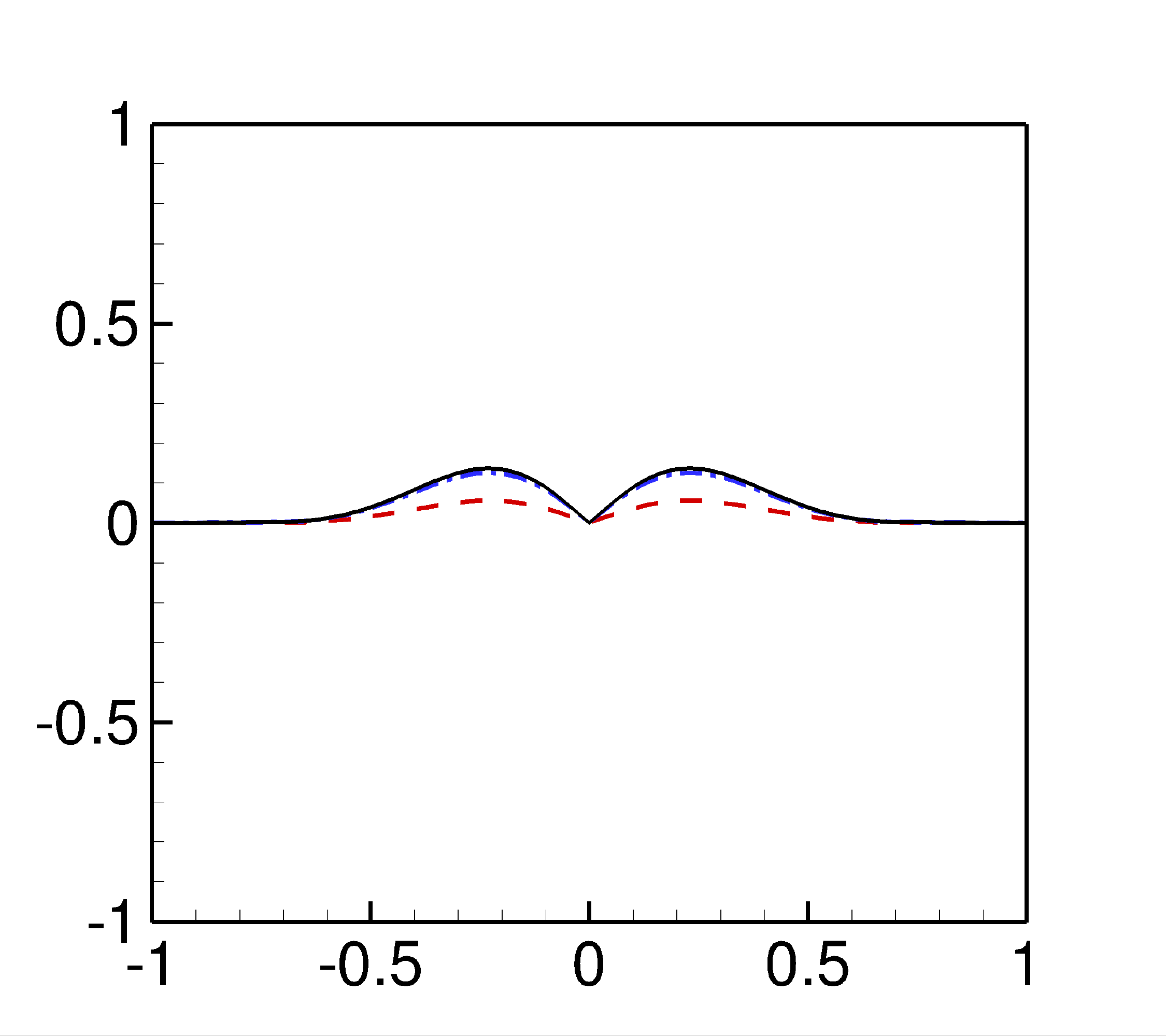}
%\put(-110,115){$(b)$}
\put(-125,55){$\hat{v}^{\dagger}$}
\put(-60,-3){$y/\delta_{ch}$}
\hspace{3mm}
 \includegraphics[height=40mm,trim={0.5cm 0.2cm 1.5cm 0cm},clip]{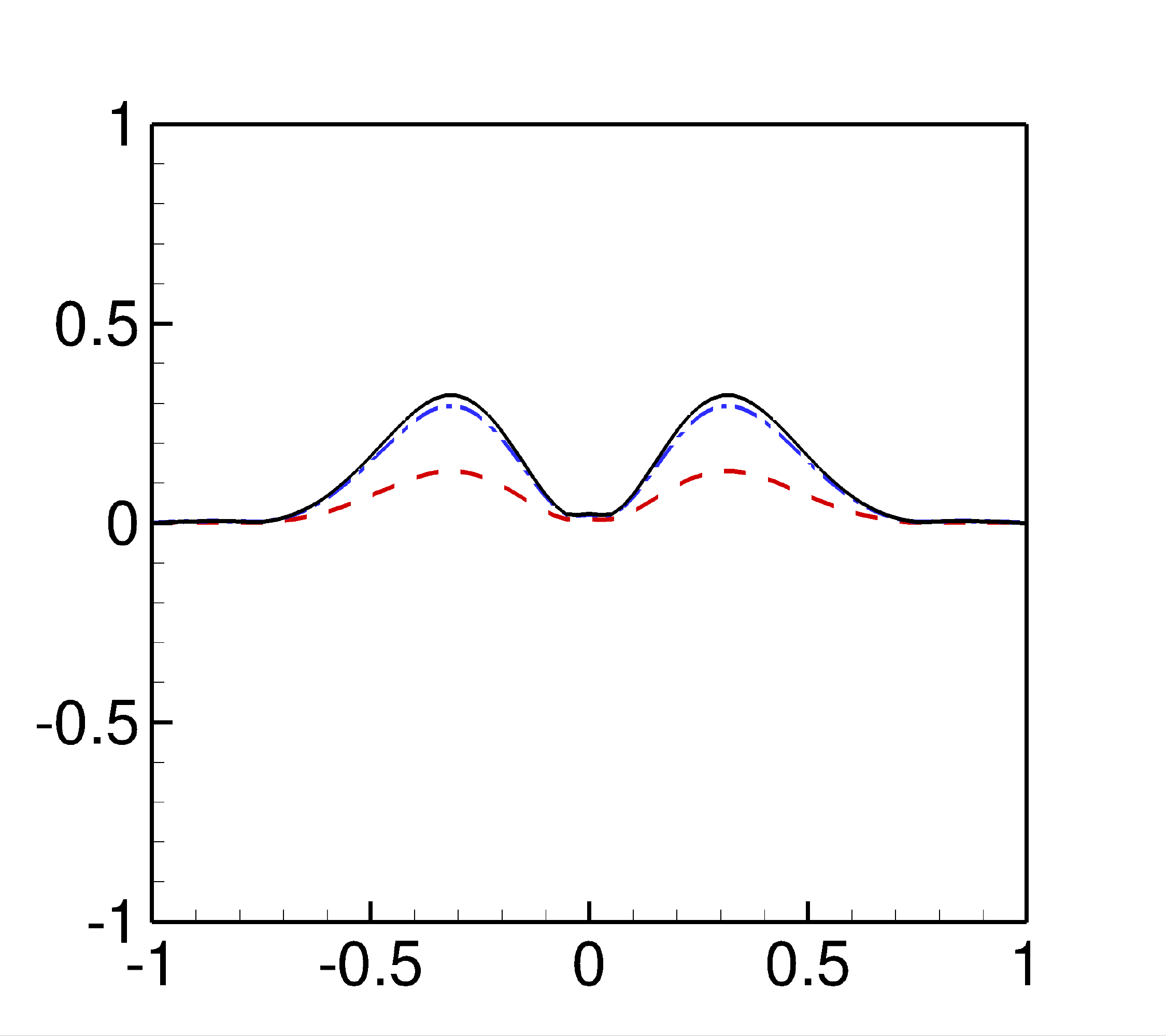}
%\put(-110,115){$(c)$}
\put(-125,55){$\hat{w}^{\dagger}$}
\put(-60,-3){$y/\delta_{ch}$}
\caption{$(a)$ Real part of the adjoint eigenmode corresponding to the second leading eigenvalue ($\alpha=1,\beta=1.5$). Shown with the contour plots at the mid-plane $z=(2\pi/3)\delta_{ch}$. $(b)$ The associated Fourier coefficients are shown for completeness. The legend is the same as figure \ref{fig:mode15}$(d)$.}
\label{fig:mode2}
\end{figure}
%direct: laminar_channel_test_4
%stability_analysis/chan/old/chan_adj/chan_2m_adj/runs/run_1_2/v/jfm_plots/

Global stability and adjoint sensitivity analyses of a laminar channel are performed. The global stability results are compared to results from parallel flow stability analysis. The Reynolds number $Re_{\tau}=44.7$ is based on the friction velocity $u_{\tau}$ and the channel half height $\delta_{ch}$. The domain length is $4\pi\delta_{ch}$ in the streamwise direction and $(4\pi/3)\delta_{ch}$ in the spanwise direction. Since the streamwise and spanwise wavenumbers $\alpha$ and $\beta$ are not specified in global stability analysis, any combination of those can be present in the global stability results. Thus we can extract $\hat{u}_i^j$ for a selective combination of $\alpha_j$ and $\beta_j$ using streamwise and spanwise Fast Fourier transforms.  The selected combination of $\alpha_j$ and $\beta_j$ can be used as the input into the parallel flow stability. The leading eigenvalues from global stability and adjoint sensitivity analyses show agreement with those from the parallel flow linear stability analysis in table \ref{tab:laminar_ch}.

The non-zero components of the first leading direct and adjoint eigenmodes are shown in figure \ref{fig:mode15}. For the contour plots, qualitative agreement is shown compared to the results from \cite{regan2017global,regan2019adjoint}. A quantitative comparison between the global stability and parallel flow linear stability results is shown in figure \ref{fig:mode15}$(c)$. Good agreement is obtained for $|\hat{w}|$. The results of the second leading direct and adjoint eignmodes are shown in figures \ref{fig:mode17} and \ref{fig:mode2} respectively. Both qualitative and quantitative agreement are obtained.

  \begin{table}
  \begin{center}
\def~{\hphantom{0}}
    \begin{tabular}{lcc}
     & $\alpha=1$, $\beta=0$ & $\alpha=1$, $\beta=1.5$ \\
     \hline
    \cite{juniper2014modal} & $-2.336 \times 10^{-2} + i9.776 \times 10^{-1}$ & $-2.561 \times 10^{-2} + i9.776 \times 10^{-1}$ \\
    Present, direct & $-2.338 \times 10^{-2} \pm i9.776 \times 10^{-1}$ & $-2.563 \times 10^{-2} \pm i9.776 \times 10^{-1}$ \\
    Present, adjoint & $-2.338 \times 10^{-2} \pm i9.776 \times 10^{-1}$ & $-2.563 \times 10^{-2} \pm i9.776 \times 10^{-1}$\\
    % $\alpha$ & $\beta$ & \cite{juniper2014modal} & Direct & Adjoint \\
    % \hline
    % 1 & 0 & $-2.336 \times 10^{-2} + i9.776 \times 10^{-1}$ & $-2.344 \times 10^{-2} \pm i9.783 \times 10^{-1}$ & $-2.344 \times 10^{-2} \pm i9.783 \times 10^{-1}$\\
    % 1 & 1.5 & $-2.561 \times 10^{-2} + i9.776 \times 10^{-1}$ & $-2.571 \times 10^{-2} \pm i9.784 \times 10^{-1}$ & $-2.571 \times 10^{-2} \pm i9.784 \times 10^{-1}$\\
    \hline
    \end{tabular}
    \caption{\label{tab:laminar_ch} The leading eigenvalues from global stability and adjoint sensitivity analyses for the laminar channel flow at $Re_{\tau}=44.7$ compared to \cite{juniper2014modal}. The selected combinations of $\alpha$ and $\beta$ are used as input to parallel flow stability analysis of Poiseuille flow. Note that the eigenvalues are normalized by the center-line velocity and the channel half height. }
     \label{tab:laminar_ch}
    \end{center}
\end{table}

%Therefore, the present global stability and adjoint sensitivity solvers on the 3-D structured platform are validated. We will now use it to study roughness-induced transition. 

\section{Problem formulation}\label{sec:setup}
In this section, the simulation set-up is shown, the base flow computation is described, and a study of grid convergence and domain length sensitivity is performed.

\subsection{Flow configuration}
\begin{figure}
%\centering
\includegraphics[width=100mm]{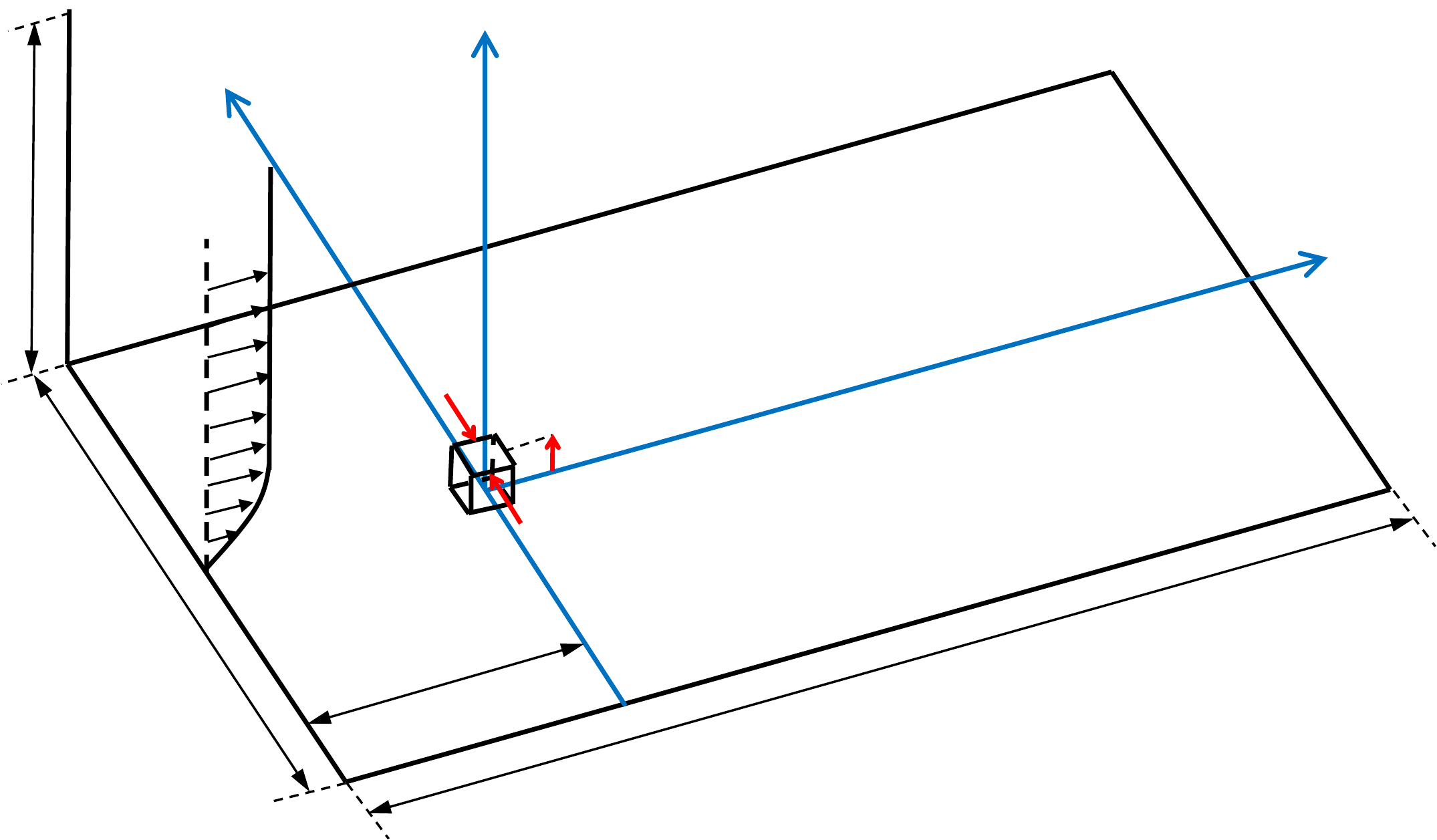}
\put(-115,20){$L_x$}
\put(-265,45){{$L_z$}}
\put(-295,125){$L_y$}
\put(-205,35){$l$}
\put(-255,125){$U_e$}
\put(-172,78){\color{red}$h$}
\put(-196,90){\color{red}$d$}
\put(-205,60){$O$}
\put(-34,118){$x$}
\put(-232,147){{$z$}}
\put(-200,155){$y$}
\includegraphics[width=40mm]{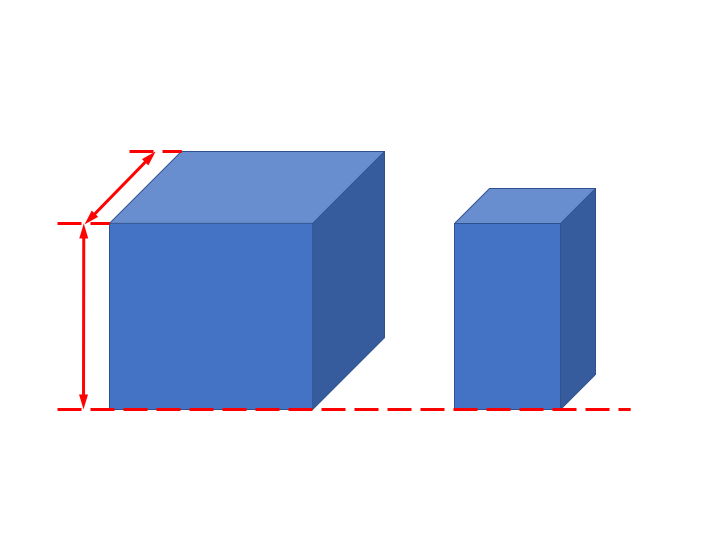}
\put(-110,33){\color{red}$h$}
\put(-105,55){\color{red}$d$}
\put(-90,8){$\eta=1$}
\put(-45,8){$\eta=0.5$}
\caption{Sketch of the flow configuration and roughness geometries. } 
\label{fig:config}
\end{figure}
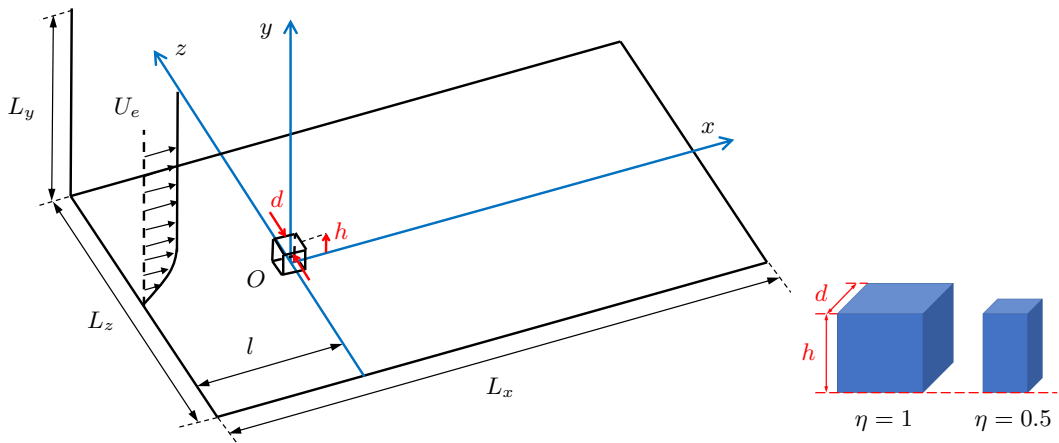

The flow configuration, the computational domain and the roughness geometries are depicted in figure \ref{fig:config}. At the inflow, a laminar Blasius boundary layer profile is prescribed. The cuboid with the height $h$ and width $d$ is centered at the origin of the Cartesian coordinate system. The ratio of the roughness height to the displacement thickness of the boundary layer $h/\delta^*$ is fixed at $2.86$. Two aspect ratios $\eta=d/h=1$ and $0.5$ are investigated. The roughness height is $h=1$, the reference length in the simulations. The streamwise extent of the computational box $L_x$ is $45h$ for global stability analyses, and is extended in the DNS to examine the transition process farther downstream. The spanwise extent is $L_z=10h$ to ensure that the roughness element behaves as isolated, and the wall-normal extent is $L_y=15h$. The distance from the inlet of the computational domain to the center of the roughness element is denoted by $l=15h$. The Blasius laminar boundary layer solution is specified at the inflow boundary, and convective boundary conditions are used at the outflow boundary. Periodic boundary conditions are used in the spanwise direction. No-slip boundary conditions are imposed on the flat plate and the roughness surfaces. The boundary conditions $U_e=1$, $\partial v/\partial y$ and $\partial w/\partial y$ are used at the upper boundary. Uniform grids are used in the streamwise and spanwise directions, and the grid in the wall-normal direction is clustered near the flat plate. Details of the grid information are shown in \S \ref{grid}.

%figure of vorticity at Reh600

\subsection{Base flow computation}
%SFD method, convergence history

%Obtaining stationary solutions of the Navier-Stokes equations is a condition precedent to stability analysis. 

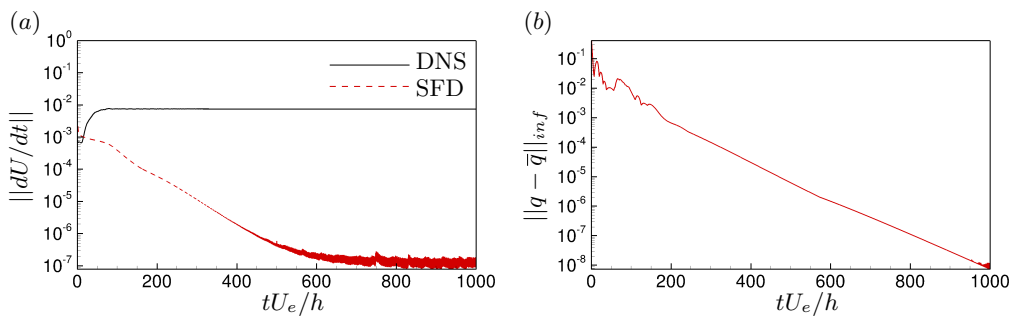
\begin{figure}
\includegraphics[width=68mm,trim={0.2cm 0.2cm 0.5cm 0cm},clip]{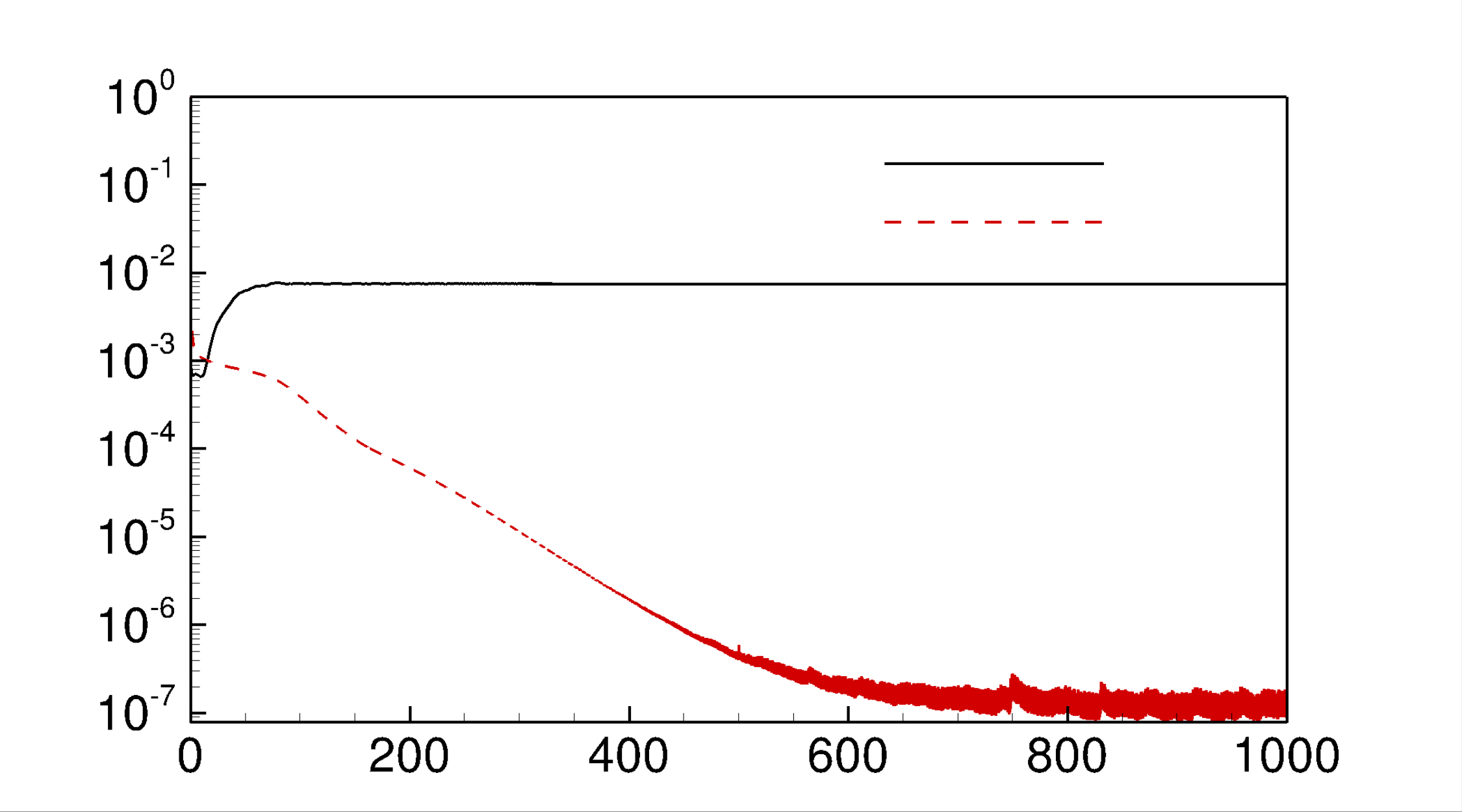}
\put(-195,35){\rotatebox{90}{$||dU/dt||$}}
\put(-100,-5){$tU_e/h$}
\put(-195,100){$(a)$}
 \put(-42,85){DNS}
 \put(-42,75){SFD}
\includegraphics[width=68mm,trim={0.2cm 0.2cm 0.5cm 0cm},clip]{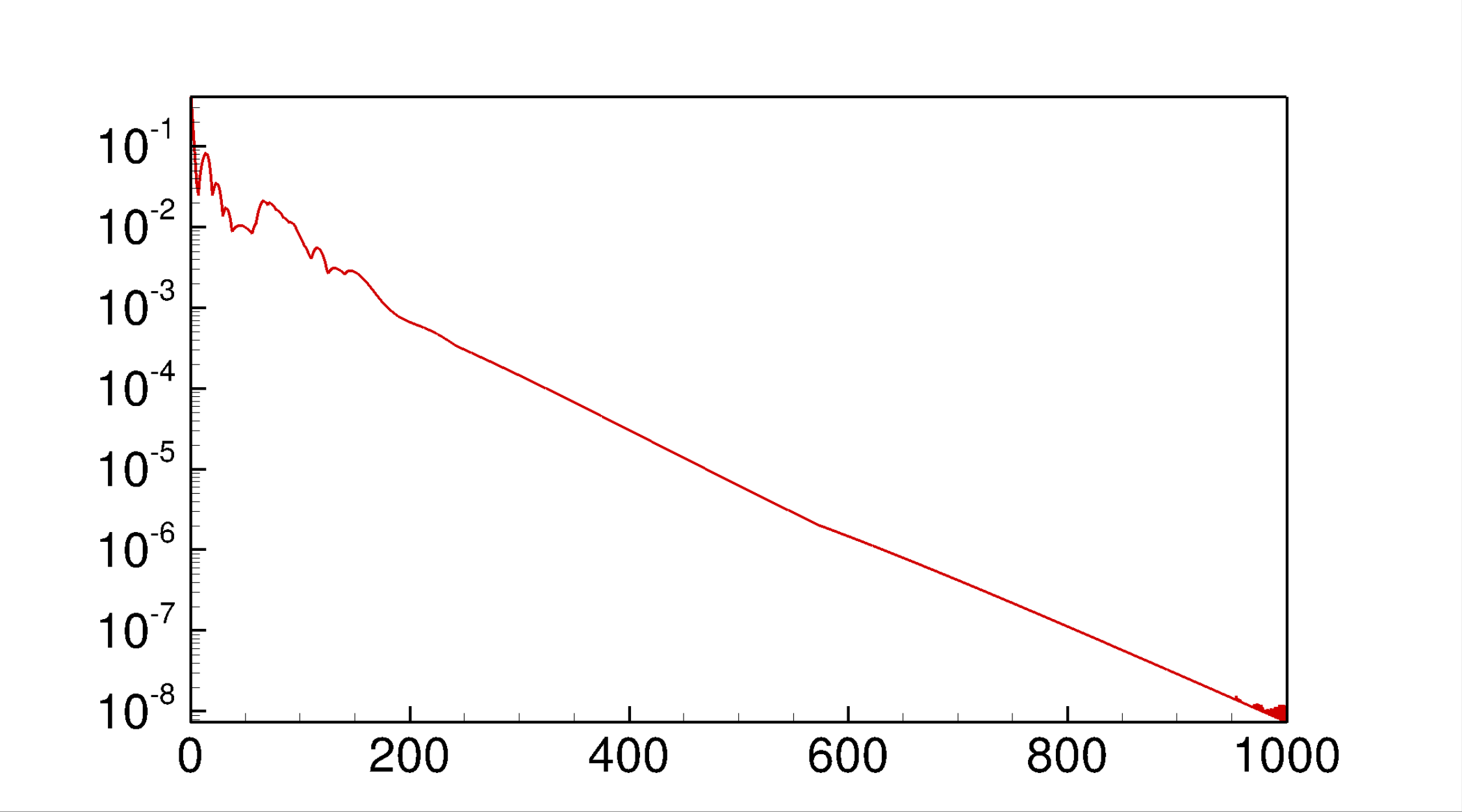}
\put(-195,28){\rotatebox{90}{$||q-\overline{q}||_{inf}$}}
\put(-100,-5){$tU_e/h$}
\put(-195,100){$(b)$}
\caption{Time evolution of $(a)$ $||dU/dt||$ and $(b)$ the residual $||q-\overline{q}||_{inf}$ using the SFD method to converge towards the steady state for Case ($Re_h,\eta$)=($600,1$). } 
\label{fig:sfd}
\end{figure}
%case: laminar_BL_Diaz_Re600_finer_iny_sfd, 432 procs

Linear stability analysis requires a stationary base flow. The time-invariant state can be either the equilibrium, or the time-averaged (mean) flow. For flows at moderate Reynolds numbers, the equilibrium state can be obtained using the selective frequency damping (SFD) method \citep{aakervik2006steady} or the BoostConv algorithm \citep{citro2017efficient}. For turbulent flows at higher Reynolds numbers, the equilibrium state is difficult to obtain; instead, the time-averaged mean flow can be used as the base state for stability analysis to seek meaningful physical interpretation \citep{turton2015prediction, tammisola2016coherent}. \cite{barkley2006linear} shows that linear stability analysis on cylinder wake flow using the mean flow as the base state is able to track the Strouhal number of vortex shedding, but yields a marginally stable state with a small growth rate. In the present work, we use SFD to compute the base flow, compare this base flow to the time-averaged mean flow, and compare their global stability results in \S \ref{sec:results}.

SFD introduced by \cite{aakervik2006steady} is a useful technique to artificially settle the flow towards a steady equilibrium. The main idea is to apply a temporal low-pass filter to damp the oscillations due to the unsteady part of the solutions, and is achieved by introducing a linear forcing term on the right-hand side of the Navier-Stokes equations. An encapsulated formulation of the SFD method developed by \cite{jordi2014encapsulated} is used in the present work. The problem is considered to have converged when $||q-\overline{q}||_{inf} \le 10^{-8}$ according to \cite{jordi2014encapsulated}, where $\overline{q}$ is the filtered state. When using SFD, the control coefficient $\chi$ and the filter width $\Delta$ play important roles in the convergence process. The control coefficient $\chi$ should be positive and larger than the growth rate of the desired mode, while the filter cut-off frequency $\omega_c=1/\Delta$ must be lower than all of the flow instabilities to ensure the unstable disturbances are well within the damped region. For example, $\chi=0.5$ and $\Delta=2$ are used for the unstable case ($Re_h,\eta$)=($600,1$), and the convergence history is shown in figure \ref{fig:sfd}.

\subsection{Grid convergence and domain length sensitivity}\label{grid}
%Re600 with SFD Resolution: 1080*120*240, 1080*240*240, 1800*240*240, 1080*480*240.
%fig 1: Show the convergence of u vs. y profiles;
%fig 2: show the leading eigenvalue convergence

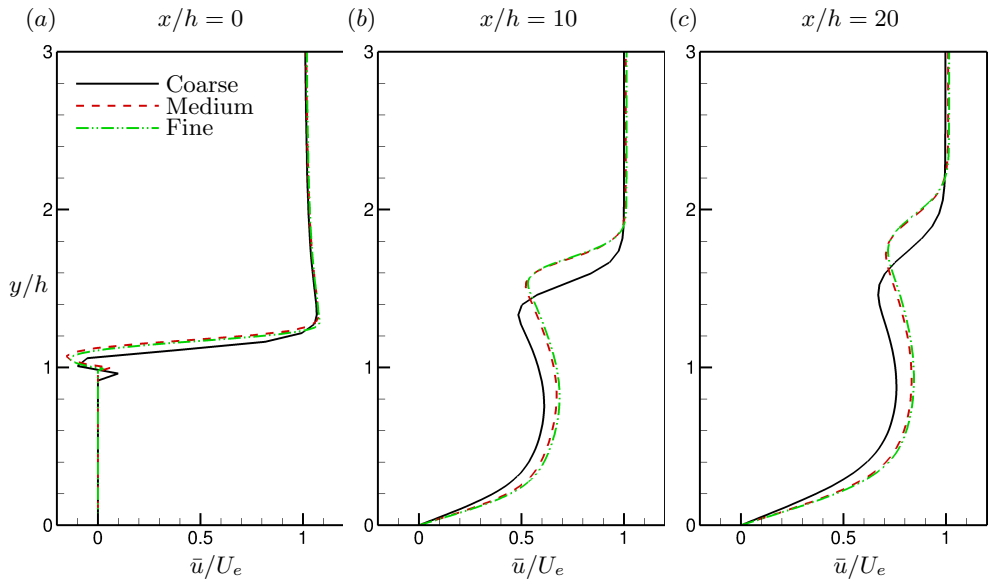
\begin{figure}
 \includegraphics[height=80mm,trim={0.5cm 0.2cm 1.5cm 0cm},clip]{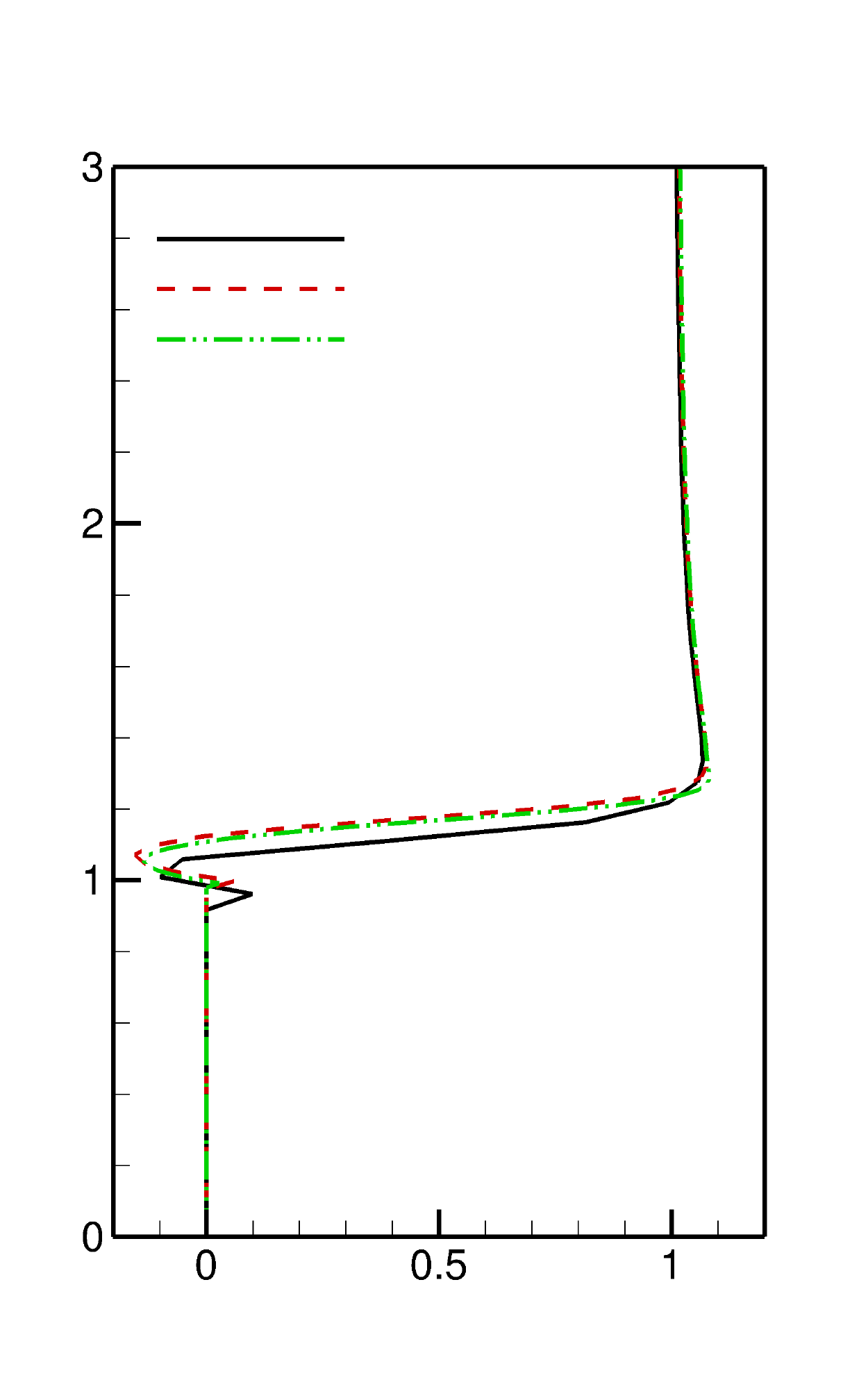}
\put(-120,210){$(a)$}
\put(-70,210){$x/h=0$}
\put(-126,110){$y/h$}
\put(-58,5){$\Bar{u}/U_e$}
\put(-67,185){Coarse}
\put(-67,177){Medium}
\put(-67,169){Fine}
 \includegraphics[height=80mm,trim={0.5cm 0.2cm 1.5cm 0cm},clip]{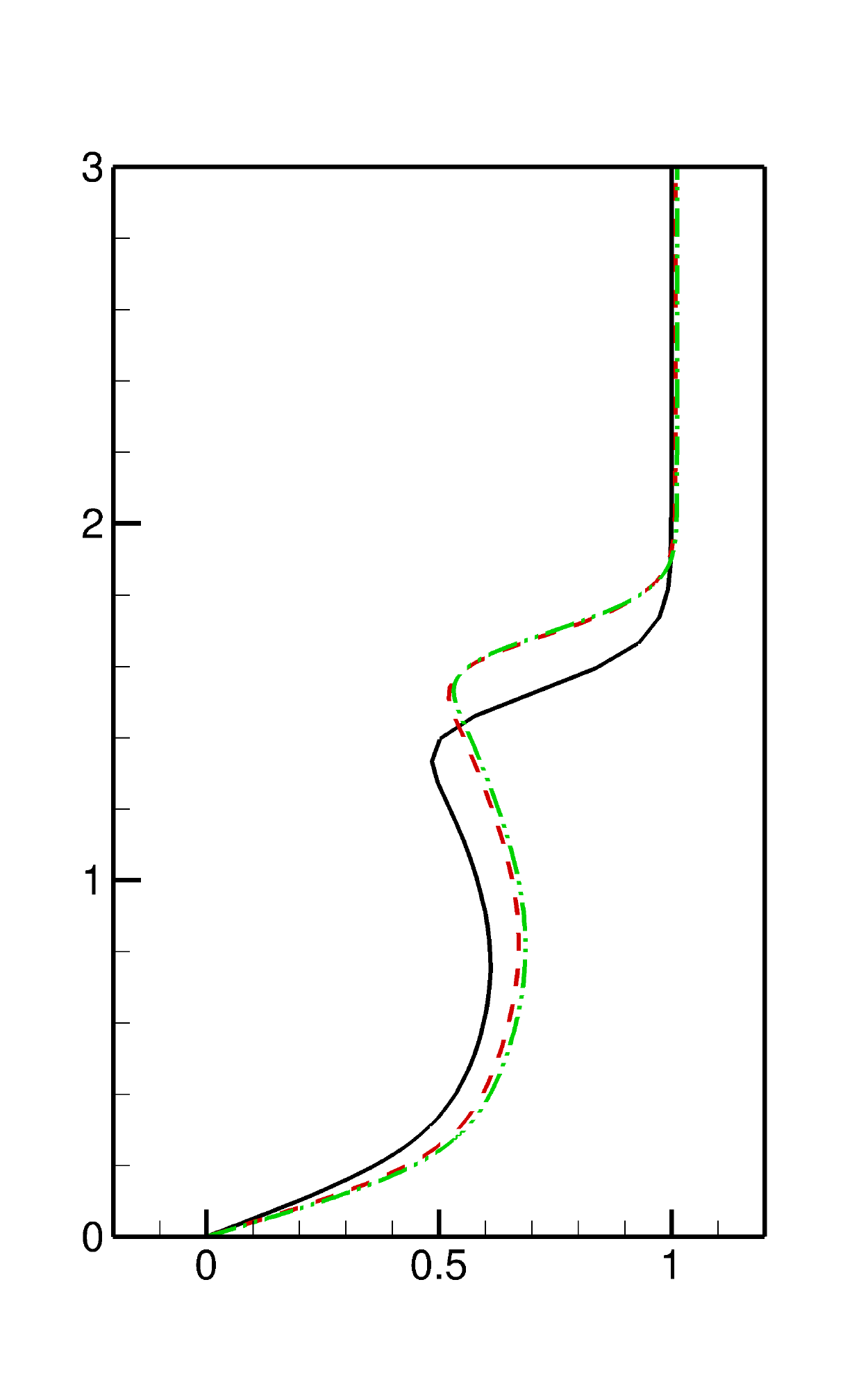}
\put(-120,210){$(b)$}
\put(-70,210){$x/h=10$}
\put(-58,5){$\Bar{u}/U_e$}
 \includegraphics[height=80mm,trim={0.5cm 0.2cm 1.5cm 0cm},clip]{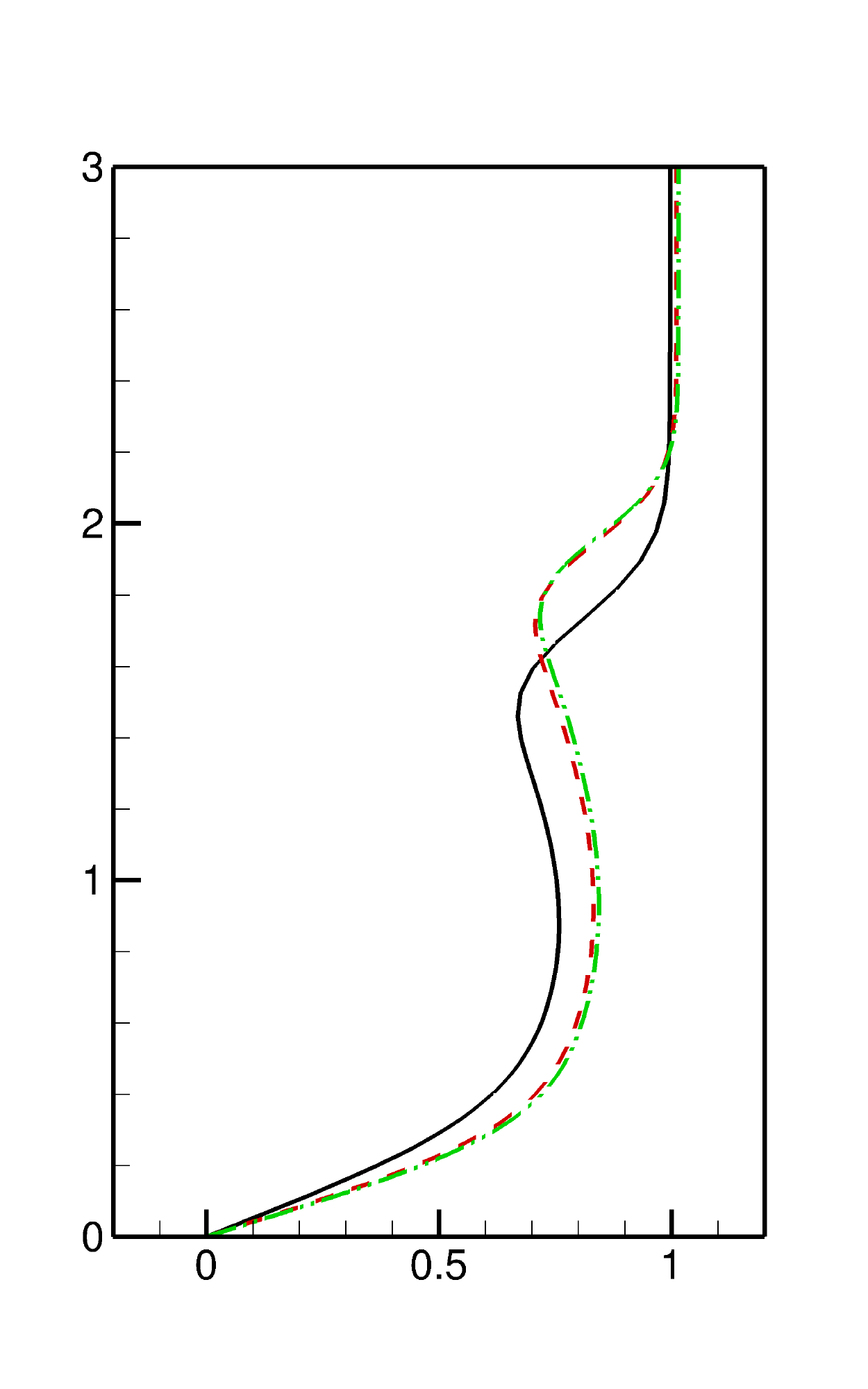}
\put(-120,210){$(c)$}
\put(-70,210){$x/h=20$}
\put(-58,5){$\Bar{u}/U_e$}
\caption{Base flow results from grid convergence study. Streamwise velocity profiles of the base flow obtained from SFD with $y$ at three streamwise stations: $(a)$ x/h=0, $(b)$ x/h=10 and $(c)$ x/h=20. } 
\label{fig:uy_profile}
\end{figure}
%laminar_BL_Diaz_Re600_double_iny_sfd_imp_correct_BC

Global stability results show strong sensitivity to grid sizes and domain lengths, highlighted by \cite{loiseau2014investigation} for roughness wake flow  and \cite{peplinski2015global} for a jet in crossflow.  A study on grid convergence and domain length sensitivity is thus performed in this section. 
Three different grids are used in the grid convergence study which are referred to as `coarse', `medium' and `fine'. Simulation details are listed in table \ref{tab:convergence}. For all cases presented in table \ref{tab:convergence}, uniform grids are used in both streamwise and spanwise directions while non-uniform grids are used in the wall-normal direction. Compared to the coarse grid, the medium grid is refined in the wall-normal direction. In the finest grid, the grid spacing in all three directions is reduced. Table \ref{tab:convergence} presents $\Delta y$ spacing at the wall (denoted by $\Delta y_{wall}$) and $\Delta y$ spacing at the roughness height location (denoted by $\Delta y_{top}$). Note that the roughness element is resolved by 43, 86 and 172 grid points in the wall-normal direction for the coarse, medium and fine cases respectively.

The streamwise velocity profiles of the base flow are examined at three different stations in figure \ref{fig:uy_profile}. The results show significant deviation of the solution for the coarse grid, while the differences between the medium and fine grids are small, indicating grid convergence. The leading eigenvalues obtained from the global stability analysis also show convergence in table \ref{tab:convergence}, suggesting that the medium grid is adequate for global stability analyses on the present case. 

\begin{table}
\begin{center}
\def~{\hphantom{0}}
    \begin{tabular}{lccccccc}
    Case &  $Re_h$ & $N_x \times N_y \times N_z$ & $L_x \times L_y \times L_z$ & $\Delta x, \Delta z$ & $\Delta y_{wall}$ & $\Delta y_{top}$ & $\sigma \pm i\omega$\\
    Coarse & 600 & $1080 \times 120 \times 240$ & $45h \times 15h \times 10h$ & $0.0417h$ & $0.0068h$ & $0.048h$ & $0.1472 \pm i1.1068$ \\
    Medium & 600 & $1080 \times 240 \times 240$ & $45h \times 15h \times 10h$ & $0.0417h$ & $0.0034h$ & $0.024h$ & $0.1110 \pm i1.1213$  \\
    Fine   & 600 & $1512 \times 480 \times 336$ & $45h \times 15h \times 10h$ & $0.0298h$ & $0.0017h$ & $0.012h$ & $0.1107 \pm i1.1213$\\
    $L_{x}75$   & 600 & $1800 \times 240 \times 240$ & $75h \times 15h \times 10h$ & $0.0417h$ & $0.0034h$ & $0.024h$ & $0.1110 \pm i1.1213$
    \end{tabular}
    \caption{\label{tab:convergence} Simulation parameters for grid convergence and domain length sensitivity study, and comparison of the direct leading eigenvalue for Case ($Re_h,\eta$)=($600,1$).}
    \end{center}
\end{table}

%Domain length: 
%table: show the leading eigenvalue convergence, and 
%fig 3: the eigenmode for Lx=45 and Lx=75.

\begin{figure}
\includegraphics[width=93mm,trim={0.5cm 2cm 0.5cm 4cm},clip]{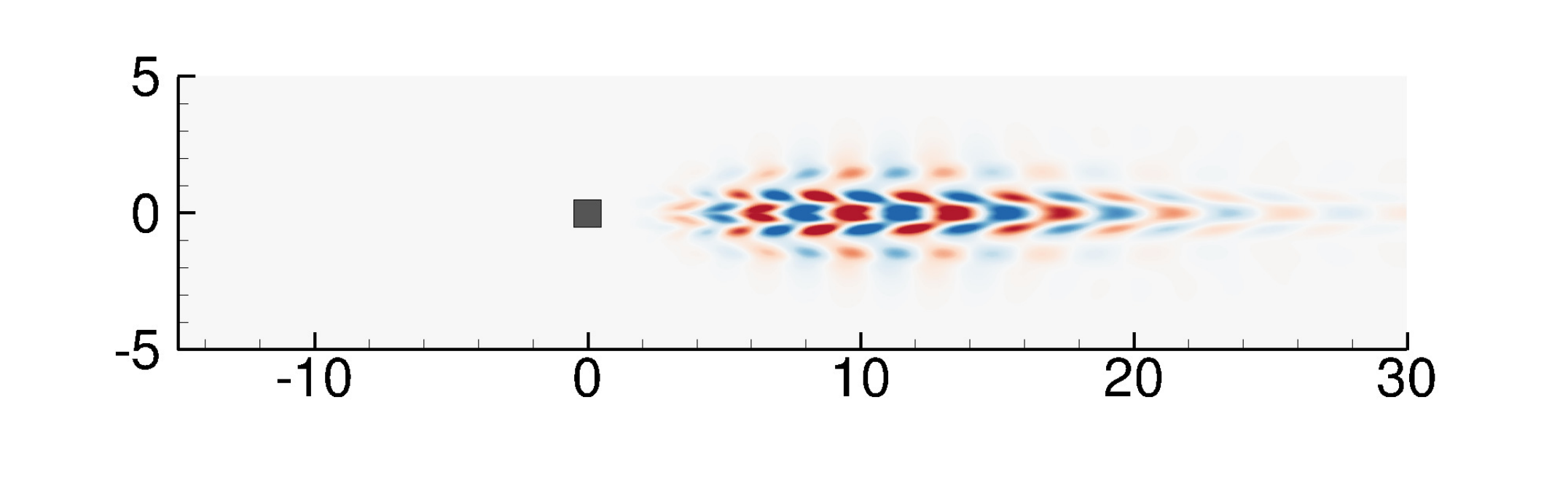}
\put(-255,40){\rotatebox{90}{$z/h$}}
\put(-130,3){$x/h$}
\put(-257,80){$(a)$}
 \hspace{3mm}
\includegraphics[width=150mm,trim={2cm 0.2cm 0.5cm 2cm},clip]{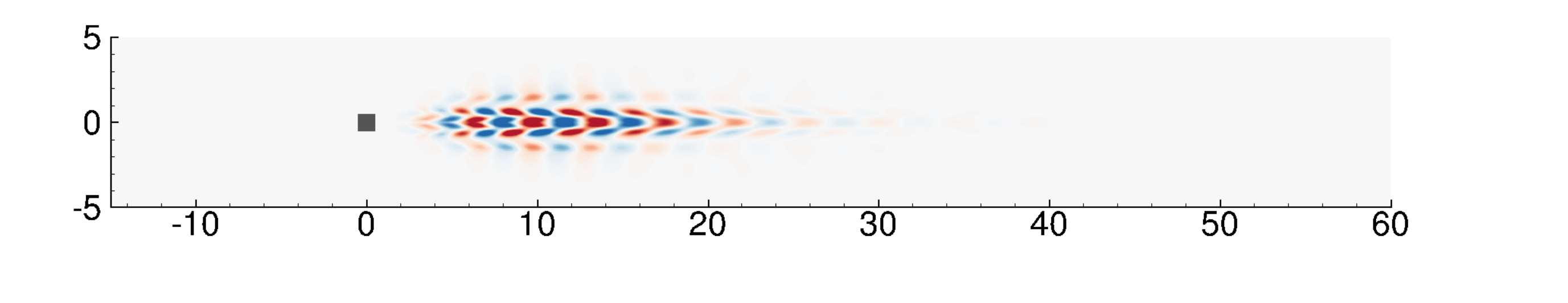}
\put(-416,42){\rotatebox{90}{$z/h$}}
\put(-215,8){$x/h$}
\put(-420,80){$(b)$}
 \caption{Contour plots of the streamwise velocity component of the leading unstable global mode at slice $y=0.5h$ for Case ($Re_h,\eta$)=($600,1$): $(a)$ short domain $L_x=45h$ (Case Medium) and $(b)$ long domain $L_x=75h$ (Case $L_{x}75$).  The contour levels depict $\pm 10 \%$ of the mode's maximum streamwise velocity. } 
\label{fig:mode_Lx75}
\end{figure}
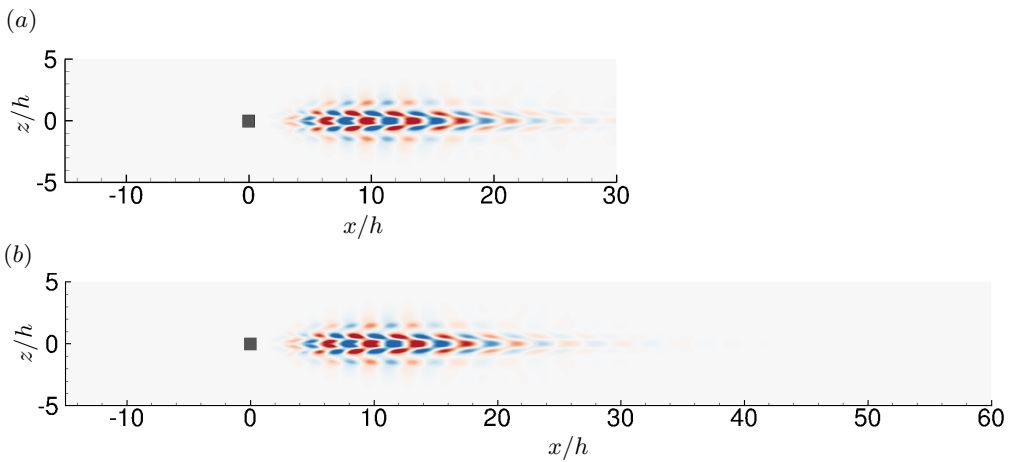

The influence of streamwise domain length on the global stability results is examined in the simulation with $L_x=75h$ (denoted by Case $L_x75$). Simulation details are listed in table \ref{tab:convergence}. Note that the grid sizes in Case $L_{x}75$ are comparable to the medium grid, already proven to sufficiently resolve the flow. The leading eigenvalue shows good agreement with that of Case Medium in table \ref{tab:convergence}, suggesting that the streamwise domain length $L_x=45h$ is adequate for the present case. The leading eigenmodes in Case Medium and Case $L_x75$ are also depicted in figure \ref{fig:mode_Lx75}. The results are identical between both cases. The global mode decays appreciably before reaching the outflow boundary, which guarantees convergence in the global stability results. Based on these conclusions, the medium grid and the domain length $L_x=45h$ are used for the cases presented in \S \ref{base} and \S \ref{GLSA}.

% laminar_BL_Diaz_Re600_medium_Lx75_correct_BC

\section{Results}\label{sec:results}
\subsection{Base flow}\label{base}
% Re600, short domain\\
% Baseflow visualization
% iso-contour ubar, cross sections at different x locations, contours of us (dubdy, dubdz)
% (compare to the mean flow, Re dependence)

\begin{figure}
\includegraphics[width=150mm,trim={2cm 0.2cm 0.5cm 0cm},clip]{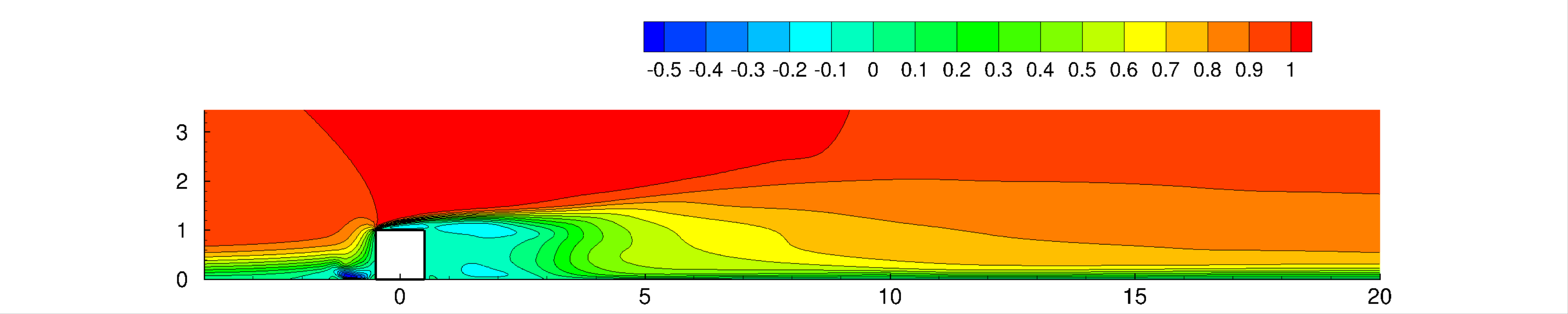}
\put(-415,25){\rotatebox{90}{$y/h$}}
\put(-220,-5){$x/h$}
 \put(-400,70){$(a)$}
 \put(-288,67){\scriptsize{$\overline{u}/U_e$}}
 \hspace{5mm}
\includegraphics[width=150mm,trim={2cm 0.2cm 0.5cm 0cm},clip]{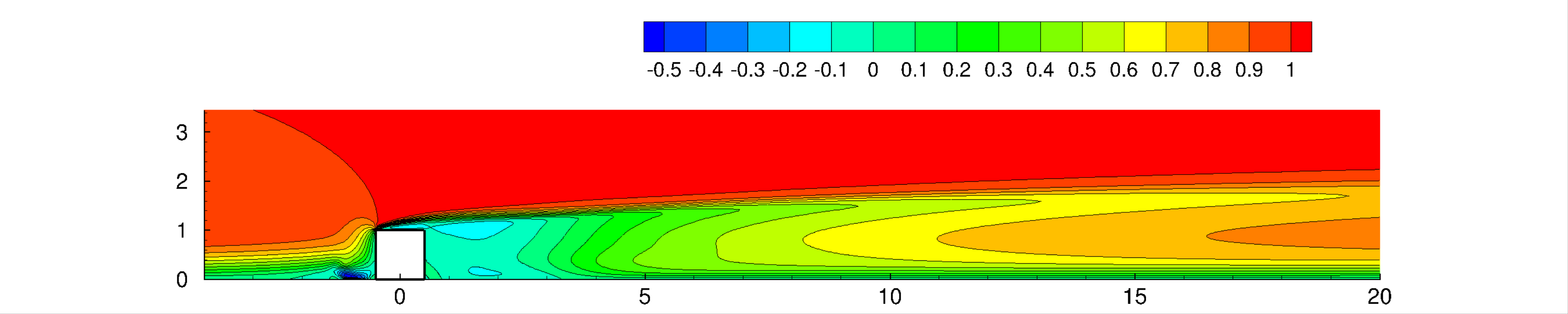}
 \put(-400,70){$(b)$}
\put(-415,25){\rotatebox{90}{$y/h$}}
\put(-220,-5){$x/h$}
\put(-288,67){\scriptsize{$\overline{u}/U_e$}}
\hspace{5mm}
\includegraphics[width=150mm,trim={2cm 0.2cm 0.5cm 0cm},clip]{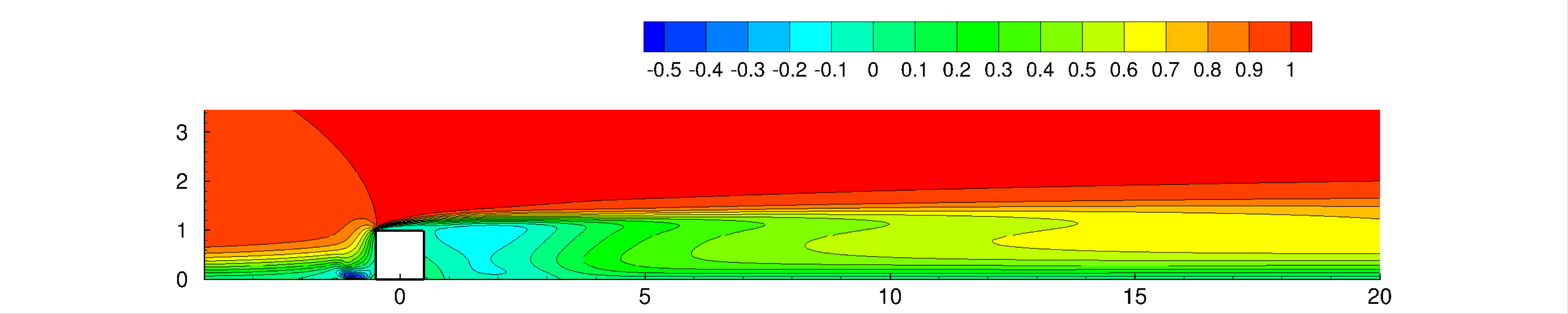}
 \put(-400,70){$(c)$}
\put(-415,25){\rotatebox{90}{$y/h$}}
\put(-220,-5){$x/h$}
\put(-288,67){\scriptsize{$\overline{u}/U_e$}}
\hspace{5mm}
\includegraphics[width=150mm,trim={2cm 0.2cm 0.5cm 0cm},clip]{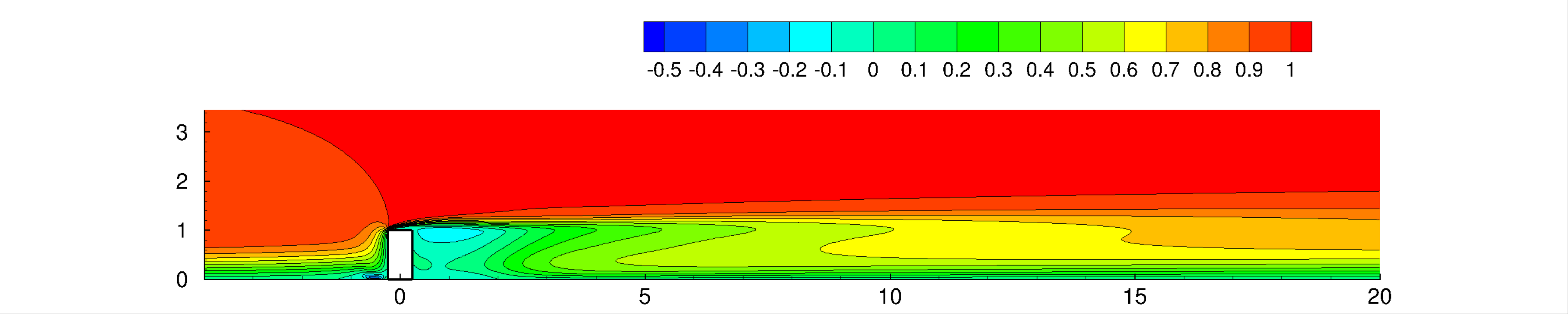}
 \put(-400,70){$(d)$}
\put(-415,25){\rotatebox{90}{$y/h$}}
\put(-220,-5){$x/h$}
\put(-288,67){\scriptsize{$\overline{u}/U_e$}}
\hspace{5mm}
\includegraphics[width=150mm,trim={2cm 0.2cm 0.5cm 0cm},clip]{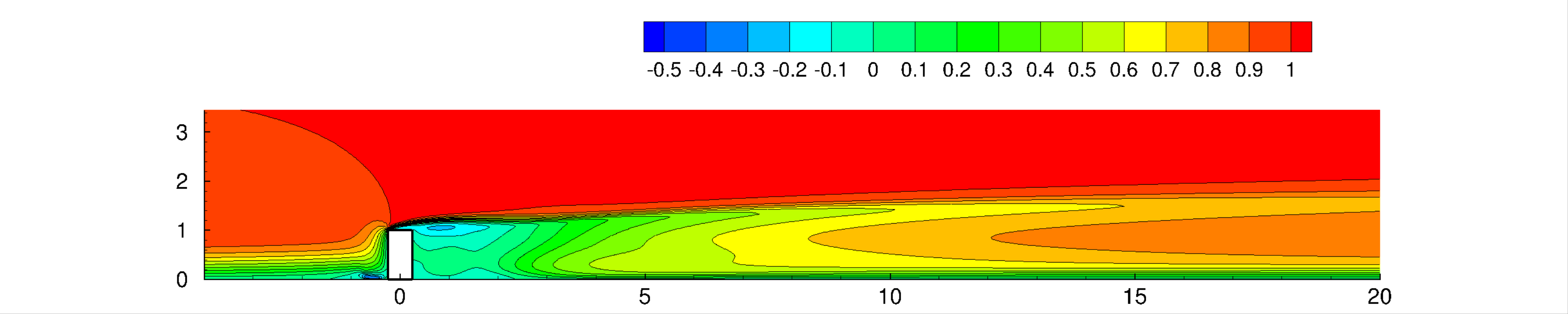}
 \put(-400,70){$(e)$}
\put(-415,25){\rotatebox{90}{$y/h$}}
\put(-220,-5){$x/h$}
\put(-288,67){\scriptsize{$\overline{u}/U_e$}}
 \caption{Contour plots at the spanwise mid-plane of $(a)$ time-averaged streamwise velocity field for Case ($Re_h,\eta$)=($600,1$), streamwise velocity field of the base flow obtained from SFD for $(b)$ Case ($Re_h,\eta$)=($600,1$), $(c)$ Case ($Re_h,\eta$)=($475,1$), $(d)$ Case ($Re_h,\eta$)=($600,0.5$) and $(e)$ Case ($Re_h,\eta$)=($800,0.5$). } 
\label{fig:contour_zslice}
\end{figure}
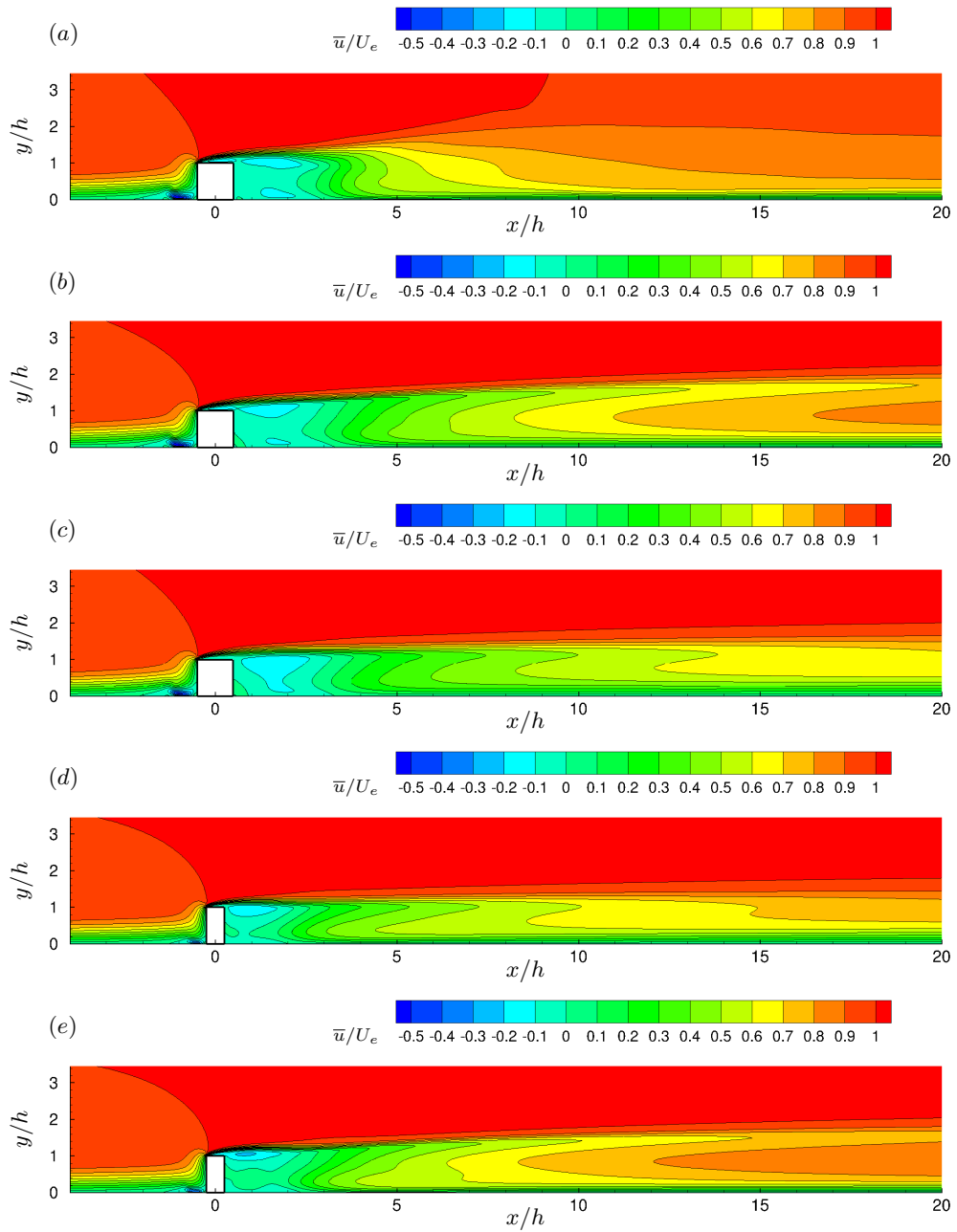
%laminar_BL_eta05_Re800_sfd

%The base flow is of great importance for global linear stability analysis. In order to better understand the global stability results, t
The differences between the base flow (obtained from the SFD method) and the time-averaged mean flow (obtained from DNS) are first investigated. %\cite{baker1979laminar} suggested that the topography and stability of horseshoe vortex system upstream of the cylinder are dependent on the Reynolds number and the ratio $d/\delta^*$. The horseshoe vortex system upstream the cube for Case ($Re_h,h/\delta^*,\eta$)=($600,2.86,1$) is stable, consistent with the observation of \cite{daniel2017direct}. 
Figures \ref{fig:contour_zslice}$(a)$ and \ref{fig:contour_zslice}$(b)$ show that the streamwise velocity field upstream of the cube is identical between the base flow and the mean flow, but the flow field downstream of the cube shows some differences. Although the reversed flow regions behind the cube are similar, with increasing downstream distance, the base flow demonstrates a stronger wall-normal gradient corresponding to the shear layer generated from the top edge of the cube. In contrast, this strong wall-normal shear is not prominent in the mean flow.

% \begin{figure}
% \includegraphics[width=60mm,trim={0.5cm 0.2cm 0.5cm 0cm},clip]{images/u_slice_z0_eta1_re600.pdf}
% \put(-167,35){\rotatebox{90}{$y$}}
% \put(-85,-5){$z$}
%  \put(-175,80){$(a)$}
%  %\put(-155,80){$x=0$}
%  %\put(-95,80){SFD}
% \includegraphics[width=60mm,trim={0.5cm 0.2cm 0.5cm 0cm},clip]{images/u_slice_mean_z0.pdf}
% \put(-167,35){\rotatebox{90}{$y$}}
% \put(-85,-5){$z$}
% %\put(-95,80){DNS}
%  %\put(-175,80){$(b)$}
%  \caption{Comparison of the base flow obtained from the SFD method on the left versus the time-averaged flow from DNS on the right for Case ($Re_h,\delta^*/h,\eta$)=($600,0.35,1$). } 
% \label{fig:contour_xslice}
% \end{figure}

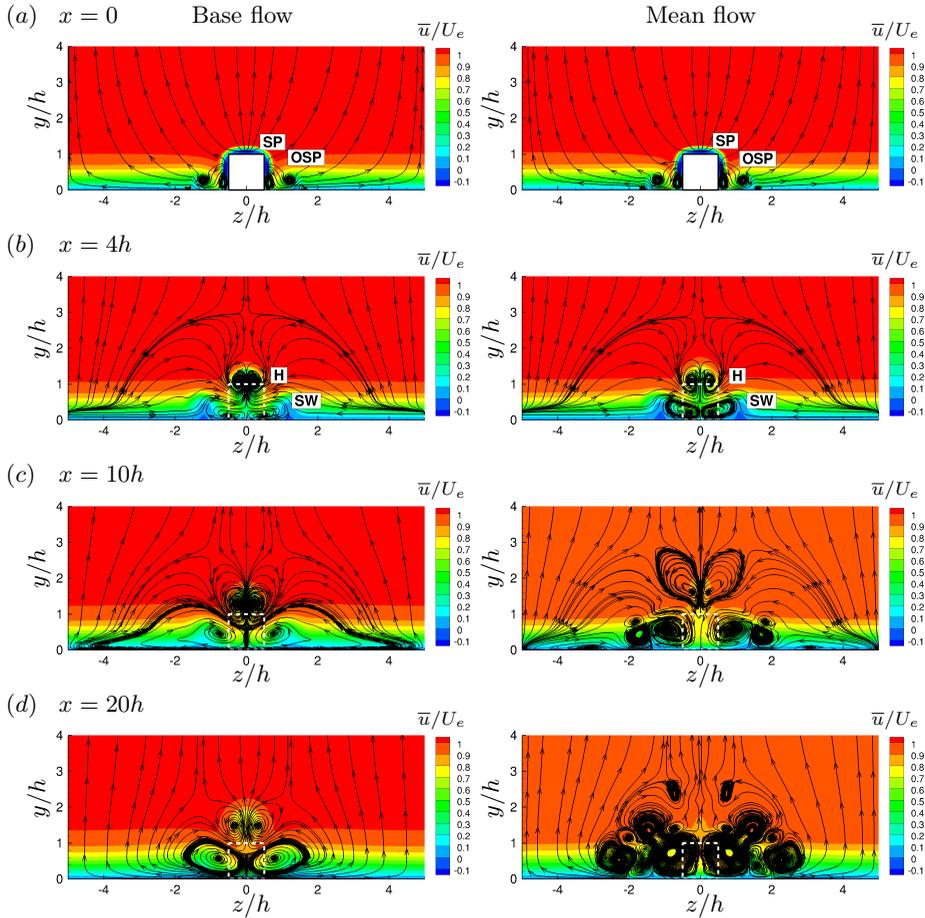
\begin{figure}
\includegraphics[width=60mm,trim={0.5cm 0.2cm 0.5cm 0cm},clip]{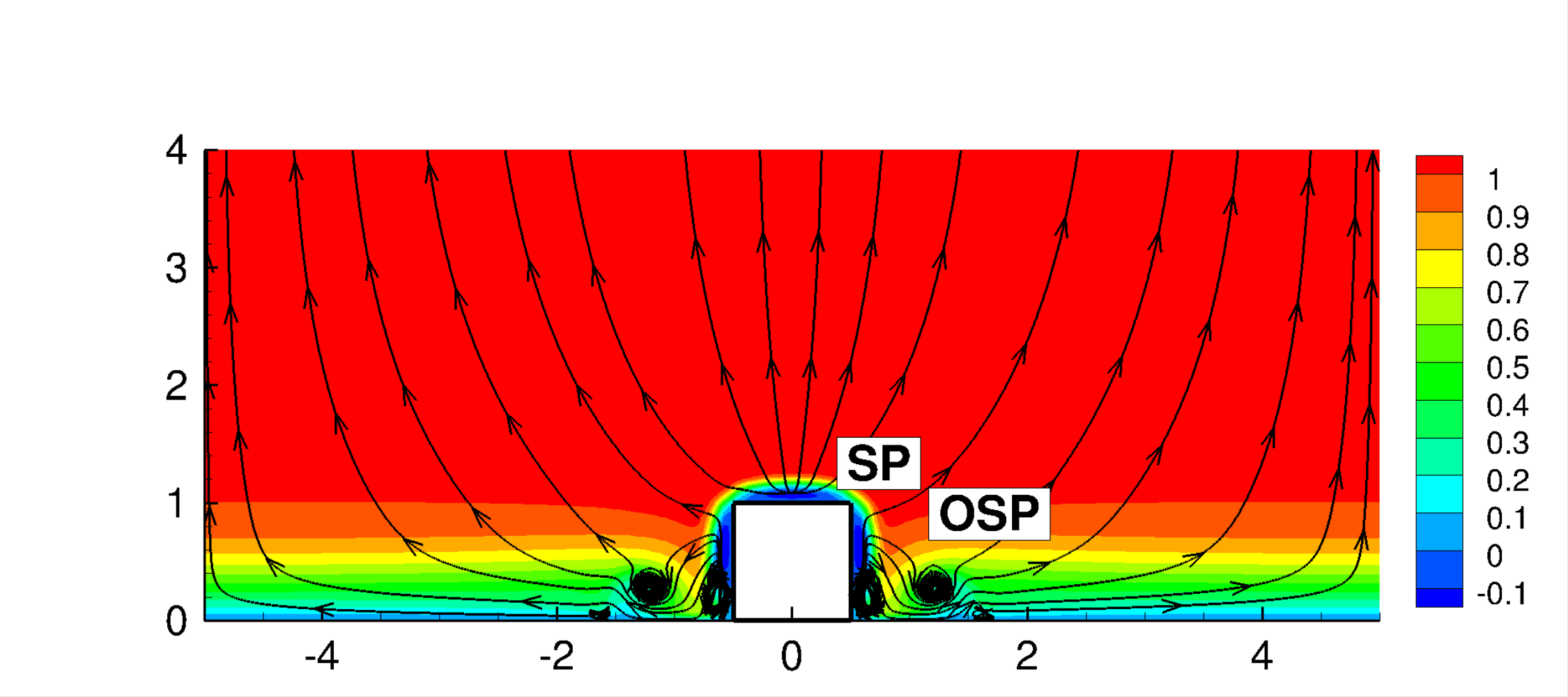}
\put(-167,30){\rotatebox{90}{$y/h$}}
\put(-90,-5){$z/h$}
 \put(-175,70){$(a)$}
 \put(-155,70){$x=0$}
 \put(-105,70){Base flow}
 \put(-20,65){\scriptsize{$\overline{u}/U_e$}}
\includegraphics[width=60mm,trim={0.5cm 0.2cm 0.5cm 0cm},clip]{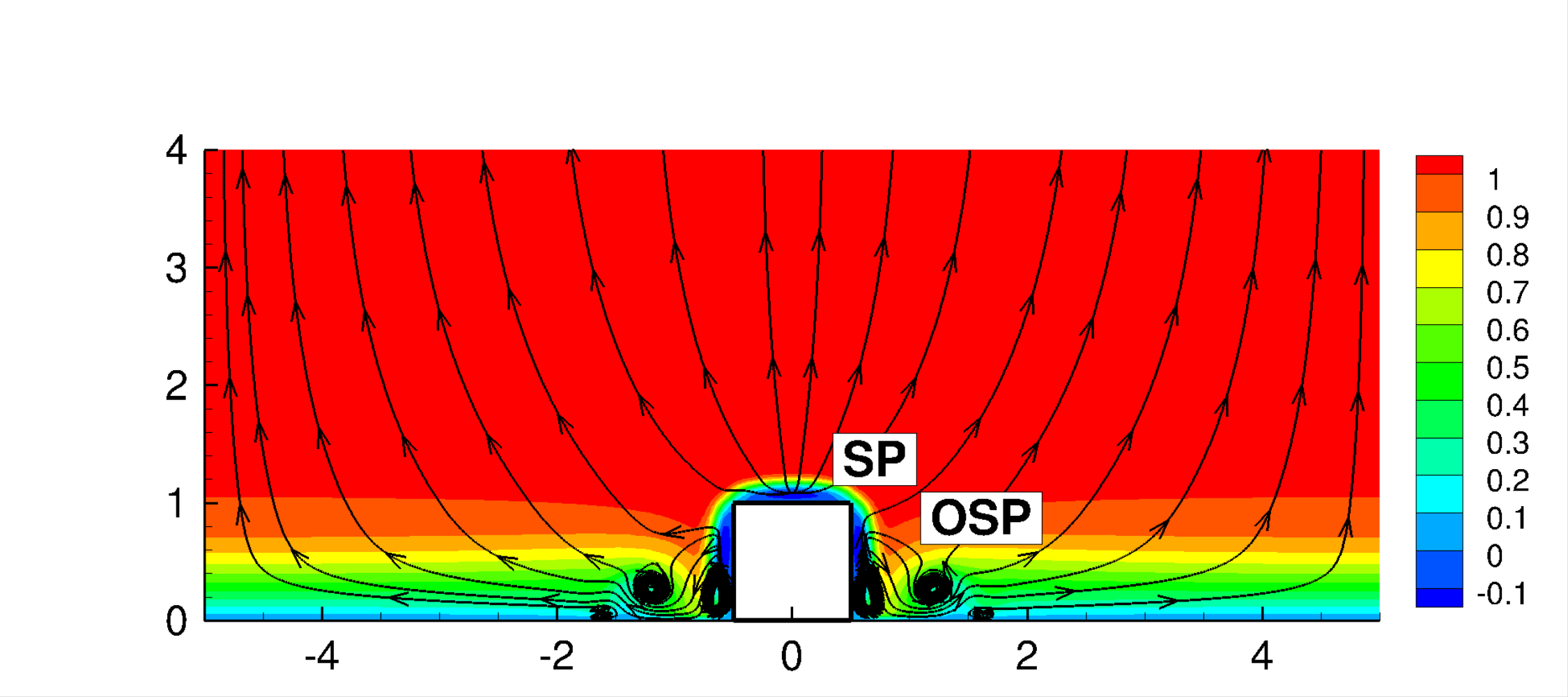}
\put(-167,30){\rotatebox{90}{$y/h$}}
\put(-90,-5){$z/h$}
\put(-105,70){Mean flow}
 \put(-20,65){\scriptsize{$\overline{u}/U_e$}}
 %\put(-175,80){$(b)$}
 \hspace{3mm}
 \includegraphics[width=60mm,trim={0.5cm 0.2cm 0.5cm 0cm},clip]{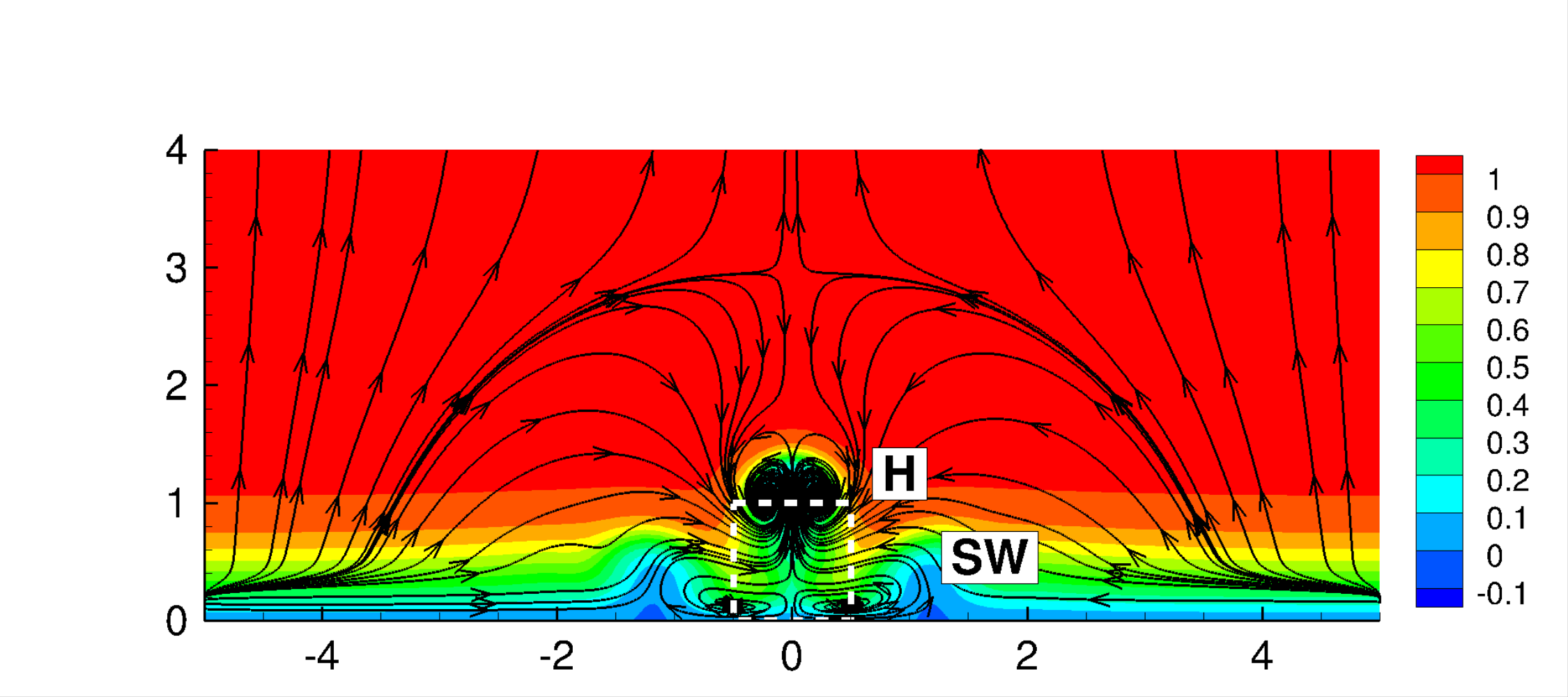}
\put(-167,30){\rotatebox{90}{$y/h$}}
\put(-90,-5){$z/h$}
 \put(-175,70){$(b)$}
 \put(-155,70){$x=4h$}
  \put(-20,65){\scriptsize{$\overline{u}/U_e$}}
\includegraphics[width=60mm,trim={0.5cm 0.2cm 0.5cm 0cm},clip]{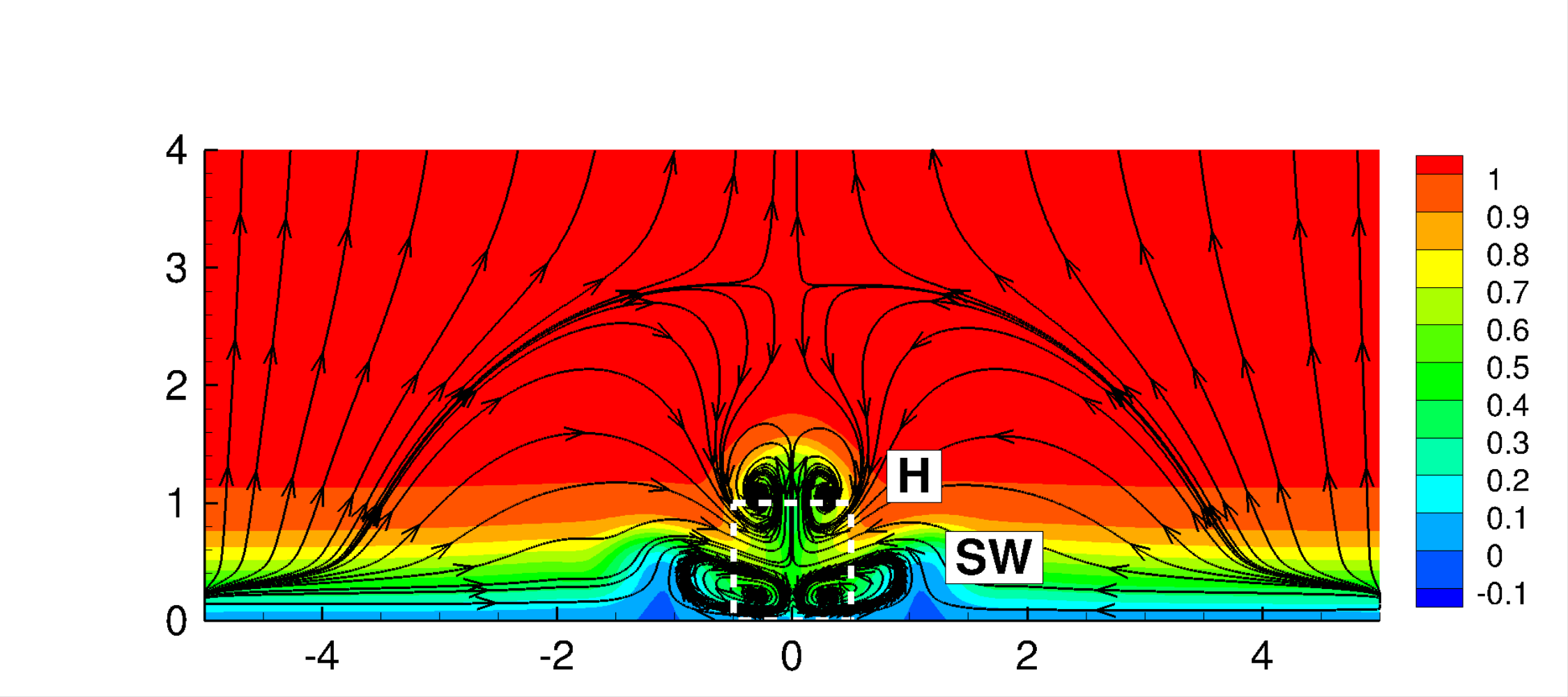}
\put(-167,30){\rotatebox{90}{$y/h$}}
\put(-90,-5){$z/h$}
% \put(-175,80){$(d)$}
 \put(-20,65){\scriptsize{$\overline{u}/U_e$}}
  \hspace{3mm}
 \includegraphics[width=60mm,trim={0.5cm 0.2cm 0.5cm 0cm},clip]{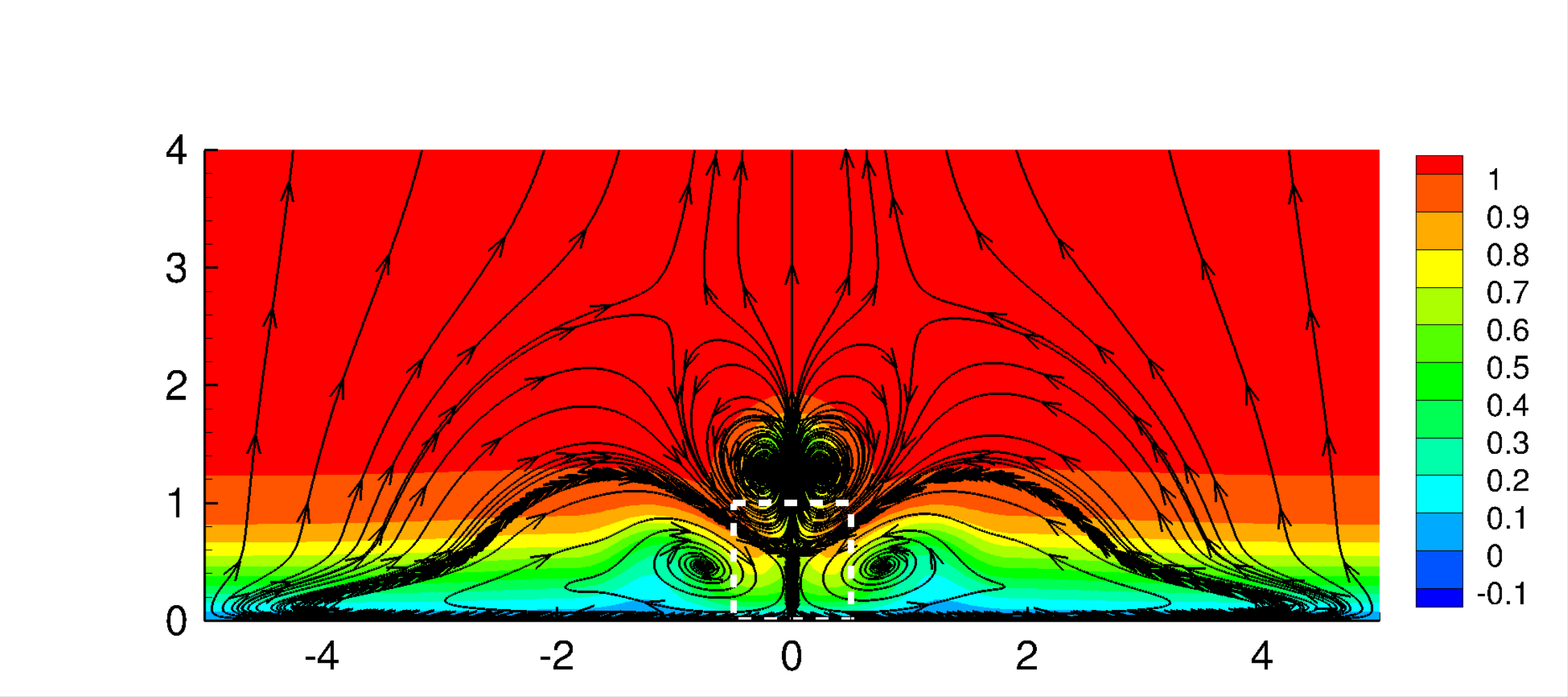}
\put(-167,30){\rotatebox{90}{$y/h$}}
\put(-90,-5){$z/h$}
 \put(-175,70){$(c)$}
 \put(-155,70){$x=10h$}
  \put(-20,65){\scriptsize{$\overline{u}/U_e$}}
\includegraphics[width=60mm,trim={0.5cm 0.2cm 0.5cm 0cm},clip]{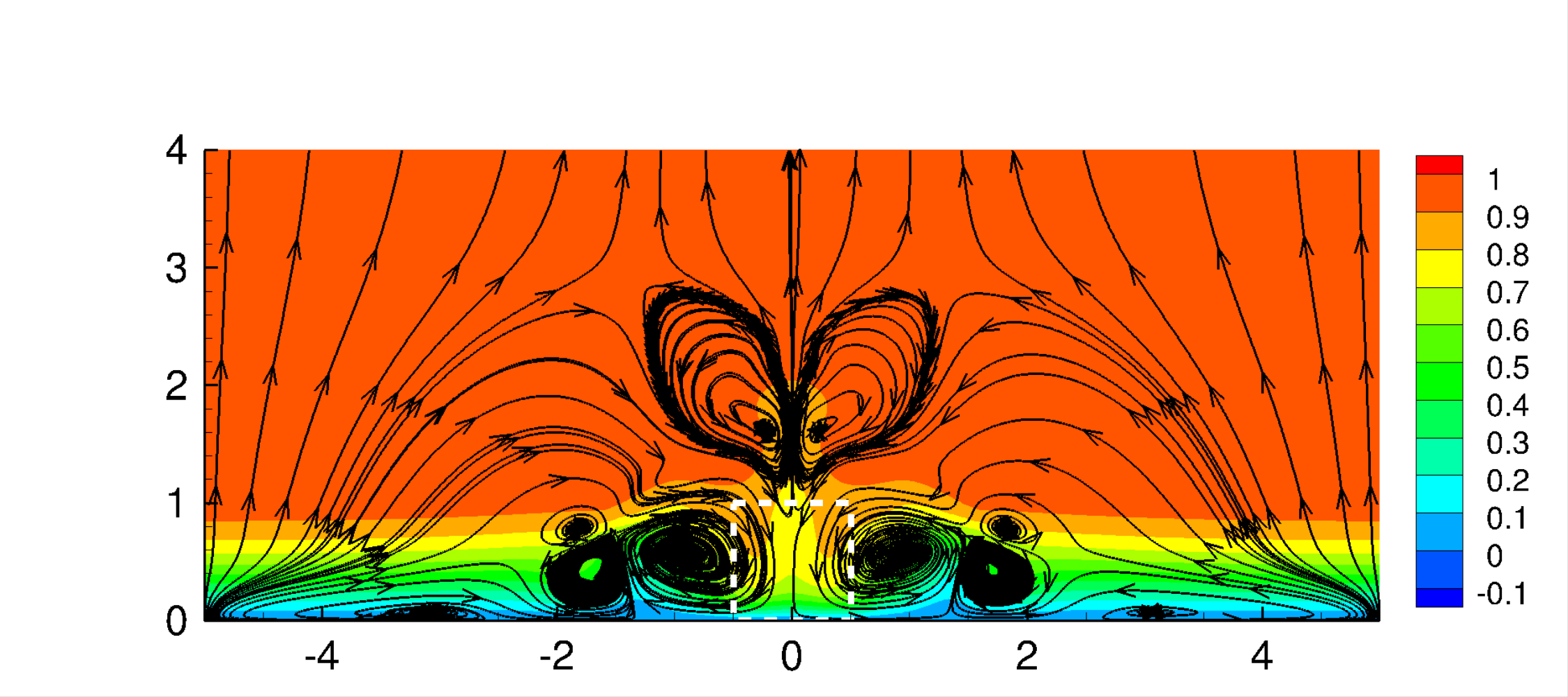}
\put(-167,30){\rotatebox{90}{$y/h$}}
\put(-90,-5){$z/h$}
 %\put(-175,80){$(f)$}
  \put(-20,65){\scriptsize{$\overline{u}/U_e$}}
  \hspace{3mm}
 \includegraphics[width=60mm,trim={0.5cm 0.2cm 0.5cm 0cm},clip]{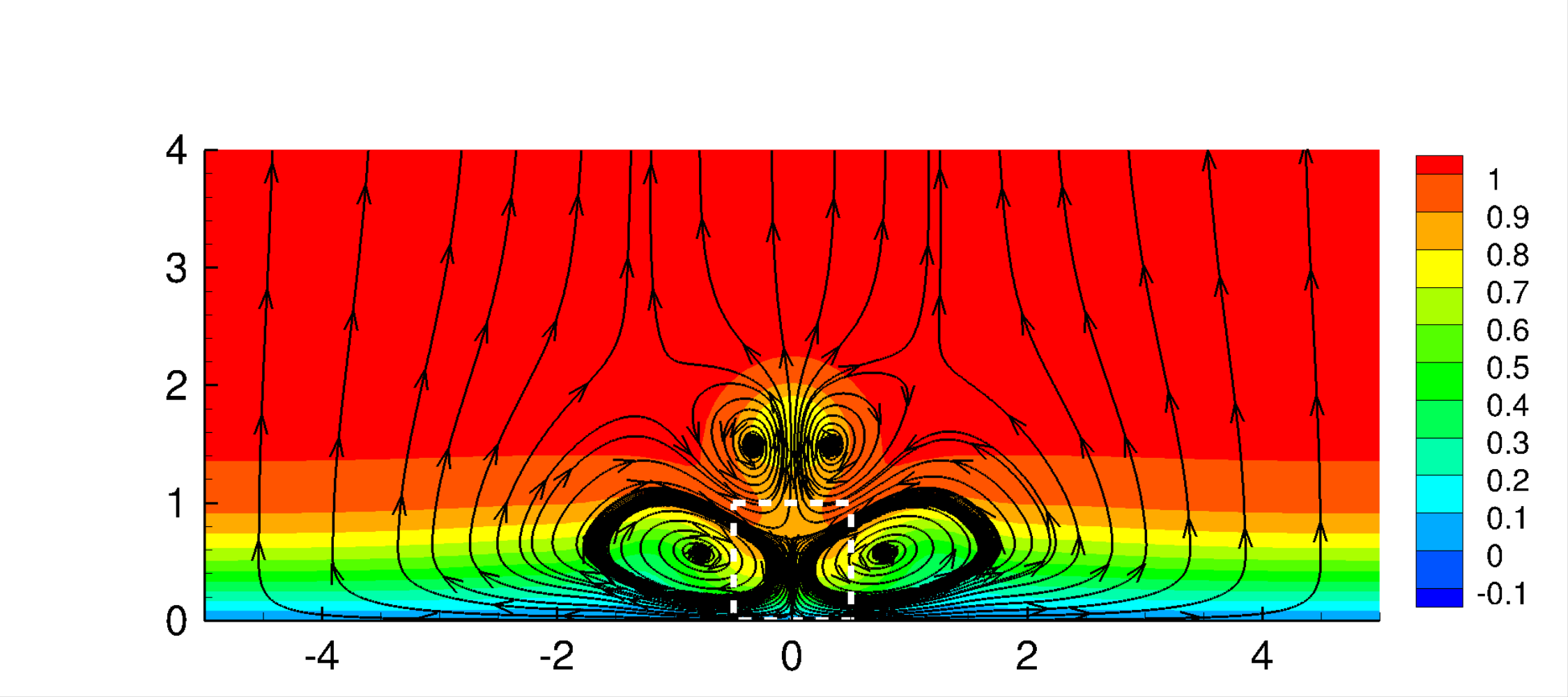}
\put(-167,30){\rotatebox{90}{$y/h$}}
\put(-90,-5){$z/h$}
 \put(-175,70){$(d)$}
 \put(-155,70){$x=20h$}
  \put(-20,65){\scriptsize{$\overline{u}/U_e$}}
\includegraphics[width=60mm,trim={0.5cm 0.2cm 0.5cm 0cm},clip]{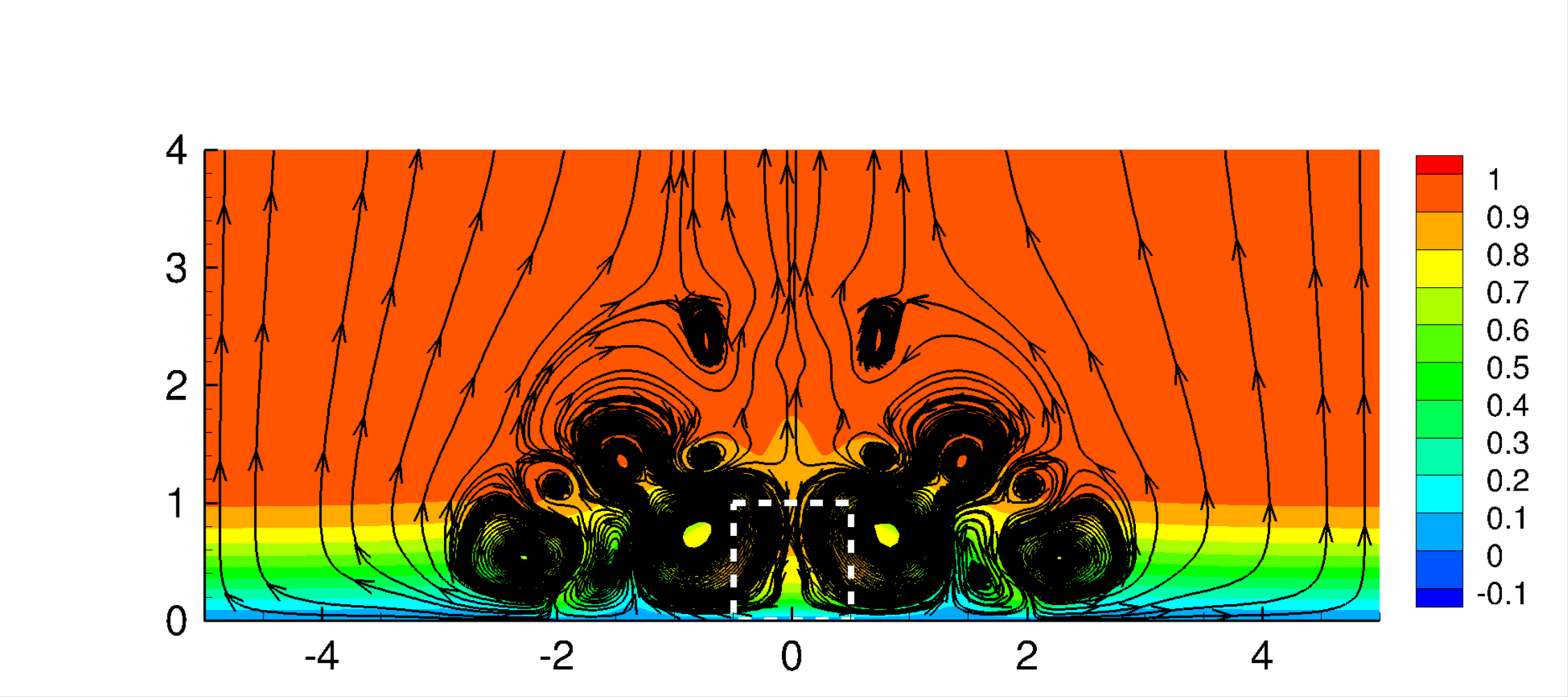}
\put(-167,30){\rotatebox{90}{$y/h$}}
\put(-90,-5){$z/h$}
% \put(-175,80){$(h)$}
 \put(-20,65){\scriptsize{$\overline{u}/U_e$}}
\caption{Comparison of the base flow obtained from the SFD method on the left versus the time-averaged flow from DNS on the right at different x locations: $(a)$ x=0, $(b)$ x=4h, $(c)$ x=10h and $(d)$ x=20h for Case ($Re_h,\eta$)=($600,1$), demonstrated by the streamlines of ($\overline{v}, \overline{w}$) with background contours of $\overline{u}$, for the base flow and the mean flow respectively. The roughness location is denoted by the dashed lines. }
\label{fig:contour_xslice}
\end{figure}
%laminar_BL_Diaz_Re600_double_iny_sfd_imp_correct_BC, laminar_BL_Diaz_Re600_DNS_correct_BC

Figure \ref{fig:contour_xslice} compares the base and time-averaged mean flows for Case ($Re_h,\eta$)=($600,1$), using streamlines and contours of streamwise velocity in different cross-flow planes. Qualitative agreement is seen between the base flow and the mean flow immediately downstream of the roughness element ($x \le 4h$). At $x=0$, as shown in figure \ref{fig:contour_xslice}$(a)$, two pairs of streamwise vortices are observed on the lateral sides of the cube in both base and mean flows. The pair very close to the cube is referred to as the symmetry plane vortices (SP) \citep{iyer2013high} or the rear pair vortices \citep{ye2016geometry, bucci2021influence}. They push low-momentum flow upwards, move closer to the symmetry plane, give rise to the low-speed region behind the roughness, and are dissipated farther downstream. %The pair beside the cube edges is the rear pair vortices which have also been observed in experiments of boundary layer flow over isolated roughness by \cite{ye2016geometry}. The rear pair vortices are generated by the fluid motion beside the roughness element and are dissipated farther downstream \citep{bucci2021influence}. 
The other counter-rotating vortex pair is formed away from the symmetry plane, which is referred to as the off-symmetry plane vortices (OSP) by \cite{iyer2013high}. They are the continuation of the vortex tubes from the horseshoe vortex system upstream. %is the horseshoe vortex pair wrapping around the roughness. 
At $x=4h$ (figure \ref{fig:contour_xslice}$(b)$), hairpin vortices (H) and secondary wall-attached vortices (SW) are observed in both the base and mean flows. As the streamwise location increases farther downstream (figures \ref{fig:contour_xslice}$(c)$ and \ref{fig:contour_xslice}$(d)$), the central low-speed region is weakened and secondary vortical structures are intensified in the mean flow, due to the enhanced nonlinear interactions. In the base flow however, only one pair of wall-attached vortices is observed below the primary hairpin vortices since the unsteady oscillations are damped out. The influence of the non-linear saturation in the mean flow results in differences in the global stability results between the base and mean flows, as further discussed in \S \ref{dynamic}. %the central low-speed region becomes weak in the mean flow. 

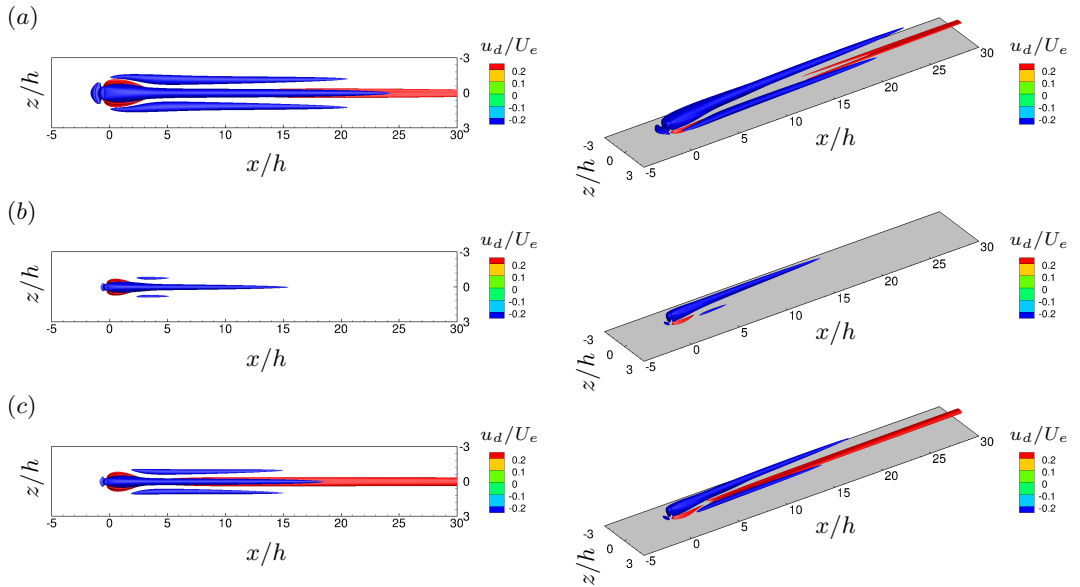
\begin{figure}
\includegraphics[width=70mm,trim={0.2cm 0.2cm 0.5cm 0cm},clip]{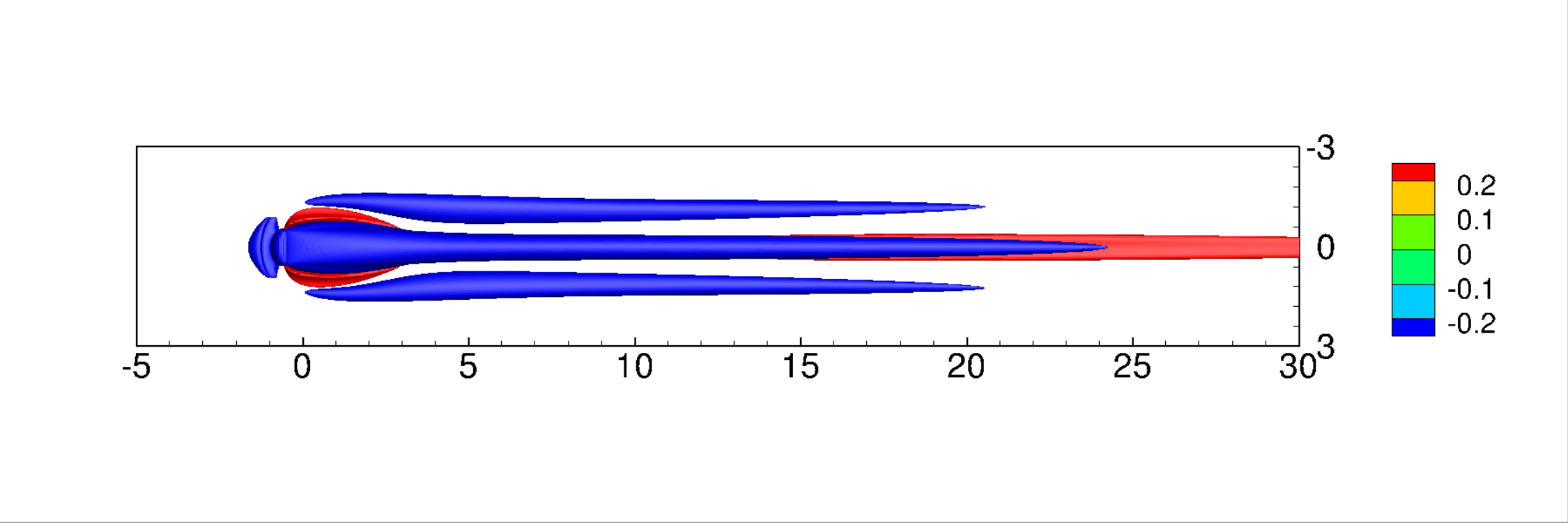}
 \put(-200,60){$(a)$}
\put(-195,28){\rotatebox{90}{$z/h$}}
\put(-110,5){$x/h$}
\put(-22,50){\scriptsize{$u_d/U_e$}}
\includegraphics[width=70mm,trim={0.2cm 0.2cm 0.5cm 0cm},clip]{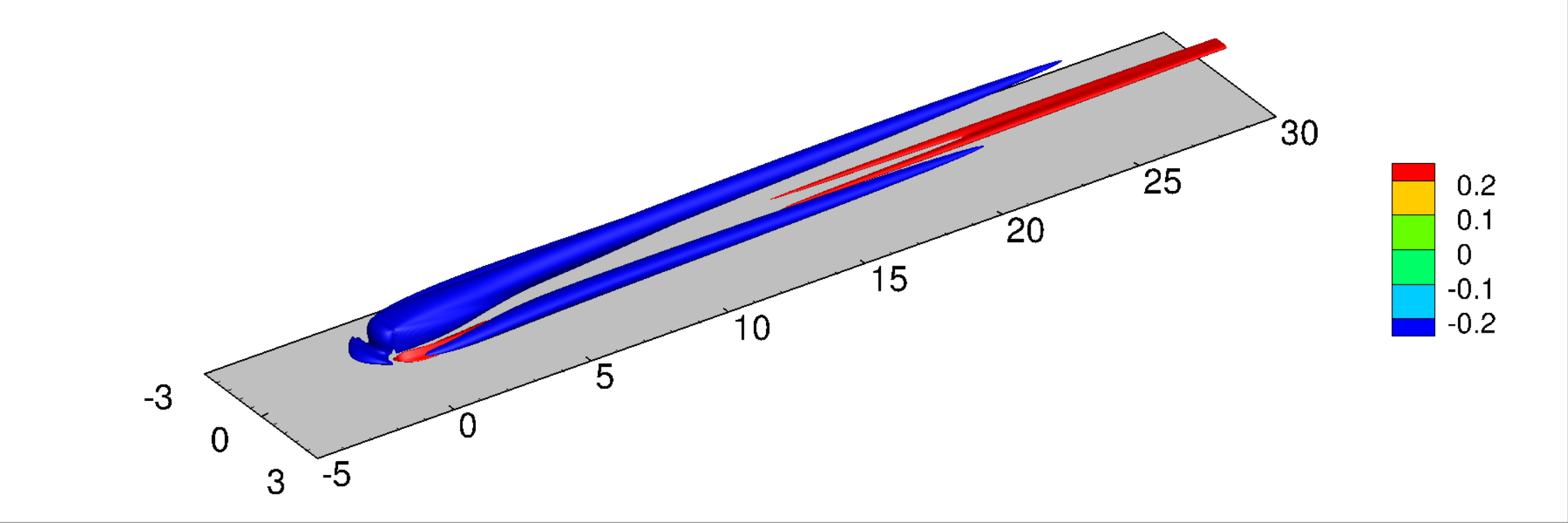}
\put(-185,-5){\rotatebox{90}{$z/h$}}
\put(-95,15){$x/h$}
\put(-22,50){\scriptsize{$u_d/U_e$}}
\hspace{3mm}
\includegraphics[width=70mm,trim={0.2cm 0.2cm 0.5cm 0cm},clip]{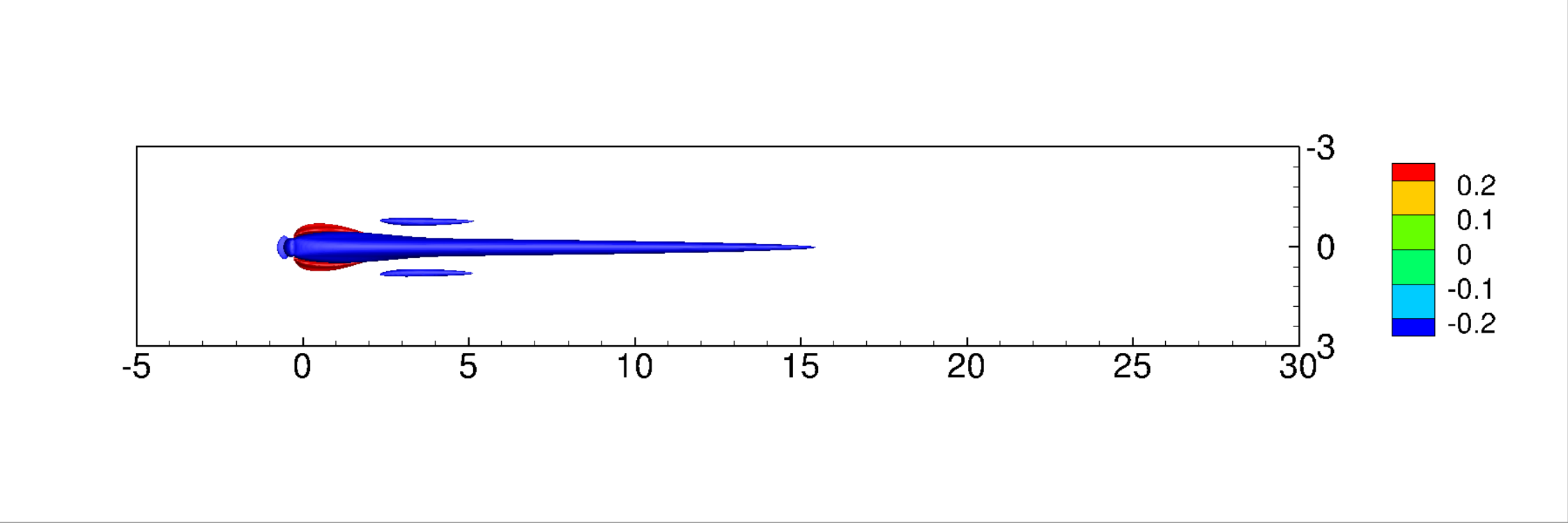}
 \put(-200,60){$(b)$}
\put(-195,28){\rotatebox{90}{$z/h$}}
\put(-110,5){$x/h$}
\put(-22,50){\scriptsize{$u_d/U_e$}}
\includegraphics[width=70mm,trim={0.2cm 0.2cm 0.5cm 0cm},clip]{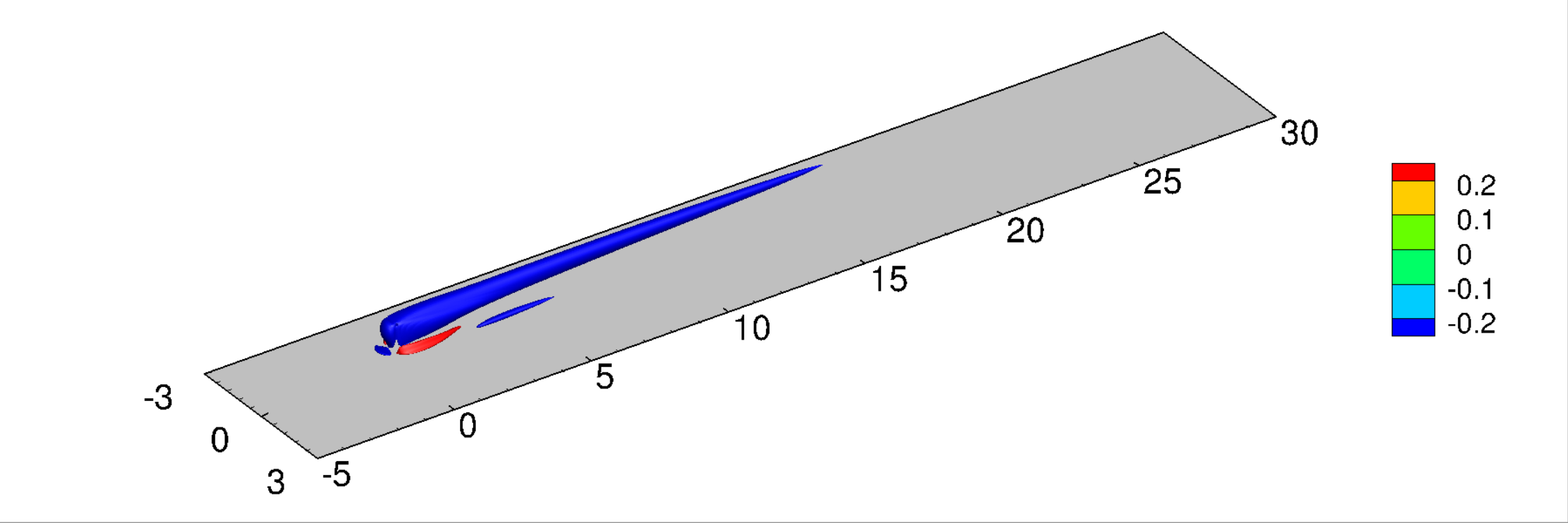}
\put(-185,-5){\rotatebox{90}{$z/h$}}
\put(-95,15){$x/h$}
\put(-22,50){\scriptsize{$u_d/U_e$}}
\hspace{3mm}
\includegraphics[width=70mm,trim={0.2cm 0.2cm 0.5cm 0cm},clip]{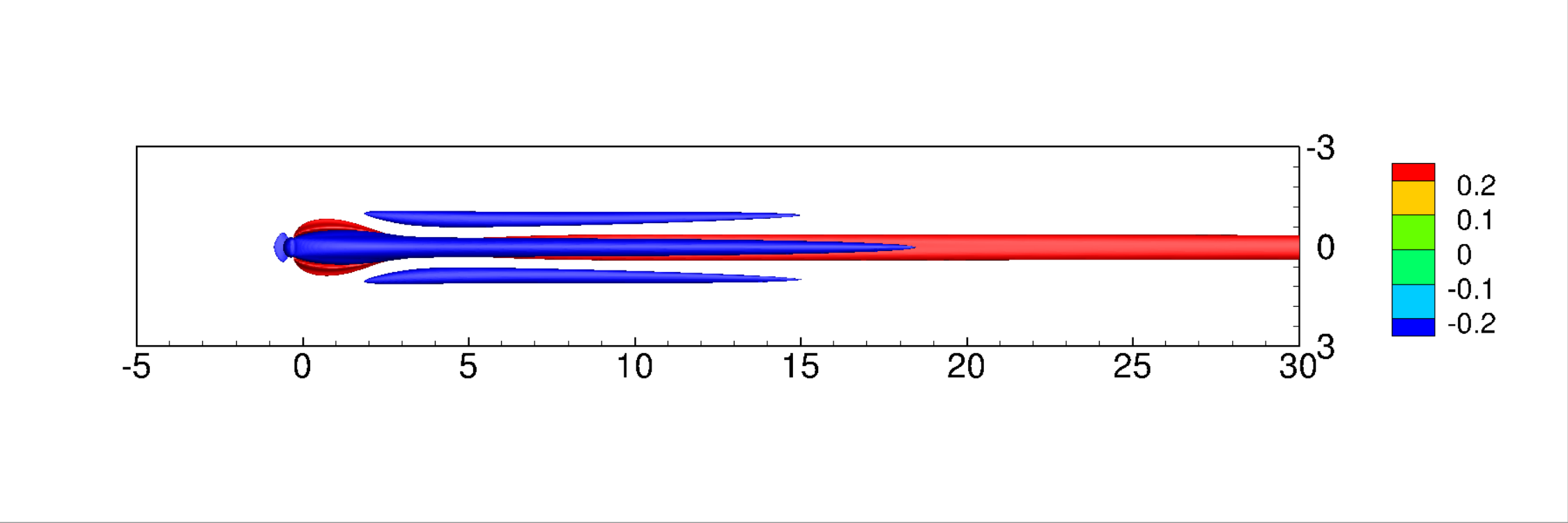}
 \put(-200,60){$(c)$}
\put(-195,28){\rotatebox{90}{$z/h$}}
\put(-110,5){$x/h$}
\put(-22,50){\scriptsize{$u_d/U_e$}}
\includegraphics[width=70mm,trim={0.2cm 0.2cm 0.5cm 0cm},clip]{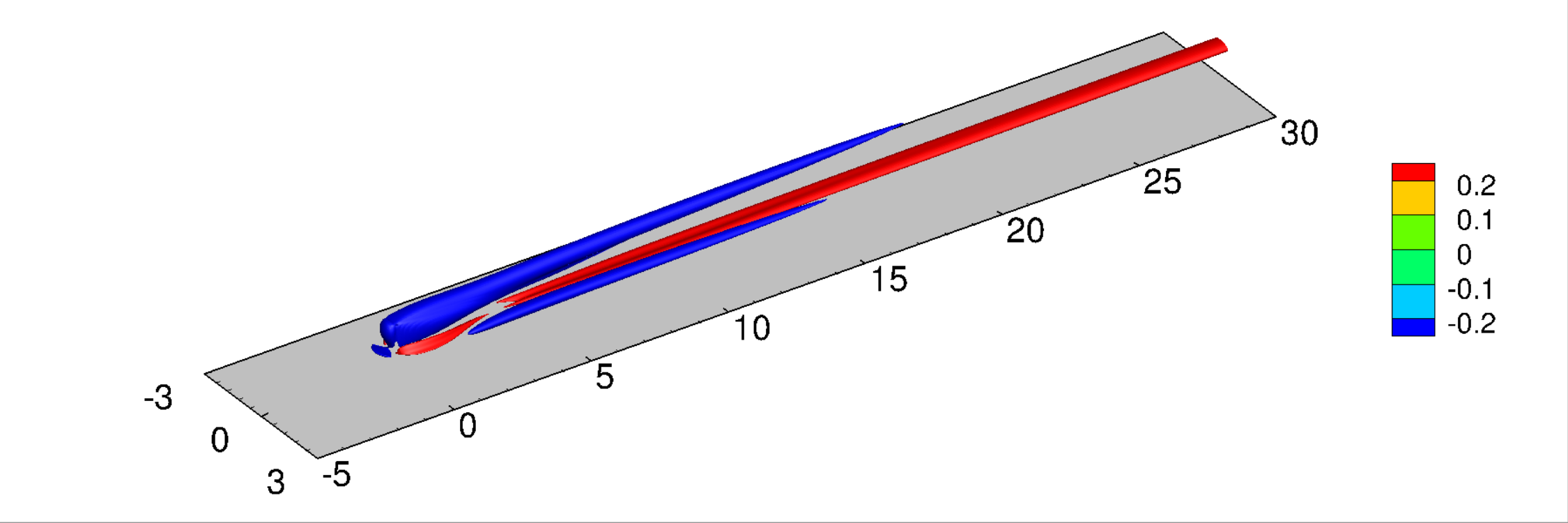}
\put(-185,-5){\rotatebox{90}{$z/h$}}
\put(-95,15){$x/h$}
\put(-22,50){\scriptsize{$u_d/U_e$}}
% \includegraphics[width=70mm,trim={0.2cm 0.2cm 0.5cm 0cm},clip]{images/umean_iso.pdf}
%  \put(-200,60){$(a)$}
% \put(-200,33){\rotatebox{90}{$z$}}
% \put(-110,5){$x$}
% \includegraphics[width=70mm,trim={0.2cm 0.2cm 0.5cm 0cm},clip]{images/umean_iso_3d.pdf}
%  \put(-200,60){$(b)$}
% \put(-185,3){\rotatebox{90}{$z$}}
% \put(-90,15){$x$}
 \caption{Top view (left) and 3-d view (right) of high- and low-speed streaks, visualized by isosurfaces of the streamwise velocity deviation of the base flow from the theoretical Blasius boundary layer solution, $u_d=\overline{u}-u_{bl}$, for $(a)$ Case ($Re_h,\eta$)=($600,1$), $(b)$ Case ($Re_h,\eta$)=($600,0.5$) and $(c)$ Case ($Re_h,\eta$)=($800,0.5$). } 
\label{fig:isocontour_ubar}
\end{figure}

%Check iso-contour for Re475, Mean flow

%laminar_BL_Diaz_Re600_double_iny_sfd_imp_correct_BC, lamianr_BL_Diaz_Re600_DNS_correct_BC(_bk)

The dependence of the base flow features on different $\eta$ and $Re_h$ is examined in figures \ref{fig:contour_zslice}$(b)$-\ref{fig:contour_zslice}$(e)$. The spanwise vortices observed upstream of the roughness element correspond to the horseshoe vortex system induced by the stagnation effect of the roughness. \cite{baker1979laminar} suggested that the stability and topology of the horseshoe vortex system is mostly dependent on $Re_h$ and $h/\delta^*$. For $\eta=1$, the location of the horseshoe vortex moves slightly farther from the front face of the roughness as $Re_h$ increases, shown in figures \ref{fig:contour_zslice}$(b)$ and \ref{fig:contour_zslice}$(c)$, consistent with the observations by \cite{daniel2017direct}. Also, the shear layer induced by the roughness lifts up and shows a stronger wall-normal gradient as $Re_h$ increases. For $\eta=0.5$, shown in figures \ref{fig:contour_zslice}$(d)$ and \ref{fig:contour_zslice}$(e)$, the regions corresponding to the upstream spanwise vortices and the dowmstream reversed flow are smaller due to thinner roughness geometry. The $Re_h$ dependence for $\eta=0.5$ is similar to what is observed for $\eta=1$.

The high- and low-speed streaks are examined in figure \ref{fig:isocontour_ubar}, using isosurfaces of the streamwise velocity deviation $u_d=\overline{u}-u_{bl}$. For Case ($Re_h,\eta$)=($600,1$), the central low-speed streak and two lateral low-speed streaks are illustrated in figure \ref{fig:isocontour_ubar}$(a)$. %The location of the central low-speed streak is corresponding to the shear layer induced by the roughness. 
The central low-speed streak, which occurs symmetrically with respect to the mid-plane, originates from the flow separation downstream of the roughness element. The lateral low-speed streaks are associated with the counter-rotating vortices. High-speed streaks close to the wall appear farther downstream. Figure \ref{fig:isocontour_ubar}$(b)$ shows that for Case ($Re_h,\eta$)=($600,0.5$), the thinner roughness geometry leads to thinner and less sustainable central and lateral low-speed streaks, and the high-speed streaks are absent farther downstream. For Case ($Re_h,\eta$)=($800,0.5$), figure \ref{fig:isocontour_ubar}$(c)$ shows that the strength of the central and lateral low-speed streaks gets amplified as $Re_h$ increases. In contrast to the other two cases, the high-speed streaks are prominent in the near-wake regions, indicating increased spanwise shear that would contribute to the sinuous instability examined in \S \ref{GLSA}. Combining the above results and the smaller $h/\delta^*$ results of \cite{loiseau2014investigation}, it can be concluded that: first, larger $h/\delta^*$, larger $\eta$ and higher $Re_h$ lead to a stronger wall-normal shear and a more sustainable central low-speed streak; Second, increasing $Re_h$ for thin roughness could result in an increased spanwise shear in the near wake region. %higher $Re_h$, higher $\eta$, and higher ratio $h/\delta^*$ could lead to a stronger shear layer and thus a more sustainable central low-speed streak. 

\subsection{Direct and adjoint analyses}\label{GLSA}
%Re600 (SFD), Re500 (SFD), Re450 (DNS), Re475 (SFD) ...
%eigenspectra, critical Re\\
%direct eigenmode (unstable, stable)\\
%production (Re dependence: add Re475 or 450)\\
%How to explain Py is stabilizing?\\
%adjoint eigenmode, wavemaker\\
%Von doenhoff-Braslow diagram

\subsubsection{Global stability analysis}\label{stability}

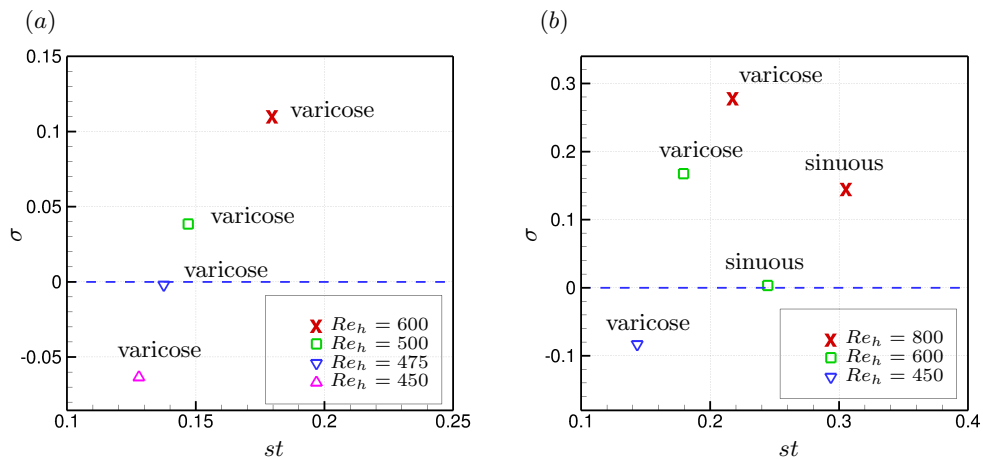
\begin{figure}
\centering
% \includegraphics[width=50mm,trim={0.5cm 0.2cm 0.5cm 0cm},clip]{images/eigenspectra_re600.pdf}
% \put(-175,65){\rotatebox{90}{$\sigma$}}
% \put(-85,0){$st$}
%\hspace{1mm}
\includegraphics[width=65mm,trim={0.5cm 0.2cm 0.5cm 0.5cm},clip]{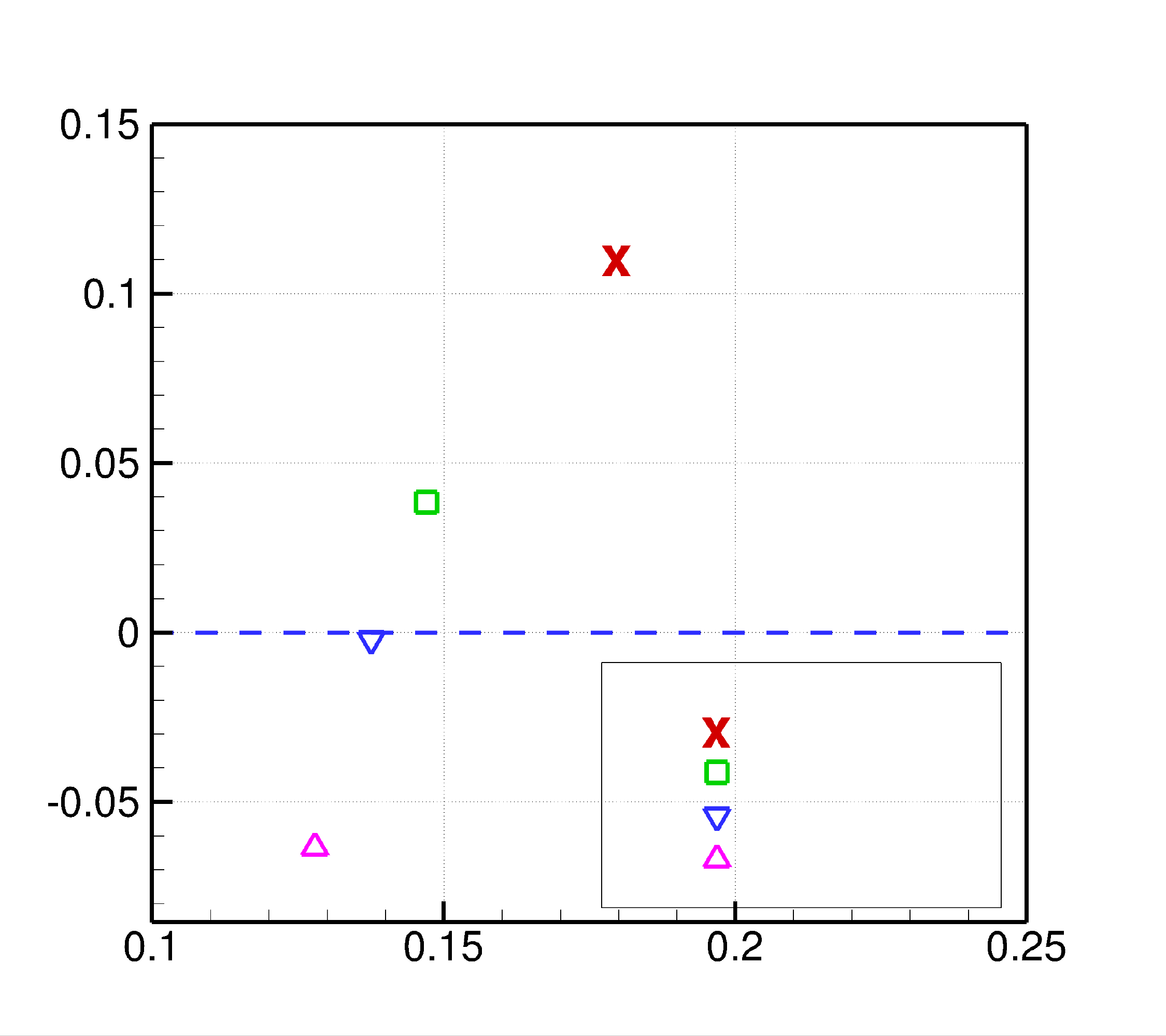}
\put(-185,80){\rotatebox{90}{$\sigma$}}
\put(-90,0){$st$}
\put(-65,47){\scriptsize{$Re_h=600$}}
\put(-65,40){\scriptsize{$Re_h=500$}}
\put(-65,33){\scriptsize{$Re_h=475$}}
\put(-65,26){\scriptsize{$Re_h=450$}}
\put(-80,127){varicose}
\put(-110,87){varicose}
\put(-120,67){varicose}
\put(-145,37){varicose}
 \put(-180,160){$(a)$}
\hspace{2mm}
\includegraphics[width=65mm,trim={0.5cm 0.2cm 0.5cm 0.5cm},clip]{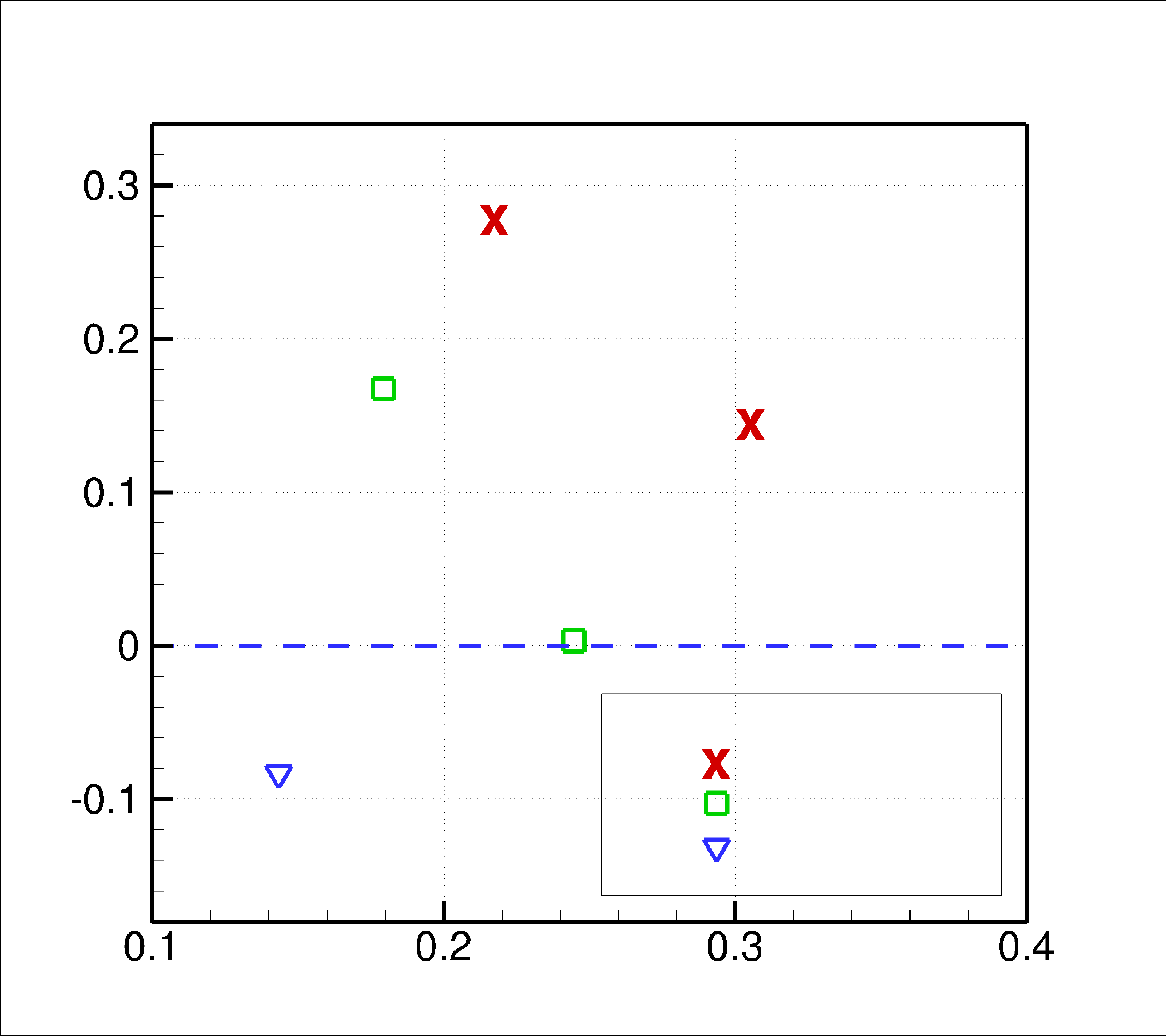}
\put(-185,80){\rotatebox{90}{$\sigma$}}
\put(-90,0){$st$}
\put(-65,42){\scriptsize{$Re_h=800$}}
\put(-65,35){\scriptsize{$Re_h=600$}}
\put(-65,28){\scriptsize{$Re_h=450$}}
\put(-105,140){varicose}
\put(-80,107){sinuous}
\put(-135,112){varicose}
\put(-110,70){sinuous}
\put(-155,47){varicose}
 \put(-180,160){$(b)$}
\caption{Leading eigenvalues of cases with $(a)$ $\eta=1$ and $(b)$ $\eta=0.5$ at different $Re_h$.} 
\label{fig:eigenspec}
\end{figure}
% Eigenvalue spectra for Case ($Re_h,h/\delta^*,\eta$)=($600,2.86,1$);
% LNS_solver/laminar_BL_Diaz_Re600_sfd_gain1; laminar_BL_eta05_Re800_sfd

Global stability analysis has been performed for cases with $\eta=1$ and $\eta=0.5$ at different $Re_h$, and the leading eigenvalues are shown in figures \ref{fig:eigenspec}$(a)$ and \ref{fig:eigenspec}$(b)$ respectively. For $\eta=1$, one leading eigenvalue is obtained at each $Re_h$, as shown in figure \ref{fig:eigenspec}$(a)$. The case at $Re_h=450$ is absolutely stable, consistent with the steady flow field observed from the DNS results. As $Re_h$ increases, both the growth rate and the temporal frequency are increased. The critical $Re_h$ can be identified when the growth rate of an eigenvalue becomes positive. The flow at $Re_h=475$ is marginally stable which suggests that the critical $Re_h$ is close to $475$ for this configuration. 

\begin{figure}
\includegraphics[width=70mm,trim={0.2cm 0.2cm 0.5cm 2cm},clip]{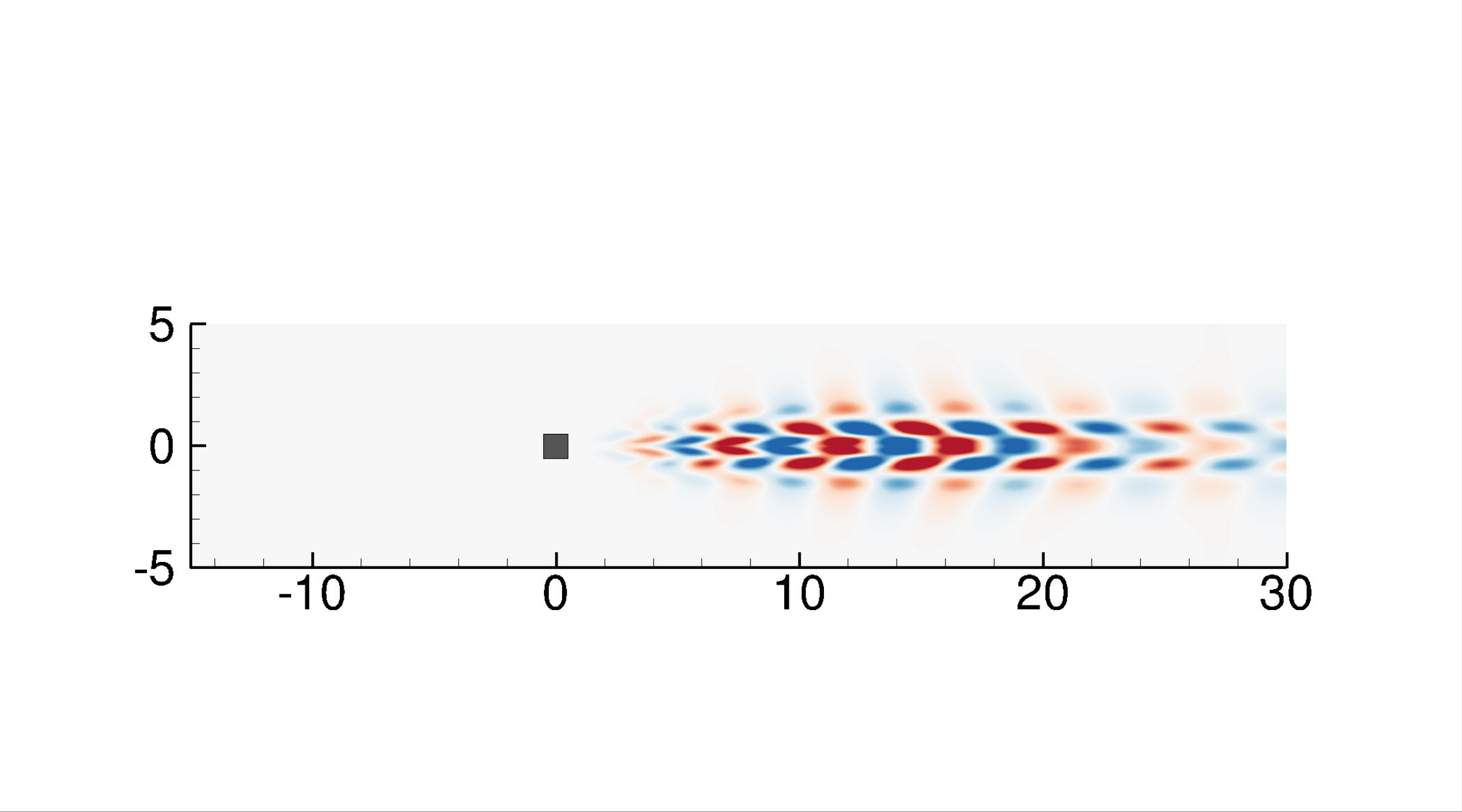}
 \put(-200,80){$(a)$}
 \put(-180,80){$\eta=1,Re_h=475$}
\put(-193,42){\rotatebox{90}{$z/h$}}
\put(-106,19){$x/h$}
\includegraphics[width=70mm,trim={0.2cm 0.2cm 0.5cm 0cm},clip]{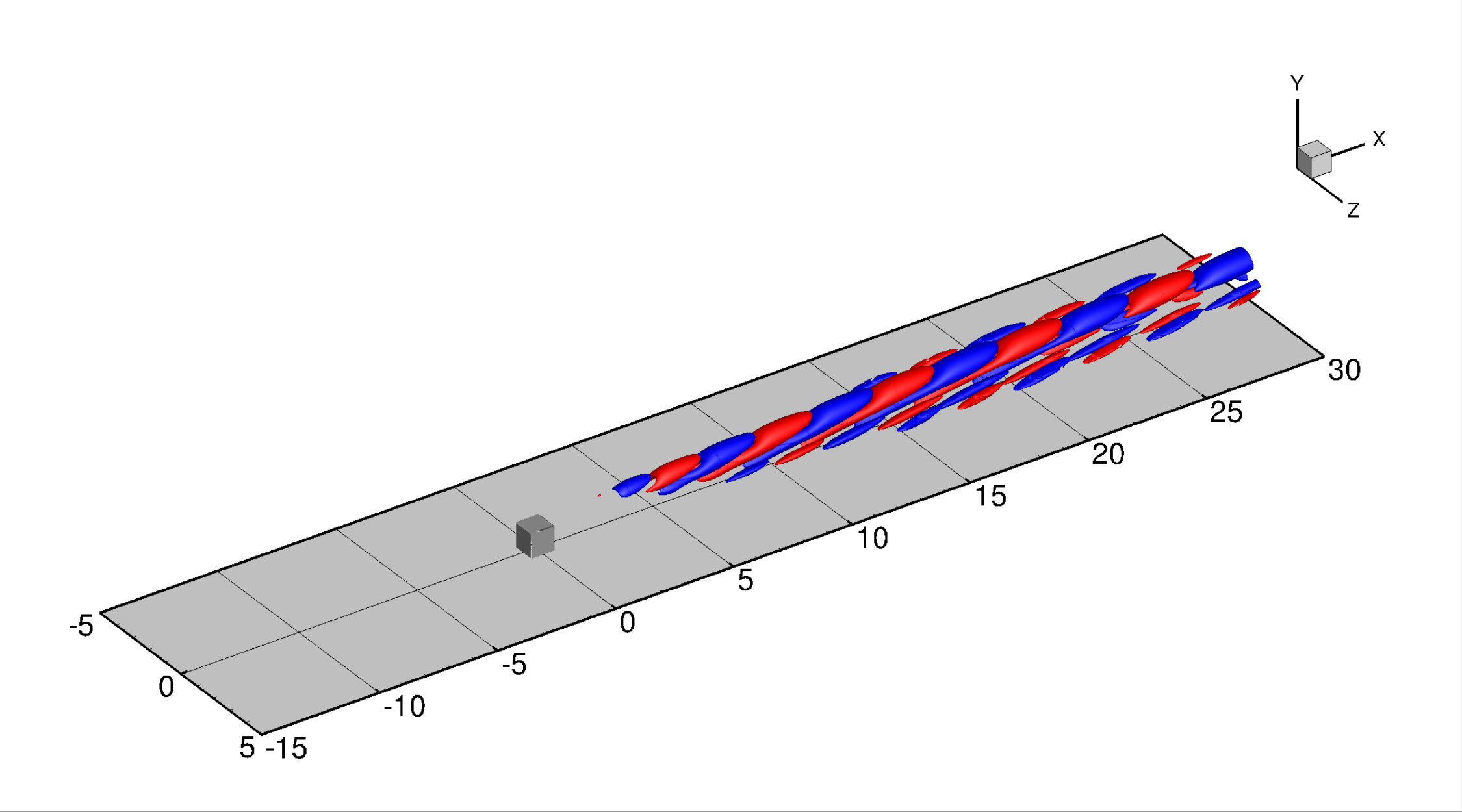}
% \put(-200,90){$(b)$}
\put(-190,3){\rotatebox{90}{$z/h$}}
\put(-90,23){$x/h$}
\hspace{3mm}
\includegraphics[width=70mm,trim={0.2cm 0.2cm 0.5cm 3.5cm},clip]{images/eigenmode_re600.pdf}
 \put(-200,80){$(b)$}
  \put(-180,80){$\eta=1,Re_h=600$}
\put(-193,42){\rotatebox{90}{$z/h$}}
\put(-106,19){$x/h$}
\includegraphics[width=70mm,trim={0.2cm 0.2cm 0.5cm 3.5cm},clip]{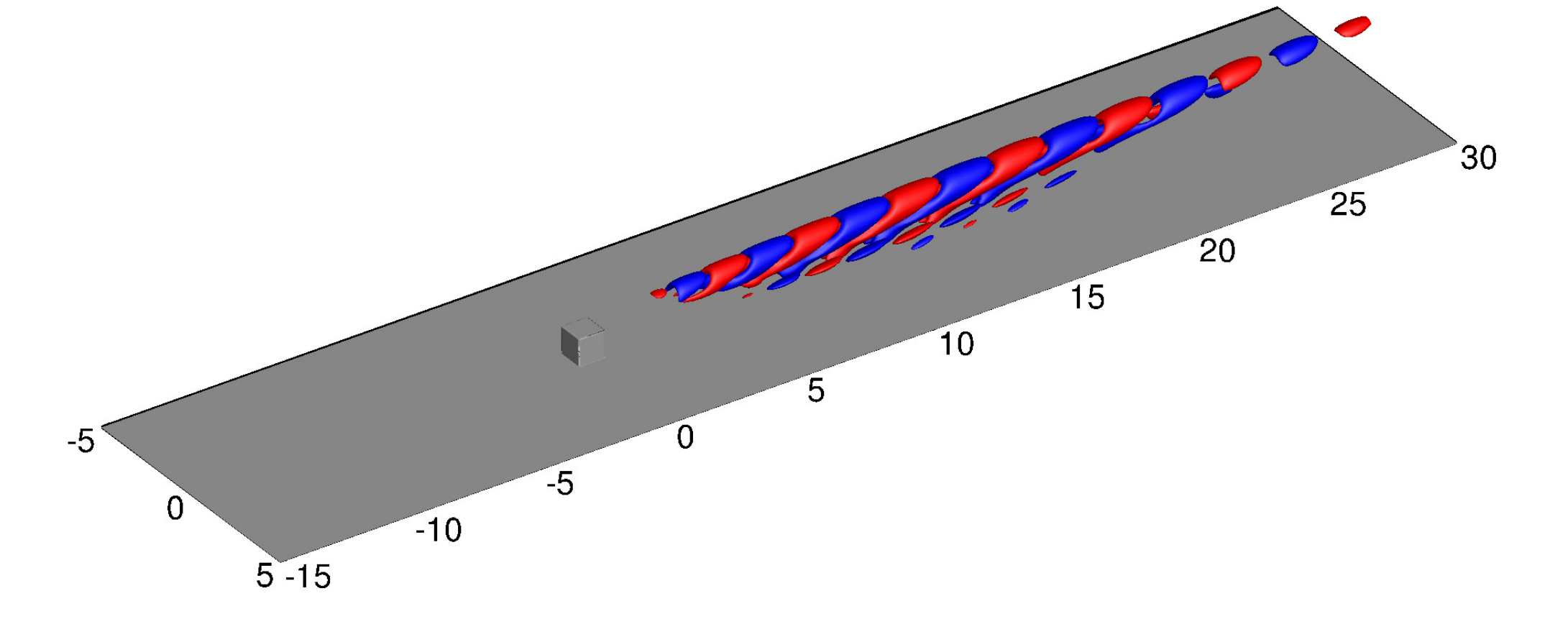}
% \put(-200,90){$(d)$}
\put(-190,3){\rotatebox{90}{$z/h$}}
\put(-90,23){$x/h$}
\hspace{3mm}
% \includegraphics[width=70mm,trim={0.2cm 0.2cm 0.5cm 3.5cm},clip]{images/eigenmode_re450_eta05.pdf}
% % \put(-200,80){$(b)$}
%   \put(-180,80){$\eta=0.5,Re_h=450$}
% \put(-193,47){\rotatebox{90}{$z$}}
% \put(-106,19){$x$}
% \includegraphics[width=70mm,trim={0.2cm 0.2cm 0.2cm 3.5cm},clip]{images/eigenmode_re450_eta05_3d.pdf}
% % \put(-200,90){$(d)$}
% \put(-195,8){\rotatebox{90}{$z$}}
% \put(-90,23){$x$}
% \hspace{3mm}
\includegraphics[width=70mm,trim={0.2cm 0.2cm 0.5cm 3.5cm},clip]{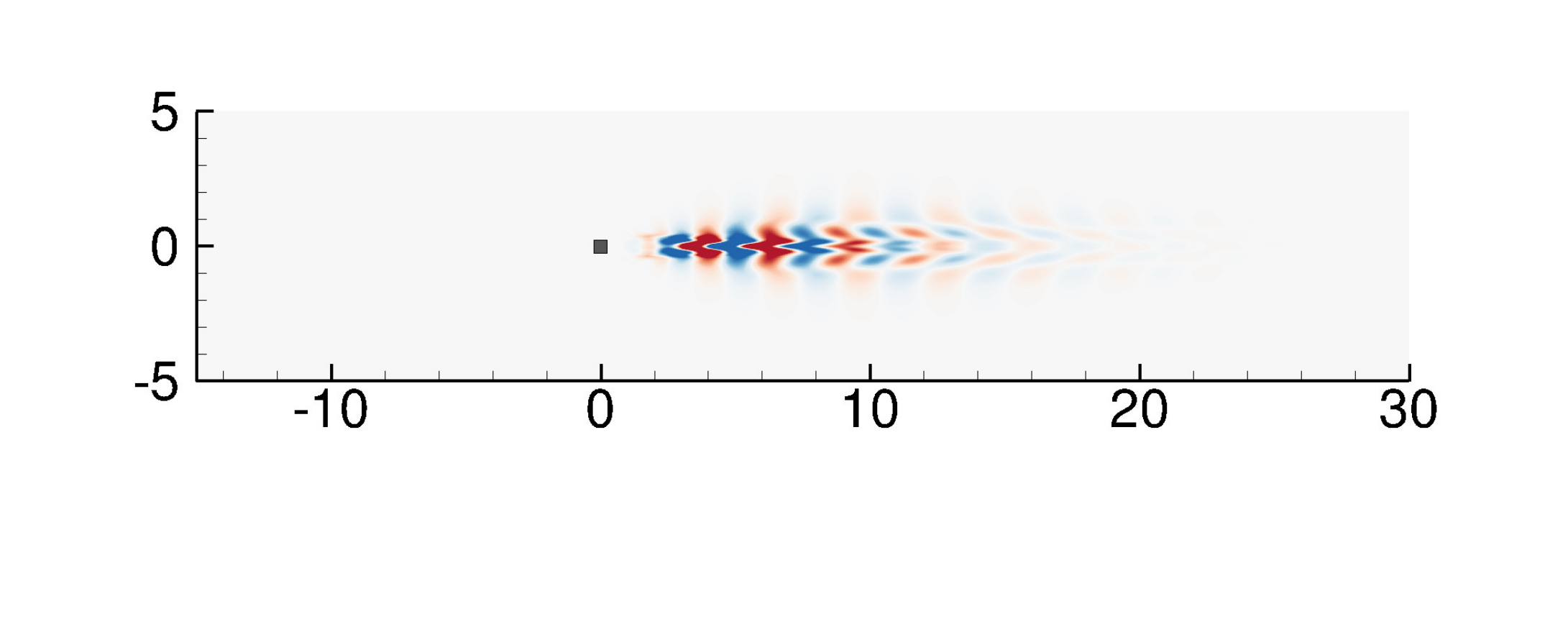}
 \put(-200,80){$(c)$}
  \put(-180,80){$\eta=0.5,Re_h=800$, varicose}
\put(-193,42){\rotatebox{90}{$z/h$}}
\put(-106,19){$x/h$}
\includegraphics[width=70mm,trim={0.2cm 0.2cm 0.5cm 3.5cm},clip]{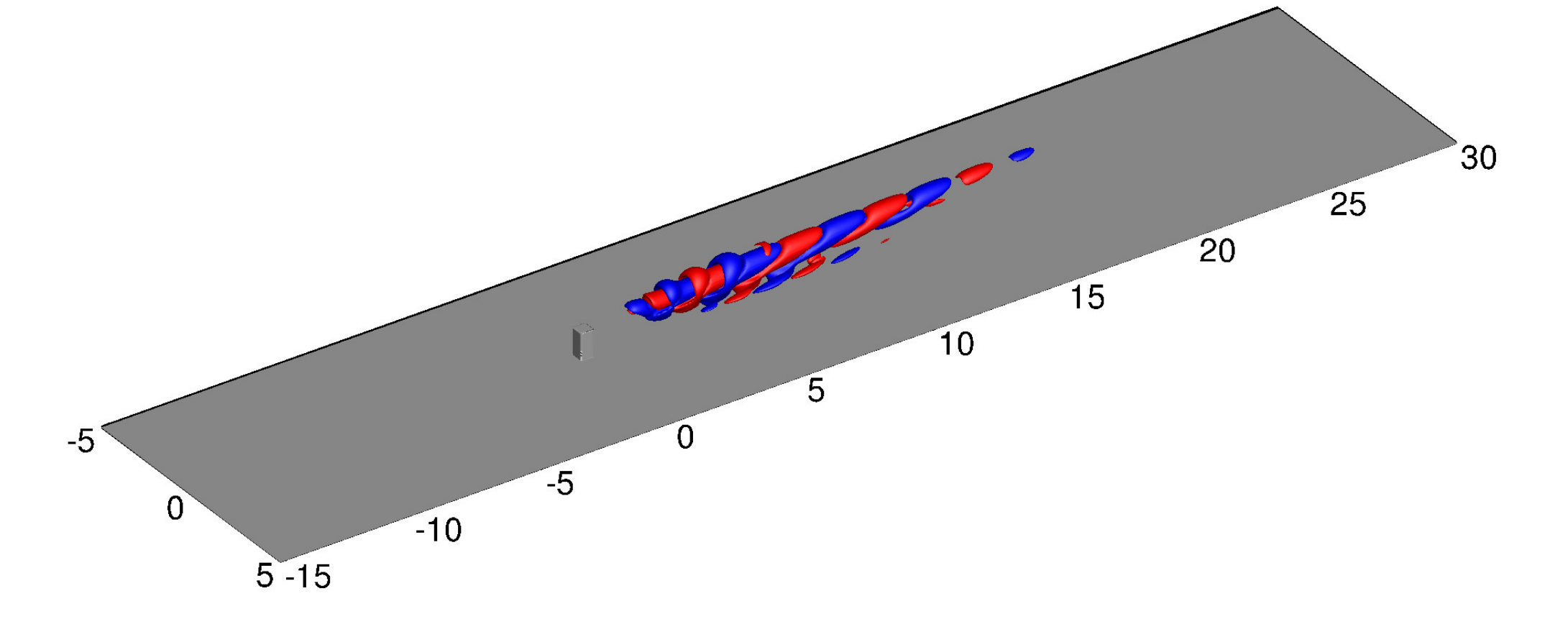}
% \put(-200,90){$(b)$}
\put(-190,3){\rotatebox{90}{$z/h$}}
\put(-90,23){$x/h$}
\hspace{3mm}
\includegraphics[width=70mm,trim={0.2cm 0.2cm 0.5cm 3.5cm},clip]{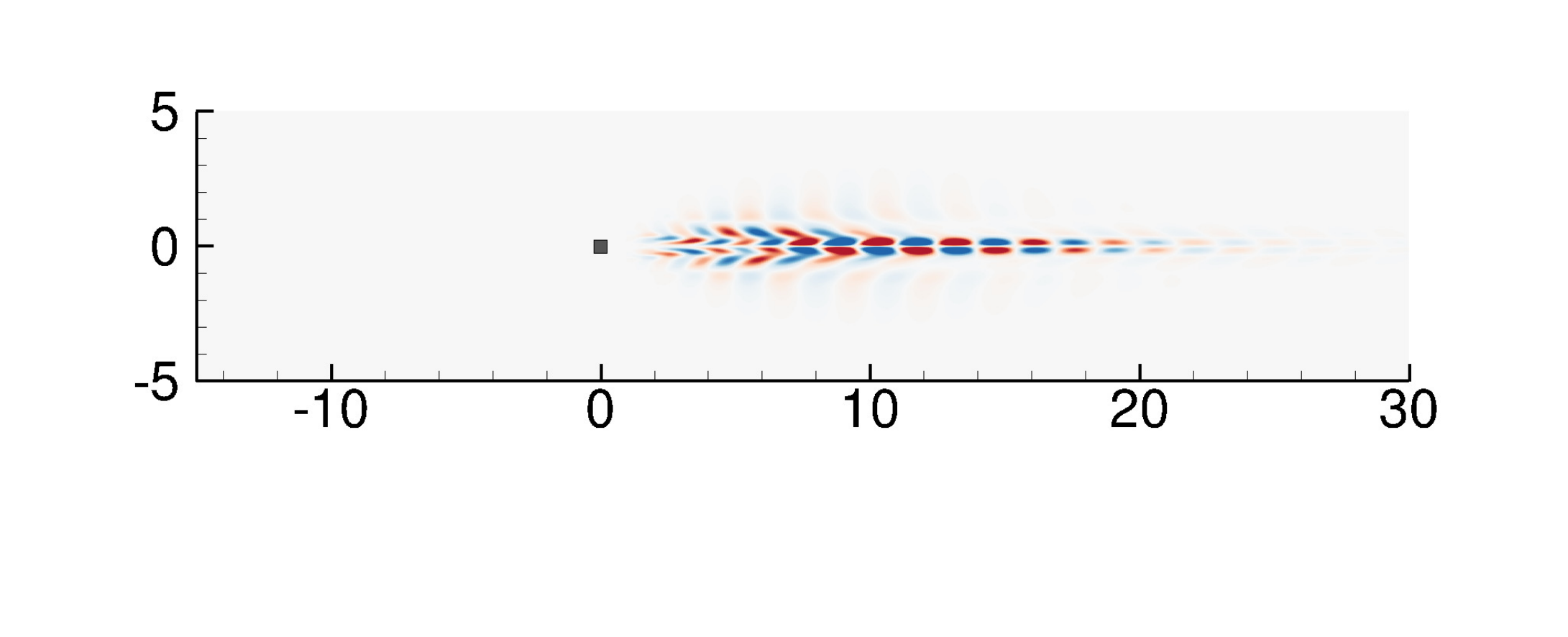}
 \put(-200,80){$(d)$}
   \put(-180,80){$\eta=0.5,Re_h=800$, sinuous}
\put(-193,42){\rotatebox{90}{$z/h$}}
\put(-106,19){$x/h$}
\includegraphics[width=70mm,trim={0.2cm 0.2cm 0.5cm 3.5cm},clip]{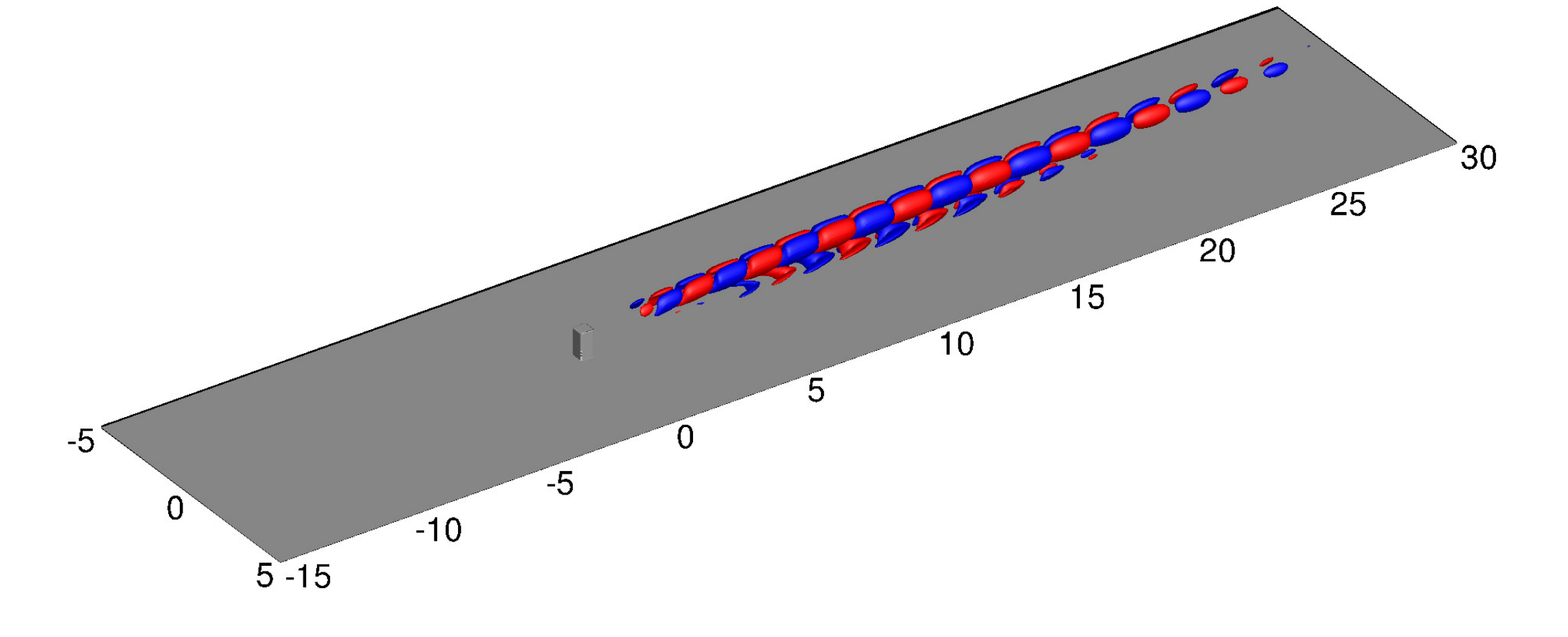}
% \put(-200,90){$(d)$}
\put(-190,3){\rotatebox{90}{$z/h$}}
\put(-90,23){$x/h$}
 \caption{Contour plots at slice $y=0.5h$ (left) and isosurfaces (right) of the streamwise velocity component of the leading unstable global modes for: $(a)$ Case ($Re_h,\eta$)=($475,1$), $(b)$ Case ($Re_h,\eta$)=($600,1$), and $(c,d)$ Case ($Re_h,\eta$)=($800,0.5$). The contour levels depict $\pm 10 \%$ of the mode's maximum streamwise velocity. } 
\label{fig:eigenmode_re600}
\end{figure}
% laminar_BL_Diaz_Re475_sfd_correct_BC
% laminar_BL_Diaz_Re600_double_iny_sfd_imp_correct_BC

For $\eta=1$, the eigenmodes of the leading eigenvalues are all varicose for the various $Re_h$ investigated. The real part of the leading eigenmodes is shown for $Re_h=475$ and $Re_h=600$ in figures \ref{fig:eigenmode_re600}$(a)$ and \ref{fig:eigenmode_re600}$(b)$. Both the leading stable and unstable global modes exhibit a varicose symmetry with respect to the spanwise mid-plane. As shown by the 3-D view of the eigenmode, the shape and location of the modes are consistent with those of the central low-speed streak observed in figure \ref{fig:isocontour_ubar}$(a)$. The varicose mode demonstrates the unstable nature of the central low-speed region induced by the roughness element. Compared to the stable mode at $Re_h=475$, the unstable mode at $Re_h=600$ %shows a more evident uplift, 
is more notably lifted, corresponding to the more raised shear layer for higher $Re_h$ observed in figure \ref{fig:contour_zslice}. 

%The case at $Re_h=475$ is marginally stable, which is consistent with the observations in the DNS simulation. The DNS results at $Re_h=475$ show that there are weak disturbances in the wake flow downstream the element, but the disturbances show a slow damping and $||dU/dt||$ keeps decreasing in a certain time history. 
%The varicose mode is related to the instability of the whole three-dimensional shear layer induced by the roughness element \citep{de2013laminar,loiseau2014investigation}

For $\eta=0.5$, a different unstable behavior is shown in figure \ref{fig:eigenspec}$(b)$. One leading stable eigenvalue is seen at $Re_h=450$ and its associated mode is varicose. Two leading eigenvalues are obtained at higher $Re_h$. The eigenvalue with larger growth rate and lower frequency is a varicose mode, and the other eigenvalue with smaller growth rate and higher frequency is a sinuous mode. For the thinner roughness geometry, the sinuous instability becomes more prominent as $Re_h$ increases. The associated varicose and sinuous eigenmodes of the leading eigenvalues for Case $(Re_h,\eta)=(800,0.5)$ are visualized in figures \ref{fig:eigenmode_re600}$(c)$ and \ref{fig:eigenmode_re600}$(d)$. While the varicose mode is associated with the central low-speed streak observed in figure \ref{fig:isocontour_ubar}$(c)$, the sinuous mode shows a larger streamwise extent along the central region. These results indicate that both varicose and sinuous oscillations exist in the wake flow, and the effect of sinuous instability could be more persistent on the transition process. It thus can be concluded that for thin roughness with large $h/\delta^*$, while the varicose instability is dominant, the sinuous instability can also be present. The onset of sinuous instability results from the interplay of small $\eta$ and increased $Re_h$, corresponding to the enhanced spanwise shear observed in the near wake of the base flow with increasing $Re_h$. 

%The varicose mode is associated with the hairpin vortex shedding in the roughness wake. 

\subsubsection{Production of disturbance kinetic energy}\label{production}

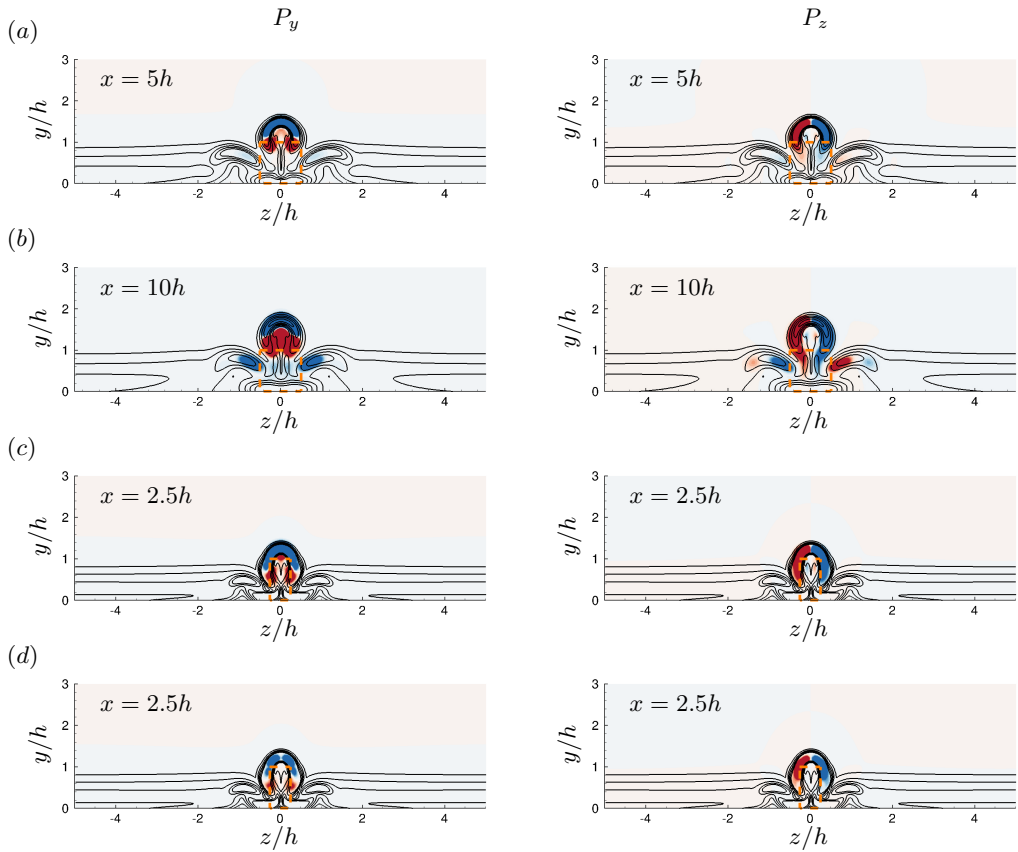
\begin{figure}
\includegraphics[width=70mm,trim={0.2cm 0.2cm 0.5cm 0cm},clip]{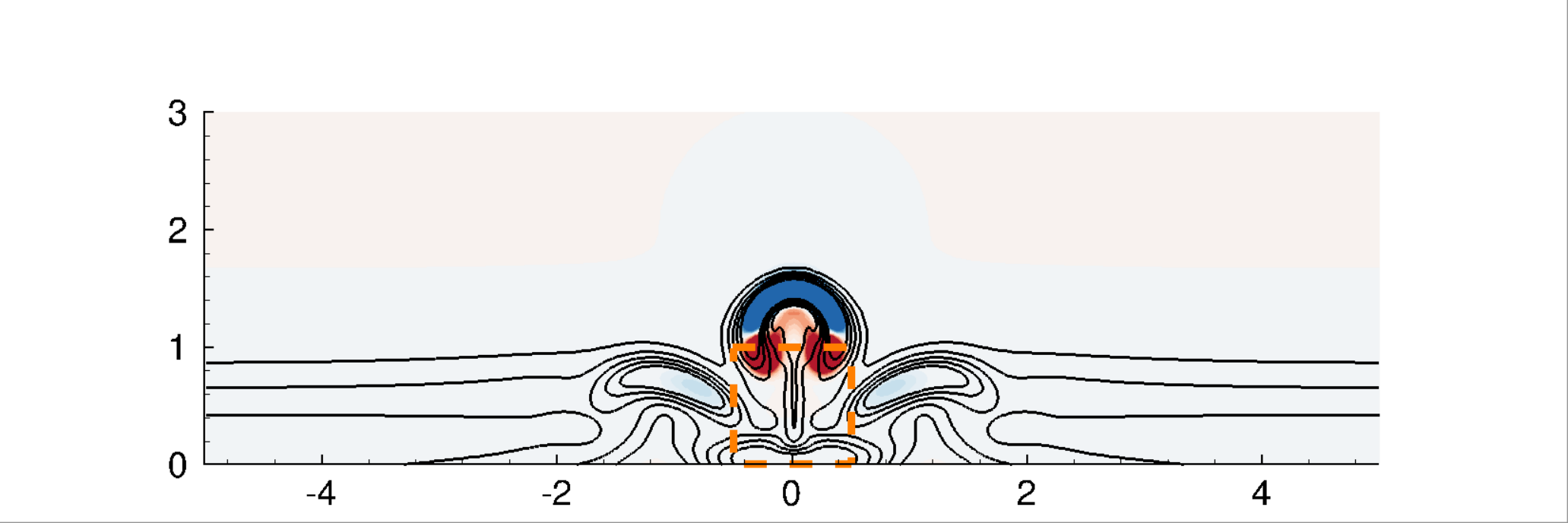}
 %\put(-400,80){$(a)$}
\put(-192,22){\rotatebox{90}{$y/h$}}
\put(-105,-8){$z/h$}
  \put(-200,60){$(a)$}
 % \put(-180,60){$(Re_h,\eta)=(600,1)$}
  \put(-165,42){$x=5h$}
  \put(-100,65){$P_y$}
\includegraphics[width=70mm,trim={0.2cm 0.2cm 0.5cm 0cm},clip]{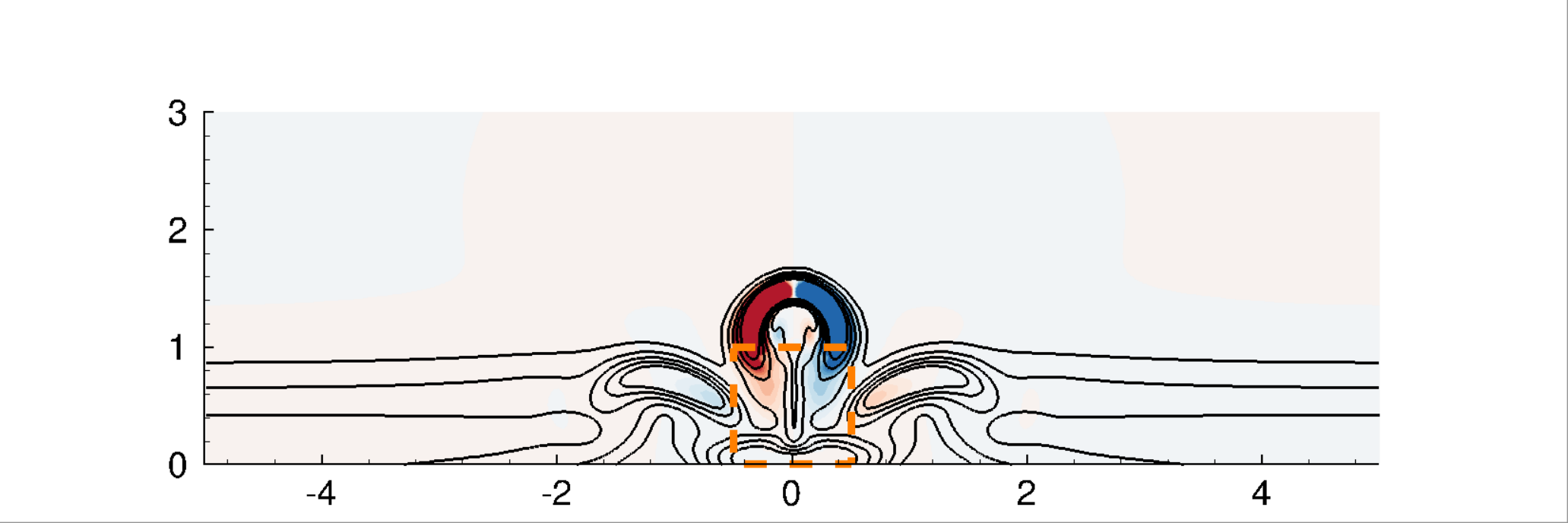}
 % \put(-200,60){$(b)$}
\put(-192,22){\rotatebox{90}{$y/h$}}
\put(-105,-8){$z/h$}
\put(-100,65){$P_z$}
\put(-165,42){$x=5h$}
\hspace{3mm}
\includegraphics[width=70mm,trim={0.2cm 0.2cm 0.5cm 0cm},clip]{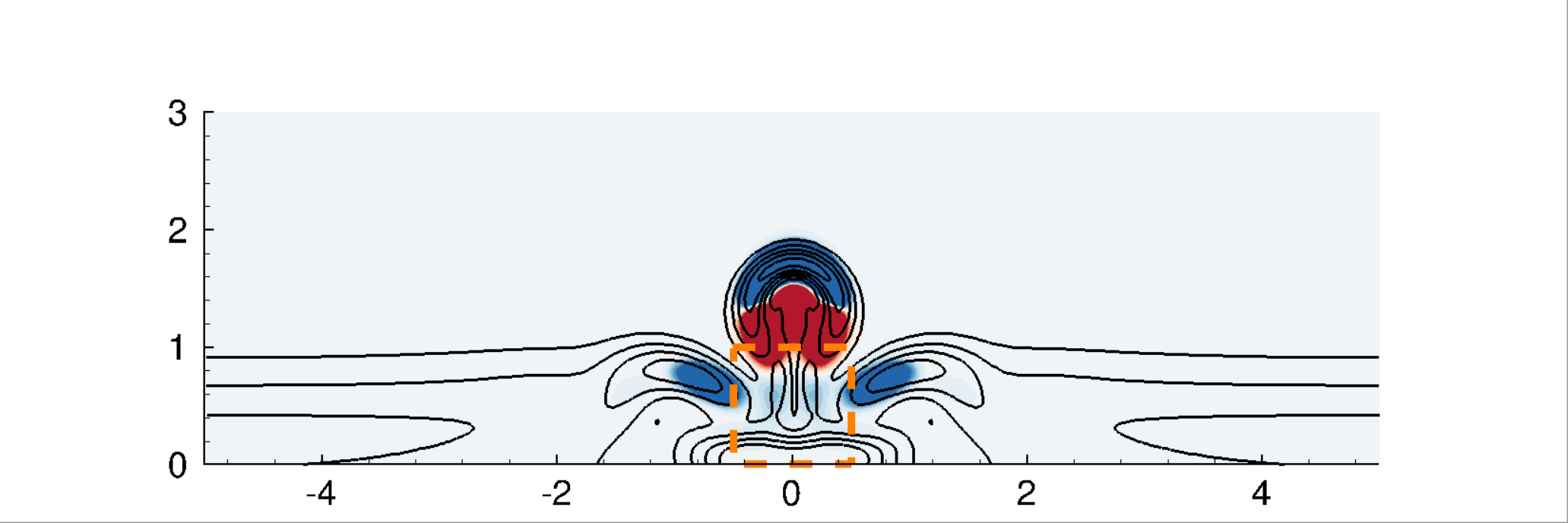}
 %\put(-400,80){$(a)$}
\put(-192,22){\rotatebox{90}{$y/h$}}
\put(-105,-8){$z/h$}
 \put(-200,60){$(b)$}
 \put(-165,42){$x=10h$}
\includegraphics[width=70mm,trim={0.2cm 0.2cm 0.5cm 0cm},clip]{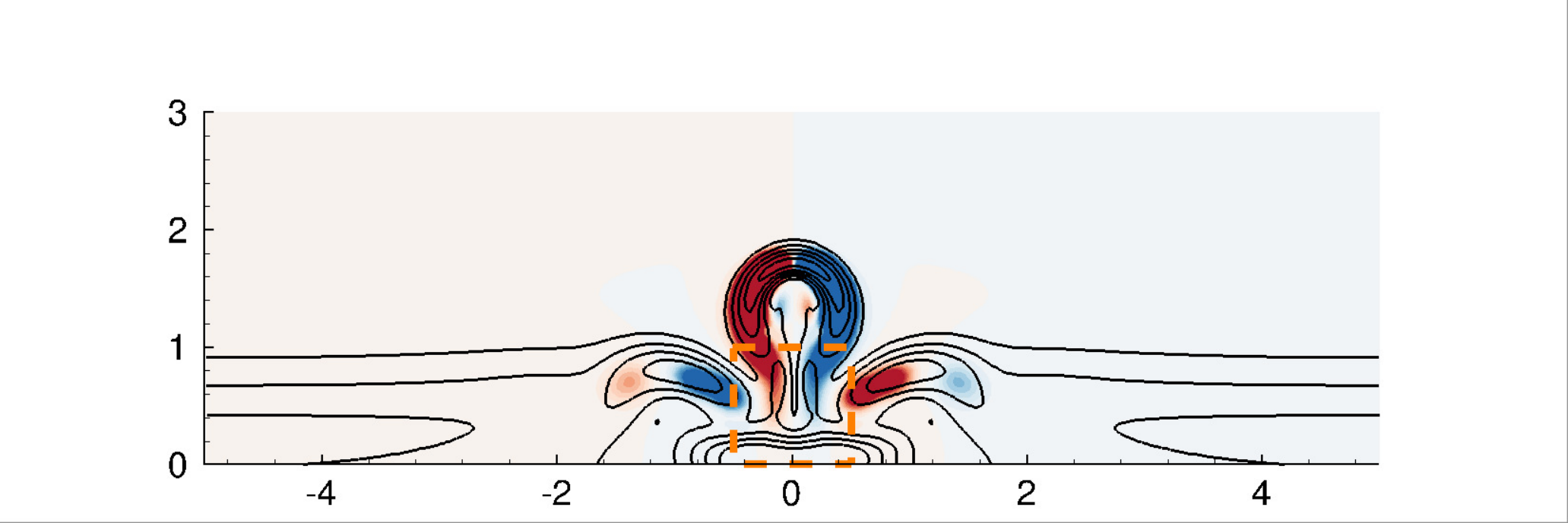}
 %\put(-400,80){$(a)$}
\put(-192,22){\rotatebox{90}{$y/h$}}
\put(-105,-8){$z/h$}
\put(-165,42){$x=10h$}
\hspace{3mm}
\includegraphics[width=70mm,trim={0.2cm 0.2cm 0.5cm 0cm},clip]{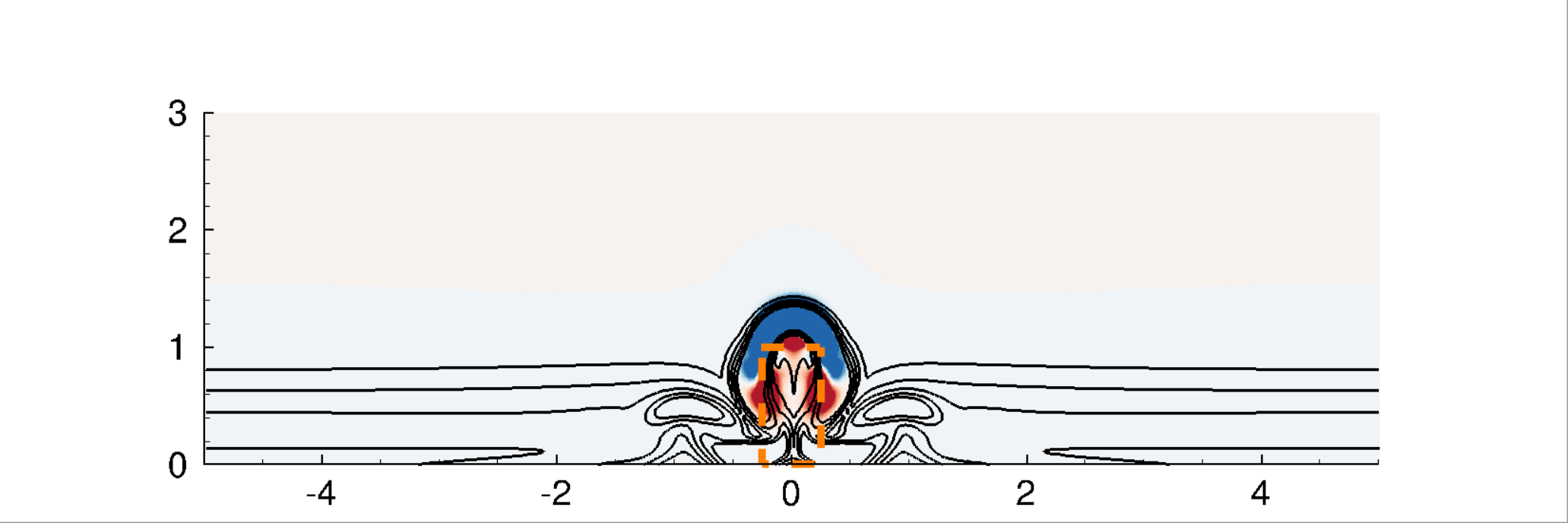}
 %\put(-400,80){$(a)$}
\put(-192,22){\rotatebox{90}{$y/h$}}
\put(-105,-8){$z/h$}
  \put(-200,60){$(c)$}
%  \put(-180,60){$(Re_h,\eta)=(800,0.5)$,varicose}
 % \put(-180,65){$x=5$}
  \put(-165,42){$x=2.5h$}
%  \put(-100,65){$P_y$}
\includegraphics[width=70mm,trim={0.2cm 0.2cm 0.5cm 0cm},clip]{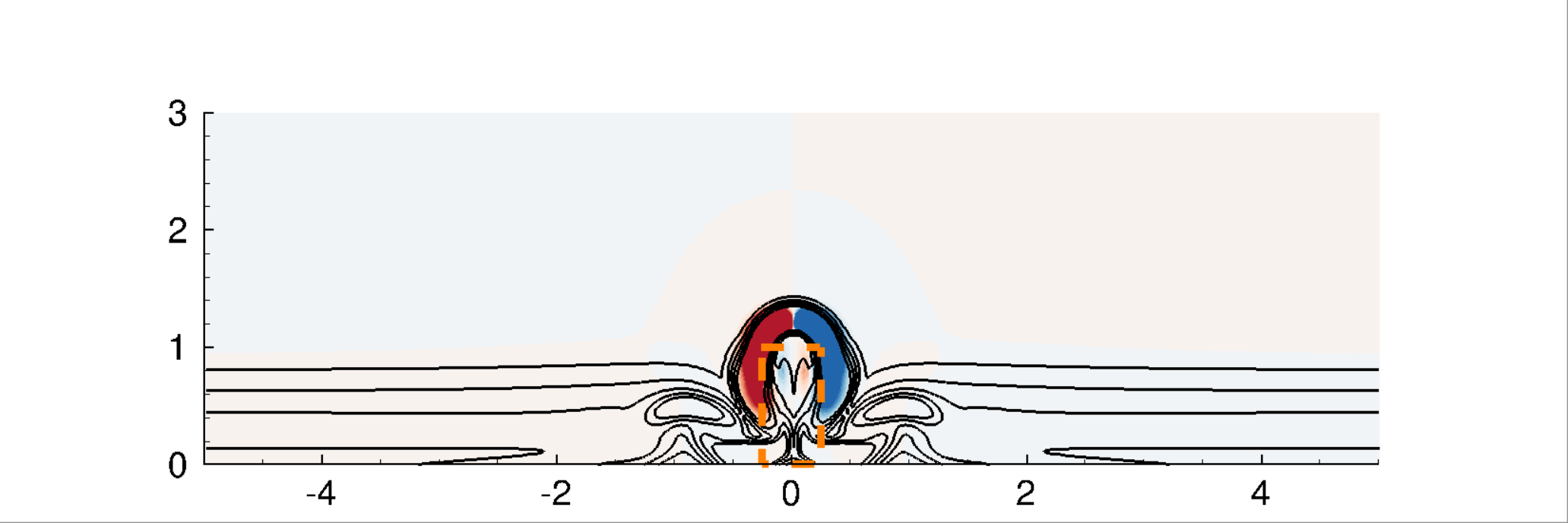}
 %\put(-400,80){$(d)$}
\put(-192,22){\rotatebox{90}{$y/h$}}
\put(-105,-8){$z/h$}
%\put(-100,65){$P_z$}
\put(-165,42){$x=2.5h$}
\hspace{3mm}
\includegraphics[width=70mm,trim={0.2cm 0.2cm 0.5cm 0cm},clip]{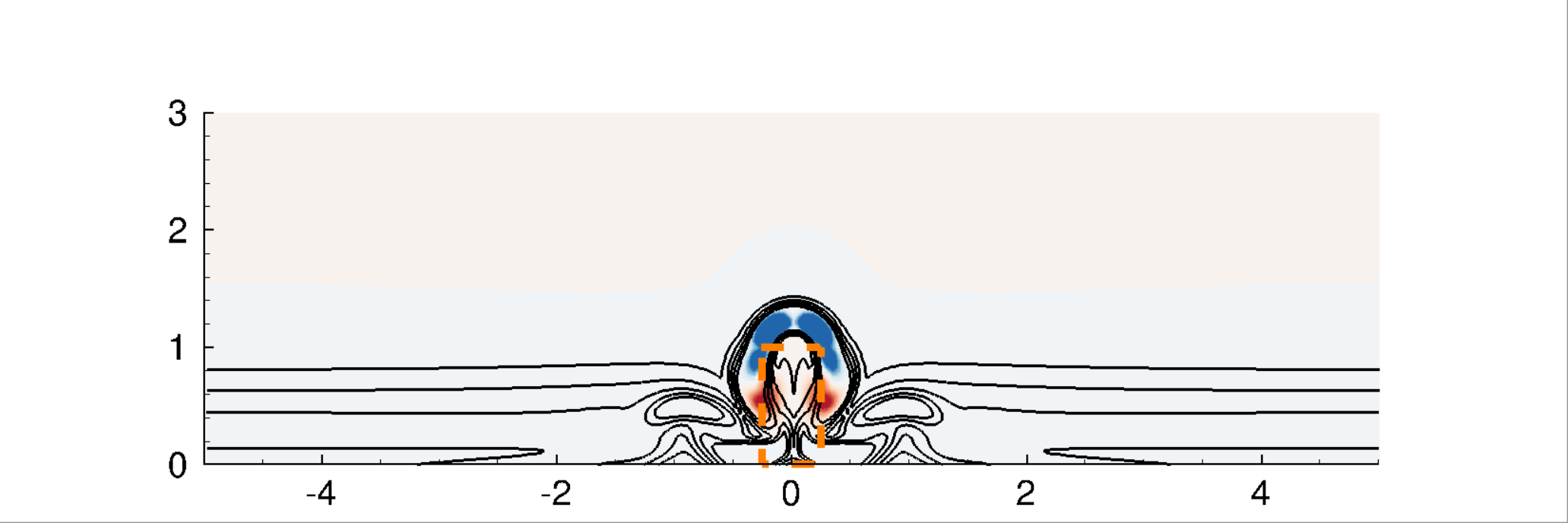}
 %\put(-400,80){$(a)$}
\put(-192,22){\rotatebox{90}{$y/h$}}
\put(-105,-8){$z/h$}
  \put(-200,60){$(d)$}
%  \put(-180,60){$(Re_h,\eta)=(800,0.5)$,sinuous}
%   \put(-180,65){$x=5$}
  \put(-165,42){$x=2.5h$}
%  \put(-100,65){$P_y$}
\includegraphics[width=70mm,trim={0.2cm 0.2cm 0.5cm 0cm},clip]{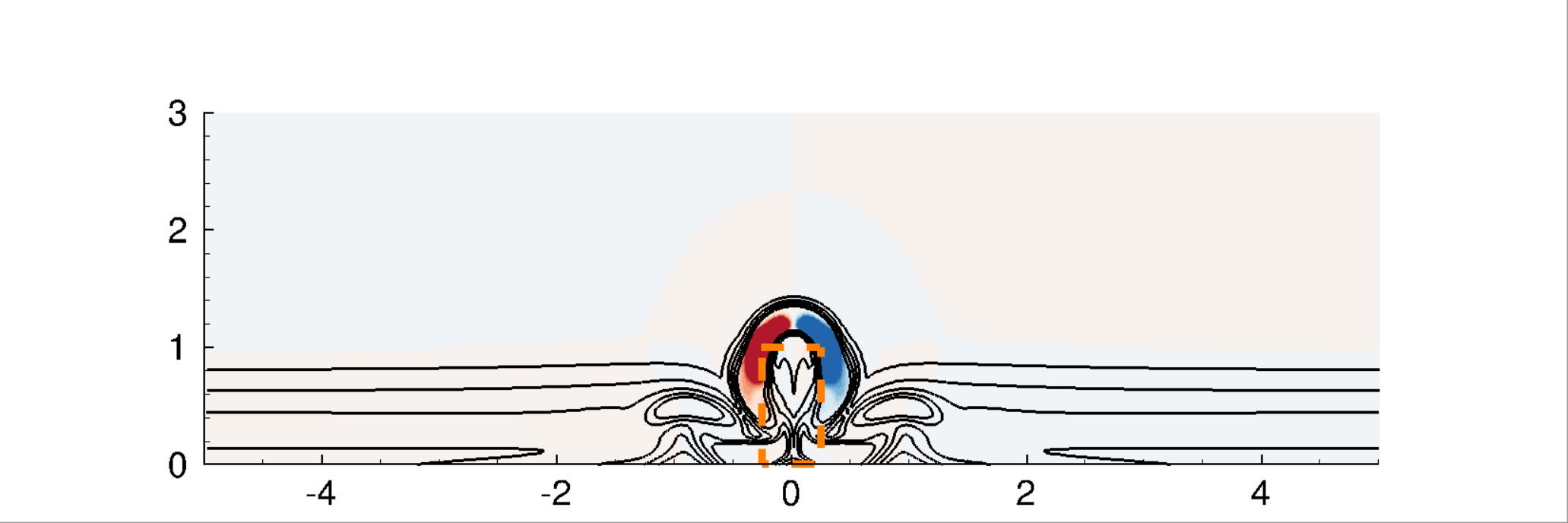}
 %\put(-400,80){$(a)$}
\put(-192,22){\rotatebox{90}{$y/h$}}
\put(-105,-8){$z/h$}
%\put(-100,65){$P_z$}
\put(-165,42){$x=2.5h$}
\caption{Contours of $P_y$ on the left and $P_z$ on the right in cross-flow planes at $(a)$ $x=5h$ and $(b)$ $x=10h$ for Case ($Re_h,\eta$)=($600,1$), $(c)$ $x=2.5h$ for the leading varicose mode of Case ($Re_h,\eta$)=($800,0.5$) and $(d)$ $x=2.5h$ for the leading sinuous mode of Case ($Re_h,\eta$)=($800,0.5$). The contour levels are shown within the range from $-1.0e^{-7}$ (blue) to $1.0e^{-7}$ (red). The localized shear is depicted by the solid lines of $u_s=((\partial \overline{u}/\partial y)^2+(\partial \overline{u}/\partial z)^2)^{1/2}$ from $0$ to $2$. The orange dashed lines show the location of the element.} 
\label{fig:I_xslice}
\end{figure}

The production of disturbance kinetic energy provides insight into how and where the global modes extract their energy from the base flow. As illustrated by \cite{de2013laminar} and \cite{loiseau2014investigation}, the main contributions to the production of disturbance kinetic energy are the two terms
\begin{equation}
P_y=-|\hat{u}||\hat{v}|\frac{\partial{U_b}}{\partial y},   P_z=-|\hat{u}||\hat{w}|\frac{\partial{U_b}}{\partial z}.
\label{eqn:eigen_adjoint}
\end{equation}
The streamwise variation and spatial distribution of these two dominant terms are examined for Cases ($Re_h,\eta$)=($600,1$) and ($Re_h,\eta$)=($800,0.5$).

% The streamwise evolution of $P_y$ and $P_z$ is shown in figure \ref{fig:prod_profiles} by integrating $P_y$ and $P_z$ in the wall-normal and spanwise directions. As expected for the varicose mode, the magnitude of $I_y$ and $I_z$ remains small in the region near the element, indicating that the varicose mode does not extract energy from the base flow in the near-wake region. They rises smoothly and reaches the peak value at $x=10$, then decreases to nearly zero along the streamwise direction. The magnitude of $I_y$ is larger than that of $I_z$ but $I_y$ is negative over the whole domain.
%This is not absolutely consistent with the observations by \cite{de2013laminar} for a roughness element in a supersonic boundary layer.

% \begin{figure}
% \centering
% \includegraphics[width=95mm,trim={0.2cm 0.2cm 0.5cm 0cm},clip]{images/prod_profiles.pdf}
% \put(-135,-8){$x$}
% \put(-165,35){\scriptsize{-$I_y$}}
% \put(-165,25){\scriptsize{$I_z$}}
% \put(-270,35){$\rotatebox{90}{production}$}
% \caption{Streamwise distribution of the production terms for the leading varicose mode in Case ($Re_h,h/\delta^*,\eta$)=($600,2.86,1$), where $I_y=\int P_y dydz$ and $I_z=\int P_z dydz$.} 
% \label{fig:prod_profiles}
% \end{figure}
% laminar_BL_diaz_Re600_double_iny_sfd_imp_correct_BC(_prod)

The spatial variations of $P_y$ and $P_z$ in cross-flow planes are depicted in figure \ref{fig:I_xslice}. In combination with the production terms, the local shear is visualized by the solid contour lines of $u_s=((\partial \overline{u}/\partial y)^2+(\partial \overline{u}/\partial z)^2)^{1/2}$ in figure \ref{fig:I_xslice}, where $\overline{u}$ is the streamwise velocity of the base flow. For Case ($Re_h,\eta$)=($600,1$), two planes at $x=5h$ and $x=10h$ are shown in figures \ref{fig:I_xslice}$(a)$ and \ref{fig:I_xslice}$(b)$. The contour lines of $u_s$ demonstrate the central low-speed streak and the lateral low-speed streaks on either side of the cube. With increasing downstream distance, both the central and lateral low-speed streaks rise, reach their maximum strength at about $x=10h$ and then fade away. The planes beyond $x=10h$ are not shown for the sake of brevity. The distributions of $P_y$ and $P_z$ show a coincidence with the location of the streaks, indicating that the varicose mode extracts the energy from the wall-normal and spanwise shear of the base flow. These results confirm that the varicose mode demonstrates the instability of the entire 3-D shear layer \citep{de2013laminar,loiseau2014investigation}.

% The streamwise evolution is consistent with the development of the streaks: the strength of $P_y$ and $P_z$ reaches its maximum around $x=10$, and then diminishes as the location goes further downstream. The strong shear layer in the wall-normal direction observed in the base flow (figure \ref{fig:contour_zslice}) results in different signs of $\partial U_b/\partial y$ along the wall-normal direction, which explains the red and blue layers of $P_y$ located within the central low-speed region. In contrast to $P_y$, the red and blue layers of $P_z$ (due to different signs of $\partial U_b/\partial z$) are shown at the location corresponding to the spanwise shear of the central low-speed streak. Figure \ref{fig:I_yslice} displays the top views of $P_y$ and $P_z$ distributions. The distribution corresponding to the two lateral streaks is observed at $y=0.75$, while the distribution corresponding to the central low-speed streak is shown clearly at $y=1.5$. 

\begin{figure}
\includegraphics[width=70mm,trim={0.2cm 0.2cm 0.5cm 0cm},clip]{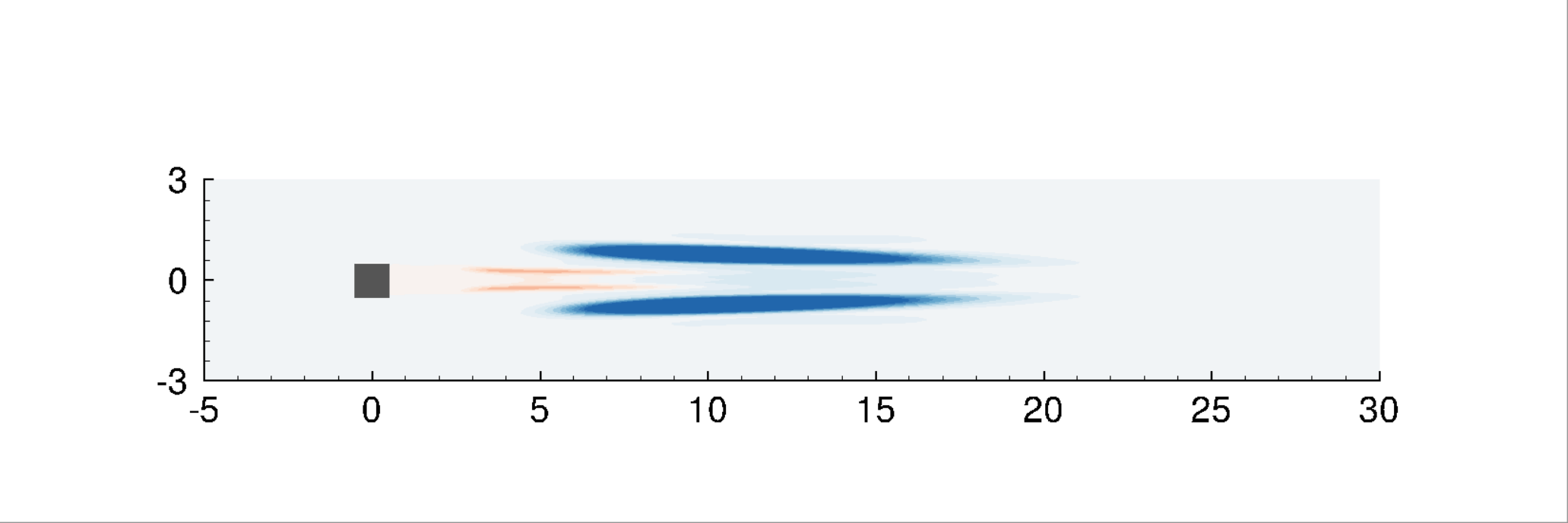}
 \put(-200,50){$(a)$}
\put(-190,23){\rotatebox{90}{$z/h$}}
\put(-105,3){$x/h$}
 \put(-180,50){$y=0.75h$}
 \put(-100,50){$P_y$}
\includegraphics[width=70mm,trim={0.2cm 0.2cm 0.5cm 0cm},clip]{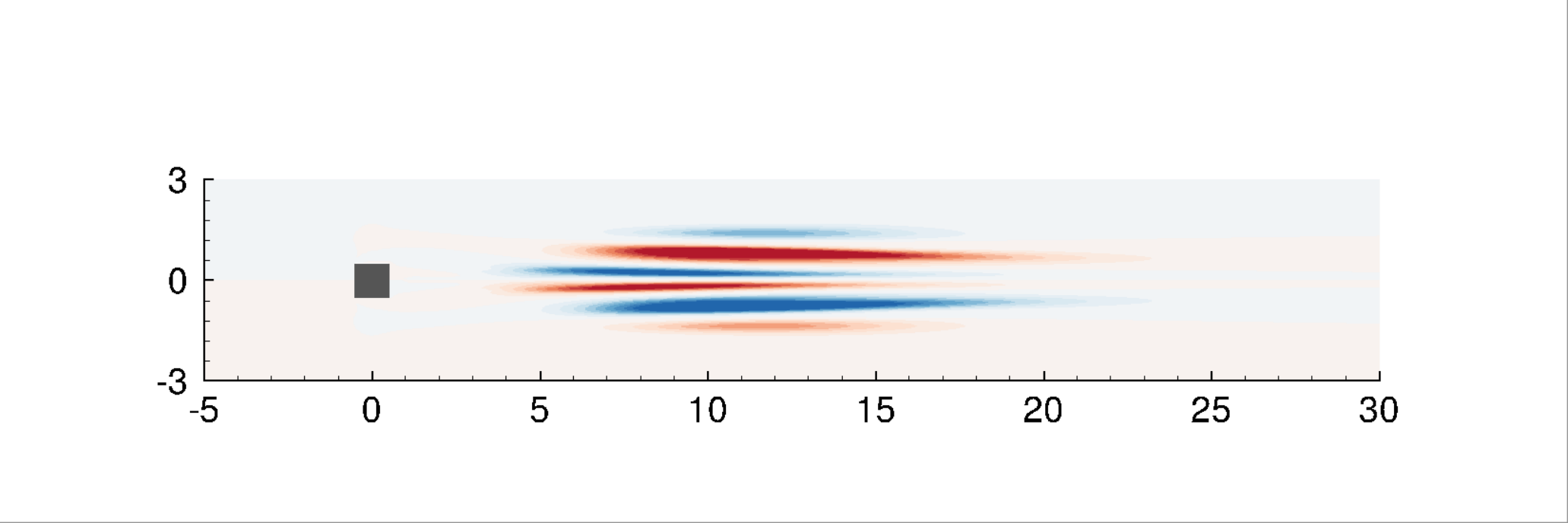}
 %\put(-200,50){$(b)$}
\put(-190,23){\rotatebox{90}{$z/h$}}
\put(-105,3){$x/h$}
\put(-100,50){$P_z$}
\hspace{3mm}
\includegraphics[width=70mm,trim={0.2cm 0.2cm 0.5cm 0cm},clip]{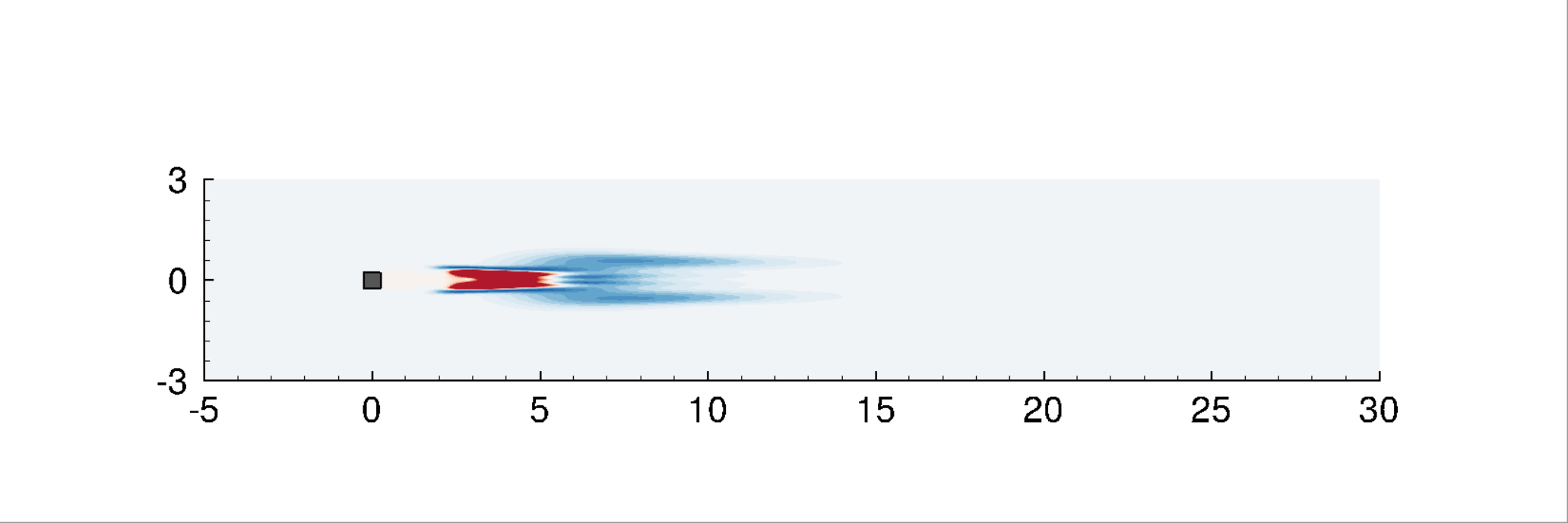}
 \put(-200,50){$(b)$}
\put(-190,23){\rotatebox{90}{$z/h$}}
\put(-105,3){$x/h$}
 \put(-180,50){$y=0.75h$}
% \put(-100,50){$P_y$}
\includegraphics[width=70mm,trim={0.2cm 0.2cm 0.5cm 0cm},clip]{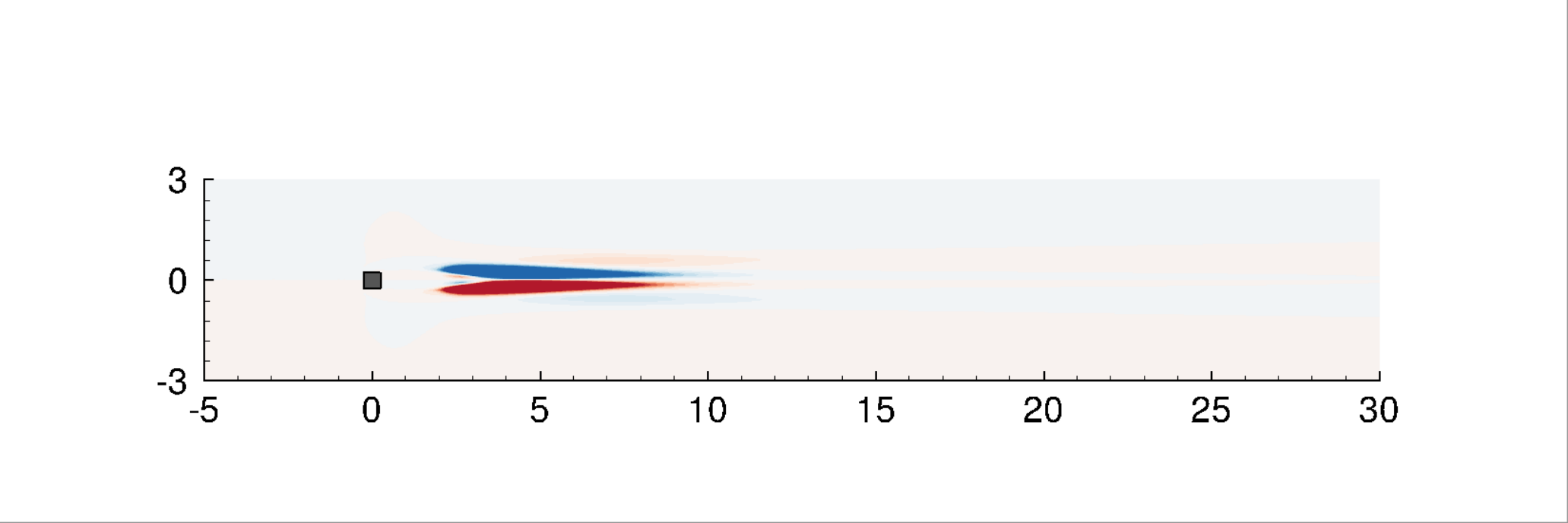}
 %\put(-200,50){$(b)$}
\put(-190,23){\rotatebox{90}{$z/h$}}
\put(-105,3){$x/h$}
%\put(-100,50){$P_z$}
\hspace{3mm}
\includegraphics[width=70mm,trim={0.2cm 0.2cm 0.5cm 0cm},clip]{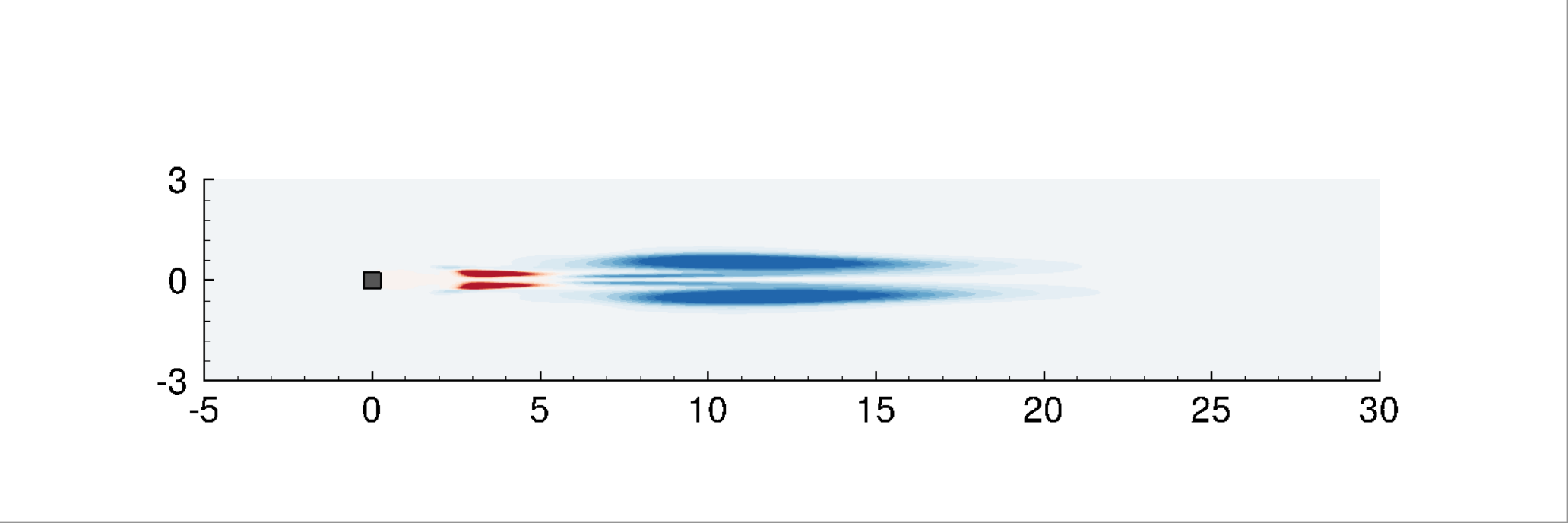}
 \put(-200,50){$(c)$}
\put(-190,23){\rotatebox{90}{$z/h$}}
\put(-105,3){$x/h$}
 \put(-180,50){$y=0.75h$}
% \put(-100,50){$P_y$}
\includegraphics[width=70mm,trim={0.2cm 0.2cm 0.5cm 0cm},clip]{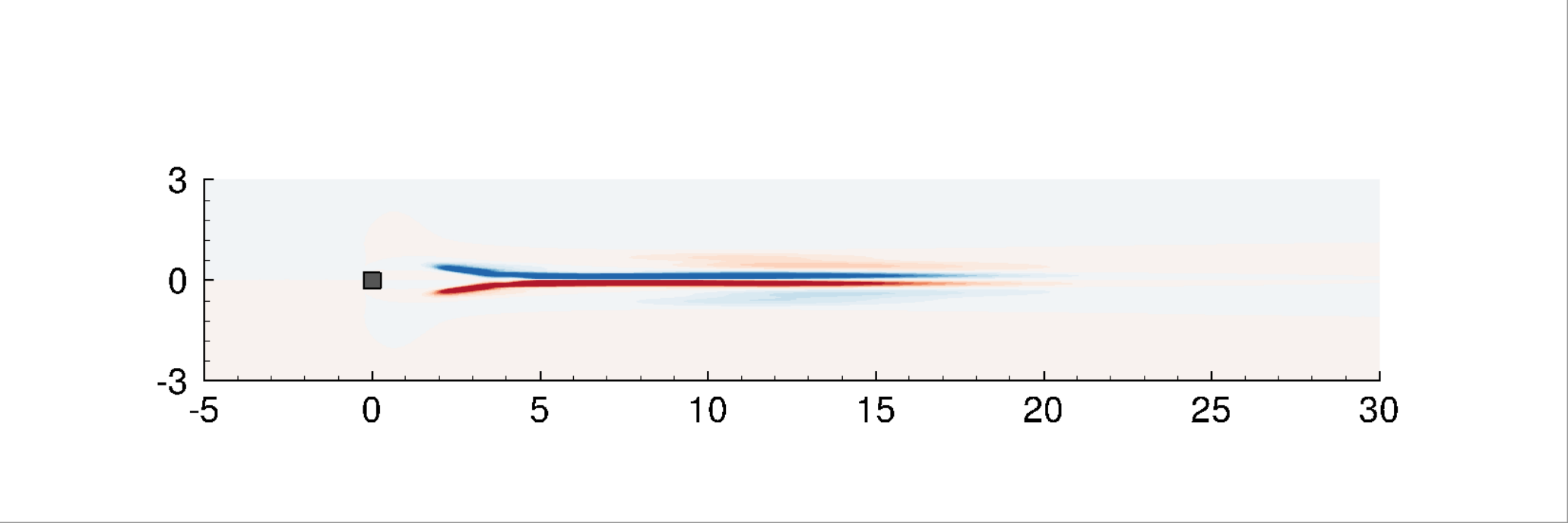}
 %\put(-200,50){$(b)$}
\put(-190,23){\rotatebox{90}{$z/h$}}
\put(-105,3){$x/h$}
%\put(-100,50){$P_z$}
 \caption{Contours of $P_y$ on the left and $P_z$ on the right in x-z planes at $y=0.75h$ for $(a)$ the leading varicose mode of Case ($Re_h,\eta$)=($600,1$), $(b)$ the leading varicose mode and $(c)$ the leading sinuous mode of Case ($Re_h,\eta$)=($800,0.5$). The contour levels are the same as figure \ref{fig:I_xslice}.} 
 \label{fig:I_yslice}
\end{figure}
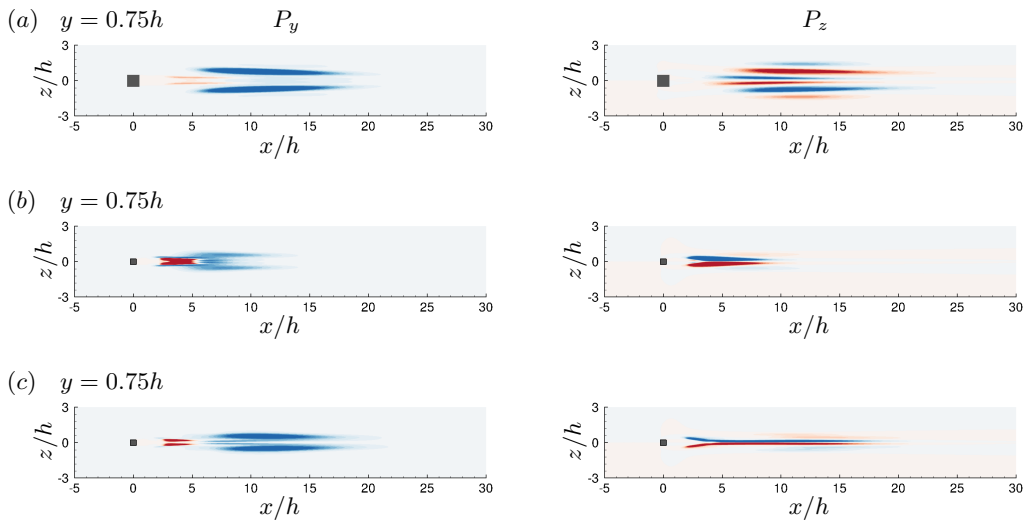

%Consistent with the observations of \cite{de2013laminar} and \cite{loiseau2014investigation}, the varicose mode demonstrates the instability of the entire 3-D shear layer.
The lateral low-speed streaks also make a contribution to the dominant production terms when $h/\delta^*$ is large. The mode extracts energy from the lateral streaks, as shown at $x=10h$ in figure \ref{fig:I_xslice}$(b)$. The top views of $P_y$ and $P_z$ for Case ($Re_h,\eta$)=($600,1$) demonstrated in figure \ref{fig:I_yslice}$(a)$ display the contributions of the two lateral streaks more clearly. The large shear ratio $h/\delta^*$ leads to stronger central and lateral streaks in the present case. Although the varicose mode extracts most of energy from the central low-speed streak, the contribution of the lateral streaks can not be neglected for cases with large shear ratios.

In contrast, the contours of $P_y$ and $P_z$ for the leading varicose and sinuous modes of Case ($Re_h,\eta$)=($800,0.5$) are shown in figures \ref{fig:I_xslice}$(c)$ and \ref{fig:I_xslice}$(d)$ respectively. The distribution of $P_y$ demonstrates that while the varicose mode extracts the energy from the top edge of the central streak, the sinuous mode extracts its energy from the lateral parts of the central streak. These results are consistent with the observation by \cite{loiseau2014investigation} for small $h/\delta^*$ cylindrical roughness. For the thinner geometry ($\eta=0.5$), there is less fluid passing above the roughness element, and a stronger spanwise shear is seen corresponding to the longer wall-normal extent for the lateral parts of the central streak, shown in figure \ref{fig:I_xslice}$(d)$. This suggests that the sinuous instability occurs due to the fact that it could extract more energy from the spanwise shear. The contour plots of $P_y$ and $P_z$ at $y=0.75h$ are shown in figures \ref{fig:I_yslice}$(b)$ and \ref{fig:I_yslice}$(c)$. %Compared to Case ($Re_h,\eta$)=($600,1$), the two lateral streaks are located more closely and have smaller sizes due to a thinner geometry. 
The $P_y$ and $P_z$ distributions of the sinuous mode show a longer streamwise extent than those of the varicose mode, implying the influence of sinuous instability on the wake flow could last farther downstream. %consistent with the length scale of the modes themselves.
Both the varicose and sinuous modes are able to extract some energy from the lateral streaks. The contribution of the lateral streaks is associated with the strength of the lateral streaks which is more likely dependent on the shear ratio.

\subsubsection{Adjoint sensitivity analysis}\label{adjoint}

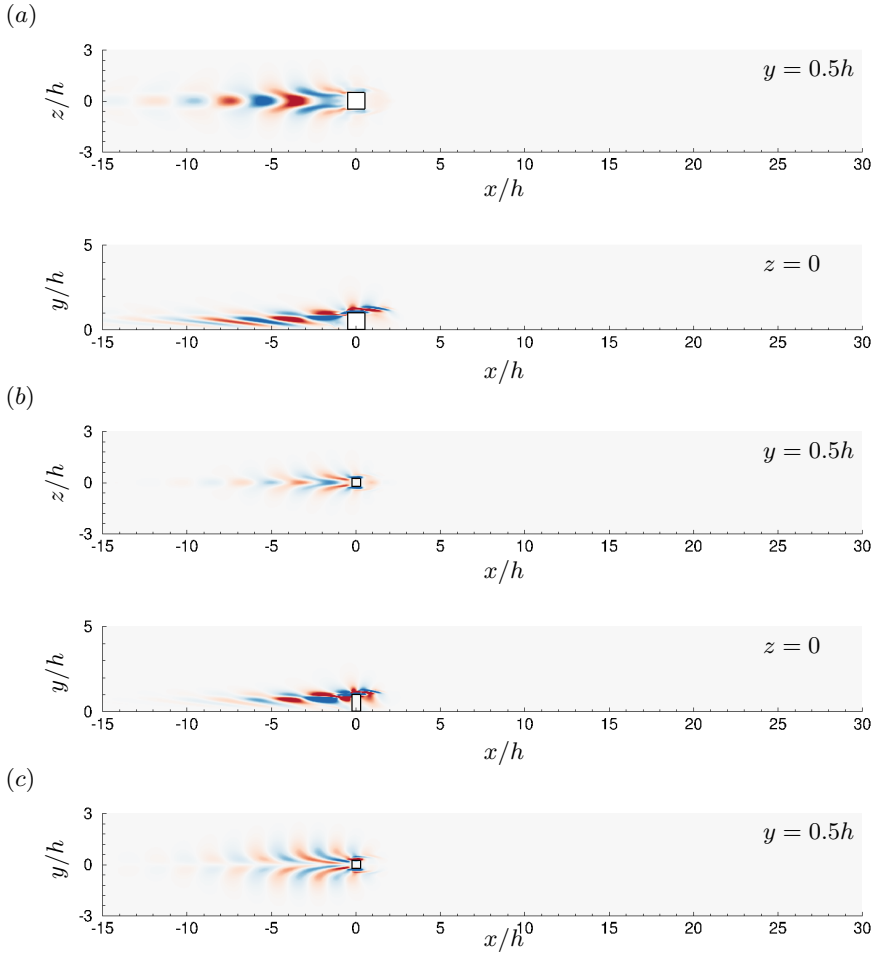
\begin{figure}
% \includegraphics[width=130mm,trim={0.2cm 0.2cm 0.5cm 0cm},clip]{images/adj_combine.pdf}
%  \put(-360,60){$(a)$}
%  \put(-340,28){\rotatebox{90}{$z$}}
%  \put(-180,-5){$x$}
\includegraphics[width=130mm,trim={0.2cm 0.2cm 0.5cm 1cm},clip]{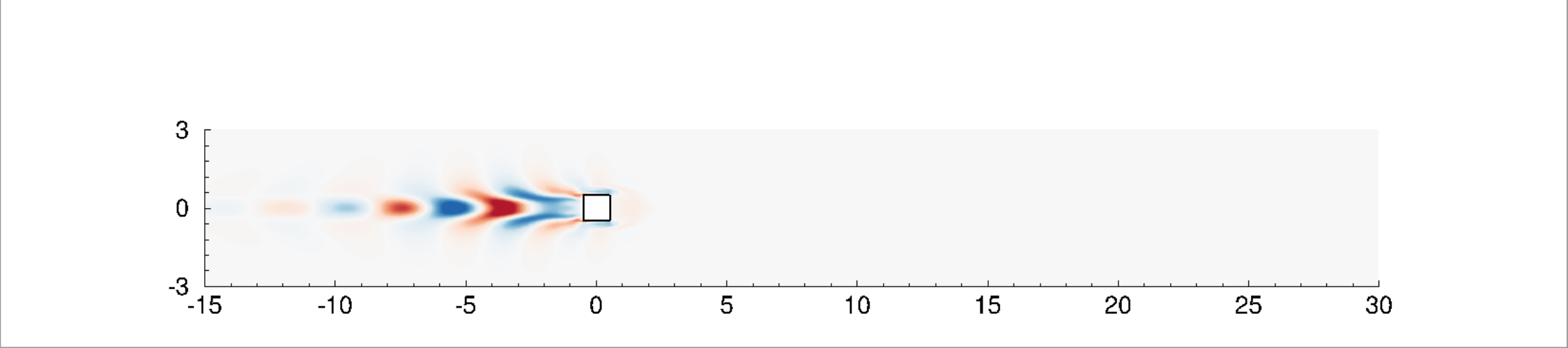}
 \put(-360,60){$(a)$}
\put(-345,23){\rotatebox{90}{$z/h$}}
\put(-180,-5){$x/h$}
\put(-75,40){$\scriptsize{y=0.5h}$}
\hspace{3mm}
\includegraphics[width=130mm,trim={0.2cm 0.2cm 0.5cm 1cm},clip]{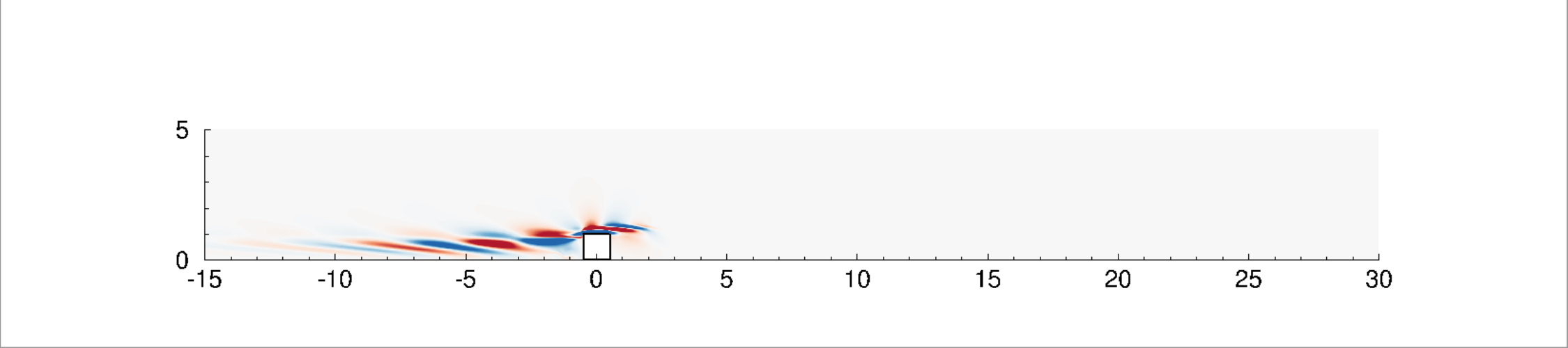}
% \put(-360,60){$(b)$}
\put(-345,25){\rotatebox{90}{$y/h$}}
\put(-180,0){$x/h$}
\put(-75,40){$\scriptsize{z=0}$}
\hspace{3mm}
% \includegraphics[width=130mm,trim={0.2cm 0.2cm 0.5cm 0cm},clip]{images/adj_eta05_combine.pdf}
%  \put(-360,60){$(b)$}
%  \put(-340,28){\rotatebox{90}{$z$}}
%  \put(-180,-5){$x$}
\includegraphics[width=130mm,trim={0.2cm 0.2cm 0.5cm 1cm},clip]{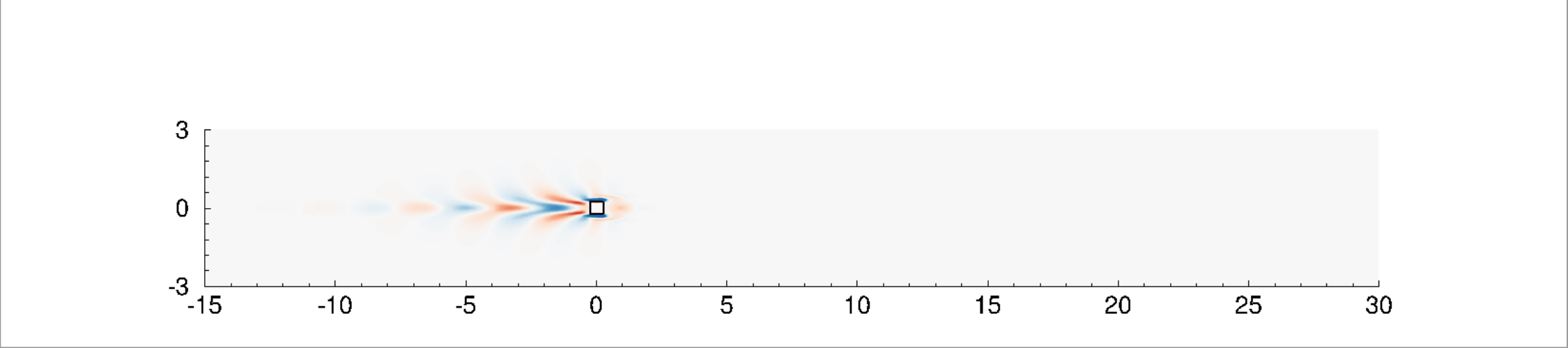}
 \put(-360,60){$(b)$}
\put(-345,23){\rotatebox{90}{$z/h$}}
\put(-180,-5){$x/h$}
\put(-75,40){$\scriptsize{y=0.5h}$}
\hspace{3mm}
\includegraphics[width=130mm,trim={0.2cm 0.2cm 0.5cm 1cm},clip]{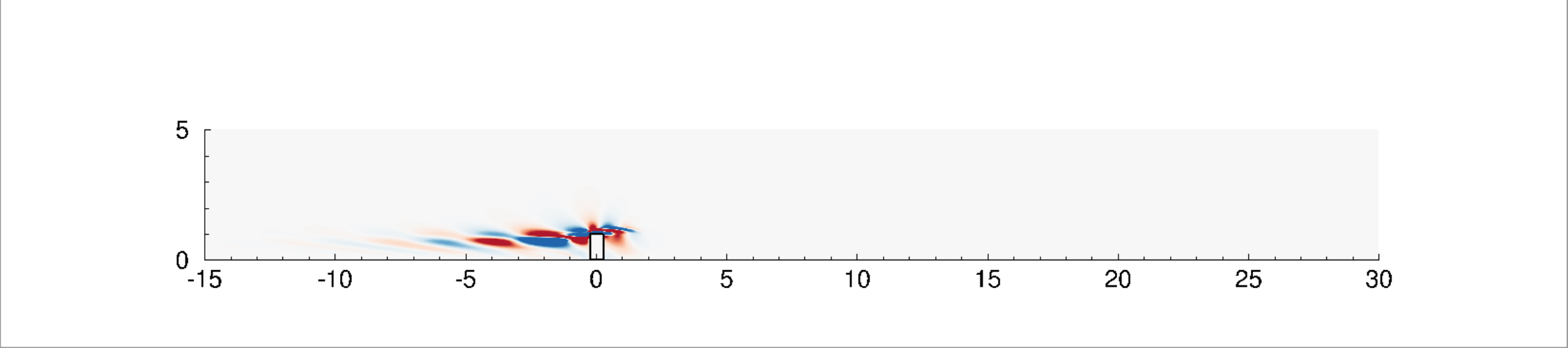}
% \put(-360,60){$(b)$}
\put(-345,25){\rotatebox{90}{$y/h$}}
\put(-180,0){$x/h$}
\put(-75,40){$\scriptsize{z=0}$}
\hspace{3mm}
\includegraphics[width=130mm,trim={0.2cm 0.2cm 0.5cm 1cm},clip]{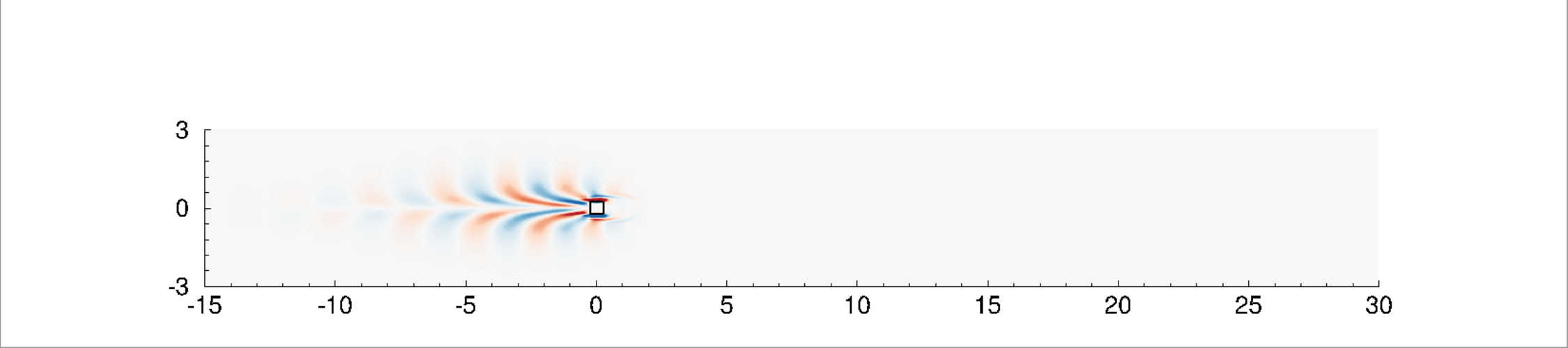}
 \put(-360,60){$(c)$}
\put(-345,25){\rotatebox{90}{$y/h$}}
\put(-180,0){$x/h$}
\put(-75,40){$\scriptsize{y=0.5h}$}
 \caption{Contour plots of $(a)$ the leading adjoint varicose mode for Case ($Re_h,\eta$)=($600,1$), $(b)$ the leading adjoint varicose mode and $(c)$ the leading adjoint sinuous mode for Case ($Re_h,\eta$)=($800,0.5$) from the top view at slice $y=0.5h$ and the side view at slice $z=0$ . } 
\label{fig:adjoint_re600}
\end{figure}
%/LNS_solver/laminar_BL_Diaz_Re600_medium_ncv60_ncd_adjoint; laminar_BL_eta05_sfd_adjoint

\begin{table}
\begin{center}
\def~{\hphantom{0}}
    \begin{tabular}{lccc}
    ($Re_h,\eta$) & ($600,1$) & ($800,0.5$) & ($800,0.5$)\\
    mode & varicose & varicose & sinuous \\
    Direct & $0.1107 \pm i1.1213$ & $0.2801 \pm i1.4831$ & $0.1468 \pm i1.7996$\\
    Adjoint & $0.1110 \pm i1.1212$ & $0.2803 \pm i1.4831$ & $0.1469 \pm i1.7995$\\
    \end{tabular}
    \caption{\label{tab:adjoint} The comparison of the leading eigenvalues of direct and adjoint modes for Cases ($Re_h,\eta$)=($600,1$) and ($Re_h,\eta$)=($800,0.5$).}
    \end{center}
\end{table}

\begin{figure}
\includegraphics[width=70mm,trim={0.2cm 0.2cm 0.5cm 2.5cm},clip]{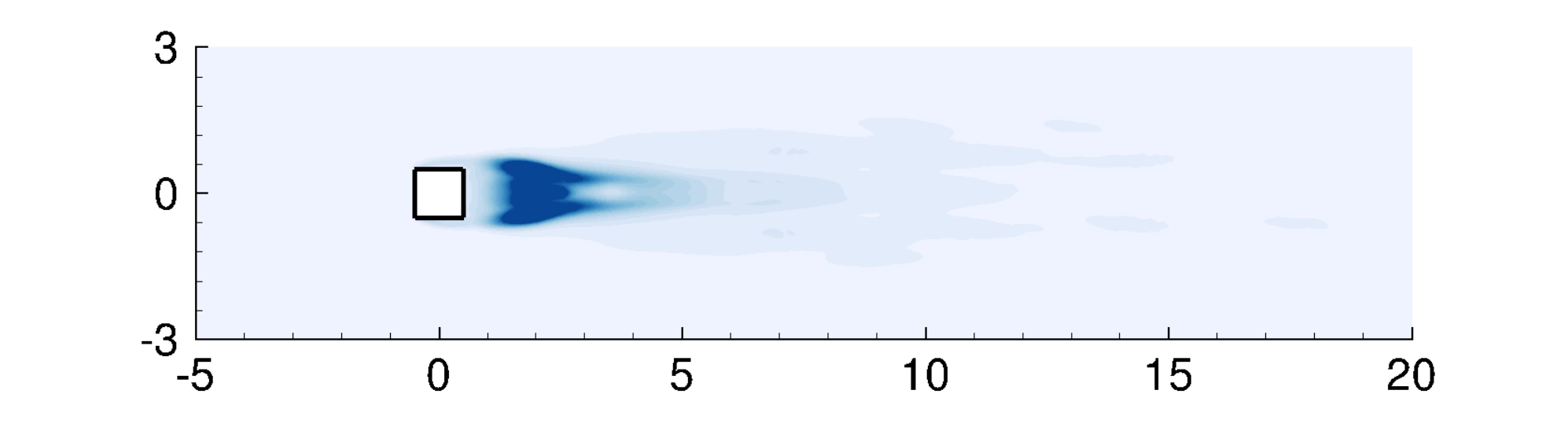} \put(-200,65){$(a)$}
 \put(-180,65){$\eta=1,Re_h=600$, varicose}
\put(-193,25){\rotatebox{90}{$z/h$}}
\put(-106,-5){$x/h$}
\includegraphics[width=70mm,trim={0.2cm 0.2cm 0.5cm 0.5cm},clip]{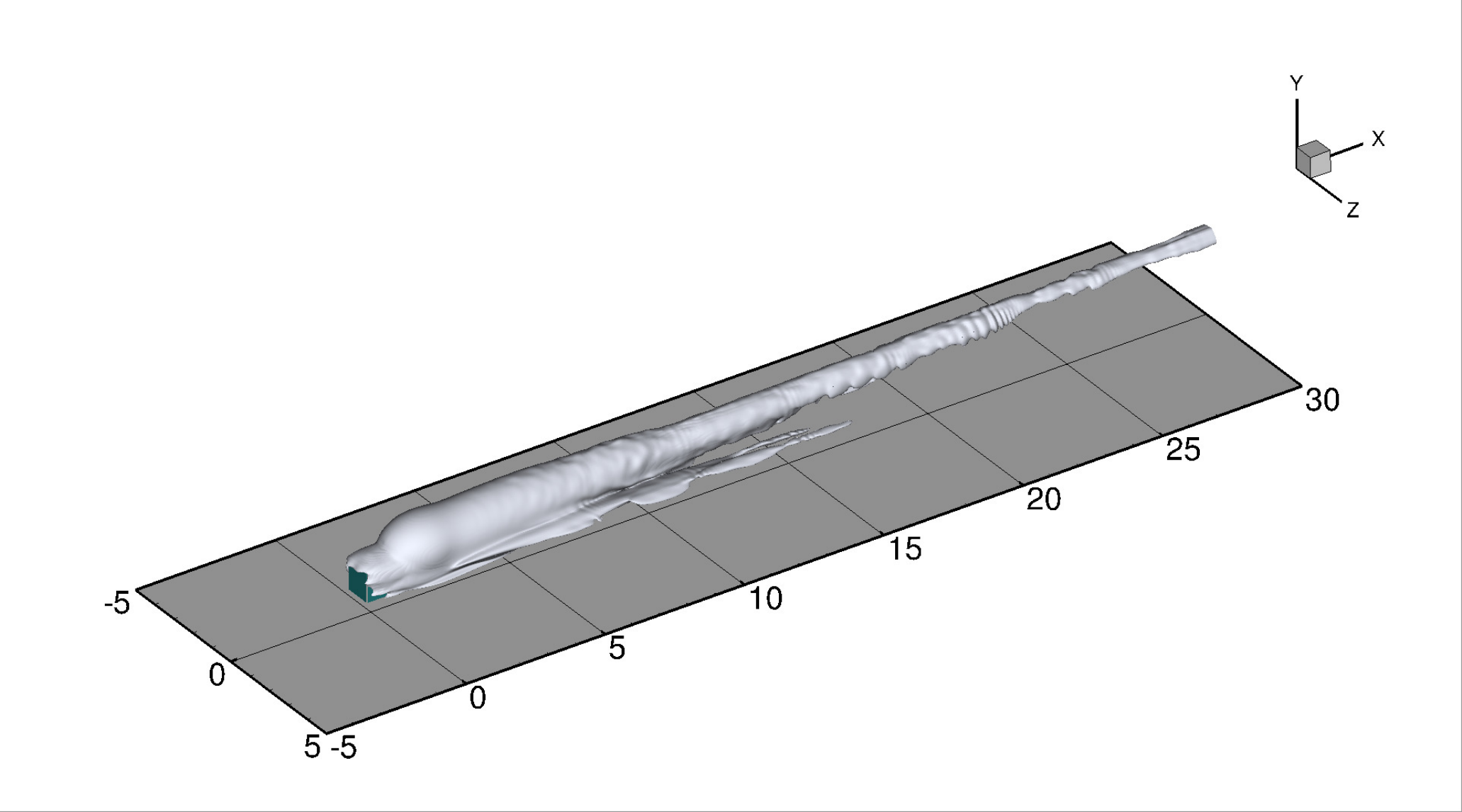}
\put(-182,5){\rotatebox{90}{$z/h$}}
\put(-85,20){$x/h$}
\hspace{3mm}
\includegraphics[width=70mm,trim={0.2cm 0.2cm 0.5cm 2.5cm},clip]{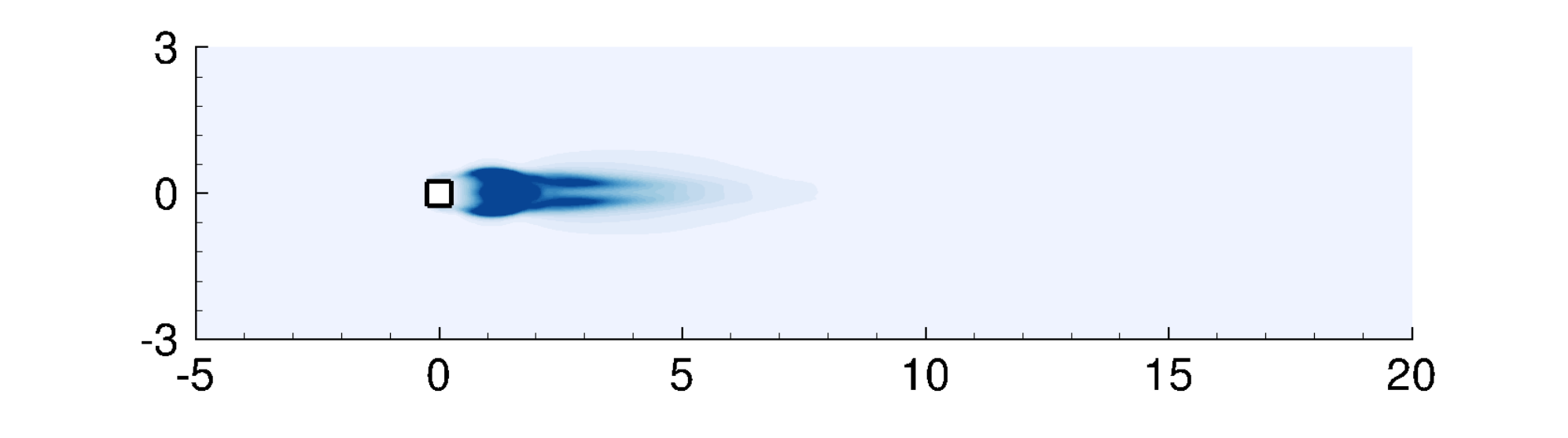} \put(-200,65){$(b)$}
 \put(-180,65){$\eta=0.5,Re_h=800$, varicose}
\put(-193,25){\rotatebox{90}{$z/h$}}
\put(-106,-5){$x/h$}
\includegraphics[width=70mm,trim={0.2cm 0.2cm 0.5cm 3.2cm},clip]{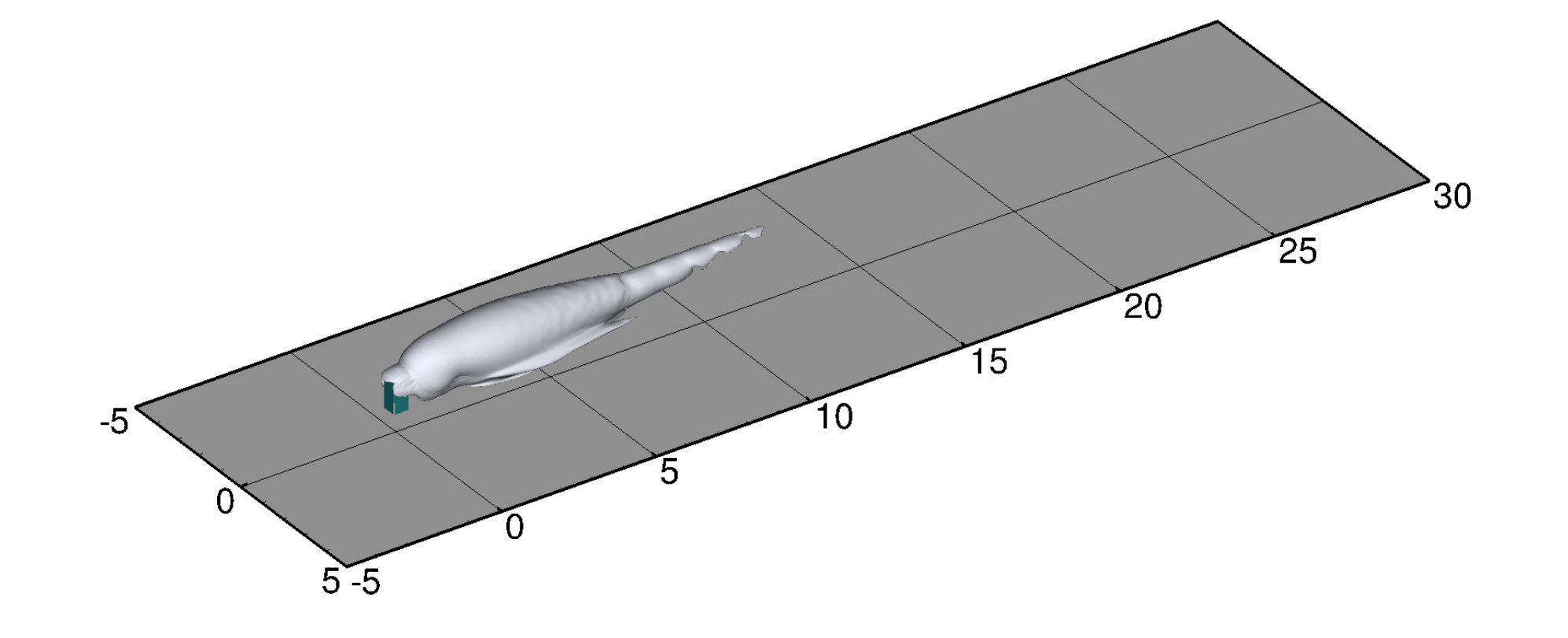}
\put(-182,5){\rotatebox{90}{$z/h$}}
\put(-85,20){$x/h$}
\hspace{3mm}
\includegraphics[width=70mm,trim={0.2cm 0.2cm 0.5cm 2.5cm},clip]{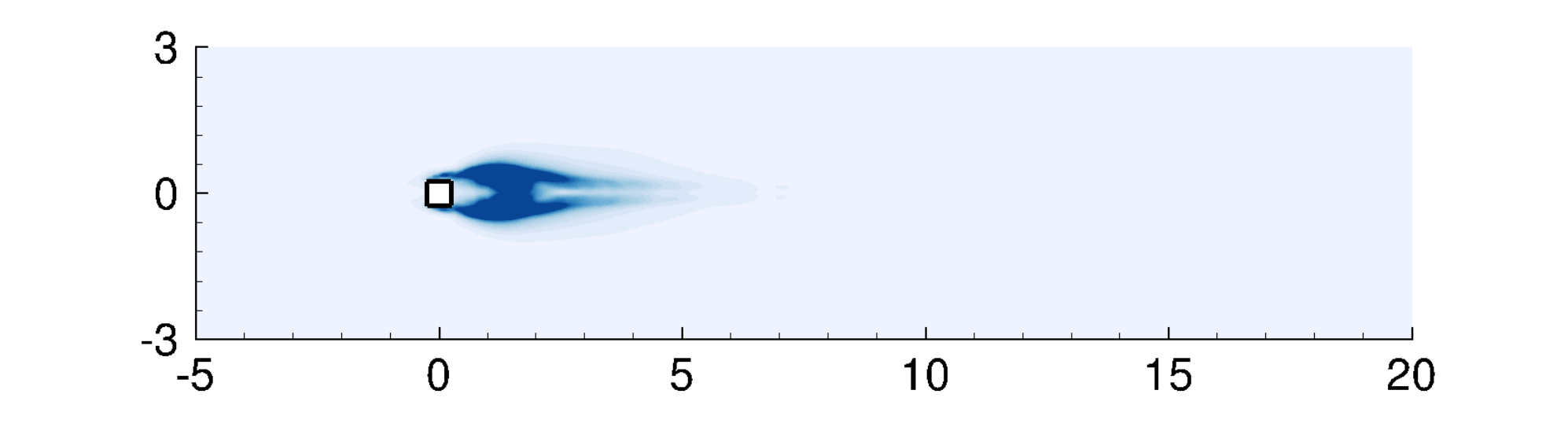}
\put(-200,65){$(c)$}
 \put(-180,65){$\eta=0.5,Re_h=800$, sinuous}
\put(-193,25){\rotatebox{90}{$z/h$}}
\put(-106,-5){$x/h$}
\includegraphics[width=70mm,trim={0.2cm 0.2cm 0.5cm 3.2cm},clip]{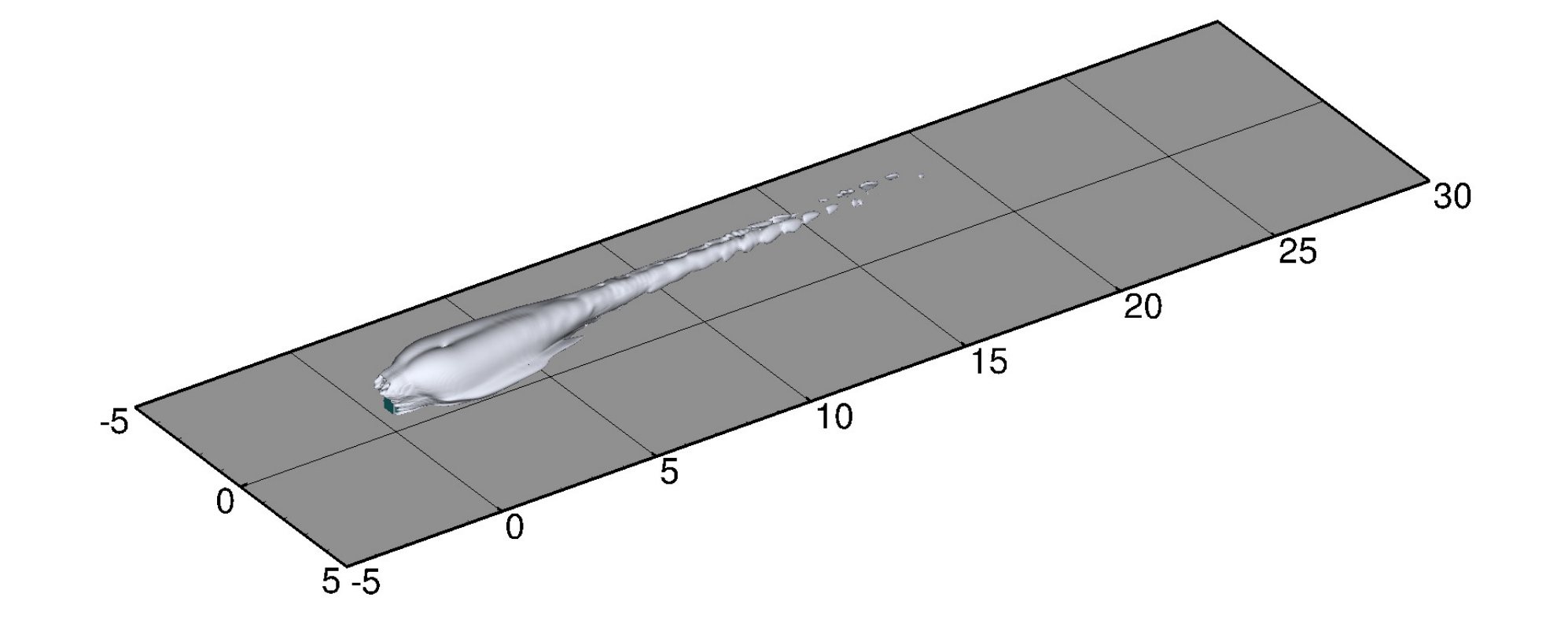}
\put(-182,5){\rotatebox{90}{$z/h$}}
\put(-85,20){$x/h$}
 \caption{Contour plots of the wavemaker from the top views at slice $y=0.5h$ (left) and isosurfaces of the wavemaker (right), for $(a)$ the leading varicose mode in Case ($Re_h,\eta$)=($600,1$), $(b)$ the leading varicose  and $(c)$ the leading sinuous modes in Case ($Re_h,\eta$)=($800,0.5$). The contour plots are displayed with a value of 0.03.} 
\label{fig:wave_3d}
\end{figure}
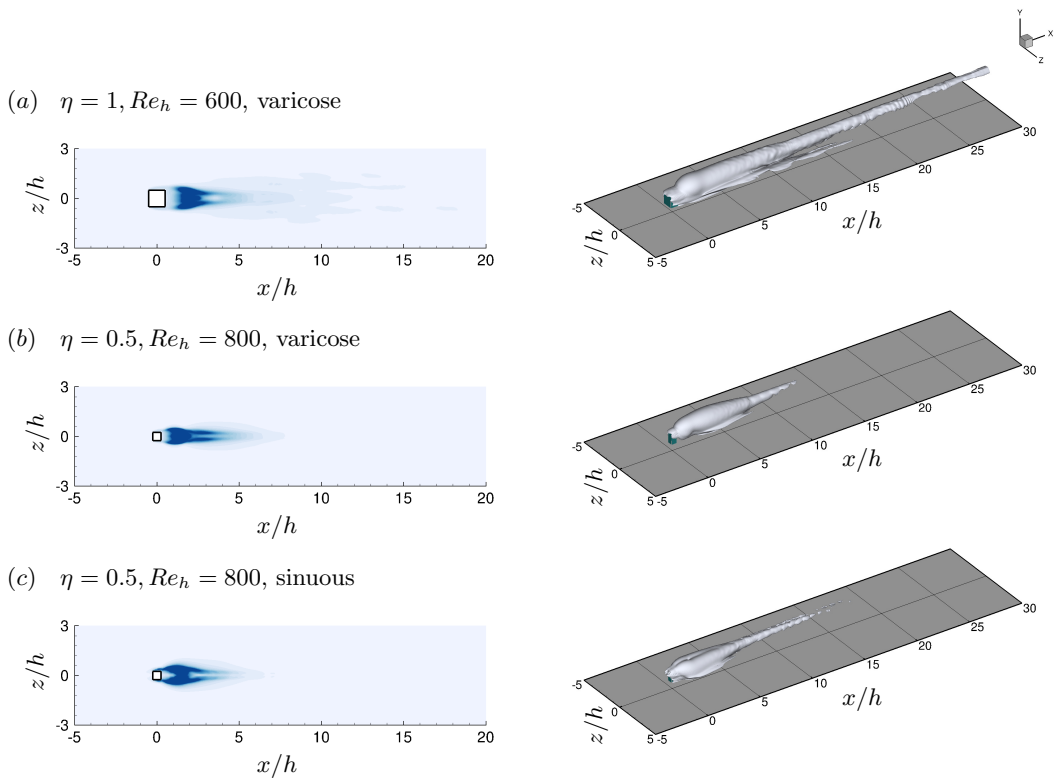
%/project/maxx0242/wavemaker
%ncd_data: eta=1,Re600: laminar_BL_Diaz_Re600_medium_ncv60_ncd_direct (base flow from laminar_BL_Diaz_Re600_double_iny_sfd_imp_correct_BC), laminar_BL_Diaz_Re600_medium_ncv60_ncd_adjoint_test2
%ncd_data: eta=0.5,Re800: laminar_BL_eta05_Re800_sfd, laminar_BL_eta05_Re800_adjoint.

The adjoint perturbation velocity field highlights the most receptive regions to momentum forcing%, which provides important information on the regions to trip the flow
. The leading adjoint eigenvalues are computed for Cases ($Re_h,\eta$)=($600,1$) and ($Re_h,\eta$)=($800,0.5$). The results show good agreement with their associated direct eigenmode counterpart in table \ref{tab:adjoint}. The streamwise velocity component of the leading adjoint modes is depicted in figure \ref{fig:adjoint_re600}. The adjoint modes are located immediately upstream of the roughness element as well as on the top edge of the separation region directly above and downstream of the roughness element. %This observation is consistent with \cite{loiseau2014investigation} and \cite{citro2015global}. %The leading varicose modes in the two cases present symmetry with respect to the spanwise mid-plane.
The receptive region for Case ($Re_h,\eta$)=($800,0.5$) is smaller than that for Case ($Re_h,\eta$)=($600,1$) due to the thinner geometry. The adjoint mode is symmetric with respect to the spanwise mid-plane, corresponding to the direct varicose mode, and is anti-symmetric corresponding to the direct sinuous mode.

Due to the large differences between the spatial distribution of direct and adjoint modes, %the non-normality of the global linear evolution operator, %the eigenvalue problem,
neither direct nor adjoint solution alone can describe the whole picture. The product for each $j$th pair of direct and adjoint global modes computed as 
\begin{equation}
W_j(x,y,z) = \frac{||\hat{u}^j||||\hat{u}^{\dagger,j}||}{max(||\hat{u}^j||||\hat{u}^{\dagger,j}||)} ,
\label{eqn:eigen_adjoint}
\end{equation}
determines the region where the eigenvalues of the LNS operator are most sensitive to localized feedback \citep{giannetti2007structural}, - also called the "wavemaker" regions. Locations where $W \approx 1$ are sensitive to localized feedback, corresponding to the instability core. The value of $W$ can be interpreted as quantification of a possible change in the eigenvalues as a result of applied forcing in the given region of the flow \citep{ilak2012bifurcation}. %The value of $W$ can be interpreted as quantification of a possible change in the eigenvalues as a result of applied perturbations in the given region \citep{ilak2012bifurcation}.

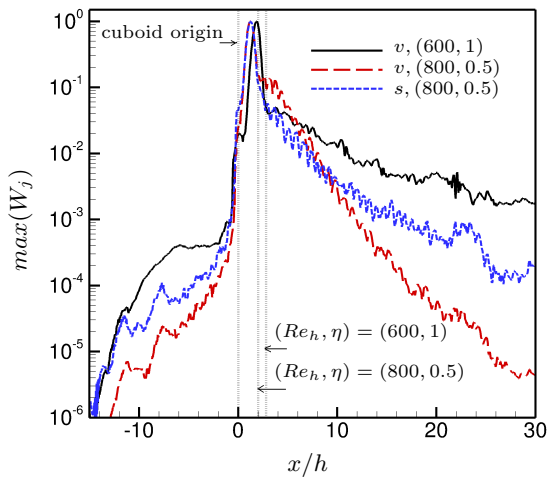
\begin{figure}
\centering
\includegraphics[width=75mm,trim={0.5cm 0.2cm 0.5cm 0.5cm},clip]{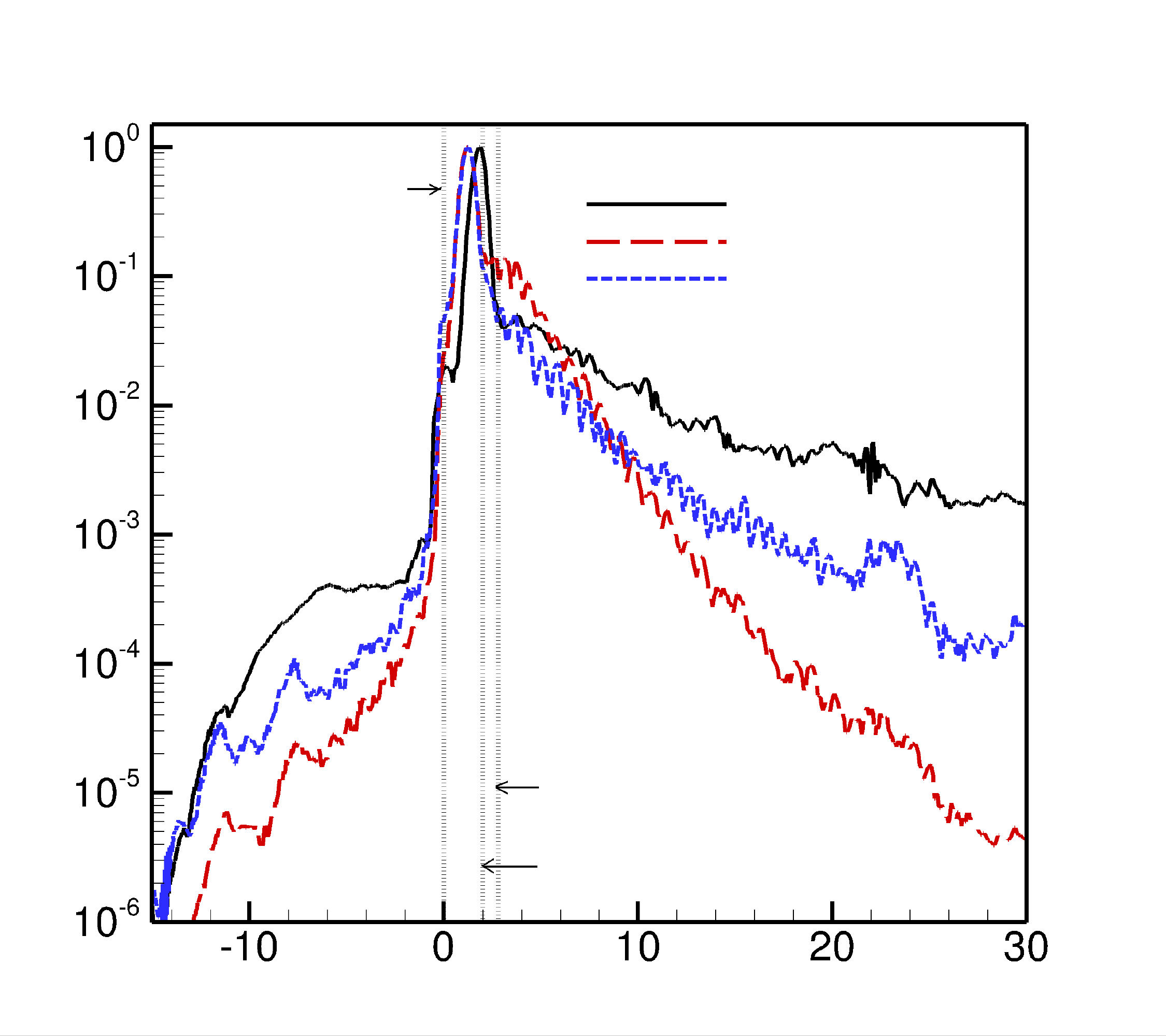}
\put(-220,75){\rotatebox{90}{$max(W_j)$}}
\put(-115,0){$x/h$}
\put(-185,162){\scriptsize{cuboid origin}}
\put(-120,50){\scriptsize{$(Re_h,\eta)=(600,1)$}}
\put(-120,35){\scriptsize{$(Re_h,\eta)=(800,0.5)$}}
\put(-75,157){\scriptsize{$v,(600,1)$}}
\put(-75,149){\scriptsize{$v,(800,0.5)$}}
\put(-75,141){\scriptsize{$s,(800,0.5)$}}
\caption{Streamwise variation of the maximum of the wavemaker for the leading varicose mode in Case ($Re_h,\eta$)=($600,1$), and the leading varicose and sinuous modes in Case ($Re_h,\eta$)=($800,0.5$). The vertical dotted lines denote the locations of cuboid origin and edges of the reversed flow regions.} 
\label{fig:max_wave}
\end{figure}

Figure \ref{fig:wave_3d} depicts the wavemaker regions for the leading modes of Cases ($Re_h,\eta$)=($600,1$) and ($Re_h,\eta$)=($800,0.5$). As shown in figure \ref{fig:wave_3d}$(a)$, the wavemaker for the varicose mode in Case ($Re_h,\eta$)=($600,1$) is prominent on the top edge of the reversed flow region and over an extended region along the central low-speed streak. %Similar phenomenon is observed by \cite{loiseau2014investigation}. 
The maximum value of the wavemaker at each $z-y$ plane is plotted along the streamwise direction in figure \ref{fig:max_wave}. It is shown that the wavemaker has its maximum value within the separation region corresponding to the instability core, and drops to the order of $10^{-1}$ as it passes through the reversed flow region. %These results are consistent with the observations for the varicose mode %of a case with parameters ($Re_h,h/\delta^*,\eta$)=($900,1.45,2$) by \cite{loiseau2014investigation}. }
%The strength of the wavemaker further downstream is about one-tenth of that in the reverse flow region. These results are consistent with the observations for the varicose mode %of a case with parameters ($Re_h,h/\delta^*,\eta$)=($900,1.45,2$) by \cite{loiseau2014investigation}. 
Similar features are also seen for the varicose mode of Case ($Re_h,\eta$)=($800,0.5$), as depicted in figure \ref{fig:wave_3d}$(b)$. However, for the thinner roughness geometry, the spatial growth shows a much smaller streamwise extent and a sharper decline is observed in figure \ref{fig:max_wave}. This could be related with the weaker and shorter central streak for the thinner roughness element.
%This could result from the weaker and shorter central streak observed in the baseflow for the thinner roughness element. %Combining these observations with the production results in \S \ref{production}, it can be concluded that the varicose instability has its core in the downstream separation region where the strong shear layer is induced by the roughness element, and spatially extends along the central low-speed region where the mode extracts most of energy from both the wall-normal and spanwise shear. 
For the sinuous mode in Case ($Re_h,\eta$)=($800,0.5$), figure \ref{fig:wave_3d}$(c)$ shows that the instability core is also located in the downstream reversed flow region, but mainly associated with the lateral sides of the reversed flow region. Unlike the varicose mode that shows one primary sensitivity region, the sinuous mode shows two lateral sensitivity regions. The wavemaker maximum of the sinuous mode shows an immediate drop behind the reversed flow region but a gradual decrease compared to that of the varicose mode. It thus can be concluded that a thinner roughness geometry results in a weaker spatial extension of the wavemaker, and the wavemaker strength of the varicose mode diminishes more quickly than that of the sinuous mode.

\subsection{Nonlinear breakdown}
To understand the effects of $\eta$ and $Re_h$ on transition, and the role of different instability characteristics in the non-linear evolution downstream of the roughness element, DNS are performed for cases with $\eta=1$ and $\eta=0.5$ at different $Re_h$. Longer streamwise domain lengths are used to examine the transition process. 

%\subsubsection{Instantaneous flow field}\label{instant}
\subsubsection{Joint effect of $\eta$ and $Re_{hh}$ on transition}\label{instant}
%Medium domain at Re600: DNS results-vorticity field, pressure field with streamlines, wx at x=0, wz at z=0\\

%Time variation of u' vs. t to show self-sustained oscillations (Re600)\\
%Sinuous: Spanwise oscillations that are antisymmetric with respect to the center line of the streak are evident on the low speed streak. 

%"The shape of the hairpin vortices becomes less pronounced as they travel downstream, which is a manifestation of transition to turbulence. " "The increasing amplitude of the sinuous component with increasing x is likely due to nonlinear interaction of unsteady disturbances, typical for laminar-turbulent breakdown." (Bucci, 2018)

\begin{figure}
\centering
\includegraphics[width=125mm,trim={0.5cm 0.2cm 0.5cm 0.5cm},clip]{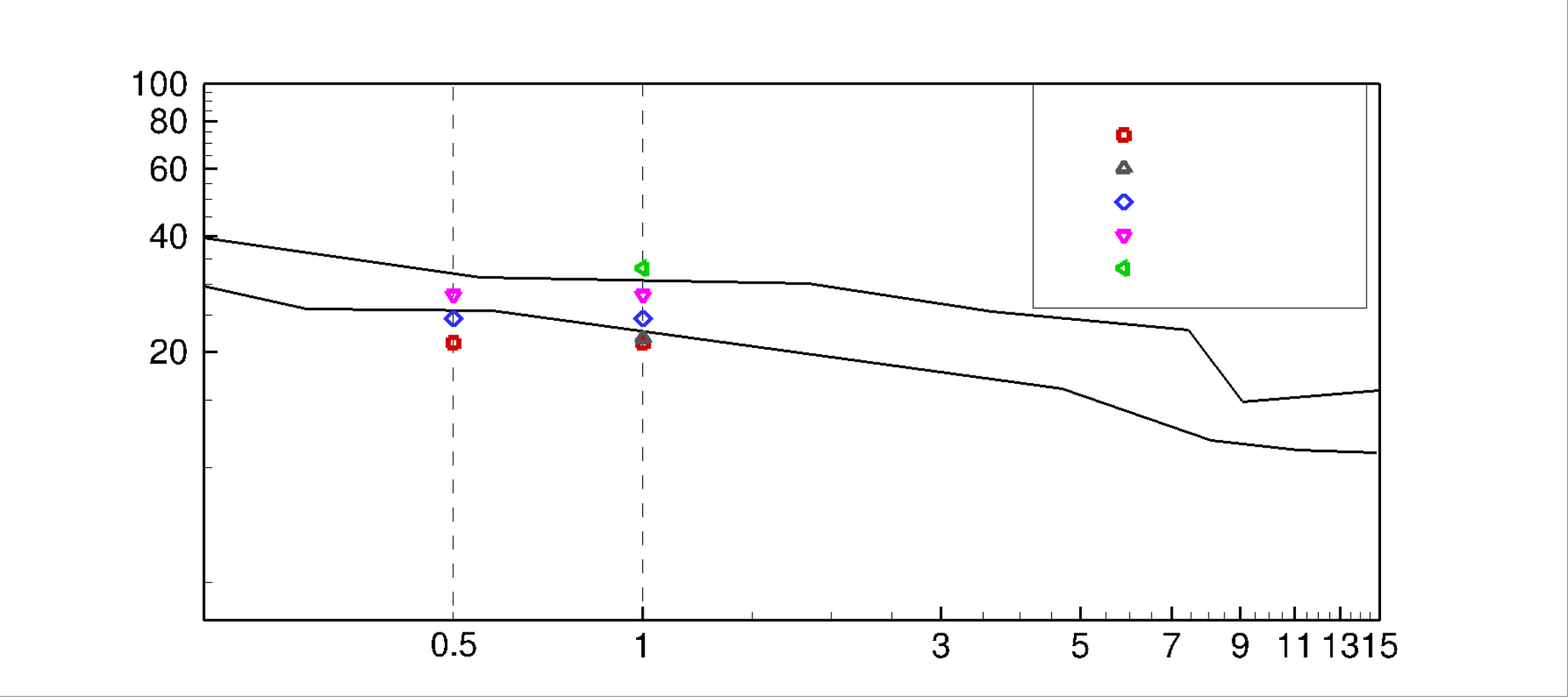}
\put(-350,75){\rotatebox{90}{$Re_{hh}^{1/2}$}}
\put(-165,0){$\eta$}
\put(-85,129){\scriptsize{$Re_h=450$}}
\put(-85,121){\scriptsize{$Re_h=475$}}
\put(-85,113){\scriptsize{$Re_h=600$}}
\put(-85,105){\scriptsize{$Re_h=800$}}
\put(-85,97){\scriptsize{$Re_h=1100$}}
\put(-165,107){(\romannumeral 3)}
\put(-165,82){(\romannumeral 2)}
\put(-165,57){(\romannumeral 1)}
\caption{Comparison to the von Doenhoff-Braslow transition diagram.} 
\label{fig:von_diagram}
\end{figure}
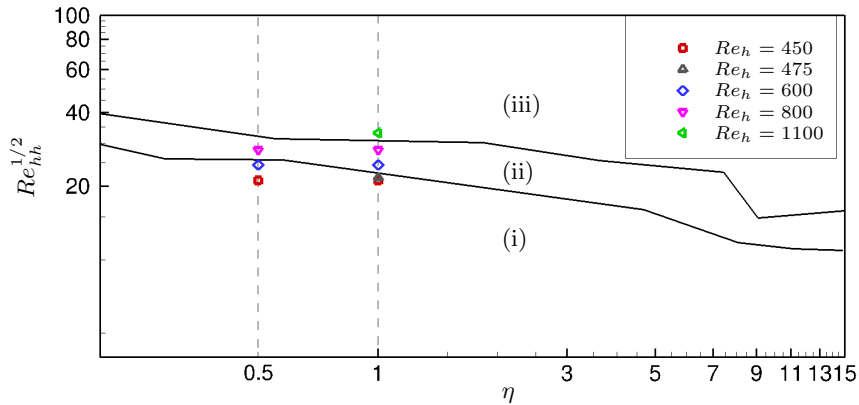

Dependence of the transition process on $\eta$ and $Re_{hh}$ is examined for the present cases by reproducing the von Doenhoff-Braslow transition diagram \citep{von1961effect} in figure \ref{fig:von_diagram}. This transition diagram shows the correlation between $\eta$ and $Re_{hh}$ and provides an approach to predict the transition characteristics. The cases located in Region %\RomanNumeralCaps{1} 
(\romannumeral 1) (below the lower curve) are expected to have a steady wake flow, the cases fitting into Region (\romannumeral 2) (between the lower and upper curves) indicate that the wake flow is unsteady and transition occurs, and the cases situated in Region (\romannumeral 3) (above the upper curve) means that transition occurs immediately downstream of the roughness. %The transition process of the present cases is examined by showing the instantaneous streamwise velocity field in the plane $y=0.5$ in figure \ref{fig:u_contour}. 
The correlation between $\eta$ and $Re_{hh}$ revealed by figure \ref{fig:von_diagram} shows that for roughness elements with $0.6 \le \eta \le 2$, $\eta$ plays a more important role on the onset of unsteadiness than the onset of immediate transition downstream of roughness, while the opposite effect is seen for roughness with $0.3 \le \eta \le 0.6$.

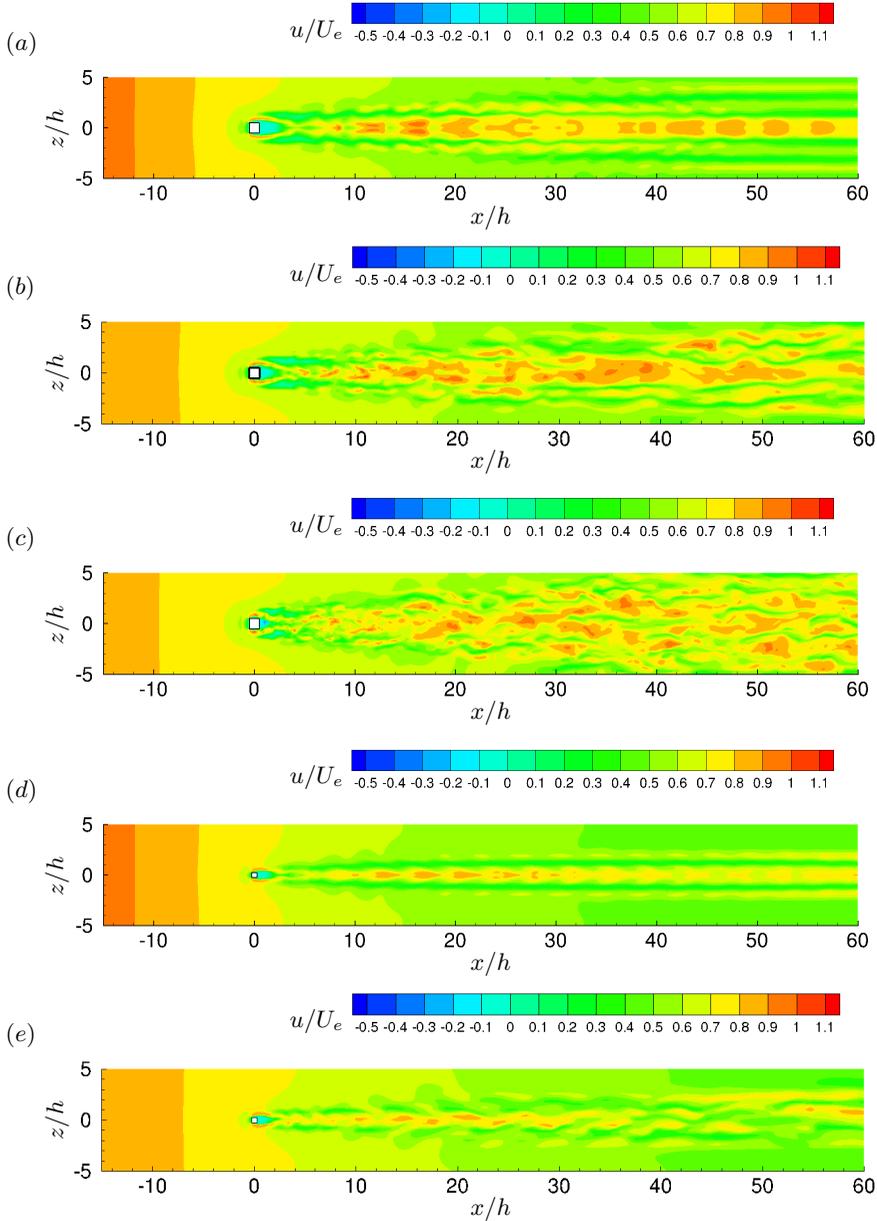
\begin{figure}
\includegraphics[width=130mm,trim={0.2cm 0.2cm 0.5cm 0.5cm},clip]{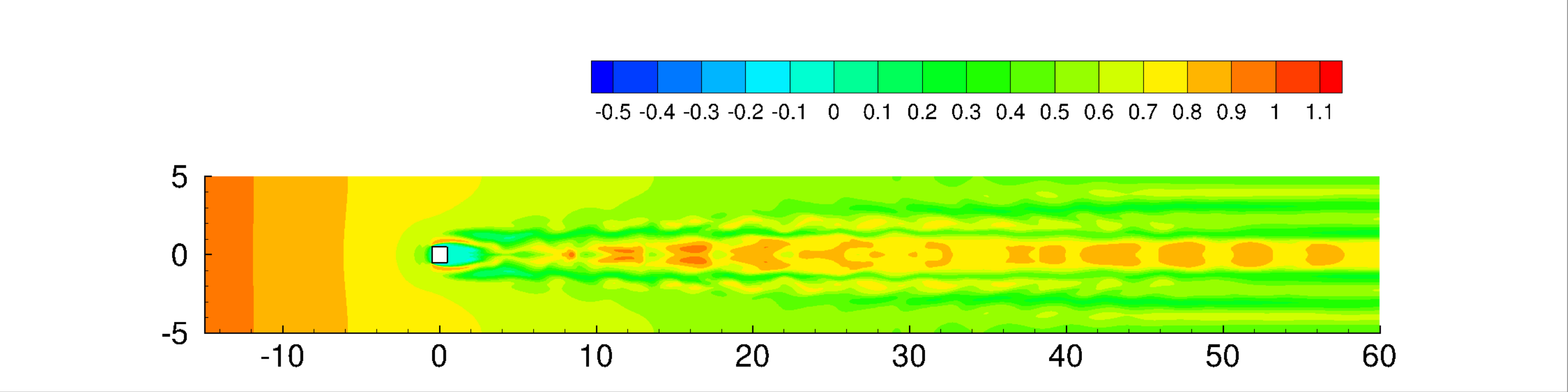}
\put(-360,60){$(a)$}
\put(-345,23){\rotatebox{90}{$z/h$}}
\put(-185,-5){$x/h$}
\put(-253,65){$u/U_e$}
\hspace{3mm}
\includegraphics[width=130mm,trim={0.2cm 0.2cm 0.5cm 0.5cm},clip]{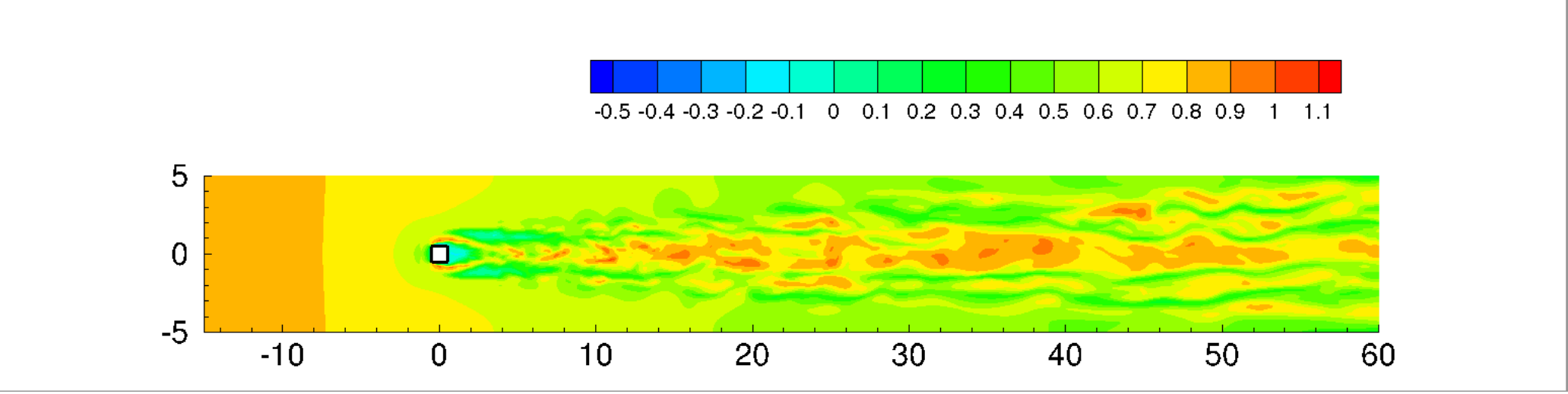}
\put(-360,60){$(b)$}
\put(-345,23){\rotatebox{90}{$z/h$}}
\put(-185,-5){$x/h$}
\put(-253,65){$u/U_e$}
\hspace{3mm}
\includegraphics[width=130mm,trim={0.2cm 0.2cm 0.5cm 0.5cm},clip]{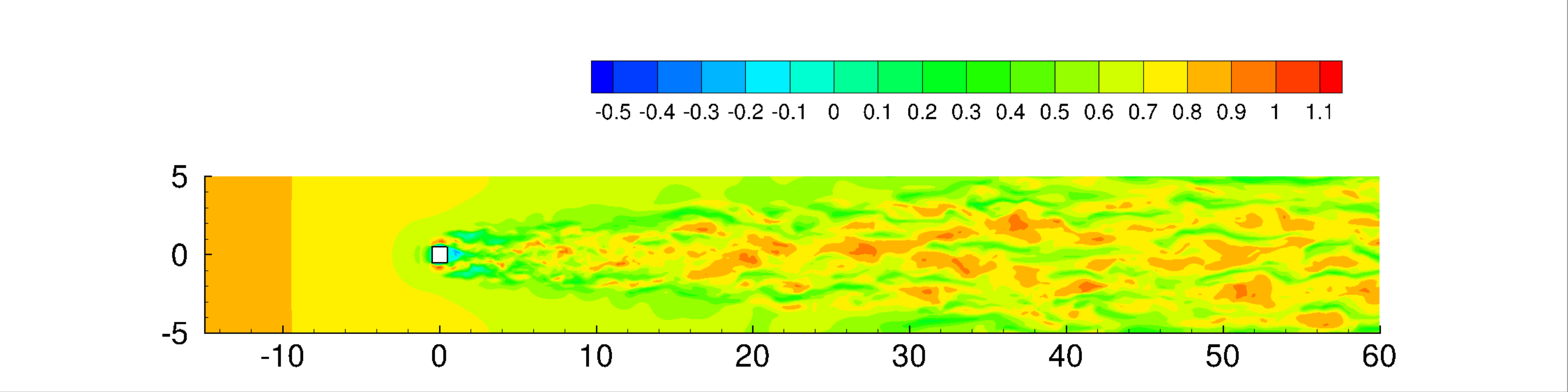}
\put(-360,60){$(c)$}
\put(-345,23){\rotatebox{90}{$z/h$}}
\put(-185,-5){$x/h$}
\put(-253,65){$u/U_e$}
\hspace{3mm}
\includegraphics[width=130mm,trim={0.2cm 0.2cm 0.5cm 0.5cm},clip]{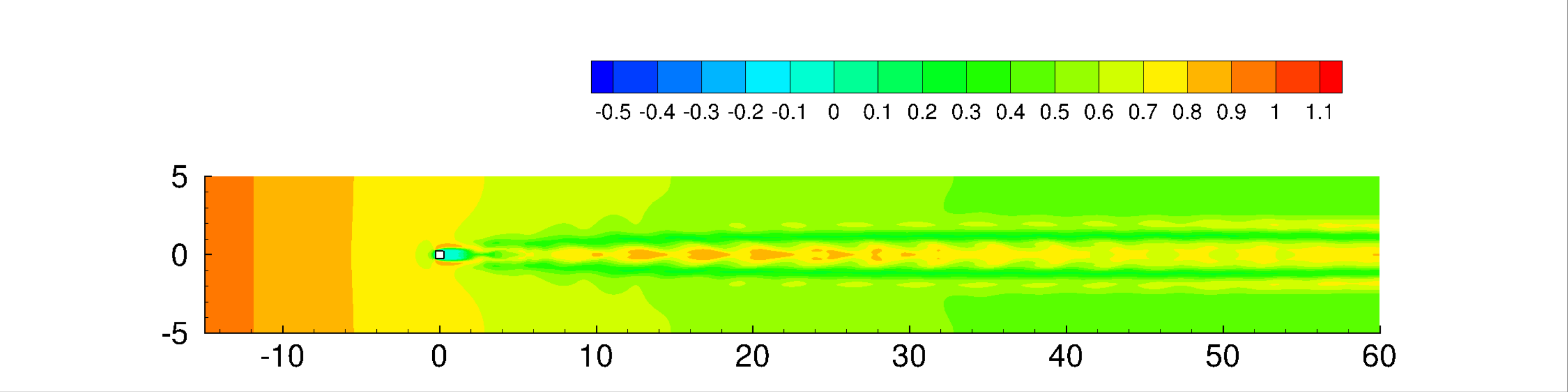}
\put(-360,60){$(d)$}
\put(-345,23){\rotatebox{90}{$z/h$}}
\put(-185,-5){$x/h$}
\put(-253,65){$u/U_e$}
\hspace{3mm}
\includegraphics[width=130mm,trim={0.2cm 0.2cm 0.5cm 0.5cm},clip]{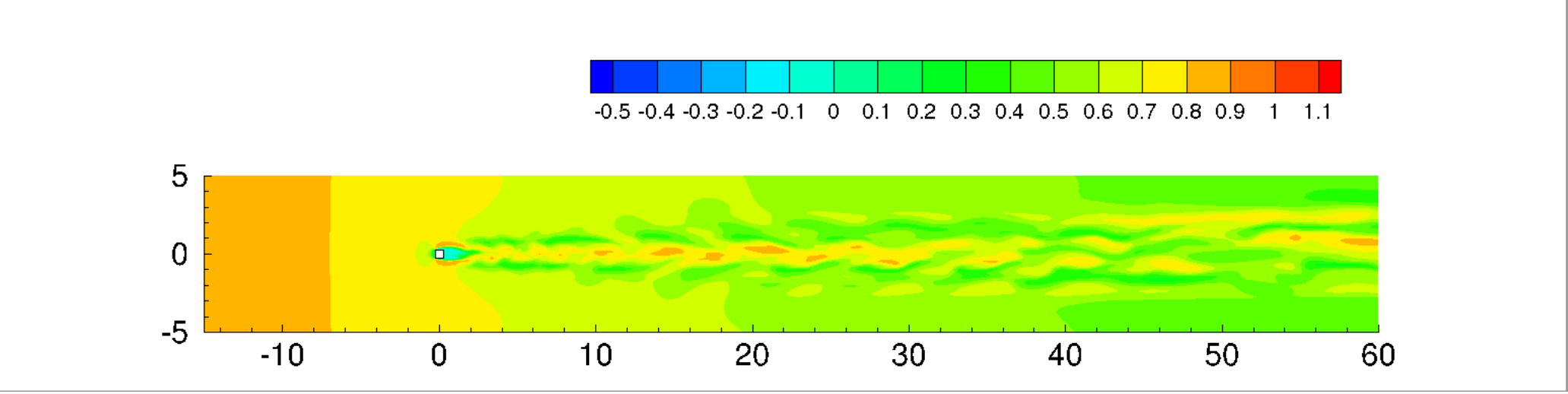}
\put(-360,60){$(e)$}
\put(-345,23){\rotatebox{90}{$z/h$}}
\put(-185,-5){$x/h$}
\put(-253,65){$u/U_e$}
 \caption{Contour plots of instantaneous streamwise velocity field at slice $y=0.5h$ for cases with $\eta=1$ at $(a)$ $Re_h=600$, $(b)$ $Re_h=800$ and $(c)$ $Re_h=1100$, and cases with $\eta=0.5$ at $(d)$ $Re_h=600$ and $(e)$ $Re_h=800$. } 
\label{fig:u_contour}
\end{figure}

%In general, the cases in the present work are well fitted in the von Doenhoff-Braslow transition diagram.
As shown in figure \ref{fig:von_diagram}, for $\eta=1$, the stable cases at $Re_h=450$ and $475$ are located in Region (\romannumeral 1), consistent with a stable wake flow observed in both the global stability and DNS results. The unstable cases at $Re_h=600$ and $800$ are within Region (\romannumeral 2) and the transition process is examined in figures \ref{fig:u_contour}$(a)$ and \ref{fig:u_contour}$(b)$ respectively. For Case ($Re_h,\eta$)=($600,1$), both the near and farther wakes are symmetric with respect to the spanwise mid-plane, corresponding to the varicose mode obtained in the global stability analysis. 

Compared to Case ($Re_h,\eta$)=($600,1$), symmetric fluid motions with smaller length scales are seen in the near-wake region for Case ($Re_h,\eta$)=($800,1$), indicating that non-linear breakdown occurs more closely downstream of the roughness as $Re_h$ increases. The perturbations farther downstream are more prominent than those at $Re_h=600$. Intense shear between streaks results in spanwise oscillations in the farther wake. Although only the varicose global instability is detected for this configuration, a sinuous like breakdown could happen when $Re_h$ is sufficiently high. This sinuous breakdown might be related or subsequently lead to secondary sinuous instabilities observed by \cite{denissen2013secondary} and \cite{vadlamani2018distributed}, which would destabilize the shear layer and promote transition to turbulence. Note that whether or not the unstable cases undergo transition to turbulence is not revealed in this transition diagram since the effects on other configuration parameters, such as spanwise spacing and $Re_{\delta}$, need to be considered. 

Case ($Re_h,\eta$)=($1100,1$) is located in Region (\romannumeral 3), suggesting that immediate transition downstream of the roughness is expected. This is verified in figure \ref{fig:u_contour}$(c)$, as $Re_h$ increases to $1100$, the non-linear breakdown occurs immediately downstream of the roughness. Also, streamwise streaks can be identified farther downstream, implying that transition to fully-developed turbulence might have occurred.

%msi:laminar_BL_Diaz_Re1100_DNS_medium_domain; lamianr_BL_eta05_Re600_DNS_medium_domain; /project/maxx0242/Q_criterion/data_files_Lx75/re800_eta05

For $\eta=0.5$, the stable case at $Re_h=450$ and the unstable case at $Re_h=800$ are situated in Regions (\romannumeral 1) and (\romannumeral 2) respectively, as shown in figure \ref{fig:von_diagram}, which is consistent with the global stability results. The unstable case at $Re_h=600$ is slightly off from Region (\romannumeral 2), which could be due to the fact that the present roughness has sharp edges and would have a lower $Re_{hh}$ for unsteadiness to occur than other smoother roughness elements. 

The effect of small $\eta$ and $Re_h$ dependence on the transition process is examined for $\eta=0.5$. Figure \ref{fig:u_contour}$(d)$ shows that the wake flow at $Re_h=600$ displays a thinner symmetric central streak compared to that of $\eta=1$. There are no sinuous oscillations observed, corresponding to the global stability results that the sinuous mode is marginally stable at $Re_h=600$. As $Re_h$ increases to $800$ (figure \ref{fig:u_contour}$(e)$), the anti-symmetric oscillations in the spanwise direction become evident in both the near and farther wakes, associated with the more prominent sinuous instability and indicating a persistent effect of sinuous oscillations farther downstream. In summary, cases with two thin roughness geometries in the present work are well fitted into the classification by this transition diagram, and the interplay between $\eta$ and $Re_{hh}$ leads to different flow behavior in the transition process, in accordance with the global instability characteristics.

% Figure \ref{fig:u_contour} shows the instantaneous streamwise velocity field in the plane $y=0.5$ for cases with $\eta=1$ at $Re_h=600$, $800$ and $1100$, and cases with $\eta=0.5$ at $Re_h=600$ and $800$. For $\eta=1$, figure \ref{fig:u_contour}$(a)$ shows that the wake flow is symmetric with respect to the spanwise mid-plane, corresponding to the varicose mode obtained in the global stability analysis. In contrast to $Re_h=600$, figure \ref{fig:u_contour}$(b)$ demonstrates fluid motions with smaller length scales in the near-wake region at $Re_h=800$, indicating that non-linear breakdown occurs more closely downstream of the roughness element as $Re_h$ increases. The perturbations further downstream are also more prominent than those at $Re_h=600$. Figure \ref{fig:u_contour}$(c)$ shows that as $Re_h$ increases to $1100$, the non-linear breakdown occurs immediately downstream of the roughness. Streamwise streaks can be identified farther downstream, implying that the transition to the fully-developed turbulence might have occurred. 

%Different flow features are observed for $\eta=0.5$. Figure \ref{fig:u_contour}$(d)$ shows that the wake flow at $Re_h=600$ displays a thinner symmetric central streak compared to that of $\eta=1$. There are no sinuous oscillations observed, consistent with the global stability results that the sinuous mode is marginally stable at $Re_h=600$. As $Re_h$ increases to $800$, the anti-symmetric spanwise oscillations are evident in the wake flow, corresponding to the more prominent sinuous instability from the global stability results.

\subsubsection{Non-linear evolution of vortical structures}\label{dynamic}
Cases ($Re_h,\eta$)=($600,1$) and ($Re_h,\eta$)=($800,0.5$) are examined to better understand how non-linear saturations are triggered and how vortical structures develop following different types of global instability.

\begin{figure}
\includegraphics[width=130mm,trim={0.2cm 0.2cm 0.5cm 0cm},clip]{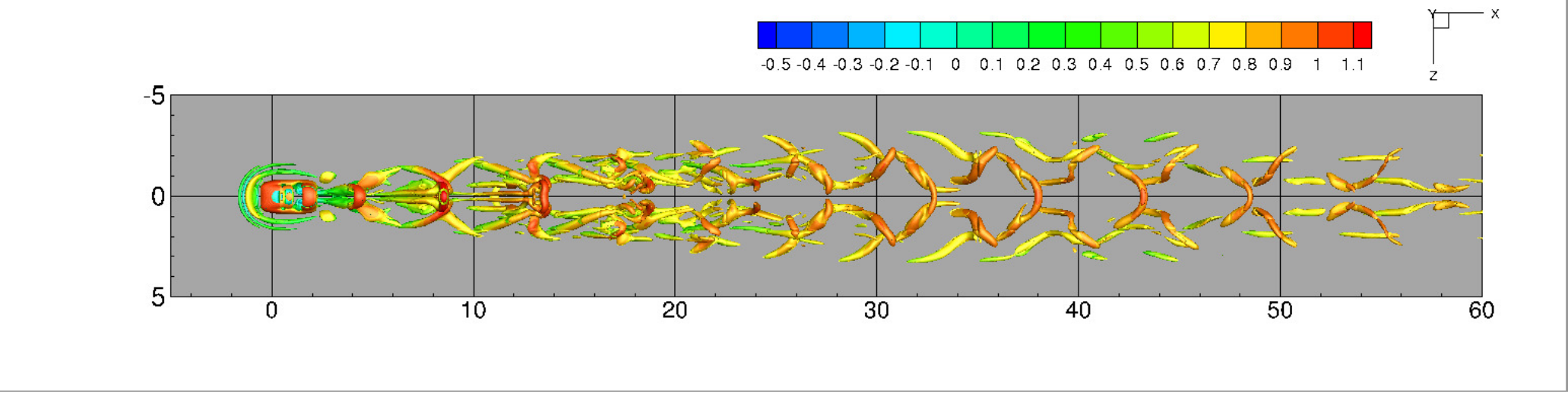}
\put(-360,80){$(a)$}
\put(-210,75){\scriptsize{$u/U_e$}}
\put(-350,37){\rotatebox{90}{$z/h$}}
\put(-175,3){$x/h$}
\hspace{3mm}
 \includegraphics[width=130mm,trim={0.2cm 0.2cm 0.5cm 0cm},clip]{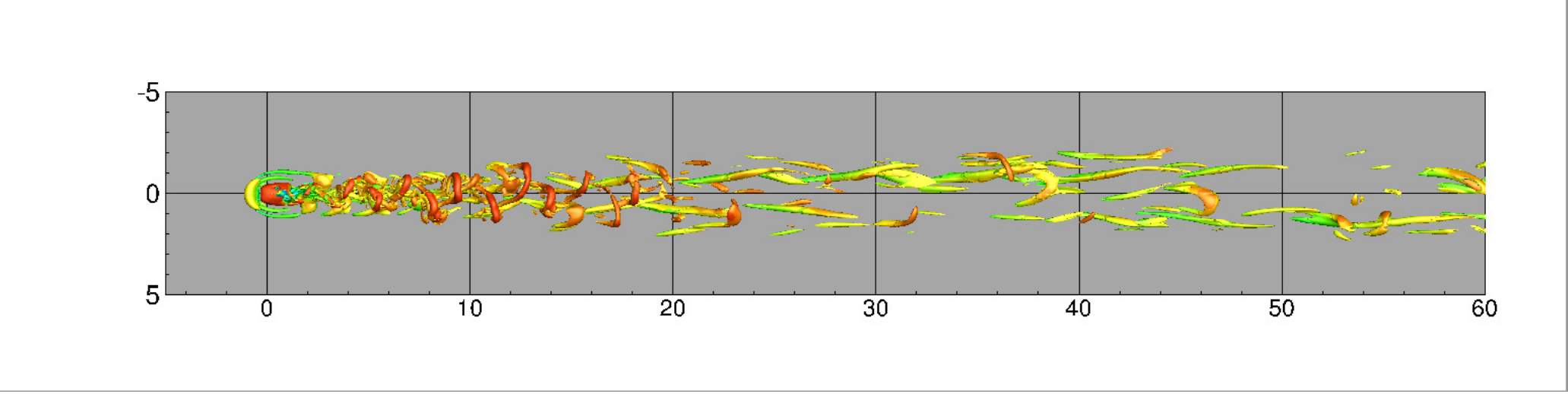}
\put(-360,80){$(b)$}
%\put(-335,115){$u/U_e$}
\put(-350,37){\rotatebox{90}{$z/h$}}
\put(-175,3){$x/h$}
 \caption{Visualizations of instantaneous vortical structures for $(a)$ Case $(Re_h,\eta)=(600,1)$ by isocontours of $Q=0.1U_e^2/h^2$ and $(b)$ Case $(Re_h,\eta)=(800,0.5)$ by isocontours of $Q=0.05U_e^2/h^2$, colored with streamwise velocity. } 
\label{fig:Q_criterion}
\end{figure}
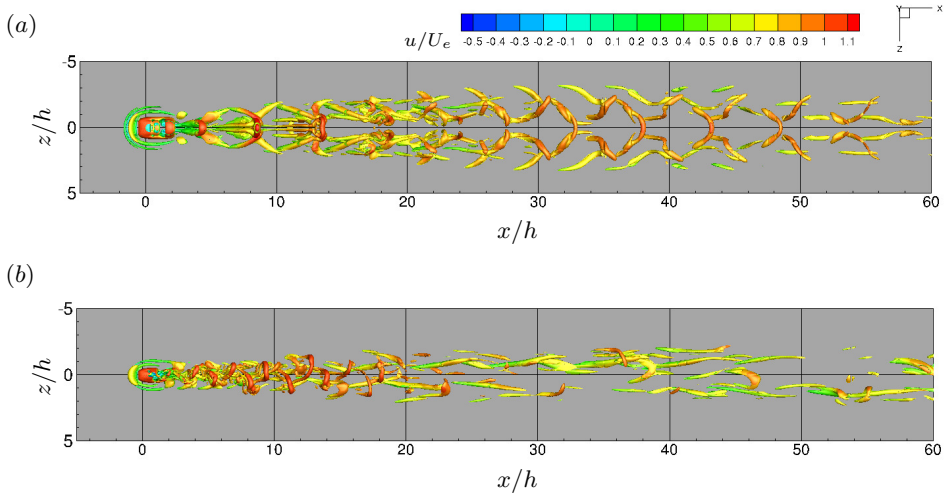

Figure \ref{fig:Q_criterion} shows the vortical structures using isocontours of Q \citep{chong1990general}. For Case $(Re_h,\eta)=(600,1)$, as shown in figure \ref{fig:Q_criterion}$(a)$, the vortical motions induced by side edges of the cube interact with the shear layer generated over the cube, giving birth to the hairpin vortices. Both the primary hairpin vortices and secondary wall-attached vortices are observed downstream of the roughness element. These vortical structures are amplified and fragment into small structures beyond $x=18h$, %indicating the non-linear saturation becomes stronger from this location
which is a manifestation of transition. The vortex "head" and "legs" are advected, stretched downstream, and diminish after $x=44h$. Even though the unsteadiness is noticeable and the unstable nature of the longitudinal streaks is revealed at $Re_h=600$, transition to turbulence may not happen since $Re_{\delta}$ for the boundary layer is low in the present case. %This could depend on other parameters, such as $Re_{\delta}$, the spanwise spacing, etc. 

In contrast, Case ($Re_h,\eta$)=($800,0.5$) shows different vortical motions in the wake flow in figure \ref{fig:Q_criterion}$(b)$. As the horseshoe vortices wrap around the roughness element and interact with the shear layer, anti-symmetric distribution of vortical structures is seen in the immediate vicinity, downstream of the roughness. This indicates that sinuous oscillations occur just downstream of the roughness element%, corresponding to the wavemaker result that the instability core is within the reverse flow region
. %The interactions between the horseshoe vortices and the shear layer give birth to the hairpin vortices which exhibit a sinuous wiggling. 
The primary hairpin vortices modulated by the sinuous oscillations of the central streak also exhibit a sinuous wiggling. %Similar behavior in this near-wake region has been observed by \cite{loiseau2014investigation} for a different set of configuration parameters with a dominant sinuous instability. 
While the hairpin vortices are advected and stretched downstream, and break down at about $x=18h$, the sinuous wiggling of the streaks continues farther downstream. %, corresponding to the longer streamwise extent of the sinuous mode observed in the global stability results. 
It is clear that for Case ($Re_h,\eta$)=($800,0.5$), both the varicose and sinuous instabilities have influences on the behavior and development of vortical structures. The varicose instability is associated with the hairpin vortices, thus has a limited streamwise extent as the hairpin vortices break down. The sinuous instability originates from the immediate vicinity of the roughness, is correlated with the wiggling of the central streak, and has a more persistent effect than the varicose instability on the wake flow.

%/project/maxx0242/Q_criterion

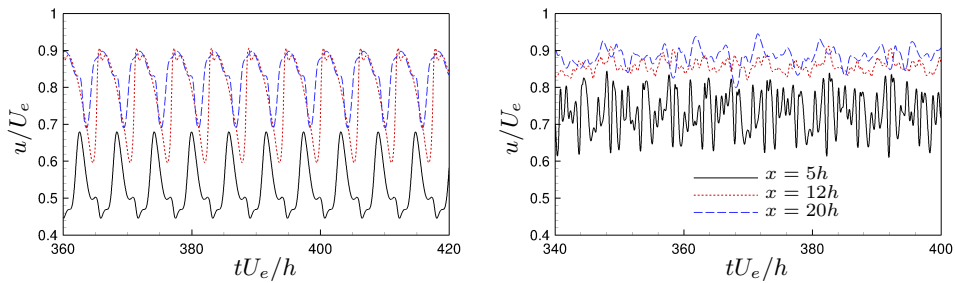
\begin{figure}
\centering
\includegraphics[width=65mm,trim={0.5cm 0.2cm 0.5cm 0cm},clip]{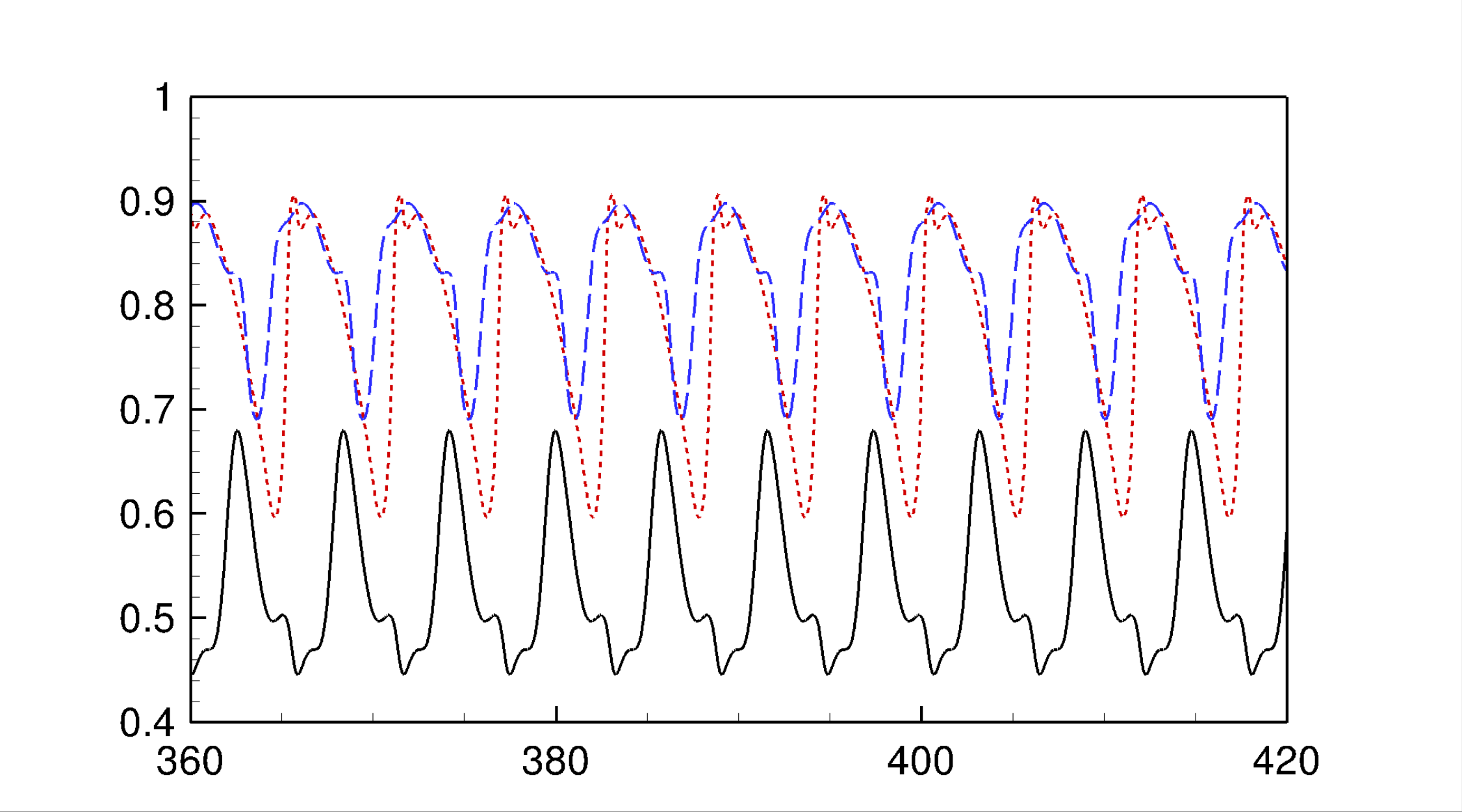}
\put(-185,40){\rotatebox{90}{$u/U_e$}}
\put(-100,-5){$tU_e/h$}
%\put(-190,90){$(a)$}
%\hspace{3mm}
% \includegraphics[width=65mm,trim={0.5cm 0.2cm 0.5cm 0cm},clip]{images/probe_u_re1100.pdf}
% \put(-185,45){\rotatebox{90}{$u$}}
% \put(-90,-5){$t$}
% %\put(-190,90){$(b)$}
% \hspace{5mm}
% \includegraphics[width=65mm,trim={0.5cm 0.2cm 0.5cm 0cm},clip]{images/probe_u_re600_eta05.pdf}
% \put(-185,45){\rotatebox{90}{$u$}}
% \put(-90,-5){$t$}
%\put(-190,90){$(a)$}
%\hspace{3mm}
\includegraphics[width=65mm,trim={0.5cm 0.2cm 0.5cm 0cm},clip]{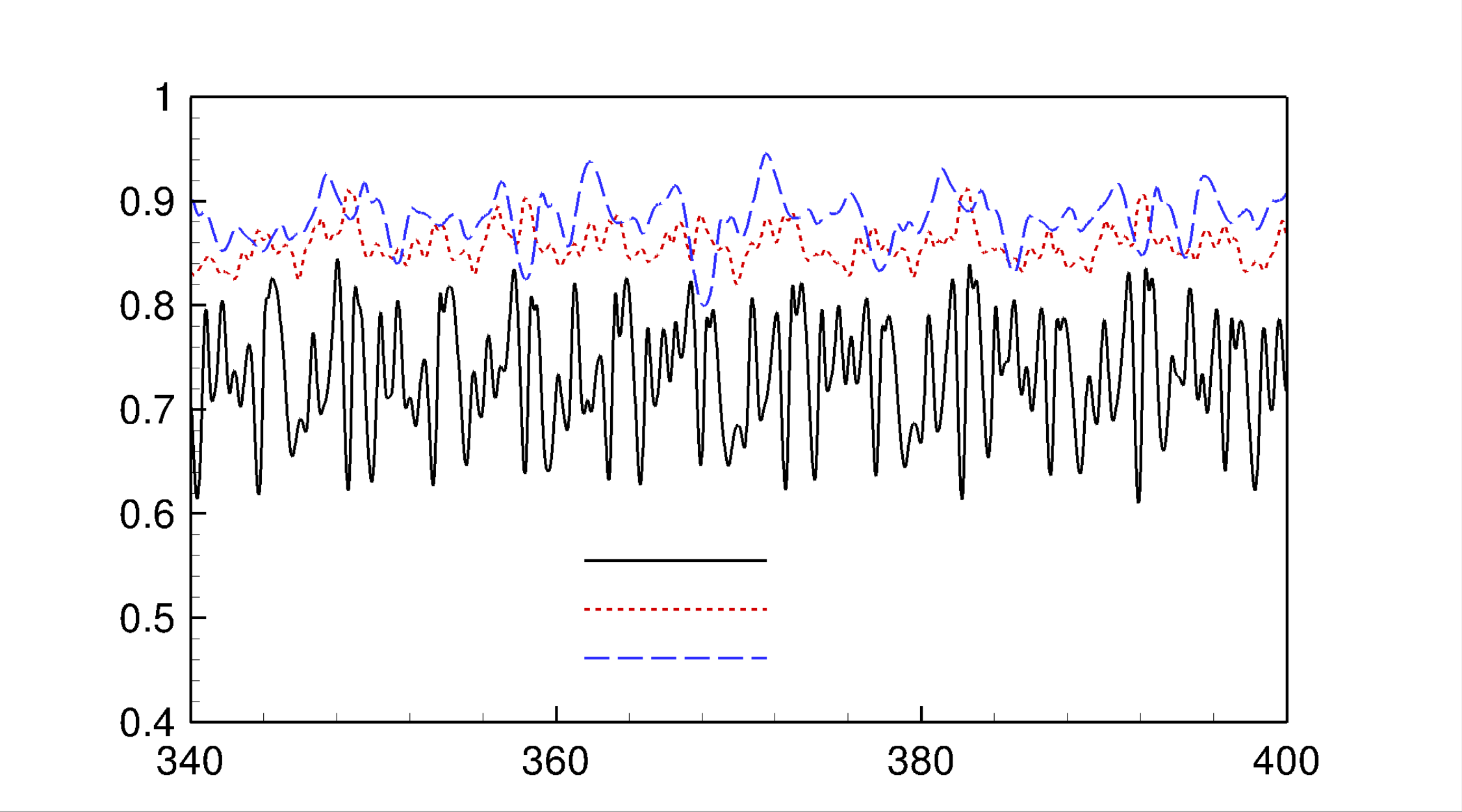}
\put(-185,40){\rotatebox{90}{$u/U_e$}}
\put(-100,-5){$tU_e/h$}
%\put(-190,90){$(b)$}
\put(-85,31){\scriptsize{$x=5h$}}
\put(-85,24){\scriptsize{$x=12h$}}
\put(-85,17){\scriptsize{$x=20h$}}
\caption{: Time history of streamwise velocity variations for $(a)$ Case ($Re_h,\eta$)=($600,1$) and $(b)$ Case ($Re_h,\eta$)=($800,0.5$) at three stations (x,y,z)=(5h,0.75h,0) (solid), (x,y,z)=(12h,0.75h,0) (dashed) and (x,y,z)=(20h,0.75h,0) (long dash).} 
\label{fig:probe_u}
\end{figure}

The time history of streamwise velocity probed at three stations is examined for Case ($Re_h,\eta$)=($600,1$) in figure \ref{fig:probe_u}$(a)$. Periodic oscillations with a circular frequency $\omega=1.088$ are seen at different streamwise stations, corresponding to the periodic shedding of hairpin vortices. These self-sustained oscillations independent of external noise, are also a sign of global instability \citep{puckert2018experiments}, and their frequency is close to the temporal frequency of the leading global varicose mode. For Case ($Re_h,\eta$)=($800,0.5$), stronger fluctuations with multiple frequencies are observed in the immediate vicinity of the roughness element, and smaller amplitude of velocity variations is seen at the farther stations.

%laminar_BL_Diaz_Re600_DNS_medium_domain/probe_290000; laminar_BL_Diaz_Re1100_DNS_medium_domain

%\subsubsection{DMD analysis}\label{shedding}
%psd to show st of hairpin vortex and DMD eigenspectra\\
%Comparison of leading eigenvalues between SFD and DMD
%DMD analysis (Re600, corresponding to the eigenmode of mean flow)\\

To understand the dynamics behind this different behavior for Cases ($Re_h,\eta$)=($600,1$) and ($Re_h,\eta$)=($800,0.5$), dynamic mode decomposition (DMD) was performed and compared to the energy spectra and global stability results. DMD is a data-driven modal decomposition technique that identifies a set of modes from multiple snapshots of the observable vectors. Each of the DMD modes has an assigned eigenvalue that describes its temporal growth/decay rate and oscillation frequency. DMD is a useful tool to isolate the regions associated with a particular frequency and provide information on system dynamics. For the present work, we use a novel DMD algorithm developed by \cite{anantharamu2019parallel} that is suitable for analysis of large datasets. The basic idea behind DMD is that the set of snapshot vectors of flow variables $\{ \psi_i \}_{i=1}^{N-1}$ can be written as a linear combination of DMD modes $\{ \phi_i \}_{i=1}^{N-1}$ as
\begin{equation}
    \psi_i = \sum_{j=1}^{N-1} c_j \lambda_j \phi_j; i=1, ..., N-1,
\end{equation}
where $\lambda_j$ are the eigenvalues of the projected linear mapping and $c_j$ are the $j$th entries of the first vector $\psi_1$. The detailed derivation of the algorithm can be obtained from \cite{anantharamu2019parallel}. To ensure the accuracy of the results, N=200 snapshots of the flow field were taken with $\Delta t U_e/h$=0.15 between them for Case $(Re_h,\eta)=(600,1)$, and N=700 snapshots were taken with $\Delta t U_e/h$=0.1 between them for Case $(Re_h,\eta)=(800,0.5)$. %We consider the comparison (i) between energy spectra and DMD spectra, and (ii) between the DMD modes and GLSA modes.

\begin{figure}
\centering
\includegraphics[width=60mm,trim={0.5cm 0.2cm 0.5cm 0cm},clip]{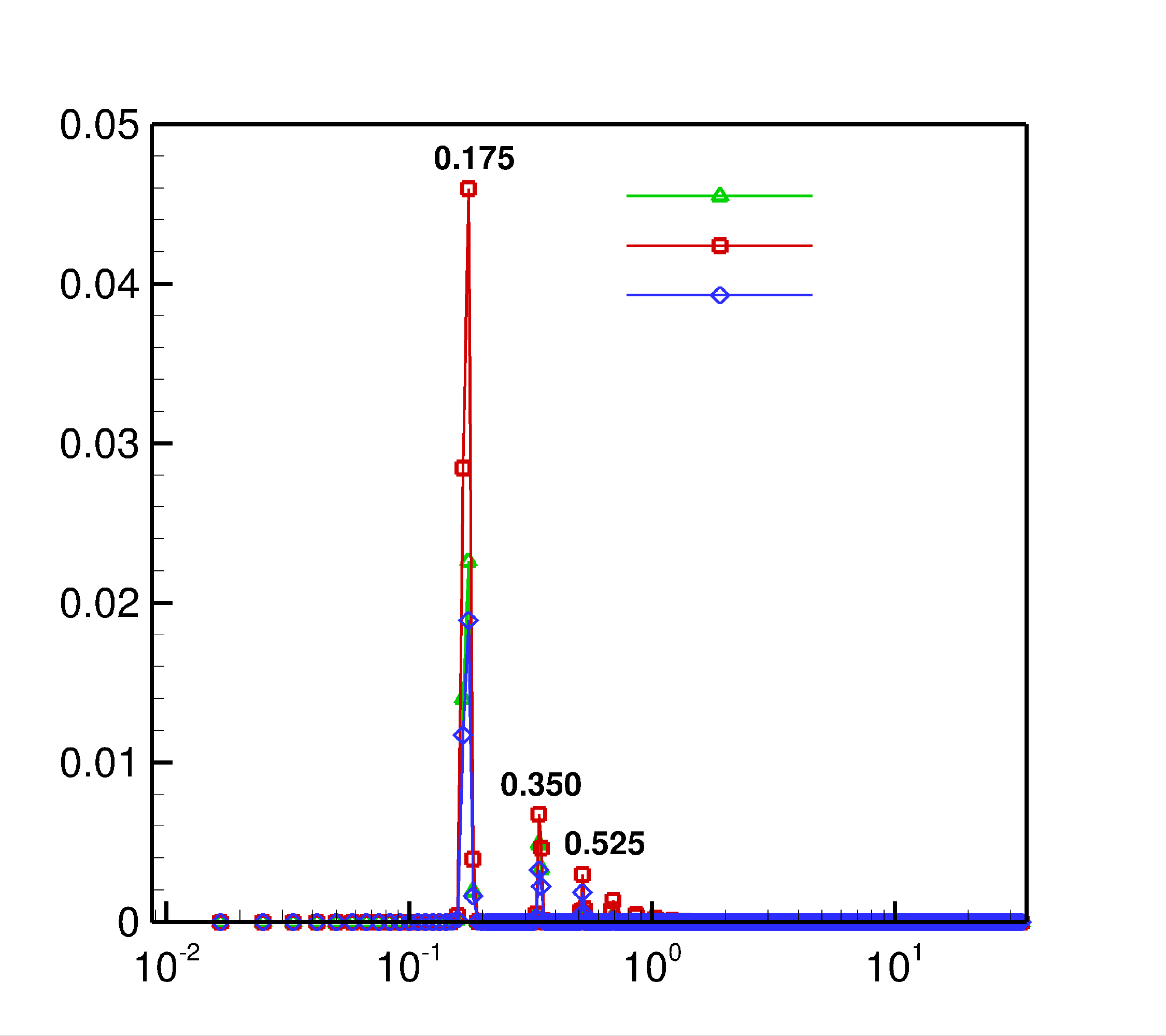}
\put(-180,65){\rotatebox{90}{$PSD_u$}}
\put(-90,-3){$St$}
\put(-47,125){\scriptsize{$x=5h$}}
\put(-47,117){\scriptsize{$x=12h$}}
\put(-47,109){\scriptsize{$x=20h$}}
\put(-180,140){$(a)$}
\hspace{3mm}
\includegraphics[width=60mm,trim={0.5cm 0.2cm 0.5cm 0cm},clip]{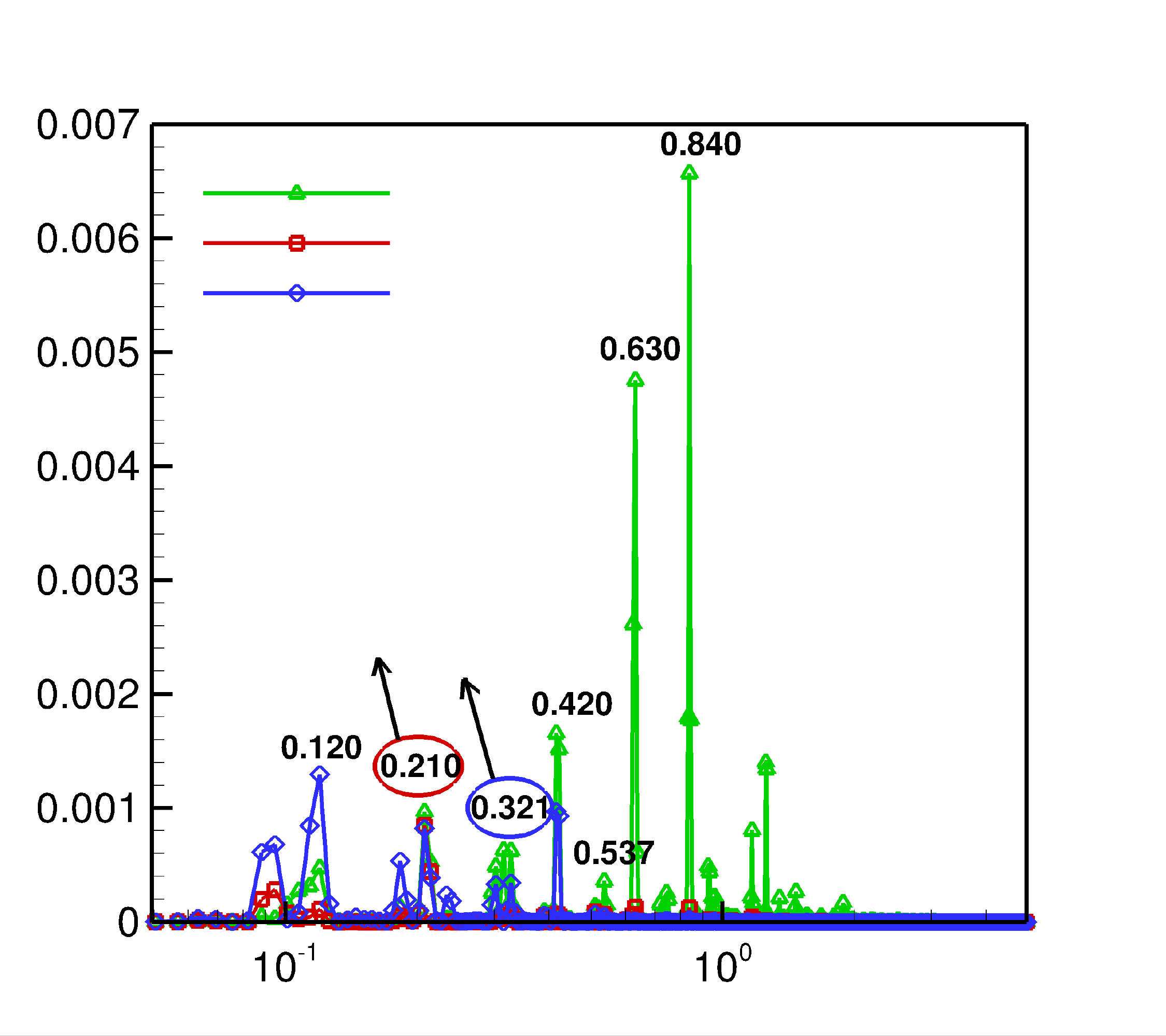}
\put(-180,65){\rotatebox{90}{$PSD_u$}}
\put(-90,-3){$St$}
\put(-112,125){\scriptsize{$x=5h$}}
\put(-112,117){\scriptsize{$x=12h$}}
\put(-112,109){\scriptsize{$x=20h$}}
\put(-180,140){$(b)$}
\put(-140,60){\color{red}{\scriptsize{varicose}}}
\put(-112,55){\color{blue}{\scriptsize{sinuous}}}
% \includegraphics[width=60mm,trim={0.5cm 0.2cm 0.5cm 0cm},clip]{images/dmd_spectra_re600.pdf}
% \put(-185,60){\rotatebox{90}{$|c_j\lambda_j^{i-1}|$}}
% \put(-97,-5){$Imag(f_j)$}
\hspace{3mm}
\includegraphics[width=60mm,trim={0.5cm 0.2cm 0.5cm 0cm},clip]{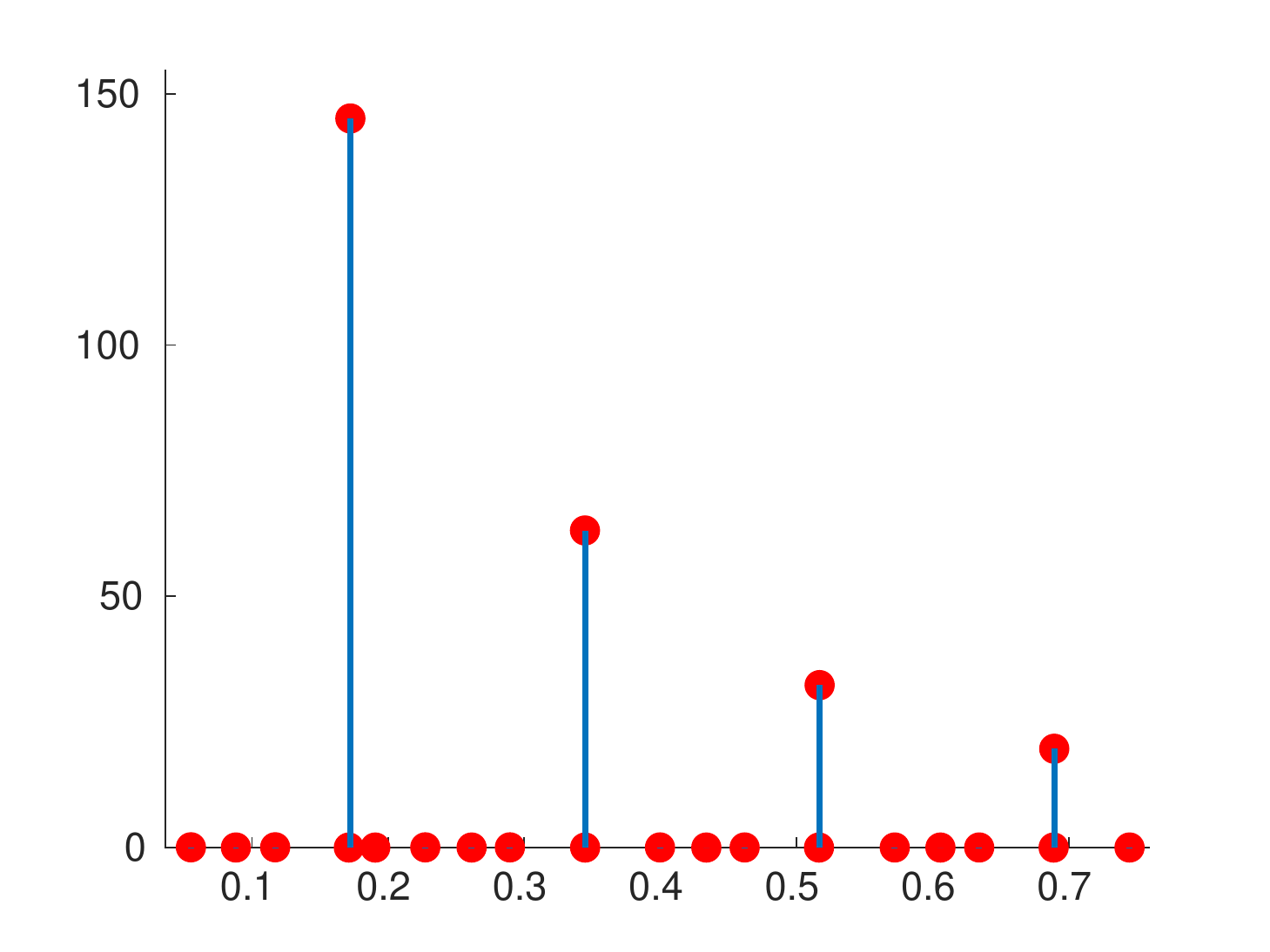}
\put(-182,60){\rotatebox{90}{$|c_j\lambda_j^{i-1}|$}}
\put(-97,-5){$Imag(f_j)$}
\put(-135,122){\scriptsize{0.174}}
\put(-99,63){\scriptsize{0.348}}
\put(-70,40){\scriptsize{0.522}}
\put(-180,130){$(c)$}
\hspace{3mm}
\includegraphics[width=60mm,trim={0.5cm 0.2cm 0.5cm 0cm},clip]{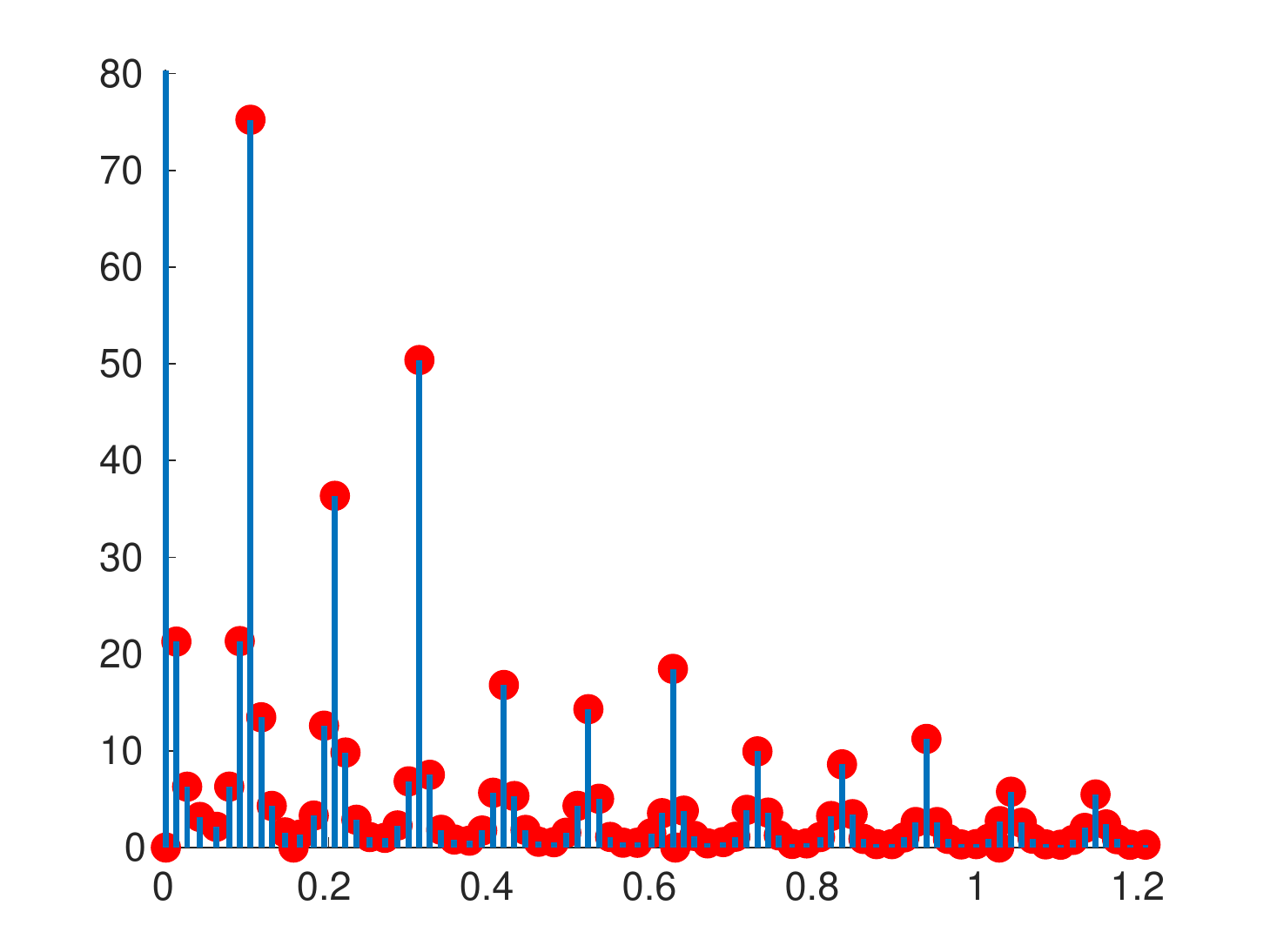}
\put(-182,60){\rotatebox{90}{$|c_j\lambda_j^{i-1}|$}}
\put(-97,-5){$Imag(f_j)$}
\put(-145,122){\scriptsize{0.115}}
\put(-137,68){\scriptsize{0.208}}
\put(-122,88){\scriptsize{0.312}}
\put(-113,44){\scriptsize{0.416}}
\put(-101,37){\scriptsize{0.520}}
\put(-89,44){\scriptsize{0.623}}
\put(-180,130){$(d)$}
\caption{Comparison between the energy spectra of streamwise velocity at different $x$ stations for $(a)$ Case ($Re_h,\eta$)=($600,1$) and $(b)$ Case ($Re_h,\eta$)=($800,0.5$), and the DMD spectra of the last snapshot for $(c)$ Case ($Re_h,\eta$)=($600,1$) and $(d)$ Case ($Re_h,\eta$)=($800,0.5$). The power spectra density (PSD) has been non-dimensionalised as $PSD=E/(U_e h)$. } 
\label{fig:psd}
\end{figure}
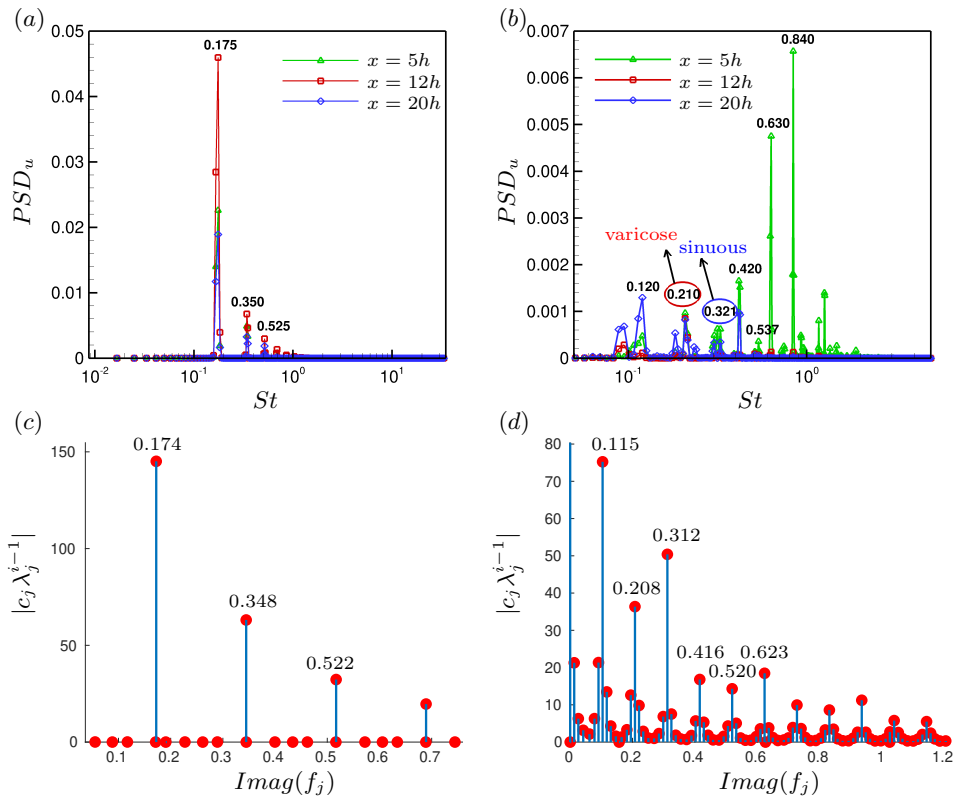

Also, power spectral density (PSD) are examined at the mid-plane for different streamwise stations downstream of the roughness. For Case ($Re_h,\eta$)=($600,1$), the PSD shows a primary peak at the Strouhal number $St=0.175$ in figure \ref{fig:psd}$(a)$, corresponding to the shedding frequency of the main hairpin vortices and the secondary wall-attached vortices observed in figure \ref{fig:Q_criterion}$(a)$. The interaction between different vortical structures results in the higher harmonics at $St=0.35$ and $St=0.525$. Note that similar peaks are also identified in the DMD spectra (figure \ref{fig:psd}$(c)$). 

% navierstokes_solver/runs/psd_computation_bl

% \begin{figure}
% \includegraphics[width=60mm,trim={0.5cm 0.2cm 0.5cm 0cm},clip]{images/psd_re600_medium.pdf}
% \put(-180,65){\rotatebox{90}{$PSD_u$}}
% \put(-90,-3){$St$}
% % \put(-47,125){$x=5$}
% % \put(-47,117){$x=12$}
% % \put(-47,109){$x=20$}
% \hspace{3mm}
% \includegraphics[width=60mm,trim={0.5cm 0.2cm 0.5cm 0cm},clip]{images/psd_eta05_re800_medium.pdf}
% \put(-180,65){\rotatebox{90}{$PSD_u$}}
% \put(-90,-3){$St$}
% % \put(-112,125){$x=5$}
% % \put(-112,117){$x=12$}
% % \put(-112,109){$x=20$}
% % \put(-140,60){\color{red}{\scriptsize{varicose}}}
% % \put(-112,55){\color{blue}{\scriptsize{sinuous}}}
% \caption{Energy spectra of streamwise velocity at different $x$ stations for $(a)$ Case ($Re_h,h/\delta^*,\eta$)=($600,2.86,1$) and $(b)$ Case ($Re_h,h/\delta^*,\eta$)=($800,2.86,0.5$). The power spectra density (PSD) has been non-dimensionalised as $PSD=E/(U_e h)$.} 
% \label{fig:psd}
% \end{figure}
% % navierstokes_solver/runs/psd_computation_bl

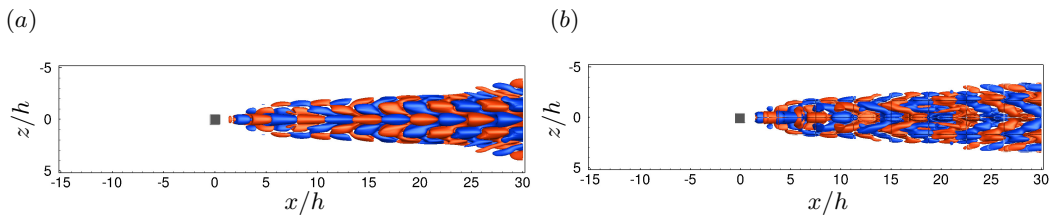
\begin{figure}
\includegraphics[width=70mm,trim={0.2cm 0.2cm 1.5cm 0cm},clip]{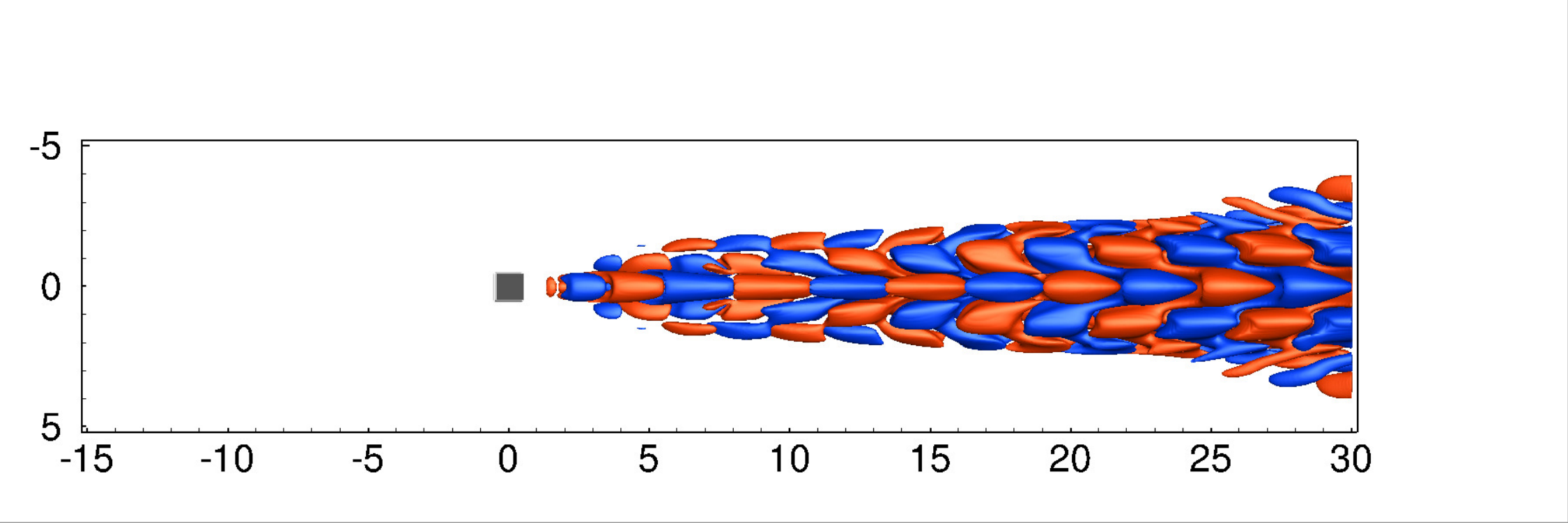}
 \put(-210,65){$(a)$}
\put(-208,23){\rotatebox{90}{$z/h$}}
\put(-105,-3){$x/h$}
\includegraphics[width=72mm,trim={0.2cm 0.1cm 0.5cm 0cm},clip]{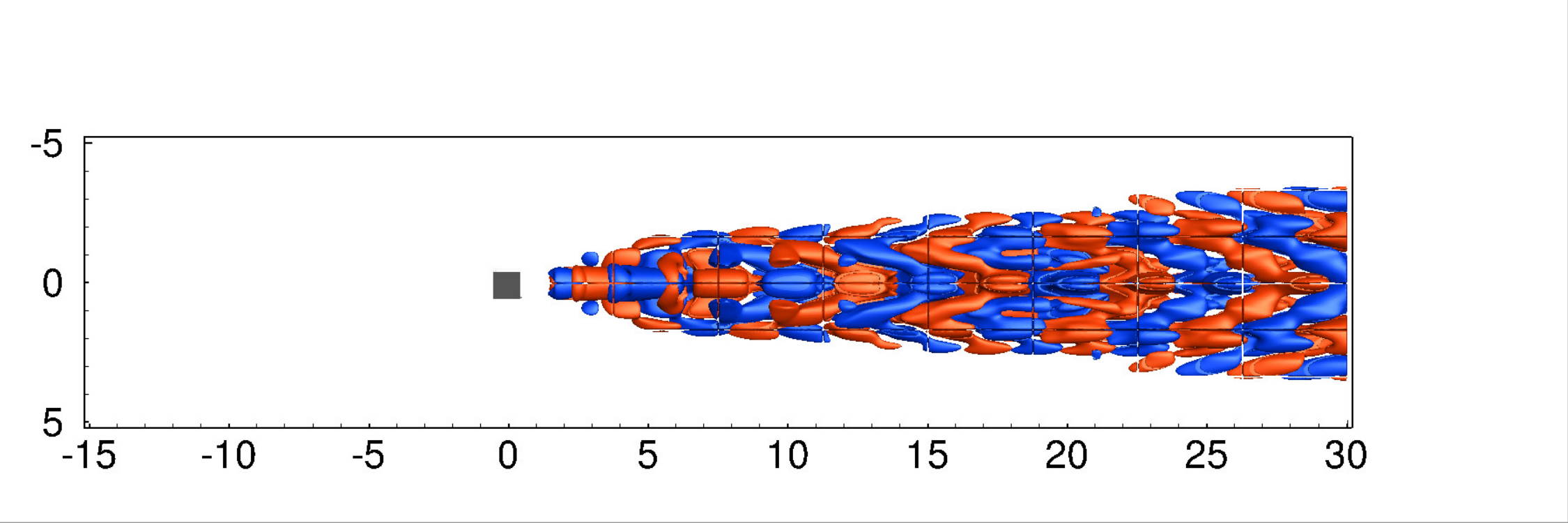}
 \put(-210,65){$(b)$}
\put(-213,23){\rotatebox{90}{$z/h$}}
\put(-113,-3){$x/h$}
 \caption{Comparison between $(a)$ the leading global unstable mode of the mean flow and $(b)$ the DMD mode at $St=0.175$ for Case ($Re_h,\eta$)=($600,1$), depicted by isocontours of the streamwise velocity component. The contour levels depict $\pm 10 \%$ of the mode's maximum streamwise velocity. } 
\label{fig:dmd_mode}
\end{figure}

Table \ref{tab:dmd} demonstrates a comparison of the eigenvalues and Strouhal numbers obtained from global stability and DMD analyses. The Strouhal numbers obtained from global stability analysis and DMD analysis show good agreement. It is worth noting that using the mean (time-averaged) flow as the base state for global stability analysis can still capture the shedding frequency of hairpin vortices, but the mode is marginally stable with a small growth rate. This discrepancy in the growth rates between base flow and mean flow is similar to the observations by \cite{barkley2006linear} for the linear stability analysis of the cylinder wake flow. The state of marginal stability is due to the strong nonlinear saturation of the mean flow observed in figure \ref{fig:contour_xslice}. \cite{sipp2007global} conducted a global weakly nonlinear analysis for cylinder flow and provided theoretical explanation for the marginal stability of mean flows: the zeroth harmonic is much stronger than the second harmonic. This could explain the fact that the mean flow is marginally stable in the present work. The associated global unstable mode of the mean flow and the DMD mode are examined in figure \ref{fig:dmd_mode}. They both demonstrates varicose features and show good qualitative agreement. 

% \begin{figure}
% \centering
% % \includegraphics[width=60mm,trim={0.5cm 0.2cm 0.5cm 0cm},clip]{images/dmd_spectra_re600.pdf}
% % \put(-185,60){\rotatebox{90}{$|c_j\lambda_j^{i-1}|$}}
% % \put(-97,-5){$Imag(f_j)$}
% %\hspace{3mm}
% \includegraphics[width=60mm,trim={0.5cm 0.2cm 0.5cm 0cm},clip]{images/dmd_zoomin_re600.pdf}
% \put(-182,60){\rotatebox{90}{$|c_j\lambda_j^{i-1}|$}}
% \put(-97,-5){$Imag(f_j)$}
% \put(-135,122){\scriptsize{0.174}}
% \put(-99,63){\scriptsize{0.348}}
% \put(-70,40){\scriptsize{0.522}}
% \includegraphics[width=60mm,trim={0.5cm 0.2cm 0.5cm 0cm},clip]{images/dmd_zoomin_eta05_re800.pdf}
% \put(-182,60){\rotatebox{90}{$|c_j\lambda_j^{i-1}|$}}
% \put(-97,-5){$Imag(f_j)$}
% \put(-145,122){\scriptsize{0.105}}
% \put(-137,68){\scriptsize{0.208}}
% \put(-122,88){\scriptsize{0.312}}
% \put(-113,44){\scriptsize{0.416}}
% \put(-101,37){\scriptsize{0.520}}
% \put(-89,44){\scriptsize{0.623}}
% \caption{DMD spectra of the last snapshot for $(a)$ Case ($Re_h,h/\delta^*,\eta$)=($600,2.86,1$) and $(b)$ Case ($Re_h,h/\delta^*,\eta$)=($800,2.86,0.5$).} 
% \label{fig:dmd_spectra}
% \end{figure}
% % navierstokes_solver/runs/DMD_computation/dmd_aux_codes 

\begin{table}
\begin{center}
\def~{\hphantom{0}}
    \begin{tabular}{cccc}
    Analysis & Base state & $\sigma \pm i\omega$ & $St=\frac{\omega h}{2\pi u_h}$ \\
    Global Stability & Base Flow (SFD) & $0.1107 \pm i1.1213$ & 0.180\\
    Global Stability & Mean Flow (DNS) & $-0.0137 \pm i1.0725$ & 0.173\\
    DMD & - & $-0.227e^{-7} \pm i1.0826$ & 0.174\\
    \end{tabular}
    \caption{\label{tab:dmd} Comparison of the eigenvalues from global stability and DMD analyses for Case ($Re_h,\eta$)=($600,1$).}
    \end{center}
\end{table}

%Mean: lamainar_BL_Diaz_Re600, DMD:DMD_computation/visualize_modes_from_nsvof/runs/laminar_BL_Diaz_Re600_DNS_correct_BC/data_files

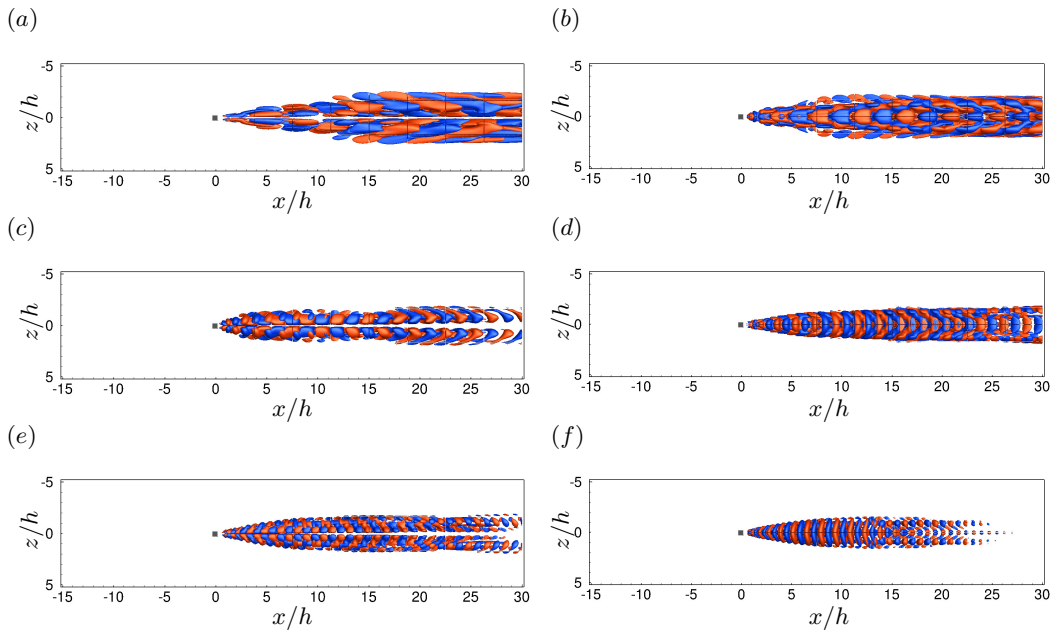
\begin{figure}
\includegraphics[width=70mm,trim={0.2cm 0.2cm 1.5cm 0cm},clip]{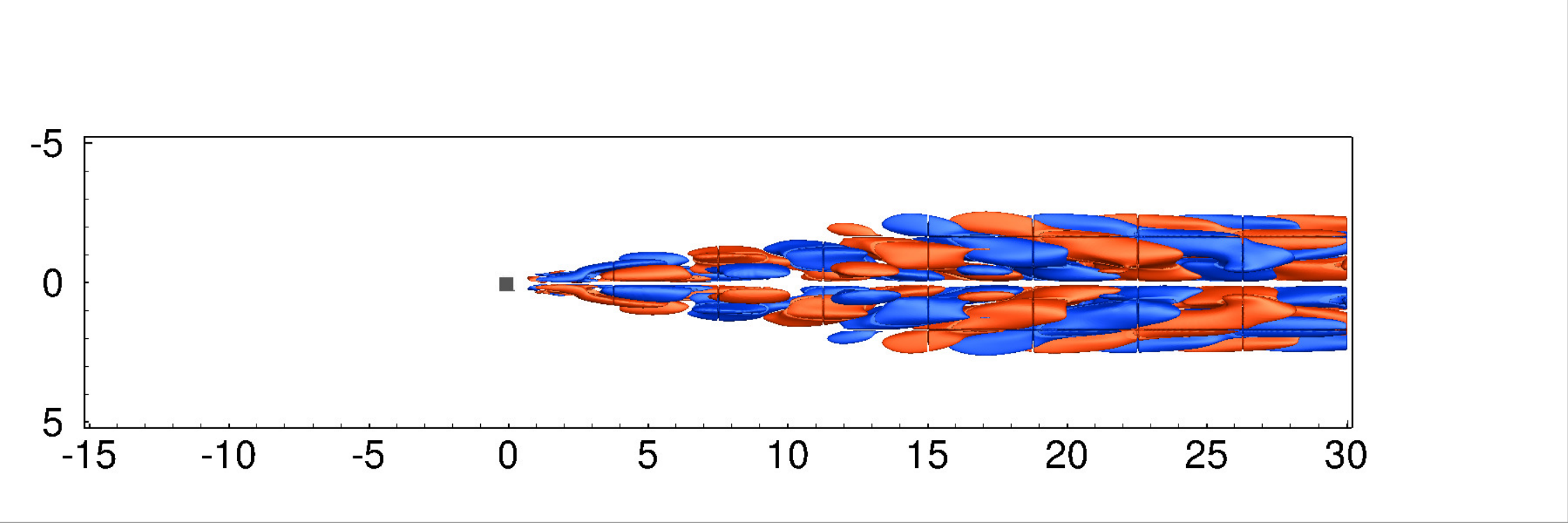}
 \put(-210,65){$(a)$}
\put(-205,25){\rotatebox{90}{$z/h$}}
\put(-110,-3){$x/h$}
\includegraphics[width=72mm,trim={0.2cm 0.1cm 0.5cm 0cm},clip]{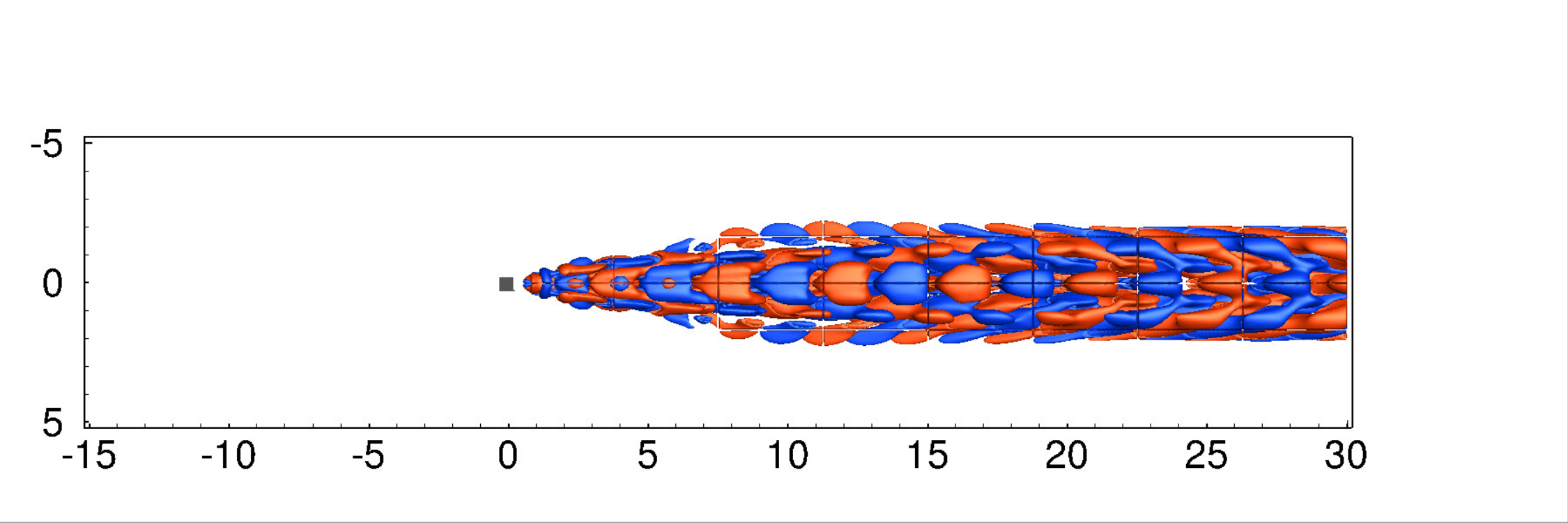}
 \put(-210,65){$(b)$}
\put(-210,25){\rotatebox{90}{$z/h$}}
\put(-113,-3){$x/h$}
\hspace{3mm}
\includegraphics[width=70mm,trim={0.2cm 0.2cm 1.5cm 0cm},clip]{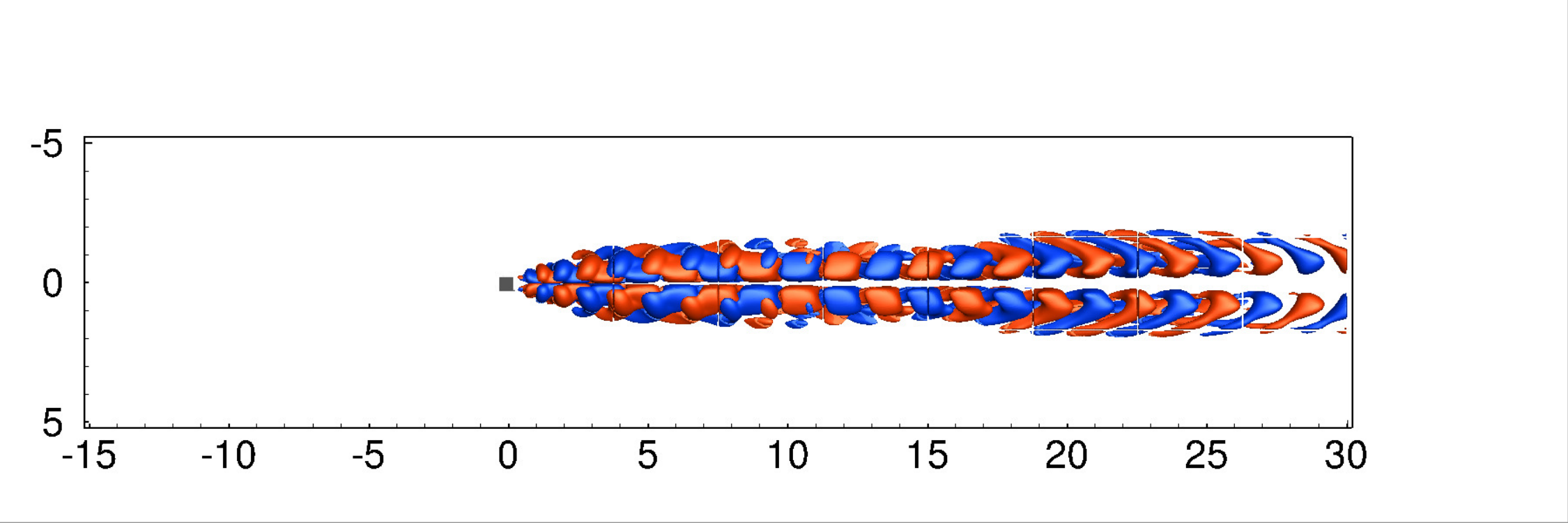}
 \put(-210,65){$(c)$}
\put(-205,25){\rotatebox{90}{$z/h$}}
\put(-110,-3){$x/h$}
\includegraphics[width=72mm,trim={0.2cm 0.1cm 0.5cm 0cm},clip]{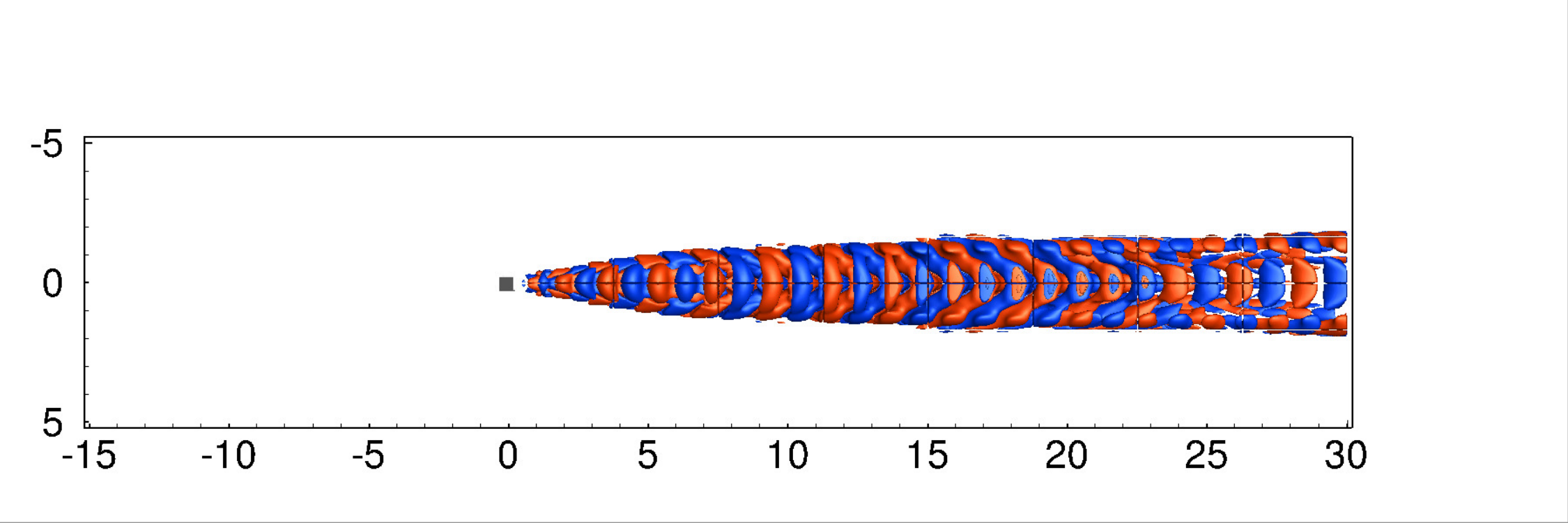}
 \put(-210,65){$(d)$}
\put(-210,25){\rotatebox{90}{$z/h$}}
\put(-113,-3){$x/h$}
\hspace{3mm}
\includegraphics[width=70mm,trim={0.2cm 0.2cm 1.5cm 0cm},clip]{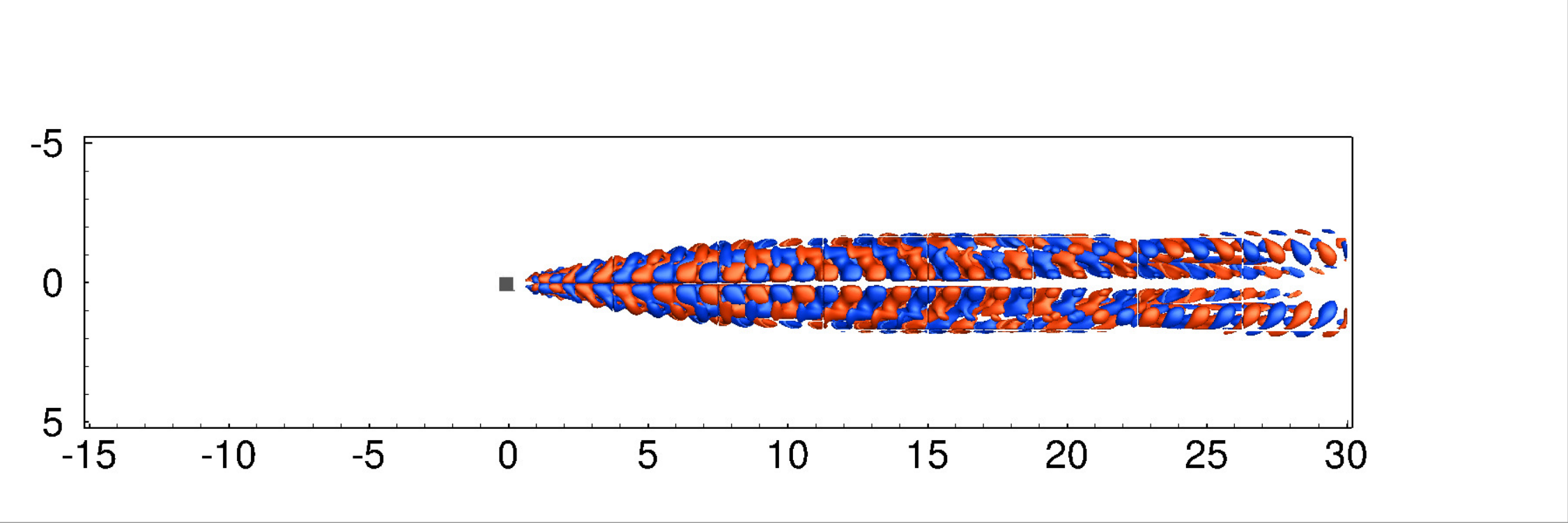}
 \put(-210,65){$(e)$}
\put(-205,25){\rotatebox{90}{$z/h$}}
\put(-110,-3){$x/h$}
\includegraphics[width=72mm,trim={0.2cm 0.1cm 0.5cm 0cm},clip]{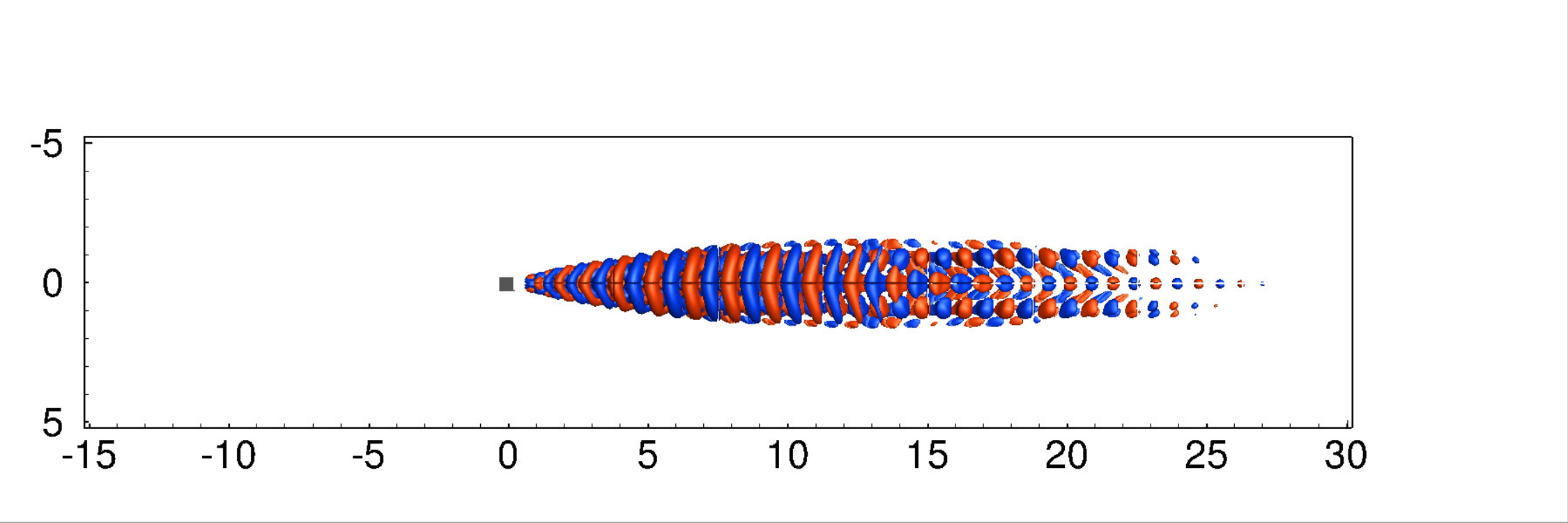}
 \put(-210,65){$(f)$}
\put(-210,25){\rotatebox{90}{$z/h$}}
\put(-113,-3){$x/h$}
 \caption{The DMD modes for Case ($Re_h,\eta$)=($800,0.5$) at $(a)$ $St=0.115$, $(b)$ $St=0.208$, $(c)$ $St=0.312$, $(d)$ $St=0.416$, $(e)$ $St=0.520$ and $(f)$ $St=0.623$, depicted by isocontours of the streamwise velocity component. The contour levels depict $\pm 10 \%$ of the mode's maximum streamwise velocity. } 
\label{fig:dmd_mode_eta05}
\end{figure}
%Mean: lamainar_BL_Diaz_Re600, DMD:DMD_computation/visualize_modes_from_nsvof/runs/laminar_BL_eta05_Re600_DNS/data_files

Compared to Case ($Re_h,\eta$)=($600,1$), a combination of multiple frequencies is distributed in the energy spectra and the DMD spectra for Case ($Re_h,\eta$)=($800,0.5$), indicating more complicated flow behavior. Figure \ref{fig:psd}$(b)$ shows that the peaks at $St=0.210$ and $St=0.321$ are close to the temporal frequency of the varicose and sinuous modes obtained from global stability analysis. Similar peaks are also seen in the DMD spectra from figure \ref{fig:psd}$(d)$. The associated DMD modes are examined in figures \ref{fig:dmd_mode_eta05}$(b)$ and \ref{fig:dmd_mode_eta05}$(c)$. The varicose and sinuous symmetries are seen for the DMD modes at $St=0.210$ and $St=0.321$, which is consistent with the global stability results. The higher harmonic peaks at $St=0.420$, $St=0.537$, $St=0.630$ and $St=0.840$ are evident in the vicinity downstream of the roughness element ($x=5h$), resulting from the interactions between the varicose and sinuous oscillations in the near-wake region. The DMD spectra show agreement with the energy spectra for the higher harmonics. The associated DMD modes at $St=0.416$ and $St=0.623$ are varicose since they are the higher multiples of the varicose mode at $St=0.208$, while the DMD mode at $St=0.520$ is sinuous due to a superposition of the varicose mode at $St=0.208$ and the sinuous mode at $St=0.312$. The results indicate that the interactions between the hairpin vortices and the general sinuous oscillations are significant in the near wake, and diminish as the vortical structures develop farther downstream. As the streamwise station increases farther downstream, a peak at a low frequency $St=0.120$ in figure \ref{fig:psd}$(b)$ gets amplified. This peak is also captured in the DMD spectra (figure \ref{fig:psd}$(d)$). The corresponding DMD mode in figure \ref{fig:dmd_mode_eta05}$(a)$ shows a sinuous symmetry. This sinuous mode is associated with the wiggling of the streaks observed farther downstream in figure \ref{fig:Q_criterion}$(b)$. It is thus clear that different roughness geometries associated with different instability characteristics lead to different wake flow behavior in the transition process. 

\subsubsection{Mean flow characteristics}\label{mean}
%Large domain at Re600 and Re1100: examine Cf and mean velocity profile\\
%max(urms) profiles vs. x at the mid-z plane\\
%Cf at z=0, 0.5 and 1: 0.5 is high ~ hairpin vortices
%Von-Donhoff Braslow diagram

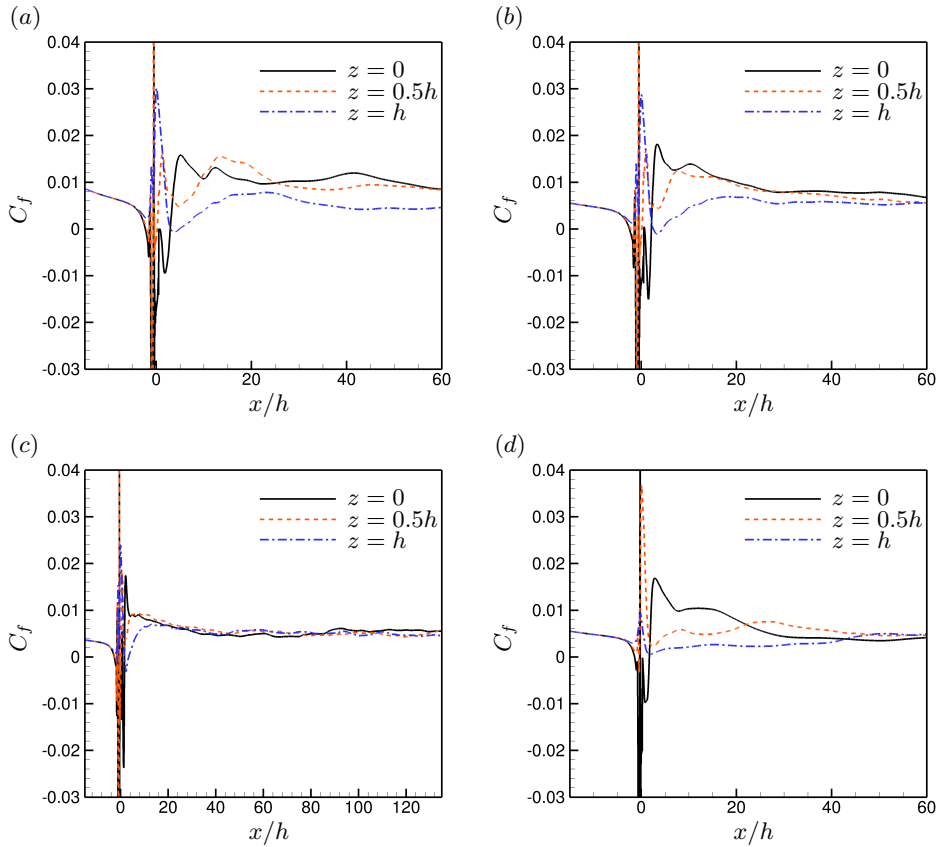
\begin{figure}
\centering
\includegraphics[width=60mm,trim={0.5cm 0.2cm 0.5cm 0cm},clip]{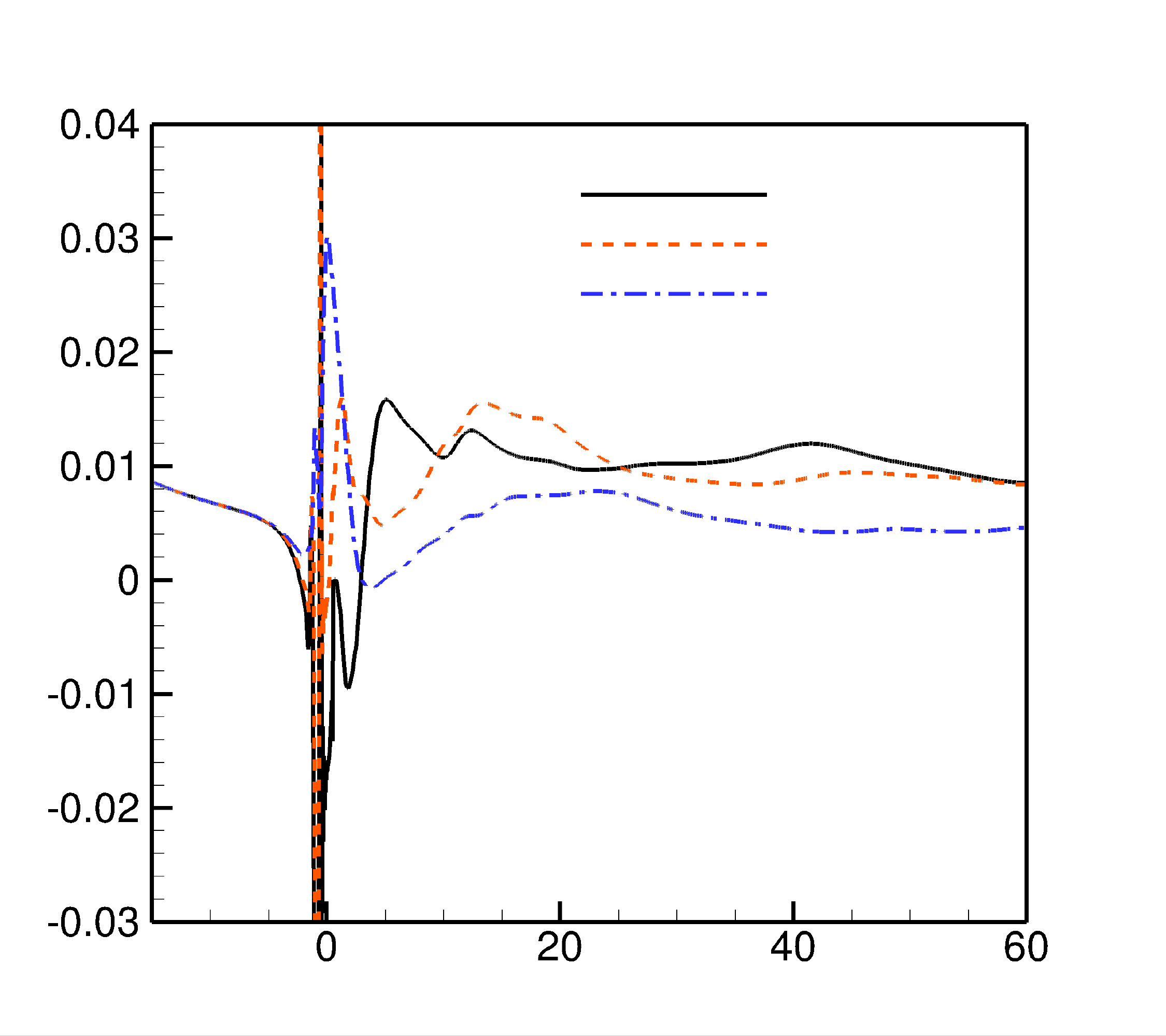}
\put(-180,70){\rotatebox{90}{$C_f$}}
\put(-90,0){$x/h$}
\put(-53,125){$z=0$}
\put(-53,117){$z=0.5h$}
\put(-53,109){$z=h$}
 \put(-180,145){$(a)$}
\hspace{3mm}
\includegraphics[width=60mm,trim={0.5cm 0.2cm 0.5cm 0cm},clip]{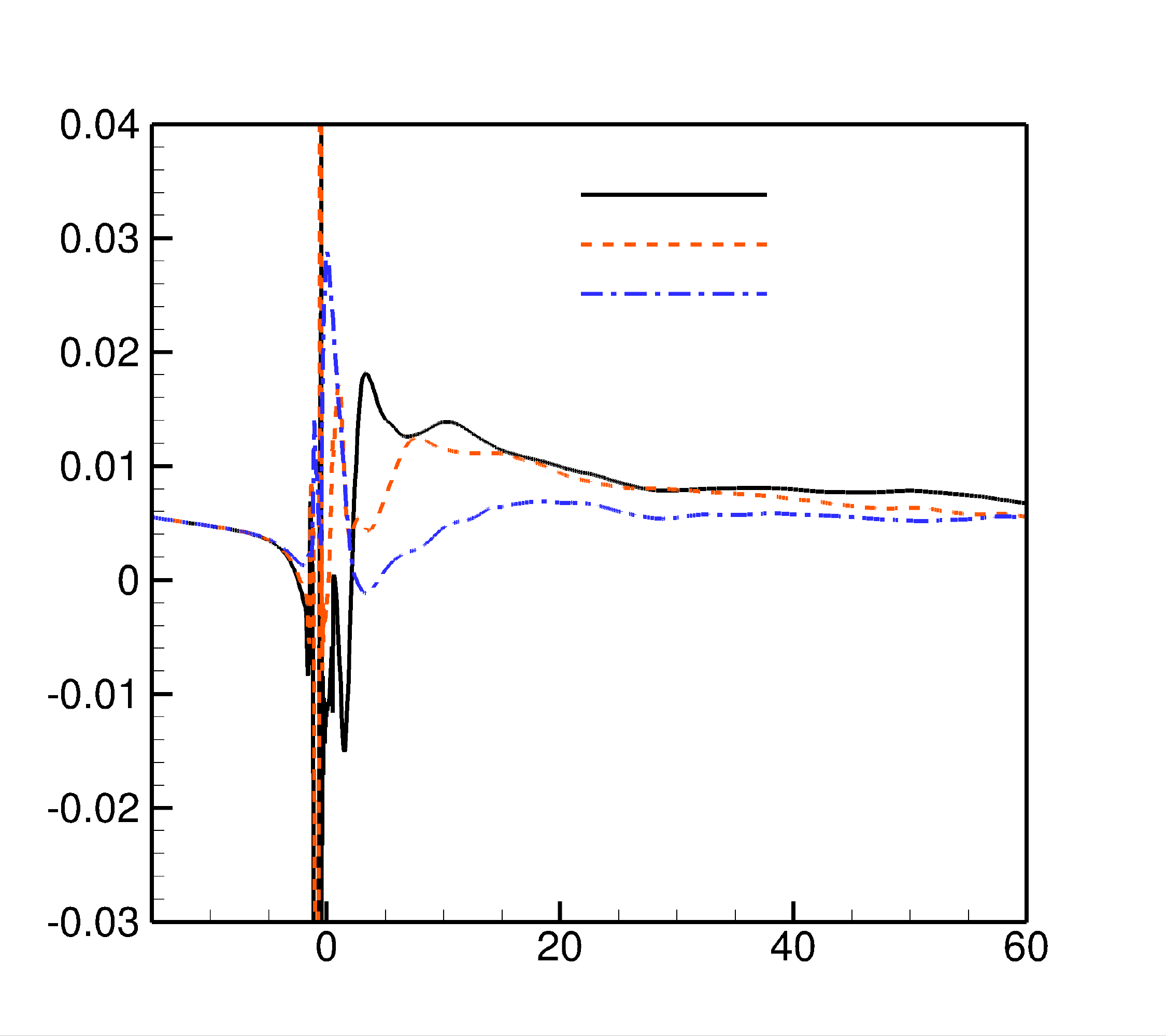}
\put(-180,70){\rotatebox{90}{$C_f$}}
\put(-90,0){$x/h$}
\put(-53,125){$z=0$}
\put(-53,117){$z=0.5h$}
\put(-53,109){$z=h$}
\put(-180,145){$(b)$}
\hspace{3mm}
\includegraphics[width=60mm,trim={0.5cm 0.2cm 0.5cm 0cm},clip]{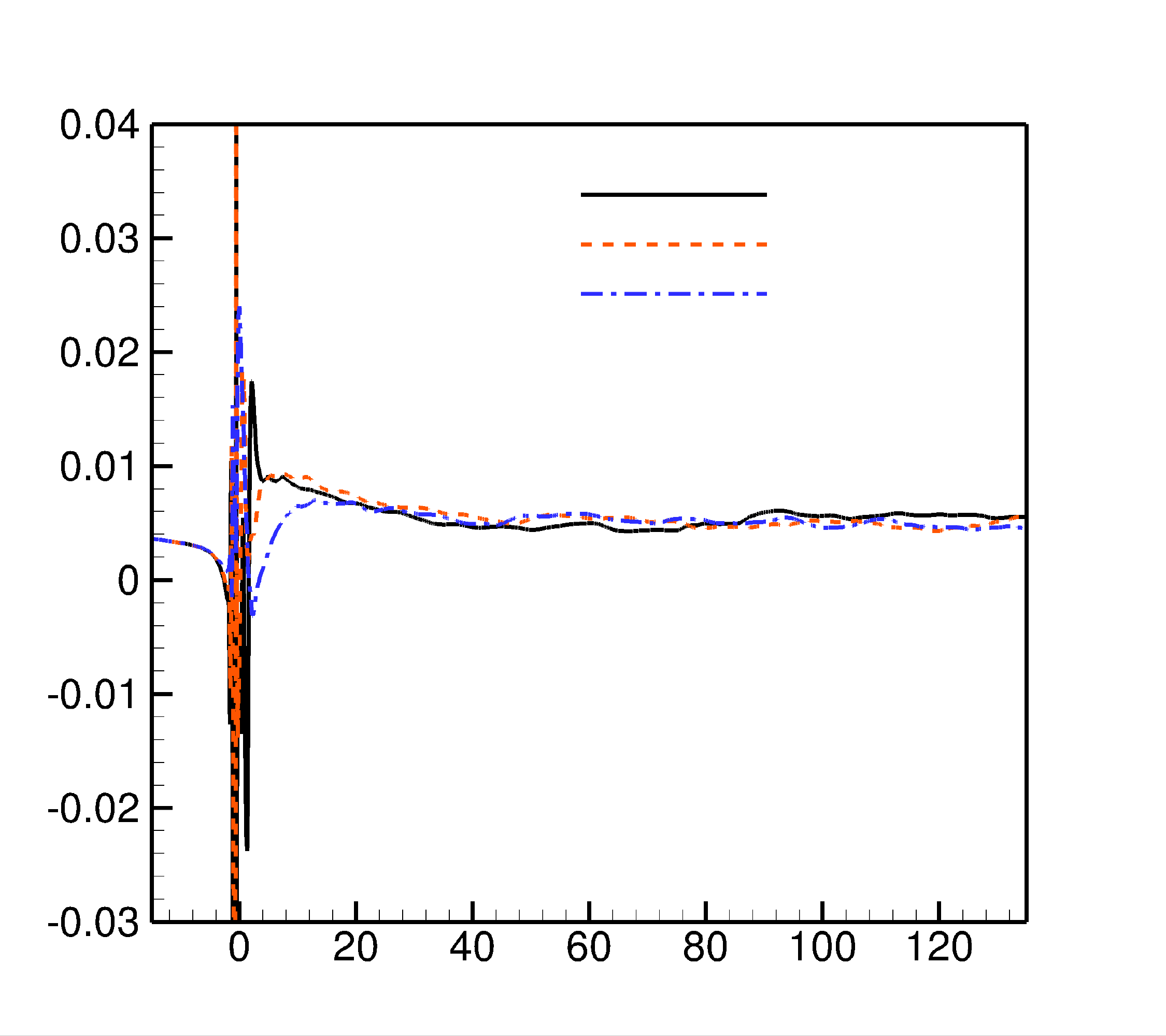}
\put(-180,70){\rotatebox{90}{$C_f$}}
\put(-90,0){$x/h$}
\put(-53,125){$z=0$}
\put(-53,117){$z=0.5h$}
\put(-53,109){$z=h$}
\put(-180,145){$(c)$}
\hspace{3mm}
\includegraphics[width=60mm,trim={0.5cm 0.2cm 0.5cm 0cm},clip]{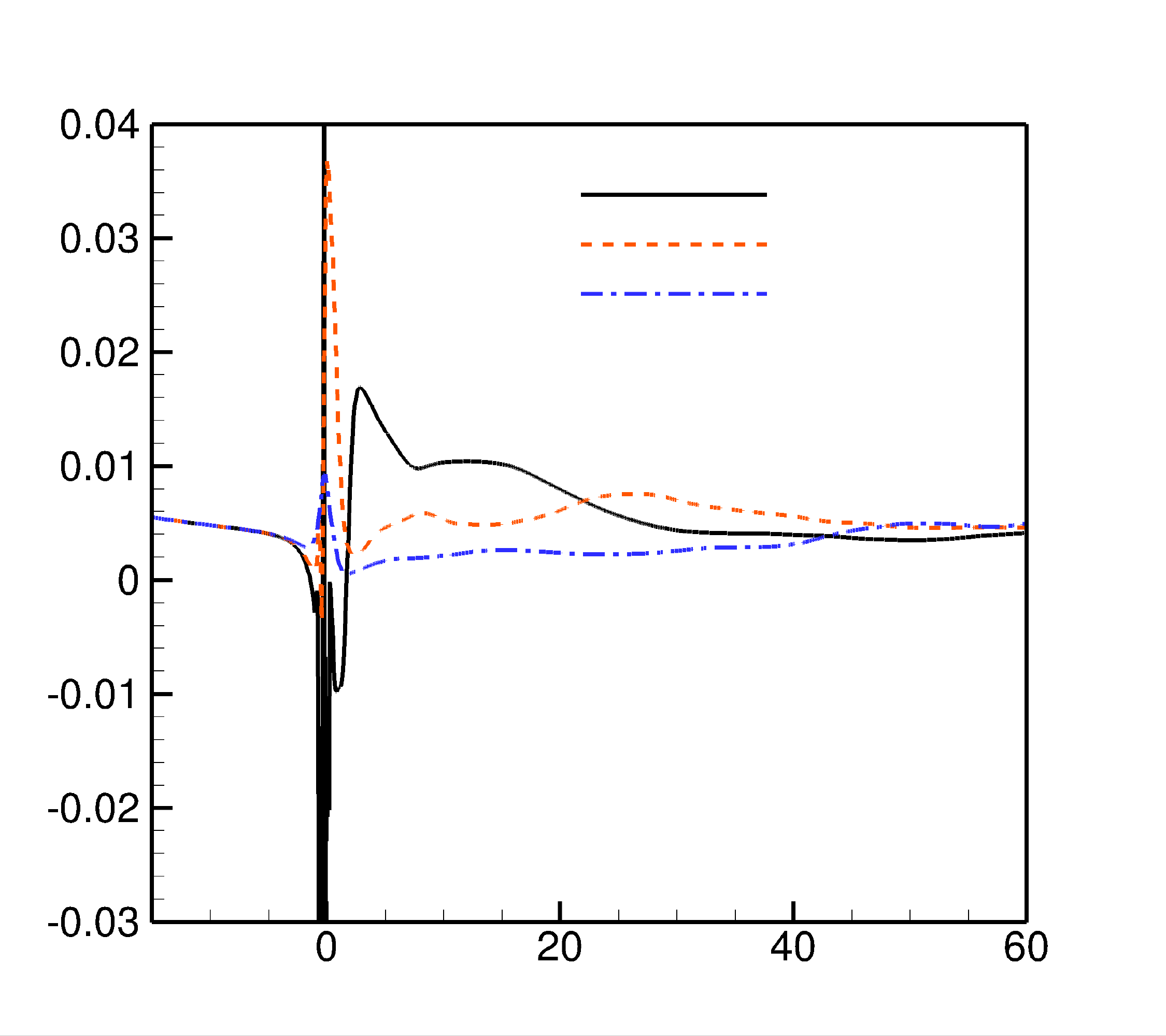}
\put(-180,70){\rotatebox{90}{$C_f$}}
\put(-90,0){$x/h$}
\put(-53,125){$z=0$}
\put(-53,117){$z=0.5h$}
\put(-53,109){$z=h$}
\put(-180,145){$(d)$}
\caption{Streamwise variation of mean skin friction at different $z$ stations for $(a)$ Case ($Re_h,\eta$)=($600,1$), $(b)$ Case ($Re_h,\eta$)=($800,1$), $(c)$ Case ($Re_h,\eta$)=($1100,1$) and $(d)$ Case ($Re_h,\eta$)=($800,0.5$).} 
\label{fig:cf}
\end{figure}

The transitional flow behavior is examined using the time-averaged flow. Figure \ref{fig:cf} shows the streamwise variation of the time-averaged skin friction at three different stations across the span for cases with $\eta=1$ and $\eta=0.5$ at different $Re_h$. For Case ($Re_h,\eta$)=($600,1$), shown in figure \ref{fig:cf}$(a)$, the $C_f$ value at $z=0$ shows a prominent increase behind the roughness location, %as the streamwise location goes farther downstream from the reversed flow region. 
corresponding to the progress from the reversed flow region to the downstream region. At $z=0.5h$ and $z=h$, the $C_f$ profiles show peaks around $x=18h$, which is associated with the evolution of the lateral wall-attached low-speed streaks observed in figures \ref{fig:contour_xslice} and \ref{fig:isocontour_ubar}. As the streamwise distance increases farther downstream, the mean skin friction at three stations does not collapse to the same level, indicating the flow experiences unsteadiness but the saturation is not sufficiently strong, and transition to turbulence may not happen eventually.

% The breakdown of hairpin vortices, as shown in figure \ref{fig:Q_criterion}, leading to an increased $C_f$ between $x=10$ and $x=20$.  The $C_f$ value at $z=1$ presents a general lower level since the probed station is at the lateral wake position and corresponding to the two low-speed regions at the sides near the wall as observed in figure \ref{fig:contour_xslice}.

As $Re_h$ increases to $800$, the peaks of $C_f$ profiles at three stations move closer to the roughness and drop to a similar level farther downstream, shown in figure \ref{fig:cf}$(b)$. This suggests that as $Re_h$ increases, the onset of unsteadiness occurs more closely to the roughness and the wake flow becomes more homogeneous. Figure \ref{fig:cf}$(c)$ shows that for $Re_h=1100$, the increase of $C_f$ profiles occur in the immediate vicinity behind the roughness element compared to those at a lower $Re_h$. As the streamwise location increases beyond $x=20h$, the $C_f$ profiles at three stations collapse and remain constant, suggesting that the wake flow becomes homogeneous and transition to turbulence may occur downstream. Note that whether or not transition to turbulence occurs could also depend on $Re_{\delta}$ of the boundary layer. For the high shear ratio $h/\delta^*$ considered in the present work, $Re_{\delta}$ corresponding to a certain $Re_h$ is relatively low. The transition to turbulence is thus unlikely to happen at moderate $Re_h$.

In contrast to $\eta=1$, the $C_f$ profiles for Case ($Re_h,\eta$)=($800,0.5$) are examined in figure \ref{fig:cf}$(d)$. Due to a thinner geometry, the $C_f$ profile at $z=0$ demonstrates a lower level, and its sharp rise occurs more closely to the roughness element compared to Case ($Re_h,\eta$)=($800,1$). The $C_f$ profiles at $z=0.5h$ and $z=h$ remain at a lower level since the width of the wake flow and the spacing of two lateral streaks are smaller for a thinner roughness geometry. %The $C_f$ profiles at three stations collapse, indicating that the wake flow becomes more homogeneous. 
Note that as the streamwise distance increases, the $C_f$ profile at $z=h$ increases and the $C_f$ values at $z=0.5h$ and $z=h$ are slightly higher than that at the mid-plane in the late stages of transition. This corresponds to the wiggling streaks observed in figure \ref{fig:Q_criterion}$(b)$, indicating that the effect of the wiggling streaks on the mean flow persists and contributes to the transition process.

% Figure \ref{fig:cf}$(b)$ shows the streamwise variation of the mean skin friction for Case ($Re_h,h/\delta^*,\eta$)=($1100,2.86,1$). In general, the $C_f$ value at $Re_h=1100$ is smaller than that at $Re_h=600$ due to the smaller viscosity. The increase of $C_f$ profile at $z=0$ occurs more closely behind the roughness element compared to that at $Re_h=600$. A relatively gentle growth is shown at $z=0.5$ and $z=1$ in a closer vicinity as well. These observations indicate that as $Re_h$ increases, the primary hairpin vortices break into small structures in a more immediate region downstream the roughness element. As the streamwise location moves above $x=20$, the $C_f$ profiles at three stations collapse and remain at a certain level, suggesting that the wake flow become homogeneous and the transition to turbulence may happen.

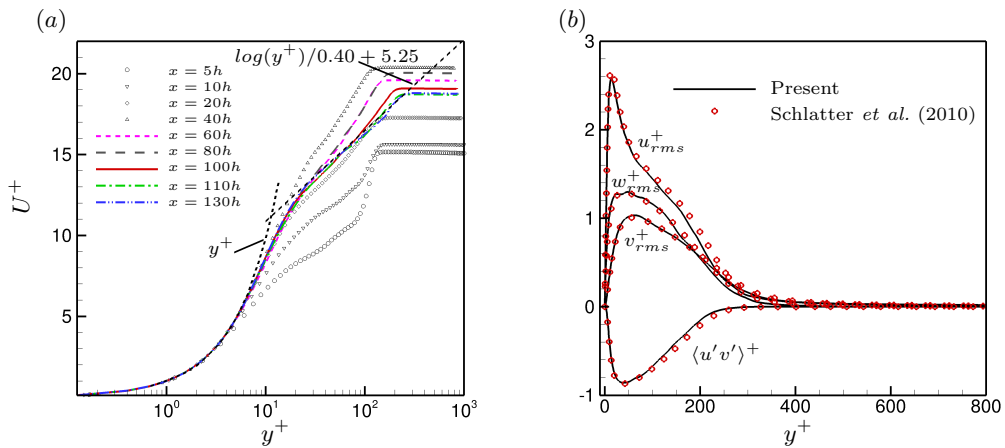
\begin{figure}
%\centering
\includegraphics[width=65mm,trim={0.5cm 0.2cm 0.5cm 0.5cm},clip]{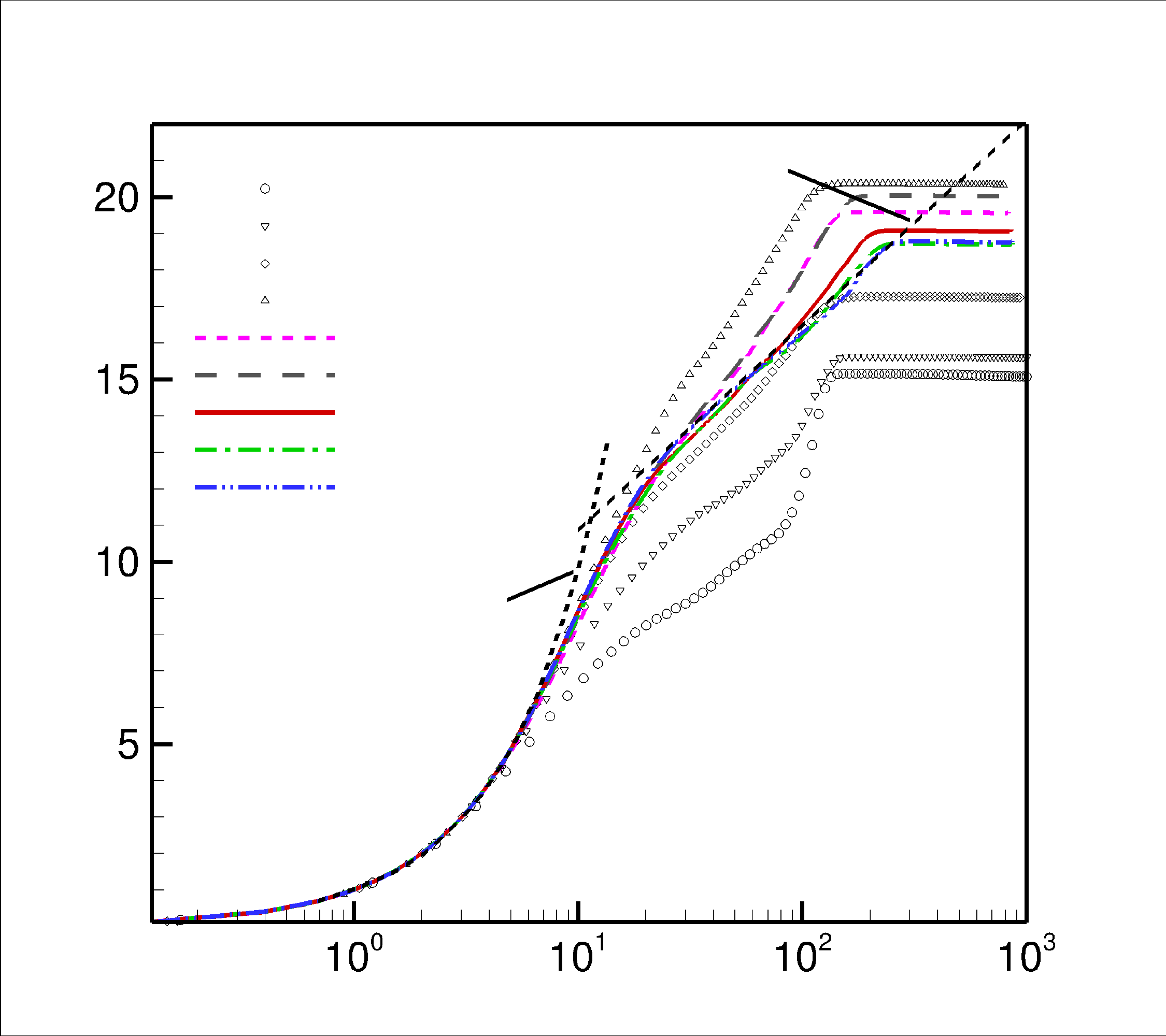}
\put(-190,85){\rotatebox{90}{$U^+$}}
\put(-95,0){$y^+$}
\put(-115,70){\scriptsize{$y^+$}}
\put(-105,142){\scriptsize{$log(y^+)/0.40+5.25$}}
\put(-130,137){\tiny{$x=5h$}}
\put(-130,131){\tiny{$x=10h$}}
\put(-130,125){\tiny{$x=20h$}}
\put(-130,119){\tiny{$x=40h$}}
\put(-130,113){\tiny{$x=60h$}}
\put(-130,107){\tiny{$x=80h$}}
\put(-130,101){\tiny{$x=100h$}}
\put(-130,94){\tiny{$x=110h$}}
\put(-130,88){\tiny{$x=130h$}}
 \put(-180,155){$(a)$}
\hspace{3mm}
\includegraphics[width=65mm,trim={0.5cm 0.2cm 0.5cm 0.5cm},clip]{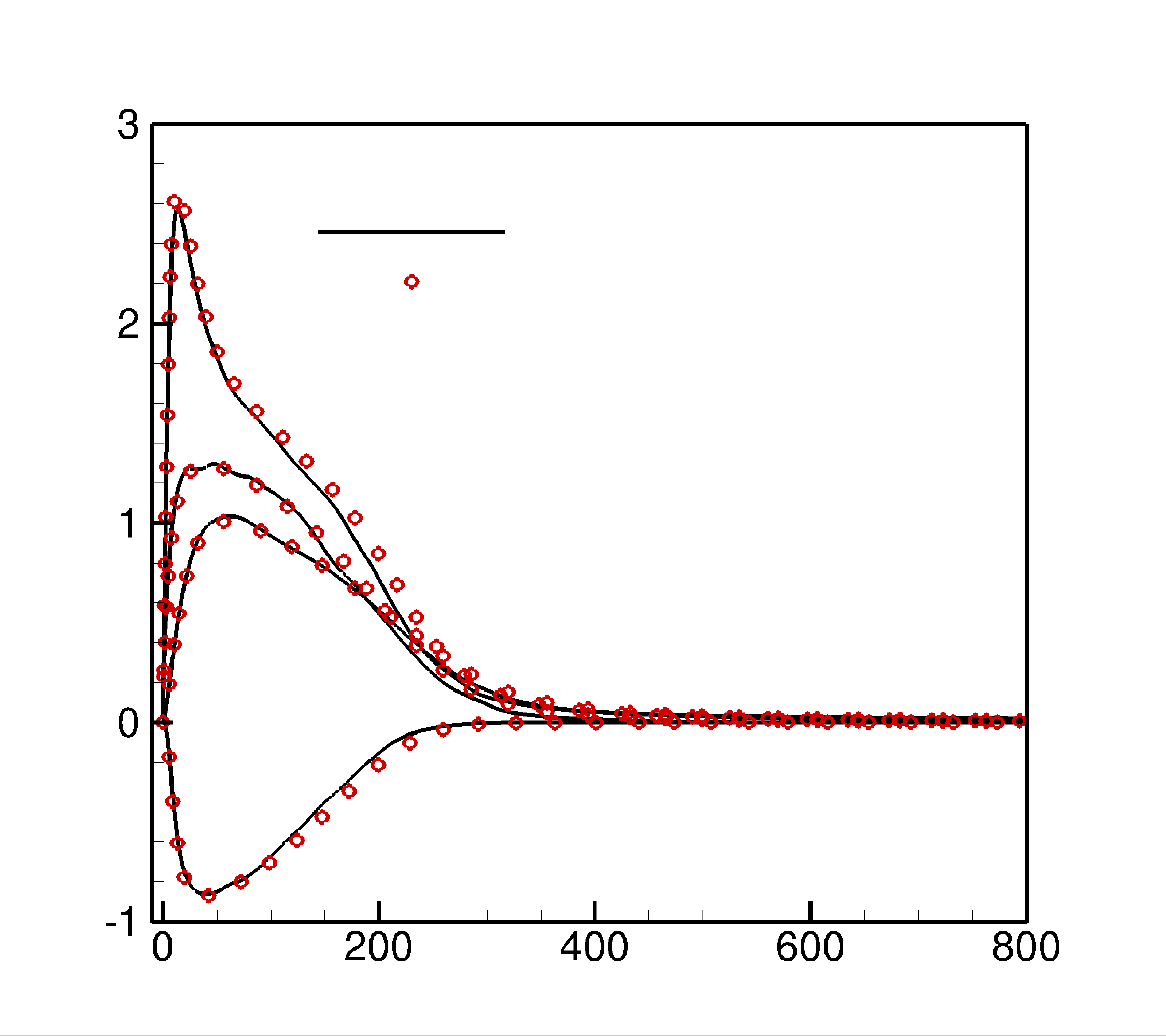}
%\put(-215,95){\rotatebox{90}{$U^+$}}
 \put(-180,155){$(b)$}
\put(-95,0){$y^+$}
\put(-150,110){\scriptsize{$u_{rms}^+$}}
\put(-160,95){\scriptsize{$w_{rms}^+$}}
\put(-155,73){\scriptsize{$v_{rms}^+$}}
\put(-130,30){\scriptsize{$\langle {u'v'}\rangle^+$}}
\put(-100,130){\scriptsize{Present}}
\put(-100,120){\scriptsize{\cite{schlatter2010simulations}}}
\caption{$(a)$ Mean velocity profiles in wall units at different streamwise stations for Case ($Re_h,\eta$)=($1100,1$) and $(b)$ Reynolds stresses for Case ($Re_h,\eta$)=($1100,1$) at $x=130h$ (corresponding to $Re_{\tau}=272$) compared with the LES of \cite{schlatter2010simulations} at $Re_{\tau}=257$.} 
\label{fig:umean_profile}
\end{figure}

The boundary layer evolution from laminar to turbulent states is examined in figure \ref{fig:umean_profile}$(a)$ using the mean velocity profiles in wall units at different streamwise locations downstream of the roughness element for Case ($Re_h,\eta$)=($1100,1$). The time-averaged streamwise velocity at the mid-plane is normalized by the local friction velocity $u_{\tau}$, where $u_{\tau}$ is computed from the $C_f$ profile at $z=0$ for each $x$ location. The wall-normal coordinate in wall units is $y^+=yu_{\tau}/\nu$. The results show that all profiles collapse well in the viscous sublayer and follow the correlation $U^+=y^+$. From $x=5h$ to $x=40h$, significant increase is observed above the viscous layer, which is due to the lift-up behavior of the shear layer. The mean velocity profile above the viscous sublayer reaches its maximum magnitude at $x=40h$ and decreases to approach the log-law profile as the $x$ location increases farther downstream. %This observation presumably results from the effect of the spanwise domain in which the flow interaction is enhanced as the wake flow reaches the periodic boundaries. 
Agreement with the logarithmic law is seen beyond $x=100h$, indicating that the inner layer is fully-developed. As the $x$ location increases even farther, the profiles at $x=110h$ and $x=130h$ show agreement in both the inner and outer layers, suggesting that fully-developed turbulent flow is established in both the inner and outer layers. The velocity fluctuations and Reynolds shear stresses at $x=130h$ are depicted in figure \ref{fig:umean_profile}$(b)$ using wall scaling. The velocity fluctuations $u_{rms}$, $v_{rms}$ and $w_{rms}$ are normalized by $u_{\tau,ave}$, and the Reynolds shear stress $\langle u'v' \rangle$ is normalized by $u_{\tau,ave}^2$, where $u_{\tau,ave}$ is the spanwise-averaged friction velocity computed from the spanwise-averaged $C_f$ at $x=130h$. The results show good agreement with the %large eddy simulation (LES) 
results of a turbulent zero-pressure gradient boundary layer from \cite{schlatter2010simulations}.

\section{Conclusions}\label{sec:conclusions}
Global stability analysis and direct numerical simulation are performed to study roughness-induced transition. Isolated cuboids with aspect ratios $\eta=1$ and $\eta=0.5$ immersed in laminar boundary layers at different $Re_h$ are investigated. The ratio $h/\delta^*$ between the roughness height and the local displacement boundary layer thickness is $2.86$, which is higher than most past studies.

%The base flow is of great importance for global stability analyses. 
Differences between the base flow computed using the SFD method and the mean flow obtained from DNS are examined. The base flow shows a stronger wall-normal shear farther downstream than the mean flow, while non-linear interactions are more evident in the mean flow. Either using the base flow or the mean flow as the base state for global stability analysis is able to capture the shedding frequency of the primary vortical structures. However, the mean flow evolves to a marginally stable state due to the strong non-linear saturation, in contrast to an unstable base flow. We thus use the base flow as the base state for global stability analysis in the present work.

The effects of $\eta$ and $Re_h$ on the base flow are investigated. As $Re_h$ increases, the downstream shear layer lifts up and shows a stronger wall-normal gradient. For a thinner roughness geometry, the central and lateral low-speed streaks are thinner and less sustainable compared to the thicker roughness at the same $Re_h$. It can be summarized that higher $Re_h$, larger $\eta$ and higher $h/\delta^*$ lead to a stronger wall-normal shear and a more sustainable central streak. Also, as $Re_h$ increases for the thinner roughness, high-speed streaks below the central streak become prominent in the near-wake region, indicating an increased spanwise shear that contributes to sinuous instability. 

Global stability analysis shows that when the shear ratio is sufficiently high ($h/\delta^*=2.86$), the varicose instability is dominant for the roughness element with small aspect ratios ($\eta \le 1$). For $\eta=1$, both the stable and unstable modes exhibit varicose symmetry. For $\eta=0.5$, the varicose instability is dominant at different $Re_h$, and the sinuous instability becomes more pronounced as $Re_h$ increases. These results highlight how the onset of sinuous instability is highly dependent on the joint effects of $h/\delta^*$, $\eta$ and $Re_h$.
At smaller $h/\delta^*$, smaller $\eta$ and higher $Re_h$, the sinuous instability is more likely to occur. 

The production of disturbance kinetic energy shows that the varicose mode extracts energy from the wall-normal and spanwise shear of the central streak%, demonstrated by the dominant terms $P_y$ and $P_z$ respectively
. In contrast, the sinuous mode extracts its energy from the lateral parts of the central streak. A longer wall-normal extent of the central streak for the thinner roughness geometry leads to a stronger spanwise shear, and the sinuous instability is able to extract more energy from the spanwise gradient of the base flow and becomes prominent. The two lateral streaks also make a contribution to energy extraction for large shear ratios. 

Global adjoint sensitivity analysis is performed to examine the receptivity and the inception of global instability. The most sensitive region to the point forcing is located immediately upstream of the roughness element and at the top edge of the separation region downstream. The wavemaker results show that the instability core is located in the reversed flow region for both varicose and sinuous modes. While the varicose mode displays one primary wavemaker region along the central streak, the sinuous mode shows two lateral wavemaker regions. The spatial growth of wavemaker is stronger for thicker roughness element. For thinner roughness, the strength of the spatial growth for varicose mode decreases faster than that for the sinuous mode.

% The varicose instability has its root in the center of the reversed flow region, experiences spatial transient growth along the central low-speed streak and extracts its energy from the whole 3-D shear layer, contributing to the birth of hairpin vortices. The sinuous instability has its root in the lateral parts of the reversed flow region, experiences spatial growth along the lateral parts of the central streak and extracts its energy from the spanwise shear, contributing to the sinuous wiggling of hairpin vortices.
% The varicose instability has its root in the center of the reversed flow region → experiences spatial transient growth along the central low-speed streak and extracts its energy from the whole 3-D shear layer → contributes to the birth of hairpin vortices. The sinuous instability has its root in the lateral parts of the reversed flow region, experiences spatial growth along the lateral parts of the central streak and extracts its energy from the span wise shear, contributes to the sinuous wiggling of hairpin vortices.

The impact of $\eta$ and $Re_h$ on the transition process associated with different instability characteristics is investigated by performing DNS. The %present cases with thin roughness geometries 
results are compared to the transition diagram by \cite{von1961effect}, and transition features are seen to agree with their classification. %well fitted into the regions classified by the diagram. 
For $\eta=1$, the peak corresponding to the shedding of the primary hairpin vortices is obtained in both the energy and DMD spectra, in accordance with the eigenfrequency of the global varicose mode. As $Re_h$ increases, transition occurs closer to the roughness element, sinuous like breakdown is seen farther downstream, destabilizing the shear layer and promoting transition to turbulence.
%Transition to turbulence may not happen for a moderate $Re_h$ since the local $Re_{\delta}$ is low.
%and the boundary layer is quite stable for itself.
When $Re_h$ is sufficiently high, a fully-developed turbulent flow is established in both the inner and outer layers farther downstream. For $\eta=0.5$, the sinuous wiggling of hairpin vortices becomes prominent in the near wake as $Re_h$ increases. Multiple peaks including the peaks corresponding to the varicose and sinuous instabilities are seen in the energy and DMD spectra. Stronger non-linear interactions between the hairpin vortices and the sinuous oscillations of the central streak are seen in the near wake. After the hairpin vortices break down, a sinuous mode associated with the wiggling of streaks persists farther downstream. %{\color{red}It is evident that when the wake flow only contains varicose instability and $Re_h$ is sufficiently high, sinuous like breakdown could happen in the farther wake, destabilizing the shear layer and promoting transition to turbulence. When the sinuous instability sets in for a thinner roughness geometry, more complicated fluid motions are seen, and the sinuous oscillations of the vortical structures persist farther downstream and contribute to the transition process. }

% The varicose and sinuous instability lead to a different behavior and development of vortical structures in the transition process. 
% While the varicose modulation for $\eta=1$ only contains one primary peak corresponding to the hairpin vortices, the combination of varicose and sinuous modulations for $\eta=0.5$ result in multiple compositions of different flow structures. 

\section*{Acknowledgements}
This work was supported by the United States Office of Naval Research (ONR) Grant N00014-17-1-2308 managed by Dr. P. Chang. Computing resources were provided by the Minnesota Supercomputing Institute (MSI) and Extreme Science and Engineering Discovery Environment (XSEDE).
%This work used the Extreme Science and Engineering Discovery Environment (XSEDE), which is supported by National Science Foundation grant number ACI-1548562..

\section*{Declaration of Interests}
The authors report no conflict of interest.
%\section*{Appendix A. Surface statistics}
%\section{Appendix A. Grid sensitivity}

\bibliographystyle{jfm}
\bibliography{jfm-instructions}

\end{document}